\documentclass[twocolumn,twocolappendix,tighten,times]{aastex6}
\usepackage{amsmath,xfrac}

\newcommand{\HI}{\ion{H}{1}}
\newcommand{\HII}{\ion{H}{2}}
\newcommand{\CI}{\ion{C}{1}}
\newcommand{\CII}{\ion{C}{2}} 
\newcommand{\CIII}{\ion{C}{3}}
\newcommand{\CIV}{\ion{C}{4}} 
\newcommand{\NII}{\ion{N}{2}} 
\newcommand{\NV}{\ion{N}{5}} 
\newcommand{\OI}{\ion{O}{1}}
\newcommand{\OIII}{\ion{O}{3}}
\newcommand{\OVI}{\ion{O}{6}} 
\newcommand{\OVIII}{\ion{O}{8}} 
\newcommand{\SiII}{\ion{Si}{2}}
\newcommand{\SiIII}{\ion{Si}{3}} 
\newcommand{\SiIV}{\ion{Si}{4}}
\newcommand{\FeII}{\ion{Fe}{2}}
\newcommand{\FeIII}{\ion{Fe}{3}} 
\newcommand{\MgII}{\ion{Mg}{2}}

\newcommand{\eqw}{\ensuremath{W_{\lambda}}}
 
\newcommand{\fuse}{{\sl FUSE}} 
\newcommand{\galex}{{\sl GALEX}} 

\newcommand{\Ha}{\ensuremath{{\rm H}\alpha}} 
\newcommand{\Hb}{\ensuremath{{\rm H}\beta}} 
\newcommand{\hst}{{\sl HST}} 
\newcommand{\kms}{\ensuremath{{\rm km\,s}^{-1}}}
\newcommand{\lya}{\ensuremath{{\rm Ly}\alpha}}
\newcommand{\lyb}{\ensuremath{{\rm Ly}\beta}}

\turnoffedit

\begin{document}

\title{Characterizing the Circumgalactic Medium of Nearby Galaxies with \hst/COS and \hst/STIS Absorption-Line Spectroscopy: II. Methods and Models\altaffilmark{1}}

\author{Brian A. Keeney\altaffilmark{2}, John T. Stocke\altaffilmark{2}, Charles W. Danforth\altaffilmark{2}, J. Michael Shull\altaffilmark{2},  Cameron T. Pratt\altaffilmark{2}, Cynthia S. Froning\altaffilmark{3}, James C. Green\altaffilmark{2}, Steven V. Penton\altaffilmark{4}, Blair D. Savage\altaffilmark{5}} 

\altaffiltext{1}{Based on observations with the NASA/ESA {\sl Hubble Space Telescope}, obtained at the Space Telescope Science Institute, which is operated by AURA, Inc., under NASA contract NAS 5-26555.}
\altaffiltext{2}{Center for Astrophysics and Space Astronomy, Department of Astrophysical and Planetary Sciences, University of Colorado, 389 UCB, Boulder, CO 80309, USA}
\altaffiltext{3}{Department of Astronomy, University of Texas at Austin, Austin, TX 78712, USA}
\altaffiltext{4}{Space Telescope Science Institute, Baltimore, MD 21218, USA}
\altaffiltext{5}{Department of Astronomy, University of Wisconsin, Madison, WI 53706, USA}

\shorttitle{UV Probes of the Circumgalactic Medium \sc{ii}}
\shortauthors{Keeney et~al.}
\submitted{To be submitted to \apjs.}

\begin {abstract}
We present basic data and modeling for a survey of the cool, photo-ionized 
Circum-Galactic Medium (CGM) of low-redshift galaxies using far-UV QSO 
absorption line probes. This survey consists of ``targeted'' and 
``serendipitous'' CGM subsamples, originally described in 
\citet[Paper~1]{stocke13}. The targeted subsample probes low-luminosity, 
late-type galaxies at $z<0.02$ with small impact parameters 
($\langle\rho\rangle = 71$~kpc), and the serendipitous subsample probes 
higher luminosity galaxies at $z\lesssim0.2$ with larger impact parameters 
($\langle\rho\rangle = 222$~kpc). \hst\ and \fuse\ UV spectroscopy of the 
absorbers and basic data for the associated galaxies, derived from ground-based 
imaging and spectroscopy, are presented. We find broad agreement with the 
COS-Halos results, but our sample shows no evidence for changing ionization 
parameter or hydrogen density with distance from the CGM host galaxy, probably 
because the COS-Halos survey probes the CGM at smaller impact parameters. 
We find at least two passive galaxies with \HI\ and metal-line absorption, 
confirming the intriguing COS-Halos result that galaxies sometimes have cool 
gas halos despite no on-going star formation. Using a new methodology for 
fitting \HI\ absorption complexes, we confirm the CGM cool gas mass of Paper~1, 
but this value is significantly smaller than found by the COS-Halos survey. 
We trace much of this difference to the specific values of the low-$z$ 
meta-galactic ionization rate assumed. After accounting for this difference, 
a best-value for the CGM cool gas mass is found by combining the results of 
both surveys to obtain $\log{(M/M_{\Sun})}=10.5\pm0.3$, or $\sim30$\% of the 
total baryon reservoir of an $L \geq L^*$, star-forming galaxy.
\end {abstract}

\maketitle

\section{Introduction}
\label{intro} 

A detailed knowledge of the Circumgalactic Medium (CGM; a.k.a gaseous
galaxy halo) is necessary for any detailed understanding of galaxy
formation and evolution. While significant advances have been made to
detect and characterize the CGM at high-$z$ \citep[e.g.,][]{steidel95, 
adelberger05, rudie13}, the advent of low redshift studies using the
ultraviolet spectrographs of the {\sl Hubble Space Telescope} (\hst) has
proven critical to recent advances in the field \citep*{tripp98, 
penton04,tumlinson11,prochaska11a,stocke13,werk13,lehner15,bowen16,werk16}, 
including the recognition that the CGM likely contains a comparable number 
of baryons as found in all the stars and gas in the disks of late-type
galaxies \citep{tumlinson11, stocke13, werk14}. The theoretical case
for the importance of a massive CGM is demonstrated by the continuing
high star formation rate (SFR) in spiral galaxies \citep{binney87, 
chomiuk11}, the detailed metallicity history of galaxies
\citep*[e.g., the ``G dwarf problem'';][]{larson72, binney87, 
chiappini01}, and the substantial deficiency of detected baryons in
spiral galaxies relative to the cosmic ratio of baryons to dark matter
\citep[e.g.,][]{mcgaugh00, klypin01}. Each of these observational
problems requires a CGM baryonic mass at least as great as the total
amount in the galaxy's disk. 

While the emission measure of this gas is too low to provide
direct imaging detections at the present time 
\citep*[although see][]{donahue95,martin15,fumagalli17},
it is possible to use background, UV-bright AGN to detect and probe
the full extent of the CGM in absorption in both the Milky Way and in other 
galaxies. In our own Galaxy's halo, the discovery of
highly-ionized high-velocity-clouds (HVCs) \citep*{sembach95, 
sembach03, collins04} using UV spectroscopy of metal ions such as \SiIII, 
\CIV, and \OVI\ has revealed a much larger reservoir of infalling gas 
\citep*{shull09, collins09} than the \HI\ 21-cm HVCs. But
only in some cases \citep{lehner11} can the distance to these
highly-ionized HVCs be determined, allowing their total mass to be
estimated. Nevertheless, the mass infall rate 
($\sim1~M_{\odot}\,{\rm yr}^{-1}$) estimated by \citet{shull09} is sufficient 
to fuel much of the on-going Milky Way SFR 
\citep[2-$4~M_{\odot}\,{\rm yr}^{-1}$;][]{diehl06, robitaille10, shull11}. 
The detection of high-covering-factor, photo-ionized gas clouds in the 
CGM of low-$z$ star-forming galaxies \citep{werk14, stocke13}
suggests that galactic ``fountains'' of recycling gas (i.e.,
what has come to be called the ``baryon cycle''), together with lower
metallicity gas from extra-galactic sources \citep[e.g., dwarf galaxy 
satellites;][or the Inter-Galactic Medium; IGM; \citealp{bowen16}]{stocke13},
is essential for the continued high star formation rate in late-type
galaxies. The possibility that these cool\footnote{The temperature nomenclature
followed in this paper is that routinely used in recent observational and 
theoretical papers on the CGM; i.e., ``cool'' refers to $T\sim10^4$~K absorbers
in photo-ionization equilibrium, ``warm'' refers to collisionally ionized absorbers
at $T\sim10^5$-$10^{6.5}$~K and ``hot'' gas is at $T>10^{6.5}$~K. Thus, the 
warm gas referenced in Galactic studies of the interstellar medium is termed 
``cool'' herein.}, photo-ionized clouds are embedded in a hotter ($T\sim10^6$~K), 
more massive diffuse gas also has been postulated based on high signal-to-noise 
(S/N) \OVI\ and \lya\ absorption-line detections \citep{stocke14, pachat16}.

In this paper we present the basic data used for a recent CGM study
\citep[][Paper~1 hereafter]{stocke13}. These data include far-UV spectroscopy
obtained with \hst's Space Telescope Imaging Spectrograph (STIS) and
Cosmic Origins Spectrograph (COS), which were used to probe the CGM of low
redshift galaxies using both ``serendipitous'' and ``targeted''
observations.  While we measure and catalog high ionization lines like the \NV\ 
and \OVI\ doublets, these absorptions are not used in our analysis herein; 
detailed discussion of \OVI\ in particular can be found in \citet{savage14}, 
\citet{stocke14}, and \citet{werk16}.

Deep \Ha\ imaging and/or long-slit spectroscopy of the targeted and serendipitous 
galaxies is also presented, and used to obtain the on-going ($\leq10^7$~yr) star 
formation rates (SFR). Near-UV photometry of these same galaxies using
the {\sl Galaxy Evolution Explorer} (\galex) satellite provides a more
comprehensive, recent SFR over a longer time frame ($\sim10^8$~yr), and is 
presented where available. Galaxy metallicities have been estimated using \HII\ 
region emission lines and/or Lick absorption-line indices.

The ultraviolet spectroscopy presented herein forms the basis 
for many of the conclusions drawn by Paper~1, including the
\lya\ covering factors (approaching 100\% inside the nearest galaxy's
virial radius) as well as cloud thicknesses (0.1 to 30 kpc) and masses
(10-$10^8~M_{\odot}$ range). A more recent study \citep{davis15} of common 
absorptions along adjacent sight lines through a foreground galaxy halo 
confirms large sizes and masses for at least some CGM absorbing clouds.
The size, mass and pressure results presented in Paper~1 were obtained using 
homogeneous, single-phase CLOUDY photo-ionization models \citep{ferland98}; 
these models are also presented and updated herein.

Section~\ref{sample} contains a brief rediscussion of the target
selection and the data analysis procedure for
the COS, STIS and \fuse\ spectra. Since both these topics have been
covered in depth in Paper~1, \citet{danforth16}, \citet{tilton12} and
\citet{keeney14}, the discussion herein will be largely a summary of
the description in those previous papers. Section~\ref{absorbers}
presents the COS spectroscopy of the absorption systems used in this
study. Section~\ref{galaxies} presents the basic galaxy data obtained
for both samples using ground-based imaging and spectroscopy,
primarily obtained at the Apache Point Observatory's 3.5m telescope
(APO hereafter) operated  by the Astrophysical Research Corporation
(ARC), as well as SDSS galaxy survey data obtained publicly through
DR12 \citep{alam15}. Our photo-ionization models are presented in 
Section~\ref{cloudy}, and Section~\ref{ensemble} presents an 
updated cool CGM gas mass calculation that conﬁrms the results of Paper~1, 
along with a comparison to similar calculations made using the COS-Halos 
results \citep{werk14,stern16,prochaska17}. The discussion of results in 
Section~\ref{discussion} includes absorber/galaxy correlations based on basic 
observables, including statistics of metal-bearing and metal-free absorbers, 
absorber kinematics relative to the associated galaxy, and absorber/galaxy 
metallicity and star formation rate comparisons. Section~\ref{discussion} 
also includes investigations into the robustness of the galaxy associations and 
whether early-type galaxies are associated with low-$z$ \lya\ absorption. 
Section~\ref{conclusion} summarizes the findings of this paper. The Appendices 
contain comments on individual absorption-line fits, galaxy measurements, and 
photo-ionization models (Appendix~\ref{indiv}), as well as the ground-based images 
of targeted and serendipitous galaxies (Appendix~\ref{imaging}).

Throughout this paper, we assume WMAP9 values \citep{hinshaw13} of 
the standard cosmological parameters: $H_0 = 69.7~{\rm km\,s^{-1}\,Mpc^{-1}}$, 
$\Omega_{\Lambda}=0.718$, and $\Omega_{\rm m}=0.282$.

\section{Sample Selection}
\label{sample}

The philosophy behind the sample selection for the COS Science Team (hereafter 
called Guaranteed Time Observers, or GTOs) QSO/galaxy pairs program was to select 
very low-redshift ($z\leq0.02$), low-luminosity spiral and irregular foreground 
galaxies at low impact parameters ($\rho \leq 1.5~R_{\rm vir}$) to bright 
(${\rm V}\leq17.5$) background QSOs, chosen from a large list of potential targets. 
The COS GTO Team then conducted a modest-sized COS survey of the CGM of these very
nearby, late-type galaxies (10 sight lines and 12 foreground galaxy CGMs
probed) by observing these bright QSOs across the full \hst\ far-UV
band (1150--1700\AA). Target QSOs were selected so that a peak signal-to-noise 
ratio ${\rm (S/N)}\sim15$-20 per resolution element of 17~\kms\ could be
obtained in 5 or fewer orbits. For each target, both G130M and G160M
exposures were obtained (see Table~1 in Paper~1 for the observing log). 

For this targeted sample, we have intentionally chosen galaxies with a range of 
luminosities ($<0.01\,L^*$ to $L^*$) and morphologies (massive spirals to dwarf 
irregulars, including star-bursting systems, passive dwarfs and 
low-surface-brightness galaxies) in order to probe the CGM of a variety of 
late-type galaxies. Although the sample has limited size, it was nevertheless 
constructed with the goal of characterizing the CGM of a wide range of late-type 
galaxies for input into models of galactic evolution. Most targeted QSO/galaxy 
pairs are single-QSO sight lines probing single-galaxy CGMs. However, one galaxy is
probed by three sight lines \citep{keeney13}, and one of these three
QSO sight lines probes two different galaxies at two distinct
redshifts. Furthermore, two adjacent sight lines (1ES~1028+511 and 
1SAX~J1032.3+5051) were chosen to probe transverse cloud sizes in the 
CGM of two low-luminosity galaxies. All of these constraints provide a 
unique, although modest-sized sample for CGM studies. The expansion of the 
COS-Halos effort to target QSO/dwarf galaxy pairs
\citep[``COS-Dwarfs'';][]{bordoloi14} now provides CGM data for a
second sample with a similar range of galaxy luminosities but
different selection methods to this sample.

Table~\ref{tab:targeted} presents the targeted sample with the following 
basic information by column: (1) the QSO sight line name; (2) the associated 
galaxy name; (3) the absorber velocity or velocities associated with the  
tabulated galaxy, in \kms; (4) the instrument(s) used for the absorber 
detections; (5) the  galaxy's recession velocity ($cz_{\rm gal}$), in \kms; 
(6) the rest-frame $g$-band galaxy luminosity in $L^*$ units 
\edit1{(we adopt $M_g^* = -20.3$, the value of \citet{montero-dorta09} at $h=0.7$)}; 
(7) the impact parameter ($\rho$), in kpc; (8) the virial radius 
($R_{\rm vir}$), in kpc; (9) the ratio of impact parameter to $R_{\rm vir}$; 
(10) the ratio of the impact parameter of the tabulated galaxy to that of the 
next-nearest galaxy 
($\eta_{\rho} \equiv \sfrac{\rho_{\rm ng}}{\rho_{\rm nng}}$); 
(11) the ratio of the normalized impact parameter (in virial radii) of the 
tabulated galaxy to that of the next-nearest galaxy 
($\eta_{\rm vir} \equiv \sfrac{(\rho/R_{\rm vir})_{\rm ng}}{(\rho/R_{\rm vir})_{\rm nng}}$); 
and (12) the ratio of normalized galaxy-absorber velocity difference of the 
tabulated galaxy to that of the next-nearest galaxy 
($\eta_{\Delta v} \equiv \sfrac{(|\Delta v|/v_{\rm esc})_{\rm ng}}{(|\Delta v|/v_{\rm esc})_{\rm nng}}$). 
The absorption velocity used in the $\Delta v$ calculation for column~12 is the 
weighted mean of the measured \lya\ component velocities, where the weights are 
set equal to the \HI\ column density of each component. Furthermore, a minimum 
$|\Delta v|$ of 15~\kms\ is assumed when calculating $\eta_{\Delta v}$; this 
value is the uncertainty in the absolute wavelength calibration of the \hst/COS 
G130M and G160M gratings \citep{green12}.

An entry in columns~10, 11 or 12 that is significantly $>1.0$ suggests that 
an alternative association with another nearby galaxy is possible. See 
Section~\ref{indiv:galaxies} of the Appendix for discussion of individual cases, 
and Section~\ref{discussion:association} for a general discussion of the question 
of association.

The derivation of virial radius (column~8) from the optical luminosity (column~6) 
is based on the abundance-matching process of Paper~1 (see their Section~3.1 for 
a detailed discussion); while Paper~1 explored several possible derivations of 
halo mass from galaxy luminosity, in this paper the ``hybrid'' prescription was 
adopted.  Because the galaxy luminosities in column~6 are derived from 
rest-frame $g$-band magnitudes (or estimates) rather than $B$-band they differ
slightly from those in Paper~1, as do the correspondingly updated halo
masses and virial radii.

To enlarge the sample over a wider range of galaxy properties, we
utilized serendipitous foreground galaxies along archival STIS
sight lines to provide information on more luminous galaxies out to
$z\leq0.2$ over similar impact parameters to the targeted sample.  
Out of the entire suite of STIS medium-resolution echelle spectroscopy, 
there are 16 sight lines that probe the extended CGM of 35 low-$z$ 
galaxies, each with a detected system of absorption lines 
\citep[see][for a definition of ``systems of absorbers'']{danforth16}. Many 
of these STIS targets are among the brightest QSOs, so that \fuse\ far-UV 
spectroscopy is available for many of them. At these low redshifts, \fuse\ 
data probes \OVI\ and \CIII\ metal lines and \lyb, and often higher Lyman lines, 
that are used to determine more accurate $N_{\rm H\,I}$ values for 
partially-saturated \lya\ lines (see Section~\ref{absorbers}).  Many of these 
same bright targets were also re-observed with COS as part of the GTO program 
\citep[see][]{savage14}.

For the serendipitous sample, we
set a limit on the impact parameter in units of the virial radius of
$\rho/R_{\rm vir} < 2.0$ and a limit on the relative velocity of the
absorber to the galaxy of $|\Delta v| < 400$~\kms. Note that Paper~1
defined both ``serendipitous'' and  ``alternate'' samples of galaxies
that were within $|\Delta v| < 400$~\kms\ of an  absorber and
$\rho/R_{\rm vir} < 1.5$ of the QSO sight line using different virial
radius definitions (see Section~4 and Tables~3 and 4 of Paper~1). 
A slightly larger range of impact parameters is adopted here to 
explore more thoroughly whether the absorbers can be associated with 
individual galaxies (see Section~\ref{discussion:association}).
We have merged  objects from both of these samples into a single
``serendipitous'' sample  with a larger virial radius cutoff
to accommodate the  updated galaxy luminosities, virial
radii, etc. This allows us to keep all but four of the
absorber-galaxy associations initially listed in the combined
``serendipitous''+``alternate'' samples of Paper~1. The QSO/galaxy systems that 
are no longer included in our analysis are
PKS~0405--123\,/\,2MASX~J04080654--1212494 (the only one from the
original serendipitous sample now excluded), PKS~2155--304\,/\,ESO~466--32,
PKS~2155--304\,/\,J215846.5--301738, and Ton~28\,/\,UGCA~201; none of these 
systems have associated  metal lines. The exclusion of these QSO/galaxy pairs 
from this study does not preclude their potential physical association since 
the CGM/IGM boundary doesn't appear to be identifiable observationally (see 
Figure~7 of Paper~1).

The sight lines used to define our serendipitous absorber
sample are presented in \citet[][see also
\citealt{tilton12}]{danforth08}, which includes very bright FUV
targets possessing both high resolution 7~\kms, moderate ${\rm
S/N}\sim5$-15 STIS E140M spectra and also \fuse\ $\sim20$~\kms\ FUV
spectra. Serendipitous absorbers have higher Lyman-series 
lines as well as the \OVI\ doublet lines falling
within the higher sensitivity regions of the \fuse\ detector, providing
detections at $\log{N_{\rm O\,VI}} \geq 13.2$ \citep{danforth05,
stocke06} when there are no obscuring Galactic absorption lines. In the
current study we have used only those $\sim500$ \lya\ absorbers with
$\log{N_{\rm H\,I}} \geq 13.0$  ($\eqw \geq 54$~m\AA) --- an 
equivalent width detectable in all STIS spectra \citep{danforth08}. The
basic data for these ``absorber-selected'' galaxies are given in
Table~\ref{tab:serendipitous}, which duplicates the information given
for the targeted sample in Table~\ref{tab:targeted}.

\floattable
\begin{deluxetable*}{llcccDcCCcCC}
\rotate
\tablecaption{Targeted Sample Summary\label{tab:targeted}}
\tablewidth{0pt}
\tabletypesize{\scriptsize}
\tablehead{ \colhead{Sight Line} & \colhead{Galaxy} & \colhead{$cz_{\rm abs}$} & \colhead{Instrument} & \colhead{$cz_{\rm gal}$} & \twocolhead{$L$} & \colhead{$\rho$} & \colhead{$R_{\rm vir}$} & \colhead{$\rho/R_{\rm vir}$} & \colhead{$\eta_{\rho}$\tablenotemark{a}} & \colhead{$\eta_{\rm vir}$\tablenotemark{b}} & \colhead{$\eta_{\Delta v}$\tablenotemark{c}} \\ & & \colhead{(\kms)} & & \colhead{(\kms)} & \twocolhead{($L^*$)} & \colhead{(kpc)} & \colhead{(kpc)} }

\decimalcolnumbers
\startdata
1ES~1028+511      & UGC~5740                 &                    728 & COS        &  649 & 0.007  & 110 &  54  & 2.04  & 4.23 & 4.00  & 0.71  \\ [-1mm]
1ES~1028+511      & SDSS~J103108.88+504708.7 &                    967 & COS        &  934 & 0.005  &  26 &  51  & 0.51  & 0.25 & 0.25  & 0.06  \\ [-1mm]
1SAX~J1032.3+5051 & UGC~5740                 &                    704 & COS        &  649 & 0.007  &  79 &  54  & 1.46  & 1.80 & 1.69  & 0.33  \\ [-1mm]
FBQS~J1010+3003   & UGC~5478                 &             1264, 1380 & COS        & 1378 & 0.059  &  57 &  78  & 0.73  & 0.26 & 0.18  & 0.01  \\ [-1mm]
HE~0435--5304     & ESO~157--49              &       1514, 1647, 1719 & COS        & 1673 & 0.12   & 172 &  93  & 1.85  & 1.28 & 0.54  & 1.41  \\ [-1mm]
HE~0439--5254     & ESO~157--49              & 1581, 1653, 1763, 1805 & COS        & 1673 & 0.12   &  93 &  93  & 1.00  & 0.93 & 0.41  & 0.17  \\ [-1mm]
HE~0439--5254     & ESO~157--50              &                   3861 & COS        & 3874 & 0.43   &  89 & 137  & 0.65  & 0.73 & 0.29  & 0.22  \\ [-1mm]
PG~0832+251       & NGC~2611                 & 5221, 5337, 5396, 5444 & COS, \fuse & 5226 & 0.42   &  53 & 136  & 0.39  & 4.08 & 1.95  & 0.27  \\ [-1mm]
PMN~J1103--2329   & NGC~3511                 &             1113, 1194 & COS        & 1114 & 0.58   &  97 & 151  & 0.64  & 2.02 & 1.31  & 6.72  \\ [-1mm]
RX~J0439.6--5311  & ESO~157--49              &       1638, 1674, 1734 & COS        & 1673 & 0.12   &  74 &  93  & 0.80  & 1.19 & 0.60  & 0.13  \\ [-1mm]
SBS~1108+560      & M~108                    &        654,  715,  778 & COS        &  696 & 0.68   &  22 & 161  & 0.14  & 0.11 & 0.03  & 0.01  \\ [-1mm]
SBS~1122+594      & IC~691                   &                   1221 & COS        & 1199 & 0.095  &  45 &  87  & 0.52  & 1.07 & 0.60  & 0.41  \\ [-1mm]
VII~Zw~244        & UGC~4527                 &                    715 & COS        &  721 & 0.003: &   7 &  47: & 0.15: & 0.02 & 0.02: & 0.04: \\
\enddata

\tablenotetext{a}{The ratio of the impact parameter (in kpc) of the tabulated galaxy to that of the next-nearest galaxy: $\eta_{\rho} = \sfrac{\rho_{\rm ng}}{\rho_{\rm nng}}$.}
\tablenotetext{b}{The ratio of the normalized impact parameter of the tabulated galaxy to that of the next-nearest galaxy: $\eta_{\rm vir} = \sfrac{(\rho/R_{\rm vir})_{\rm ng}}{(\rho/R_{\rm vir})_{\rm nng}}$.}
\tablenotetext{c}{The ratio of the normalized absorber-galaxy velocity difference of the tabulated galaxy to that of the next-nearest galaxy: $\eta_{\Delta v} = \sfrac{(|\Delta v|/v_{\rm esc})_{\rm ng}}{(|\Delta v|/v_{\rm esc})_{\rm nng}}$.}

\end{deluxetable*}

\floattable
\begin{deluxetable*}{llcccDcCCDcC}
\rotate
\tablecaption{Serendipitous Sample Summary\label{tab:serendipitous}}
\tablewidth{0pt}
\tabletypesize{\scriptsize}
\tablehead{ \colhead{Sight Line} & \colhead{Galaxy} & \colhead{$cz_{\rm abs}$} & \colhead{Instrument} & \colhead{$cz_{\rm gal}$} & \twocolhead{$L$} & \colhead{$\rho$} & \colhead{$R_{\rm vir}$} & \colhead{$\rho/R_{\rm vir}$} & \twocolhead{$\eta_{\rho}$\tablenotemark{a}} & \colhead{$\eta_{\rm vir}$\tablenotemark{b}} & \colhead{$\eta_{\Delta v}$\tablenotemark{c}} \\ & & \colhead{(\kms)} & & \colhead{(\kms)} & \twocolhead{($L^*$)} & \colhead{(kpc)} & \colhead{(kpc)} }

\decimalcolnumbers
\startdata
3C~273            & SDSS~J122815.96+014944.1  &                                     1019 &  COS, \fuse &   911 & 0.004  &  70 &  50  & 1.40  &   0.41  &   0.51  &   1.54  \\ [-1mm]
3C~273            & SDSS~J122950.57+020153.7  &                                     1585 &  COS, \fuse &  1775 & 0.006  &  81 &  53  & 1.53  &   0.31  &   0.70  &   4.55  \\ [-1mm]
3C~351            & Mrk~892                   &                              3465,  3597 & STIS        &  3581 & 0.15   & 173 &  98  & 1.77  &   0.63  &   0.81  &   0.46  \\ [-1mm]
H~1821+643        & SDSS~J182202.70+642138.8  &        36139, 36307, 36339, 36439, 36631 &  COS        & 36436 & 1.1    & 157 & 189  & 0.83  &  <0.16  & \nodata & \nodata \\ [-1mm]
Mrk~335           & SDSS~J000529.16+201335.9  &                              1961,  2281 &  COS, \fuse &  1950 & 0.008  &  97 &  55  & 1.76  &   0.22  &   0.81  &   0.85  \\ [-1mm]
Mrk~876           & NGC~6140                  &                               923,   978 &  COS, \fuse &   908 & 0.72   & 257 & 163  & 1.58  &   0.41  &   0.25  &   0.12  \\ [-1mm]
PG~0953+414       & SDSS~J095638.90+411646.1  &               42512, 42664, 42759, 42907 &  COS, \fuse & 42759 & 3.1    & 438 & 265  & 1.65  &   1.08  &   0.53  &   0.88  \\ [-1mm]
PG~1116+215       & SDSS~J111905.51+211733.0  &               17614, 17676, 17786, 18202 &  COS, \fuse & 17993 & 0.10   & 133 &  89  & 1.49  &   0.52  &   1.05  &   3.89  \\ [-1mm]
PG~1116+215       & SDSS~J111906.68+211828.7  &                             41522, 41522 &  COS, \fuse & 41428 & 1.2    & 139 & 192  & 0.72  &   0.22  &   0.19  &   0.16  \\ [-1mm]
PG~1211+143       & IC~3061                   &                                     2114 & STIS, \fuse &  2316 & 0.32   & 138 & 125  & 1.10  &   0.39  &   0.62  &   1.83  \\ [-1mm]
PG~1211+143       & SDSS~J121409.55+140420.9  &        15170, 15321, 15357, 15431, 15574 & STIS, \fuse & 15309 & 0.92   & 137 & 177  & 0.77  &   0.75  &   0.48  &   0.69  \\ [-1mm]
PG~1211+143       & SDSS~J121413.94+140330.4  &               19305, 19424, 19481, 19557 & STIS, \fuse & 19334 & 0.12   &  72 &  93  & 0.77  &   0.48  &   0.94  &   0.91  \\ [-1mm]
PG~1216+069       & SDSS~J121930.86+064334.4  &                      23880, 24059, 24141 &  COS, \fuse & 24116 & 3.5    & 505 & 275  & 1.84  &   0.98  &   0.36  &   0.22  \\ [-1mm]
PG~1216+069       & SDSS~J121923.43+063819.7  &               37049, 37138, 37363, 37455 &  COS, \fuse & 37204 & 0.74   &  93 & 164  & 0.57  &   0.13  &   0.19  &   0.05  \\ [-1mm]
PG~1259+593       & UGC~8146                  &                                      702 &  COS, \fuse &   668 & 0.046  & 114 &  74  & 1.54  &   0.17  &   0.11  &   0.65  \\ [-1mm]
PG~1259+593       & SDSS~J130101.05+590007.1  &                      13825, 13914, 14014 &  COS, \fuse & 13862 & 0.47   & 138 & 141  & 0.98  &   1.55  &   0.86  &   0.53  \\ [-1mm]
PHL~1811          & SDSS~J215456.65--091808.6 &                             15418, 15444 &  COS, \fuse & 15453 & 1.5    & 269 & 207  & 1.30  &   0.89  &   0.29  &   0.02  \\ [-1mm]
PHL~1811          & SDSS~J215517.30--091752.0 &                             21998, 22042 &  COS, \fuse & 21951 & 2.7    & 502 & 253  & 1.98  &   1.47  &   0.77  &   0.53  \\ [-1mm]
PHL~1811          & J215447.5--092254         &                                    23313 &  COS, \fuse & 23278 & 0.74   & 309 & 164  & 1.88  &   0.57  &   0.65  &   0.49  \\ [-1mm]
PHL~1811          & J215450.8--092235         &                             23652, 23710 &  COS, \fuse & 23623 & 0.26   & 237 & 116  & 2.04  &   0.74  &   0.82  &   0.95  \\ [-1mm]
PHL~1811          & 2MASS~J21545996--0922249  &                                    24226 &  COS, \fuse & 24223 & 0.56   &  35 & 150  & 0.23  &   0.39  &   0.42  &   0.21  \\ [-1mm]
PHL~1811          & J215506.5--092326         &                             39658, 39795 &  COS, \fuse & 39758 & 2.3    & 228 & 239  & 0.95  &  <0.23  & \nodata & \nodata \\ [-1mm]
PHL~1811          & J215454.9--092331         &                             52914, 52933 &  COS        & 52873 & 1.4    & 354 & 204  & 1.74  &  <0.35  & \nodata & \nodata \\ [-1mm]
PKS~0312--770     & J031201.7--765517         &                                    17824 & STIS        & 17792 & 0.34:  & 239 & 127: & 1.88: &  <0.24  & \nodata & \nodata \\ [-1mm]
PKS~0312--770     & J031158.5--764855         &                             35466, 35813 & STIS        & 35732 & 2.1:   & 381 & 231: & 1.65: &  <0.38  & \nodata & \nodata \\ [-1mm]
PKS~0405--123     & 2MASX~J04075411--1214493  &                             28947, 28958 &  COS, \fuse & 29050 & 1.2    & 378 & 192  & 1.97  &   1.41  &   0.70  &   4.39  \\ [-1mm]
PKS~0405--123     & J040743.9--121209         &                      45617, 45783, 45871 &  COS        & 45989 & 0.89   & 197 & 175  & 1.13  &  <0.20  & \nodata & \nodata \\ [-1mm]
PKS~0405--123     & J040751.2--121137         & 49910, 49946, 50001, 50059, 50104, 50158 &  COS        & 50127 & 1.8    & 117 & 222  & 0.53  &  <0.12  & \nodata & \nodata \\ [-1mm]
PKS~1302--102     & NGC~4939                  &                                     3455 &  COS, \fuse &  3112 & 3.2    & 261 & 267  & 0.98  &   0.33  &   0.23  &   0.84  \\ [-1mm]
PKS~1302--102     & 2MASX~J13052026--1036311  &                      12573, 12655, 12703 &  COS, \fuse & 12755 & 2.8    & 227 & 256  & 0.89  &   0.72  &   0.41  &   0.52  \\ [-1mm]
PKS~1302--102     & 2MASX~J13052094--1034521  &                             28179, 28439 &  COS, \fuse & 28304 & 3.4    & 353 & 273  & 1.29  &   0.91  &   0.39  &   0.08  \\ [-1mm]
PKS~2155--304     & 2MASX~J21584077--3019271  &                      16965, 17113, 17340 &  COS, \fuse & 17005 & 1.7    & 425 & 217  & 1.96  &   0.59  &   0.43  &   0.11  \\ [-1mm]
PKS~2155--304     & J215845.1--301637         &                      31635, 31697, 31754 &  COS, \fuse & 31887 & 2.2:   & 403 & 238: & 1.69: &  <0.40  & \nodata & \nodata \\ [-1mm]
Q~1230+0115       & CGCG~014--054             &                                     1497 &  COS, \fuse &  1105 & 0.004  &  70 &  50  & 1.40  &   0.32  &   0.75  &   3.34  \\ [-1mm]
Q~1230+0115       & SDSS~J123047.60+011518.6  &                             23294, 23404 &  COS, \fuse & 23327 & 0.19   &  55 & 105  & 0.52  &   0.10  &   0.14  &   0.07  \\
\enddata

\tablenotetext{a}{The ratio of the impact parameter (in kpc) of the tabulated galaxy to that of the next-nearest galaxy: $\eta_{\rho} = \sfrac{\rho_{\rm ng}}{\rho_{\rm nng}}$.}
\tablenotetext{b}{The ratio of the normalized impact parameter of the tabulated galaxy to that of the next-nearest galaxy: $\eta_{\rm vir} = \sfrac{(\rho/R_{\rm vir})_{\rm ng}}{(\rho/R_{\rm vir})_{\rm nng}}$.}
\tablenotetext{c}{The ratio of the normalized absorber-galaxy velocity difference of the tabulated galaxy to that of the next-nearest galaxy: $\eta_{\Delta v} = \sfrac{(|\Delta v|/v_{\rm esc})_{\rm ng}}{(|\Delta v|/v_{\rm esc})_{\rm nng}}$.}

\end{deluxetable*}

\clearpage

\begin{figure}[!t]
\plotone{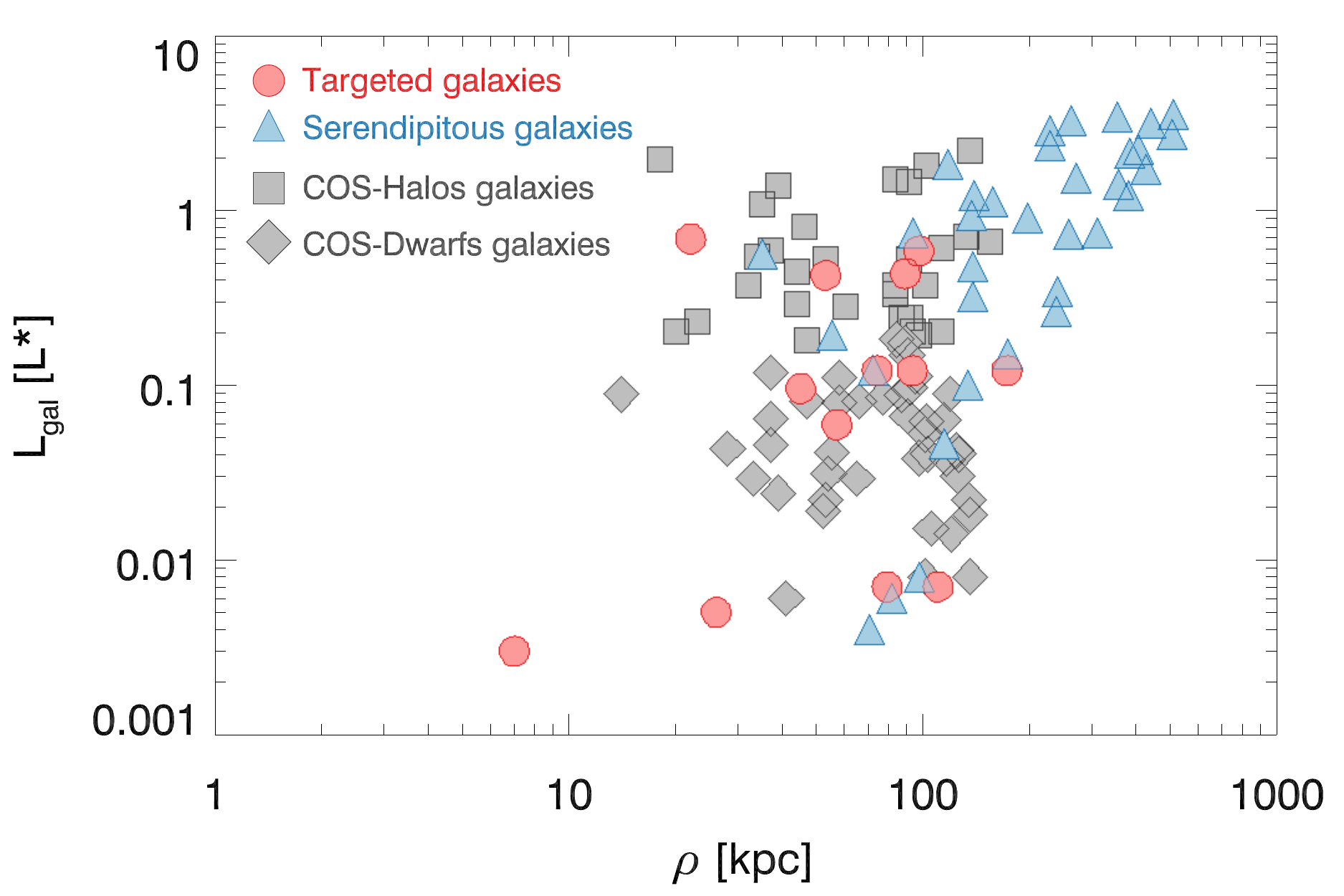}
\plotone{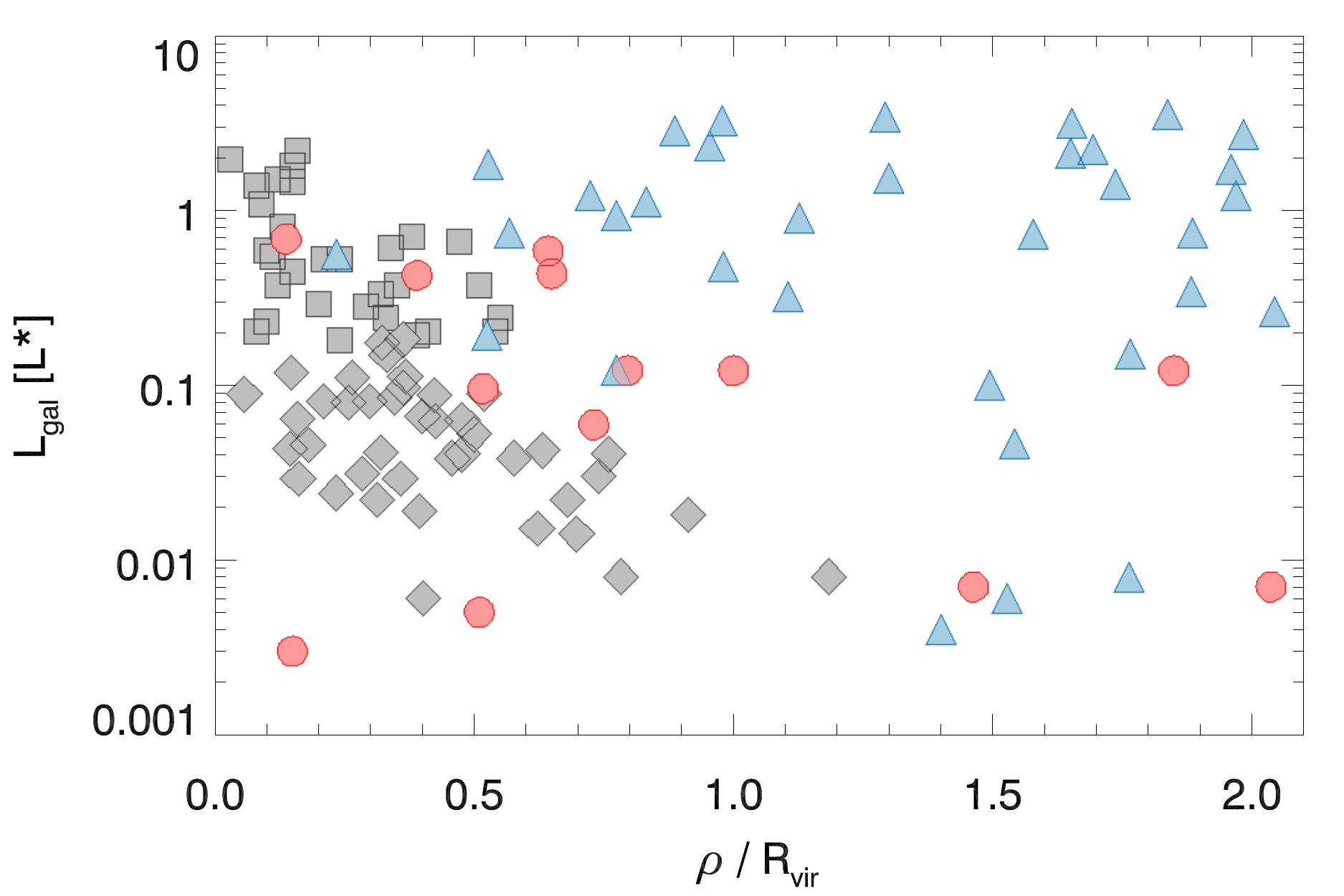}
\caption{\textit{Top:} The distribution of targeted and serendipitous galaxies in impact parameter to the QSO sight line ($\rho$) and rest-frame $g$-band luminosity ($L_{\rm gal}$). The targeted galaxies tend to have lower luminosities and smaller impact parameters. \edit1{Gray squares show the distribution of COS-Halos galaxies, and gray diamonds show the distribution of COS-Dwarfs galaxies.} \textit{Bottom:} Same as the top panel, except the impact parameter has been normalized by the virial radius of the nearest galaxy. The targeted and serendipitous galaxies are probed by the QSO sight lines over a similar range of virial radii, but the targeted galaxies are probed primarily at $\rho<R_{\rm vir}$. \label{fig:Lrho}}
\end{figure}

\edit1{
\subsection{Comparison with COS-Halos and COS-Dwarfs}
\label{sample:comparison}
}

\edit1{
The CGM sample in the present study has many similarities with the COS-Halos 
\citep{tumlinson11,tumlinson13} and COS-Dwarfs \citep{bordoloi14} surveys, but 
also important differences. Here, we compare the general properties of our 
targeted and serendipitous subsamples with those of COS-Halos and COS-Dwarfs. 
However, comparisons of specific measurements and derivations can also be found 
throughout the manuscript when appropriate, particularly in 
Sections~\ref{cloudy:coshalos} and \ref{ensemble:coshalos}.
}

\begin{figure}[!t]
\plotone{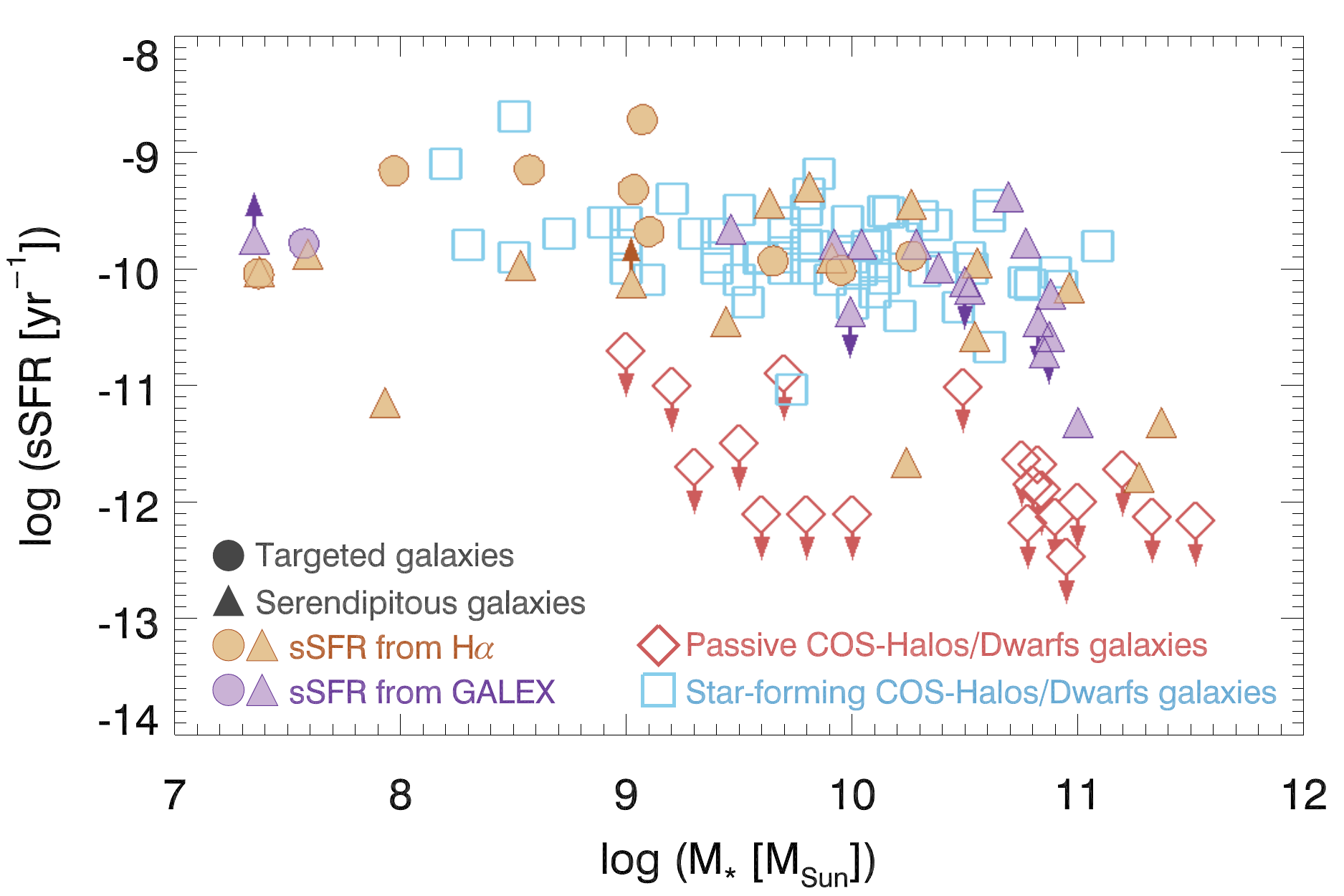}
\caption{A comparison of the stellar mass ($M_*$) and specific star formation rate (sSFR) of the targeted and serendipitous galaxies with those of the COS-Halos and COS-Dwarfs samples. Our sample includes several galaxies (both targeted and serendipitous) that are less massive than any COS-Dwarfs galaxy. \label{fig:coshalos}}
\end{figure}

\edit1{
The survey design for COS-Halos and COS-Dwarfs is similar to that for our targeted 
subsample. All examine close QSO-galaxy associations ($\rho\lesssim160$~kpc; see 
top panel of Figure~\ref{fig:Lrho}), with COS-Halos focusing on 
$L\gtrsim0.1\,L^*$ galaxies and COS-Dwarfs focusing on $L\lesssim0.1\,L^*$ 
galaxies. Our targeted subsample probes galaxies at similar impact parameters over 
nearly the full range of luminosity of the combined COS-Halos and COS-Dwarfs 
samples. However, the similarity in the distribution of basic observables breaks 
down when the impact parameter is normalized by the virial radius of the nearest 
galaxy, as shown in the bottom panel of Figure~\ref{fig:Lrho}. In this view it is 
clear that the COS-Halos galaxies are all probed at 
$\rho \lesssim \sfrac{1}{2}\,R_{\rm vir}$ and the COS-Dwarfs galaxies are 
generally probed at $\rho \lesssim R_{\rm vir}$, whereas our targeted galaxies 
are probed at $\sfrac{1}{2}\,R_{\rm vir} \lesssim \rho \lesssim 2\,R_{\rm vir}$. 
Further, even though our serendipitous galaxies were not targeted in the same way as the other surveys, when combined with our targeted galaxies they form a rather complementary sample to the COS-Halos and COS-Dwarfs galaxies in this parameter space.
}

\edit1{
However, despite these similarities in targeted survey design, the COS-Halos 
galaxies in particular were chosen to be rather isolated so that, in standard 
theoretical terminology, they would be classified as dominant or ``central'' 
halos. Our targeted and serendipitous samples are much more heterogeneous, with 
QSOs probing the CGM of spirals and irregular galaxies spanning a wide range in 
luminosity. None of the targeted or nearest serendipitous galaxies were chosen specifically to be isolated or to be ``central'' halos. 
}

Figure~\ref{fig:coshalos} compares the stellar masses and specific 
star formation rates of the targeted and serendipitous galaxies with the 
galaxies studied by COS-Halos and COS-Dwarfs. We discuss our methods for deriving 
stellar masses and specific star formation rates for the targeted and 
serendipitous galaxies in Section~\ref{galaxies}, but present 
Figure~\ref{fig:coshalos} now to demonstrate that the targeted and serendipitous 
samples include a few associated galaxies an order of magnitude less massive than 
any studied with COS-Dwarfs. However, only a few passive galaxies are
included in this sample, compared to the substantial number targeted by COS-Halos. 
Disregarding the passive galaxies for the moment, both this sample and the 
COS-Halos and COS-Dwarfs samples show a remarkable constancy of sSFR over 
4~decades of stellar mass. While most of the targeted and serendipitous galaxies 
in Figure~\ref{fig:coshalos} align well with the star-forming galaxies studied by 
COS-Halos and COS-Dwarfs, there are only four early-type serendipitous galaxies 
that overlap with the passive COS-Halos and COS-Dwarfs galaxies. We did not target 
passive galaxies specifically; this is another difference between the current 
survey and the COS-Halos/Dwarfs studies. However, there could be late-type 
galaxies that are good alternative identifications for galaxies associated with 
these absorbers (as postulated in Paper~1); this situation will be reviewed in 
detail in Section~\ref{discussion:association:passive}. As shown in Paper~1 (see 
their Figure~6), the position angle distribution of absorber locations relative to 
the associated galaxy's disk is reasonably isotropic (see 
Section~\ref{discussion:correlations}), although we did not specifically select 
our sample using that as a criterion.

\begin{figure}[!t]
\plotone{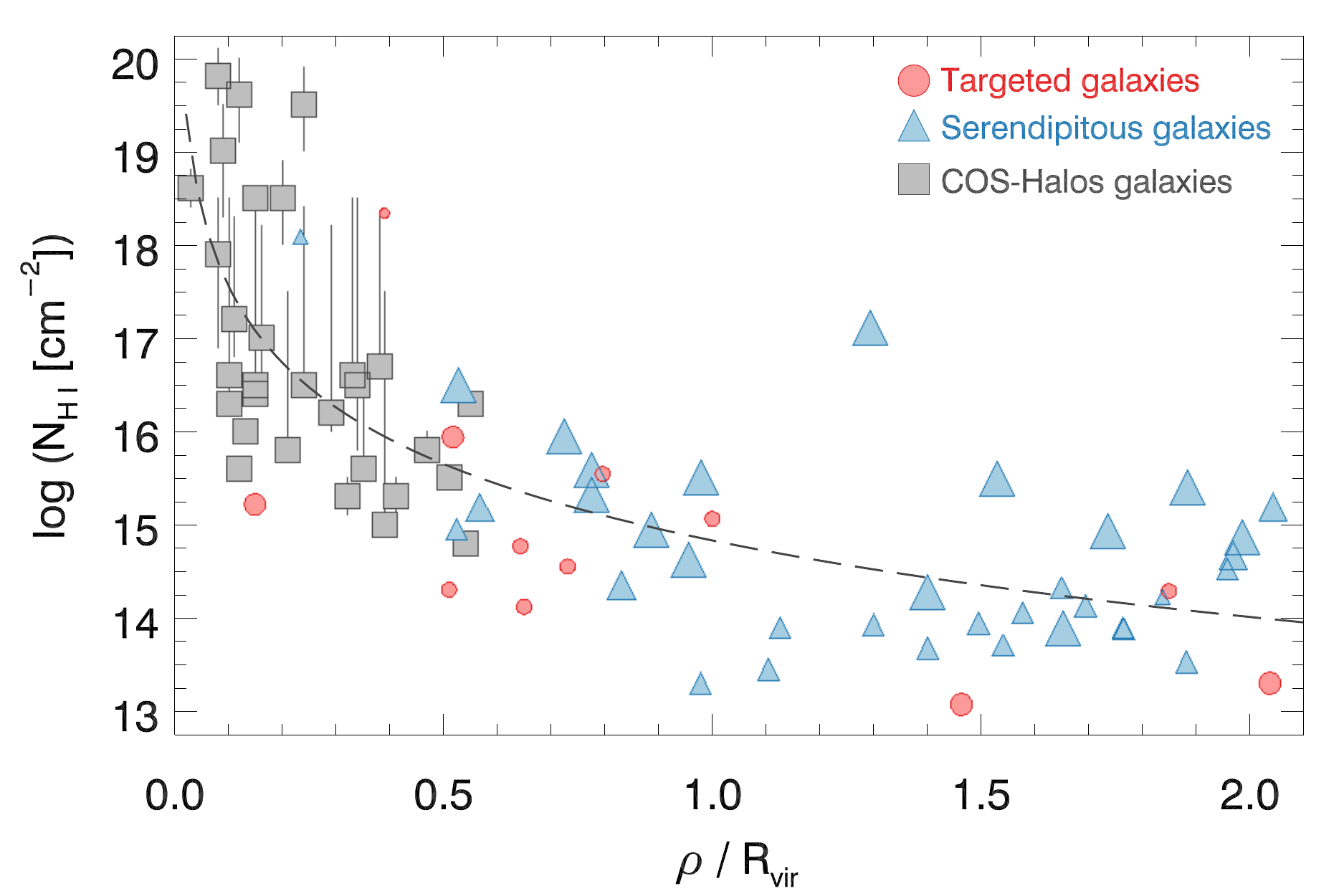}
\caption{Total \HI\ column density, $N_{\rm H\,I}$, in the CGM as a function of normalized impact parameter for the targeted and serendipitous galaxies, compared with measurements from COS-Halos \citep{werk14}. All of the measurements are consistent with a single power-law fit (dashed line), which is described in detail in Section~\ref{cloudy:coshalos}.
\label{fig:N_HI}}
\end{figure}

\edit1{
While the current sample does not lack in close QSO/galaxy pairs (see 
Figure~\ref{fig:Lrho}), COS-Halos has many more probes through the 
close-in CGM where most of the cool gas mass is concentrated 
\citep[][and Paper 1]{werk14}, allowing the COS-Halos studies to better determine 
physical conditions in the region at $\rho \leq \sfrac{1}{2}\,R_{\rm vir}$. 
Remarkably, however, when the COS-Halos measurements of total CGM \HI\ column 
density, $N_{\rm H\,I}$, are plotted next to the values measured for this sample 
as a function of $\rho/R_{\rm vir}$ as in Figure~\ref{fig:N_HI}, they are found 
to be consistent with a common power-law distribution (dashed line in 
Figure~\ref{fig:N_HI})\footnote{The \HI\ column density measurements for the 
targeted and serendipitous samples are described in Section~\ref{absorbers} and 
the power-law fit in Section~\ref{discussion:correlations}.}. Thus, there seems to be broad agreement between our measurements and those of COS-Halos on basic observables despite the different ranges of galaxy luminosity and impact parameter probed by these samples.
}

\edit1{
Unfortunately, this agreement does not generally extend to derived quantities. 
The difference in the range of impact parameters probed by this study compared 
to COS-Halos can be at least partially responsible for different trends with 
radius; e.g., whereas the photo-ionization modeling of \citet{werk14} found 
correlations of ionization parameter and pressure with normalized impact 
parameter, this study does not find such correlations in the current dataset (see 
Sections~\ref{cloudy} and \ref{discussion:correlations}). It is possible that the 
more diverse galaxy selection is a factor in this difference; i.e., the targeted 
and serendipitous galaxies in this study are often not the gravitational center of 
their circumgalactic environment, whereas the COS-Halos galaxies have a much 
better claim to being ``central'' halos, at the gravitational center of the 
region. Operationally, this means that the impact parameters in this study may not 
be the most appropriate indicator of distance from the local gravitational center 
to which the CGM clouds find themselves responding. For example, if the internal 
cloud pressures are at least partially due to the external pressure of a 
surrounding hot medium, it is distance from the physical centroid of that 
surrounding medium that should be the reference point from which to determine 
physical quantities. Given their luminous and isolated selection criteria, the 
COS-Halos galaxies are likely to be a much closer match to that ``local'' physical 
centroid than the targeted and serendipitous galaxies in the present sample.
}

\edit1{
A major difference in inferred CGM properties between our analysis and that of the 
COS-Halos team is in the total mass of the cool CGM 
\citep[see][and Section~\ref{ensemble:coshalos}]{werk14,stern16,prochaska17}; 
our total mass estimate is a factor of $\sim6$ lower than the latest estimate 
from \citet{prochaska17}. While there are significant differences in associated 
galaxy properties and absorber impact parameters as detailed above, a large 
portion of the difference between these values is attributable to the different 
choice of meta-galactic ionizing UV radiation field employed in the 
photo-ionization modeling. COS-Halos uses the higher intensity \citet{haardt01} 
radiation field, while we use the lower \citet{haardt12} spectrum. We quantify the 
difference this makes on the CGM cool mass calculation in 
Section~\ref{ensemble:coshalos}, and use both surveys to derive a new, best 
estimate for the CGM cool gas mass somewhat higher than found in Paper~1 and lower 
than the COS-Halos values.
}

\bigskip
\section{\hst\ and \fuse\ UV Spectroscopy}
\label{absorbers}

In this Section we present the COS, STIS and \fuse\ spectra of the
absorbers associated with both the targeted and the serendipitously
detected galaxy halos. The COS spectroscopy of many of these targets
was described in detail in \citet{danforth16}, including specifics
of the data handling (e.g., flux and wavelength calibrations,
absorber identifications) and basic observables such as equivalent widths. 
A detailed discussion of the error analysis of these measurements can be found 
in \citet{keeney13}. The STIS (E140M data only) and \fuse\ spectroscopy were 
described in detail in \citet{tilton12}. Only one of the targeted sight lines 
(PG~0832+251) was observed with \fuse, but many of the serendipitous sight lines 
were. Because the COS spectra have significantly higher S/N compared to the STIS 
data for these serendipitous sight lines, we have used the COS spectra where
they exist. Since most of the STIS targets are among the brightest
QSOs in the sky, several have been reobserved using COS, including in
the GTO program of bright targets reported in \citet{savage14} and 
\citet{danforth16}.

\begin{figure*}[!t]
\gridline{\fig{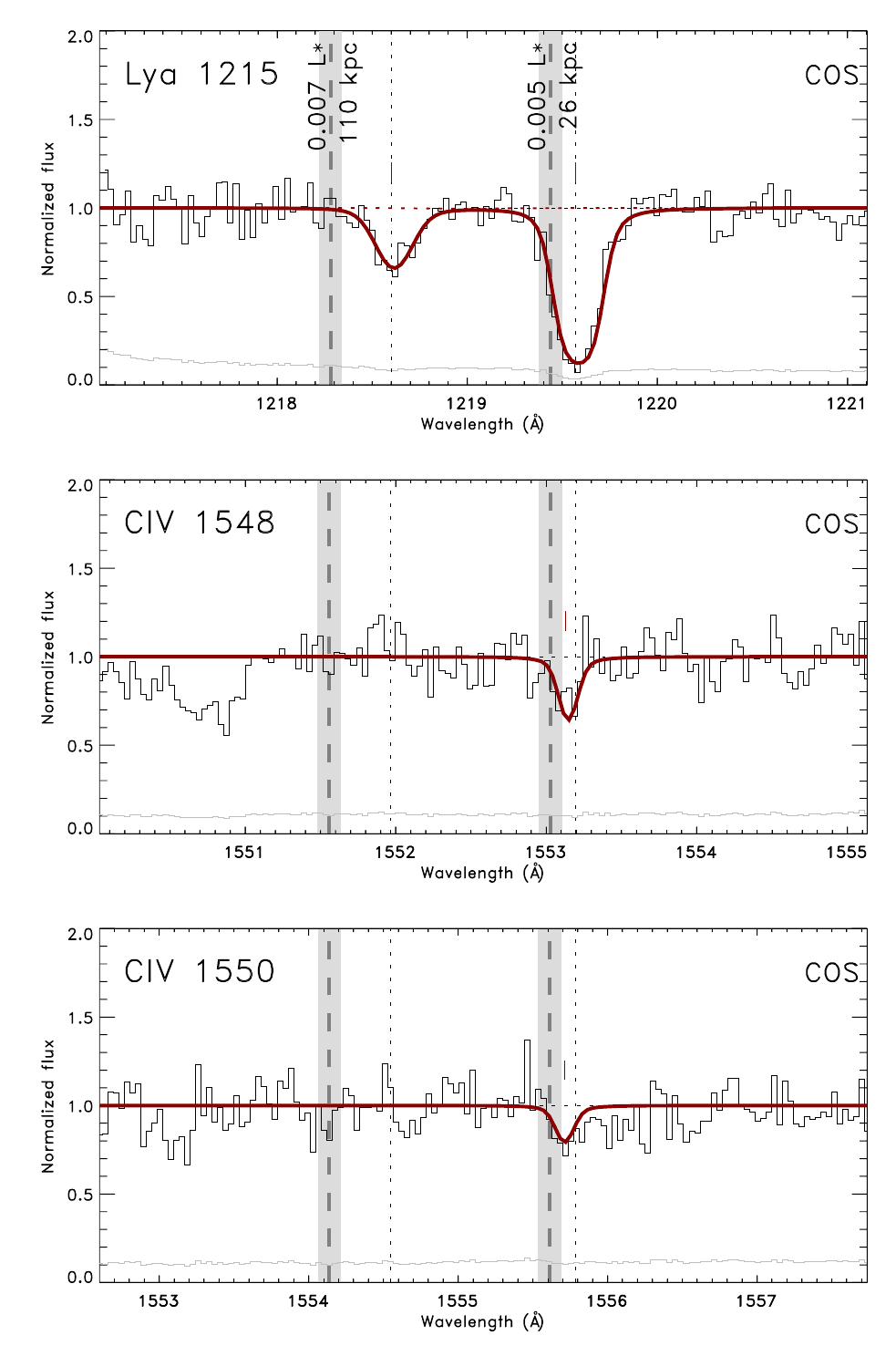}{0.5\textwidth}{4.1: 1ES~1028+511 absorbers at $cz_{\rm abs}=728$ \& 967~\kms\ (Systems~1 \& 2).}}
\gridline{\fig{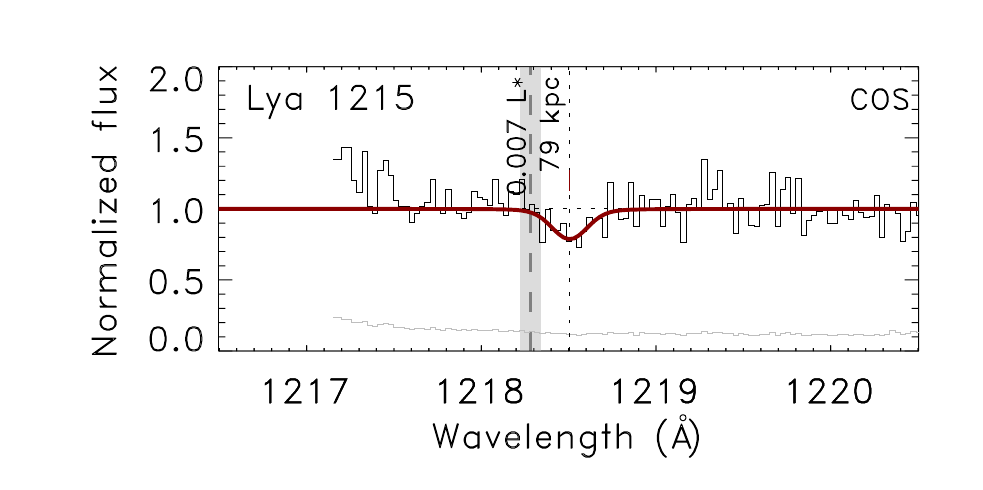}{0.5\textwidth}{4.2: 1SAX~J1032.3+5051 absorber at $cz_{\rm abs}=704$~\kms\ (System 3).}}
\caption{Absorption line fits for \HI\ and associated metal lines detected by COS, STIS or \fuse\ for all absorbers listed in Table~\ref{tab:absprop}. The species name and the provenance of the spectrum are labeled in each individual panel. Dotted vertical lines show \lya\ component velocities and dashed vertical lines with gray shaded areas show the nearest galaxy velocity and $1\sigma$ uncertainty; \edit1{the luminosity of this nearest galaxy and its impact parameter from the QSO sight line are also labeled}. The horizontal dotted line shows the continuum fit, dotted red lines show Voigt profile fits to individual velocity components, and the solid red line shows the total fit to all components. When absorption is confused by the presence of gas at another redshift, the relevant portion of the spectrum is still shown but not fit by a Voigt profile. The complete figure set (45~images) is available in the online journal.
\label{fig:stackplot}}
\end{figure*}

\bigskip
\figsetstart
\figsetnum{4}
\figsettitle{Absorption Line Fits}

\figsetgrpstart
\figsetgrpnum{4.1}
\figsetgrptitle{Systems~1\& 2}
\figsetplot{./fig01_1es1028_847_stack.pdf}
\figsetgrpnote{1ES~1028+511 absorbers at $cz_{\rm abs}=728$ \& 967~\kms\ (Systems~1 \& 2).}
\figsetgrpend

\figsetgrpstart
\figsetgrpnum{4.2}
\figsetgrptitle{System~3}
\figsetplot{./fig02_1sj1032_704_stack.pdf}
\figsetgrpnote{1SAX~J1032.5051 absorber at $cz_{\rm abs}=704$~\kms\ (System~3).}
\figsetgrpend

\figsetend

\clearpage
\floattable
\begin{deluxetable*}{cllCCCl}
\tablecaption{Multi-Component Absorption Line Fits\label{tab:absprop}}
\tablewidth{0pt}
\tabletypesize{\small}
\tablehead{ \colhead{System} & \colhead{Sight Line} & \colhead{Ion} & \colhead{$cz_{\rm abs}$} & \colhead{$b$} &  \colhead{$\log{N}$} & \colhead{Comments} \\ \colhead{\#} & & & \colhead{(\kms)} & \colhead{(\kms)} & \colhead{($N$ in cm$^{-2}$)} }

\colnumbers
\startdata
  1 & 1ES~1028+511      & \HI\    &   728 &  27.3\pm 5.9 &  13.30\pm0.06 & Voigt profile fit to \lya\ \\
  2 & 1ES~1028+511      & \HI\    &   967 &  21.8\pm 2.5 &  14.30\pm0.15 & Voigt profile fit to \lya\ \\*
    &                   & \CIV\   &   954 &  12.3\pm 6.6 &  13.34\pm0.10 & Simultaneous fit to doublet \\
  3 & 1SAX~J1032.3+5051 & \HI\    &   704 &  29.6\pm13.1 &  13.08\pm0.27 & Voigt profile fit to \lya\ \\
  4 & FBQS~J1010+3003   & \HI\    &  1264 &  36.9\pm11.9 &  13.27\pm0.09 & Voigt profile fit to \lya\ \\
  5 & FBQS~J1010+3003   & \HI\    &  1380 &  36.8\pm 2.9 &  14.47\pm0.08 & Voigt profile fit to \lya\ \\
  6 & HE~0435--5304     & \HI\    &  1514 &  29.4\pm 3.7 &  13.80\pm0.04 & Voigt profile fit to \lya\ \\
  7 & HE~0435--5304     & \HI\    &  1647 &  67.5\pm15.9 &  14.00\pm0.11 & BLA; Voigt profile fit to \lya\ \\
  8 & HE~0435--5304     & \HI\    &  1719 &  25.3\pm11.8 &  13.49\pm0.30 & Voigt profile fit to \lya\ \\
  9 & HE~0439--5254     & \HI\    &  1581 &   5.6\pm14.3 &  12.76\pm0.24 & Voigt profile fit to \lya\ \\
\enddata

\tablecomments{Table~\ref{tab:absprop} is published in its entirety in machine-readable format. A portion is shown here for guidance regarding its form and content.}

\end{deluxetable*}

In Figure~\ref{fig:stackplot} we show the associated \HI\ and metal lines 
for all the targeted and serendipitous absorbers in our sample. The origin of 
the spectrum (COS, STIS or \fuse) for each particular ion is indicated on the 
individual plots. The sight lines and adopted \lya\ velocity components are given 
in the caption for each plot and the vertical dotted line shows the adopted
velocities. The horizontal dotted line shows the continuum fit
\citep[see][]{danforth16}. Voigt profile fits to individual velocity 
components are shown as dotted red lines, and the total ensemble fit is
shown as a solid red line; the best-fit Voigt profile parameters are 
tabulated in Table~\ref{tab:absprop}. The associated galaxy redshifts 
are indicated on the \lya\ plots as vertical dashed lines with a grey
shaded area indicating the $1\sigma$ uncertainties on the galaxy's
redshift. \edit1{The nearest galaxy luminosity and its impact parameter to the QSO sight line are also indicated}. Ions in spectral regions not covered by the FUV
spectrographs due to the redshift of the absorber are not shown. When
the line is obscured or confused by the presence of a Galactic
absorption line (e.g., the Galactic DLA) or a strong absorption from
gas at another redshift, the relevant portion of the spectrum is still
shown but not fit by a Voigt profile. In some cases the associated metal 
lines do not line up precisely with the \HI\ lines (particularly the \OVI\ 
doublet) but the velocity offsets are quite small ($<30$~\kms) once 
uncertainties in the absolute wavelength calibration of the COS, STIS 
and \fuse\ data are accounted for. While the higher ions, particularly the 
\NV\ and \OVI\ doublets, are measured and tabulated here, they are not used 
in this study of the cool CGM.

Table~\ref{tab:absprop} contains the following information about each absorber 
as follows by columns: (1) A running system ID number, with absorption 
associated with our targeted sample having numbers $<100$ and absorption 
associated with our serendipitous sample having numbers $>100$; (2) sight 
line target name; (3) the ion being fit; (4) absorber velocity
(\kms); (5) Doppler $b$-value in \kms\ and 1$\sigma$ errors; (6) $\log{N}$
and 1$\sigma$ errors; and (7) comments pertaining to the fits to these 
absorptions. The presence of non-Poisson noise in COS data requires assuming 
a $b$-value to set a rigorous upper limit on equivalent width or column 
density \citep{keeney12}; $b$-values with no uncertainties are those assumed 
when calculating upper limits from COS data.

Since most of the serendipitous absorbers are found in the spectra
of very bright QSOs, there have been analyses made of these same
absorbers by other authors (see Section~\ref{indiv:absorbers} of the Appendix 
for details). In general, these analyses are either identical to within the quoted 
velocity and column density errors with what we show in Figure~\ref{fig:stackplot} 
and Table~\ref{tab:absprop} or the previous analyses suggest a more complicated 
set of components to the more complex \HI\ absorbers. These differences are to 
be expected when the lines (\lya\ specifically) are saturated. In general, 
these differences are not important for the results we present here and
those presented in Paper~1 but do enter the CGM ensemble mass calculation 
through a ``shadowing'' factor as well as the individual cloud masses of
the components (see Paper~1 and Section~\ref{ensemble}). 
In some cases, whether the multiple-component fits to \lya\ allow a very 
broad ($b > 40$~\kms), probably collisionally-ionized component 
\citep[a broad \lya\ absorber or BLA;][]{savage14} or not is an important detail 
\citep[see also][]{stocke14}. The 28~BLAs in Table~\ref{tab:absprop} are indicated
with a comment in column~7 and reproduced in Table~\ref{tab:bla} for 
convenience.

\clearpage
\begin{deluxetable*}{cllCCCl}
\tablecaption{BLA Absorbers\label{tab:bla}}
\tablewidth{0pt}
\tabletypesize{\small}
\tablehead{ \colhead{System} & \colhead{Sight Line} & \colhead{Ion} & \colhead{$cz_{\rm abs}$} & \colhead{$b$} &  \colhead{$\log{N}$} & \colhead{Comments} \\ \colhead{\#} & & & \colhead{(\kms)} & \colhead{(\kms)} & \colhead{($N$ in cm$^{-2}$)} }

\colnumbers
\startdata
  7 & HE~0435--5304     & \HI\    &  1647 &  67.5\pm15.9 &  14.00\pm0.11 & BLA; Voigt profile fit to \lya\ \\
 11 & HE~0439--5254     & \HI\    &  1763 & 105.3\pm56.7 &  13.49\pm0.20 & BLA; Voigt profile fit to \lya\ \\
 15 & PG~0832+251       & \HI\    &  5337 &  48.0:       &  14.96:       & BLA; Voigt profile fit to \lya, $cz$ fixed \\*
    &                   & \CI\    &       &  10          & <13.16        & \\*
    &                   & \CII\   &  5346 &  12.5\pm 4.5 &  13.59\pm0.07 & 1334~\AA\ line only \\*
    &                   & \CIV\   &  5327 &  20.5\pm 2.7 &  14.41\pm0.07 & Simultaneous fit to doublet \\*
    &                   & \SiII\  &       &              & <12.71        & \\*
    &                   & \SiIII\ &  5346 &  30.1\pm11.0 &  12.98\pm0.14 & \\*
    &                   & \SiIV\  &  5327 &  60.4\pm43.9 &  13.24\pm0.27 & Simultaneous fit to doublet \\*
    &                   & \OI\    &       &  10          & <13.73        & \\*
    &                   & \NV\    &       &              & <13.77        & Upper limit from intervening line \\*
    &                   & \FeII\  &       &  10          & <13.65        & \\*
    &                   & \FeIII\ &       &  10          & <14.17        & \\
 19 & PMN~J1103--2329   & \HI\    &  1194 &  48.6\pm 7.0 &  14.76\pm0.16 & BLA; Voigt profile fit to \lya\ \\*
    &                   & \CI\    &       &  20          & <13.33        & \\*
    &                   & \CII\   &       &  20          & <13.22        & \\*
    &                   & \CIV\   &  1189 &  23.6\pm 1.8 &  14.27\pm0.04 & Simultaneous fit to doublet \\*
    &                   & \SiII\  &       &  20          & <12.26        & \\*
    &                   & \SiIII\ &  1195 &  28.9\pm 0.5 &  12.95\pm0.01 & \\*
    &                   & \SiIV\  &  1191 &  25.1\pm 4.8 &  13.01\pm0.05 & Simultaneous fit to doublet \\*
    &                   & \NV\    &  1201 &  52.5\pm10.3 &  13.66\pm0.06 & Simultaneous fit to doublet \\*
    &                   & \FeII\  &       &  20          & <13.84        & \\
 20 & RX~J0439.6--5311  & \HI\    &  1638 &  44.4\pm30.7 &  13.87\pm0.81 & BLA; Voigt profile fit to \lya\ \\
 26 & SBS~1122+594      & \HI\    &  1221 &  44.8\pm 5.2 &  15.94\pm0.47 & BLA; Voigt profile fit to \lya\ \\*
    &                   & \CI\    &       &  30          & <13.36        & \\*
    &                   & \CII\   &  1198 &  34.6\pm 7.1 &  13.80\pm0.06 & 1334~\AA\ line only \\*
    &                   & \CIV\   &  1194 &  39.6\pm 1.5 &  14.74\pm0.03 & Simultaneous fit to doublet \\*
    &                   & \SiII\  &       &  30          & <12.83        & \\*
    &                   & \SiIII\ &  1215 &  46.3\pm 3.3 &  13.39\pm0.03 & \\*
    &                   & \SiIV\  &  1208 &  39.9\pm 3.6 &  13.54\pm0.03 & Simultaneous fit to doublet \\*
    &                   & \OI\    &       &  30          & <13.88        & \\*
    &                   & \NV\    &       &  45          & <13.64        & \\*
    &                   & \FeII\  &       &  30          & <13.82        & \\
103 & 3C~351            & \HI\    &  3465 &  41.1\pm 4.8 &  13.47\pm0.05 & BLA; Voigt profile fit to \lya \\
107 & H~1821+643        & \HI\    & 36339 &  46.5\pm 1.9 &  14.12\pm0.03 & BLA; simultaneous fit to \lya+Ly$\beta$ \\
108 & H~1821+643        & \HI\    & 36439 &  74.0\pm10.9 &  13.60\pm0.10 & BLA; simultaneous fit to \lya+Ly$\beta$ \\*
    &                   & \OVI\   & 36415 &  50.3\pm 8.6 &  13.68\pm0.13 & 1032~\AA\ line only \\
112 & Mrk~876           & \HI\    &   923 &  48.1\pm 4.2 &  13.90\pm0.10 & BLA; Voigt profile fit to \lya\ \\*
    &                   & \OVI\   &   932 &  25.0\pm 6.0 &  13.55\pm0.08 & \citet{tilton12} \\
117 & PG~0953+414       & \HI\    & 42907 &  53.1\pm 6.1 &  13.15\pm0.04 & BLA; simultaneous fit to \lya-Ly$\gamma$ \\*
    &                   & \OVI\   & 42873 &  11.8\pm 6.8 &  13.08\pm0.09 & Simultaneous fit to doublet \\
122 & PG~1116+215       & \HI\    & 41522 &  59.8\pm 5.7 &  13.78\pm0.10 & BLA; simultaneous fit to \lya-Ly6 \\
124 & PG~1211+143       & \HI\    &  2114 &  65.2\pm 3.9 &  13.45\pm0.03 & BLA; Voigt profile fit to \lya\ \\*
    &                   & \OVI\   &       &              & <13.87        & Upper limit from intervening line \\
127 & PG~1211+143       & \HI\    & 15357 & 127.8\pm 9.0 &  14.18\pm0.06 & BLA; simultaneous fit to \lya-Ly$\delta$ \\
134 & PG~1216+069       & \HI\    & 23880 &  63.3\pm14.0 &  13.14\pm0.07 & BLA; Voigt profile fit to \lya\ \\
136 & PG~1216+069       & \HI\    & 24141 &  42.3\pm 2.8 &  14.15\pm0.02 & BLA; Voigt profile fit to \lya\ \\
145 & PHL~1811          & \HI\    & 15418 &  72.8\pm 5.1 &  13.65\pm0.03 & BLA; Voigt profile fit to \lya\ \\*
154 & PHL~1811          & \HI\    & 39795 &  60.9\pm 8.0 &  13.19\pm0.04 & BLA; simultaneous fit to \lya-Ly$\delta$ \\*
    &                   & \OVI\   & 39826 &  22.4\pm 3.6 &  13.59\pm0.04 & Simultaneous fit to doublet \\*
156 & PHL~1811          & \HI\    & 52914 &  41.7\pm 0.9 &  14.22\pm0.03 & BLA; simultaneous fit to \lya-Ly5 \\*
    &                   & \OVI\   & 52922 &  18.1\pm 2.2 &  14.11\pm0.03 & 1038~\AA\ line only \\
160 & PKS~0312--770     & \HI\    & 35813 &  41.0\pm 2.0 &  13.79\pm0.02 & BLA; \citet{tilton12} \\
162 & PKS~0405--123     & \HI\    & 28958 &  69.0\pm 3.0 &  13.93\pm0.04 & BLA; \citet{savage14} \\*
    &                   & \OVI\   & 28955 &  23.0\pm 6.0 &  13.70\pm0.07 & \citet{savage14} \\
165 & PKS~0405--123     & \HI\    & 45871 &  47.6\pm 0.8 &  13.72\pm0.01 & BLA; Voigt profile fit to \lya\ \\*
    &                   & \OVI\   &       &  48          & <13.02        & \\
167 & PKS~0405--123     & \HI\    & 49910 &  54.0\pm13.0 &  13.11\pm0.16 & BLA; \citet{savage14} \\*
172 & PKS~0405--123     & \HI\    & 50158 &  41.0\pm 3.0 &  13.90\pm0.08 & BLA; \citet{savage14} \\
179 & PKS~2155--304     & \HI\    & 16965 &  52.3\pm 0.9 &  14.34\pm0.01 & BLA; Voigt profile fit to \lya\ \\
180 & PKS~2155--304     & \HI\    & 17113 &  67.9\pm 1.7 &  14.08\pm0.01 & BLA; Voigt profile fit to \lya\ \\*
    &                   & \OVI\   & 17149 &  24.3\pm 7.1 &  13.60\pm0.09 & 1032~\AA\ line only \\
182 & PKS~2155--304     & \HI\    & 31635 &  47.1\pm 2.3 &  13.98\pm0.03 & BLA; Voigt profile fit to \lya\ \\*
    &                   & \OVI\   &       &  47          & <13.50        & \\
184 & PKS~2155--304     & \HI\    & 31754 &  43.2\pm 8.6 &  13.28\pm0.12 & BLA; Voigt profile fit to \lya\ \\*
    &                   & \OVI\   &       &  43          & <13.48        & \\
\enddata

\end{deluxetable*}

\begin{deluxetable*}{cllCCCl}
\tablecaption{\OVI\ Absorbers with No Associated \HI\label{tab:ovi}}
\tablewidth{0pt}
\tabletypesize{\small}
\tablehead{ \colhead{System} & \colhead{Sight Line} & \colhead{Ion} & \colhead{$cz_{\rm abs}$} & \colhead{$b$} &  \colhead{$\log{N}$} & \colhead{Comments} \\ \colhead{\#} & & & \colhead{(\kms)} & \colhead{(\kms)} & \colhead{($N$ in cm$^{-2}$)} }

\colnumbers
\startdata
155 & PHL~1811          & \OVI\   & 39930 &  33.2\pm 4.6 &  13.61\pm0.04 & No associated \HI; simultaneous fit to doublet \\
166 & PKS~0405--123     & \OVI\   & 49780 &  56.0\pm 2.0 &  13.85\pm0.01 & No associated \HI; \citet{savage14} \\
\enddata

\end{deluxetable*}

Additionally, there are two \OVI-only absorbers (i.e., no 
associated \HI) that are demonstrably collisionally-ionized gas
\citep{savage10, savage14, stocke14}. These were not included in the
discussions in Paper~1 and it is not clear what associations these
absorbers may or may not have with the individual, associated galaxies
in this study. Regardless, given the likely high temperature ($T>10^5$~K) 
of these clouds, we do not consider them further in this study of the 
cool, photo-ionized CGM. Speculation as to the physical conditions and 
association of these absorbers to individual galaxies and galaxy groups 
can be found in \citet{savage14} and \citet{stocke14}. The Voigt profile 
fits for these two absorbers are detailed in Table~\ref{tab:ovi}. All 
column densities, including those in logarithms, have units of 
${\rm cm}^{-2}$.

All absorption from \HI\ and associated metal lines are fit with the minimum 
number of velocity components necessary to obtain an acceptable fit 
\edit1{(characterized by a reduced-$\chi^2\approx1$)}, 
and all lines of a particular ion with significant absorption are fit 
simultaneously (i.e., all available Ly-series lines or both lines in the \SiIV, 
\CIV, or \OVI\ doublets). 
\edit1{During these simultaneous fits between different instruments, the 
data are aligned in velocity space prior to the fit. During the fit, we preserve 
the relative spacing between velocity components but allow the absolute
velocities to shift by $<30$~\kms\ to account for uncertainties in velocity
registration between instruments.} 
The velocities reported in Table~\ref{tab:absprop} are those of the strongest line 
being fit (e.g., \lya\ or \OVI\ $\lambda$1032). 
\edit1{All Voigt profiles are convolved with the instrumental line spread 
functions as part of the fitting procedure: we assume Gaussian line spread 
functions with $\mathcal{R}=17,500$ and 45,800 for \fuse\ and STIS data, 
respectively, and the empirical line spread functions of \citet{kriss11} for 
COS data.}

Generally speaking, the serendipitous absorbers have higher-order
Ly-series lines available to inform the \HI\ fits, either in the COS
or STIS data themselves when $z_{\rm abs}\gtrsim0.1$, or from
\fuse\ at lower redshifts. These higher-order lines are crucial for assessing the
component structure and Voigt profile parameters of \HI\ systems with
saturated, or even partially saturated, \lya\ absorption. When
\lya\ absorption is present and saturated and no other \HI\ lines are
available (as is typical for the targeted absorbers) choosing the
number of velocity components to include in the fit is quite
subjective and can have significant effects on the derived Voigt profile
parameters. Furthermore, since these saturated \lya\ absorbers reside 
on the flat part of the curve of growth, it is often true that a high 
$b$-value, low column density fit and a low $b$-value, high column density 
fit have very similar $\chi^2$ values. \edit1{Due to low-S/N or severe blending, 
the difference between the low and high column density solutions can span orders 
of magnitude; we adopt the lower column density solution whenever possible as a 
consequence of our minimum-component fitting philosophy.} 

These systematic uncertainties imply that the component structure, and inferred 
$b$-values and column densities, are much more reliable for absorbers with 
multiple Ly-series lines informing the \HI\ fits. Uncertainties in 
Tables~\ref{tab:absprop}-\ref{tab:ovi} are formal fitting uncertainties only and 
do not attempt to quantify the systematic issues described above. 
\edit1{We characterize the amount of subjectivity in individual measurements 
qualitatively using absorber grades as described in Sections~\ref{cloudy} and 
\ref{ensemble}.}

In the Appendix, Section~\ref{indiv:absorbers}, we briefly discuss previous
work done on the absorbers in the targeted and serendipitous samples.
Since publication in Paper~1, all of the targeted and serendipitous absorbers 
have been re-fit after updating our data reduction pipeline. Updates were 
necessary to fix errors in parsing data quality flags at the exposure level, and 
to re-bin the data by 3~pixels so it is approximately Nyquist sampled after all 
exposures have been co-added (see \citealp{danforth16} for details). 
Consequently, some of the component-level fits have changed from our earlier 
papers \citep[Paper~1 and][]{danforth16}. A discussion on the impact that these 
multiple absorber fits have on the CGM absorber properties we have derived is 
found in Section~\ref{ensemble}.

\vspace{0.5in}
\section{Galaxies Associated with these Absorbers}
\label{galaxies}

In this Section we describe the ground-based observations and analysis
of the targeted and serendipitous galaxies associated with the
absorbers as listed in Tables~\ref{tab:targeted} and
\ref{tab:serendipitous}. The basic information obtained on these
galaxies from our ground-based observations are listed in
Tables~\ref{tab:targ_galprop} and \ref{tab:ser_galprop}  for the
targeted and serendipitous samples, respectively. The following
information is provided in these tables by column: (1) the QSO target
sight line name; (2) the galaxy name; (3) the galaxy's recession
velocity ($cz_{\rm gal}$) in \kms;  (4) the galaxy luminosity in $L^*$
units derived from rest-frame $g$-band magnitudes (excepting as
discussed below); (5) the galaxy impact parameter ($\rho$) in kpc; (6)
the galaxy's virial radius in kpc; (7) and (8) the logarithms of the
galaxy halo and stellar masses in $M_{\Sun}$ units; (9) the galaxy's
metallicity relative to the  solar value, with 2$\sigma$
(95\% confidence level) uncertainties; (10) the galaxy's inclination
angle on the sky where well-defined; (11) and (12) the SFR in
$M_{\Sun}\,{\rm yr}^{-1}$ as given by the galaxy's \Ha\ and \galex\
FUV  luminosities; and (13) and (14) the specific SFR (sSFR) in units
of ${\rm yr}^{-1}$, as derived from the \Ha\ and FUV SFRs,
respectively, and the galaxy stellar mass from column~8.

Many of the quantities in Tables~\ref{tab:targ_galprop} and 
\ref{tab:ser_galprop} were either derived directly from the SDSS galaxy 
redshift survey database \citep[DR12;][]{alam15} or from our own observations. 
Most redshifts are derived from the SDSS galaxy redshift survey and have associated 
errors of $\sim30$~\kms; other sources of galaxy redshifts and associated errors, 
including \HI\ 21-cm emission redshifts, are described with the individual galaxy 
descriptions in Section~\ref{indiv:galaxies} of the Appendix where applicable. 
Tabulated stellar masses (column~8) use Equation~8 of \citet{taylor11}, 
which needs only rest-frame $(g-i)$ color and $i$-band absolute magnitude, so it 
can be applied uniformly to our whole sample. We correct to rest-frame colors and 
magnitudes using the $K$-corrections of \citet{chilingarian10} and 
\citet{chilingarian12}. The virial radius (column~6) and halo mass (column~7) are 
derived from the galaxy luminosity (column~4) as described in Section~\ref{sample} 
and Paper~1.

\begin{figure}[!t]
\plotone{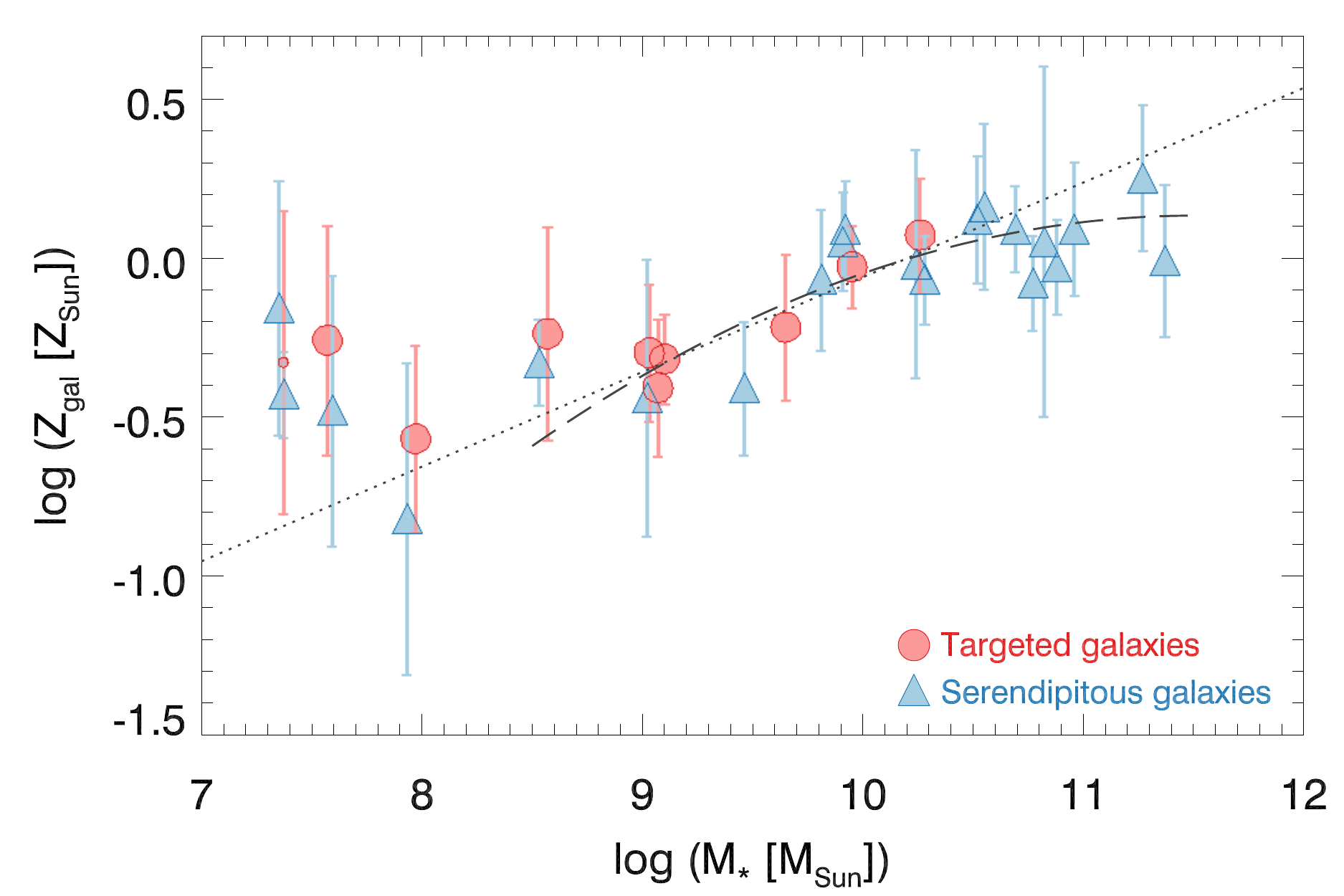}
\caption{Relationship between the metallicity and stellar mass of targeted and serendipitous absorbers, with the relationships of \citet[dashed line]{tremonti04} and \citet[dotted line]{lee06} overlaid. These are a reasonable match to our data.
\label{fig:MZ}}
\end{figure}

\edit1{
The galaxy metallicity (column~9) is derived using the O3N2 index of 
\citet{pettini04} unless otherwise noted below, with the exception of values that 
have no listed uncertainty; these metallicity estimates are derived from the 
galaxy's stellar mass using the mass-metallicity relations of \citet{tremonti04} 
when $\log{M_*}>10.2$ or \citet{lee06} when $\log{M_*}<10.2$. Figure~\ref{fig:MZ} 
shows the measured galaxy metallicity as a function of stellar mass for our 
targeted and serendipitous galaxies. The mass-metallicity relations of 
\citet[dashed line]{tremonti04} and \citet[dotted line]{lee06} are overlaid, and 
are clearly a good match to our measurements. Thus, we feel that they provide 
reasonable estimates of the galaxy metallicity in cases where no direct 
measurements are available.
}

\floattable
\begin{deluxetable*}{llcDccDDCcccDD}
\rotate
\tablecaption{Targeted Galaxy Properties\label{tab:targ_galprop}}
\tablewidth{0pt}
\tabletypesize{\scriptsize}
\tablehead{ \colhead{Sight Line} & \colhead{Galaxy} & \colhead{$cz_{\rm gal}$} & \twocolhead{$L$} & \colhead{$\rho$} & \colhead{$R_{\rm vir}$} & \twocolhead{$\log{M_{\rm h}}$} & \twocolhead{$\log{M_*}$} & \colhead{[O/H]\tablenotemark{a}} & \colhead{$i$} & \colhead{SFR(\Ha)} & \colhead{SFR(FUV)} & \multicolumn{4}{c}{$\log{\rm sSFR}$} \\ & & \colhead{(\kms)} & \twocolhead{($L^*$)} & \colhead{(kpc)} & \colhead{(kpc)} & & & & & & \colhead{(deg)} & \colhead{($M_{\Sun}\,{\rm yr}^{-1}$)} & \colhead{($M_{\Sun}\,{\rm yr}^{-1}$)} & \twocolhead{(\Ha)} & \twocolhead{(FUV)} }

\decimalcolnumbers
\startdata
1ES~1028+511      & UGC~5740                 &  649 & 0.007  & 110 &  54  &  9.94  &  7.97  & -0.57\pm0.59 &    47   &     0.052-0.063 &   \nodata    &   -9.17  &     .    \\ [-1mm]
1SAX~J1032.3+5051 &                          &      &  .     &  79 \\ [-1mm]		       								       
1ES~1028+511      & SDSS~J103108.88+504708.7 &  934 & 0.005  &  26 &  51  &  9.87  &  7.57  & -0.26\pm0.72 & \nodata & $>0.00004$      &  0.006       & >-11.97  &   -9.79  \\ [-1mm]
FBQS~J1010+3003   & UGC~5478                 & 1378 & 0.059  &  57 &  78  & 10.42  &  8.57  & -0.24\pm0.67 & \nodata &     0.067-0.26  &   \nodata    &   -9.16  &     .    \\ [-1mm]
HE~0435--5304     & ESO~157--49              & 1673 & 0.12   & 172 &  93  & 10.65  &  9.03  & -0.30\pm0.43 &    80   &      0.23-0.50  &   0.02-0.08  &   -9.33  &  -10.12  \\ [-1mm]
HE~0439--5254     &                          &      &  .     &  93 \\ [-1mm]
RX~J0439.6--5311  &                          &      &  .     &  74 \\ [-1mm]
HE~0439--5254     & ESO~157--50              & 3874 & 0.43   &  89 & 137  & 11.16  &  9.07  & -0.41\pm0.43 &    83   &      0.79-2.2   &   \nodata    &   -8.73  &     .    \\ [-1mm]
PG~0832+251       & NGC~2611                 & 5226 & 0.42   &  53 & 136  & 11.15  & 10.26  & +0.07\pm0.36 &    74   &      0.99-2.3   &   0.11-0.49  &   -9.90  &  -10.57  \\ [-1mm]
PMN~J1103--2329   & NGC~3511                 & 1114 & 0.58   &  97 & 151  & 11.29  &  9.65  & -0.22\pm0.46 &    73   &      0.28-0.52  &   \nodata    &   -9.94  &     .    \\ [-1mm]
SBS~1108+560      & M~108                    &  696 & 0.68   &  22 & 161  & 11.37  &  9.95  & -0.03\pm0.26 &    77   &      0.42-0.85  &  0.58:-0.89: &  -10.02  &  -10.00: \\ [-1mm]
SBS~1122+594      & IC~691                   & 1199 & 0.095  &  45 &  87  & 10.57  &  9.10  & -0.32\pm0.28 & \nodata &      0.13-0.26  &   0.09-0.30  &   -9.69  &   -9.62  \\ [-1mm]
VII~Zw~244        & UGC~4527                 &  721 & 0.003: &   7 &  47: &  9.76: &  7.37: & -0.33\pm0.95 & \nodata &     0.002       &  0.004       &  -10.05: &   -9.77: \\
\enddata

\tablecomments{All masses are in units of $M_{\Sun}$ and specific star formation rate (sSFR) has units of $\mathrm{yr}^{-1}$.}
\tablenotetext{a}{Logarithm of the galaxy's oxygen abundance relative to solar, with $2\sigma$ (95\% confidence) error bars.}

\end{deluxetable*}

\newcommand{\tnb}{\tablenotemark{b}}

\floattable
\begin{deluxetable*}{llcDccDDCcccDD}
\rotate
\tablecaption{Serendipitous Galaxy Properties\label{tab:ser_galprop}}
\tablewidth{0pt}
\tabletypesize{\scriptsize}
\tablehead{ \colhead{Sight Line} & \colhead{Galaxy} & \colhead{$cz_{\rm gal}$} & \twocolhead{$L$} & \colhead{$\rho$} & \colhead{$R_{\rm vir}$} & \twocolhead{$\log{M_{\rm h}}$} & \twocolhead{$\log{M_*}$} & \colhead{[O/H]\tablenotemark{a}} & \colhead{$i$} & \colhead{SFR(\Ha)} & \colhead{SFR(FUV)} & \multicolumn{4}{c}{$\log{\rm sSFR}$} \\ & & \colhead{(\kms)} & \twocolhead{($L^*$)} & \colhead{(kpc)} & \colhead{(kpc)} & & & & & & \colhead{(deg)} & \colhead{($M_{\Sun}\,{\rm yr}^{-1}$)} & \colhead{($M_{\Sun}\,{\rm yr}^{-1}$)} & \twocolhead{(\Ha)} & \twocolhead{(FUV)} }
\decimalcolnumbers
\startdata
3C~273            & SDSS~J122815.96+014944.1 &   911 & 0.004  &  70 &  50  &  9.83  &   7.37  & -0.43\pm0.27      &    63   &     0.001-0.002  &  0.003-0.006 &  -10.03  &   -9.58  \\ [-1mm]
3C~273            & SDSS~J122950.57+020153.7 &  1775 & 0.006  &  81 &  53  &  9.92  &   7.93  & -0.82\pm0.98\tnb  & \nodata &   0.0006         & $>0.0005$    &  -11.15  & >-11.23  \\ [-1mm]
3C~351            & Mrk~892                  &  3581 & 0.15   & 173 &  98  & 10.72  &   9.25  & -0.16\pm0.27      &    65   &      0.02-0.25   &   0.03-0.05  &   -9.85  &  -10.56  \\ [-1mm]
H~1821+643        & SDSS~J182202.70+642138.8 & 36436 & 1.1    & 157 & 189  & 11.58  &  10.04  &     -0.04         &    43   &      0.58-1.4    &    1.3-1.7   &   -9.89  &   -9.80  \\ [-1mm]
Mrk~335           & SDSS~J000529.16+201335.9 &  1950 & 0.008  &  97 &  55  &  9.98  &   7.59  & -0.48\pm0.85      & \nodata &     0.005        &   \nodata    &   -9.88  &     .    \\ [-1mm]
Mrk~876           & NGC~6140                 &   908 & 0.72   & 257 & 163  & 11.39  &   9.81  & -0.07\pm0.44      &    44   &           3.1    &   \nodata    &   -9.31  &     .    \\ [-1mm]
PG~0953+414       & SDSS~J095638.90+411646.1 & 42759 & 3.1    & 438 & 265  & 12.02  &  10.96  & +0.09\pm0.42      &    41   &      0.93-6.1    &   \nodata    &  -10.17  &     .    \\ [-1mm]
PG~1116+215       & SDSS~J111905.51+211733.0 & 17993 & 0.10   & 133 &  89  & 10.60  &   9.44  &     -0.23         &    28   &      0.03-0.09   &   0.06-0.09  &  -10.46  &  -10.48  \\ [-1mm]
PG~1116+215       & SDSS~J111906.68+211828.7 & 41428 & 1.2    & 139 & 192  & 11.60  &  11.27  & +0.25\pm0.46      &    40   &      0.05-0.29   &   \nodata    &  -11.80  &     .    \\ [-1mm]
PG~1211+143       & IC~3061                  &  2316 & 0.32   & 138 & 125  & 11.04  &   9.91  & +0.05\pm0.31      &    87   &      0.33-1.0    &  0.58:-3.9:  &   -9.91  &   -9.32: \\ [-1mm]
PG~1211+143       & SDSS~J121409.55+140420.9 & 15309 & 0.92   & 137 & 177  & 11.50  &  10.28  & -0.07\pm0.28      &    35   &     0.06:-0.77:  &   0.84-3.0   &  -10.39: &   -9.80  \\ [-1mm]
PG~1211+143       & SDSS~J121413.94+140330.4 & 19334 & 0.12   &  72 &  93  & 10.65  &   9.02  & -0.44\pm0.87      &    47   &   $>0.08$        &   \nodata    & >-10.13  &     .    \\ [-1mm]
PG~1216+069       & SDSS~J121930.86+064334.4 & 24116 & 3.5    & 505 & 275  & 12.07  &  10.69  & +0.09\pm0.27      &    31   &      7.0:-27:    &     10-20    &   -9.26: &   -9.39  \\ [-1mm]
PG~1216+069       & SDSS~J121923.43+063819.7 & 37204 & 0.74   &  93 & 164  & 11.40  &  10.26  &     +0.01         &    46   &       1.6-6.3    &   \nodata    &   -9.46  &     .    \\ [-1mm]
PG~1259+593       & UGC~8146                 &   668 & 0.046  & 114 &  74  & 10.35  &   8.53  & -0.33\pm0.27      &    89   &     0.011-0.035  &   0.09-0.56  &   -9.98  &   -8.78  \\ [-1mm]
PG~1259+593       & SDSS~J130101.05+590007.1 & 13862 & 0.47   & 138 & 141  & 11.20  &  10.24  & -0.02\pm0.72      &    54   &     0.008-0.04   &   \nodata    &  -11.67  &     .    \\ [-1mm]
PHL~1811          & SDSS~J215456.65--091808. & 15453 & 1.5    & 269 & 207  & 11.70  &  10.52  & +0.12\pm0.406     &    48   &   $>0.08$        &    1.2-2.2   & >-11.62  &  -10.18  \\ [-1mm]
PHL~1811          & SDSS~J215517.30--091752. & 21951 & 2.7    & 502 & 253  & 11.96  &  11.37  & -0.01\pm0.480     &    22   &      0.07-1.1    & $<0.11$      &  -11.32  & <-12.33  \\ [-1mm]
PHL~1811          & J215447.5--092254        & 23278 & 0.74   & 309 & 164  & 11.40  &   9.92  & +0.09\pm0.30      &    26   &     0.94:-4.0:   &    1.0-1.3   &   -9.32: &   -9.80  \\ [-1mm]
PHL~1811          & J215450.8--092235        & 23623 & 0.26   & 237 & 116  & 10.95  &   9.46  & -0.41\pm0.42      &    50   &     0.37:-3.3:   &   0.39-0.62  &   -8.94: &   -9.67  \\ [-1mm]
PHL~1811          & 2MASS~J21545996--0922249 & 24223 & 0.56   &  35 & 150  & 11.28  &  10.24  &     +0.01         &    53   &     \nodata      &   \nodata    &     .    &     .    \\ [-1mm]
PHL~1811          & J215506.5--092326        & 39758 & 2.3    & 228 & 239  & 11.89  &  10.88  & -0.03\pm0.30      &    43   &      1.9:-21:    &    2.4-4.6   &   -9.56: &  -10.22  \\ [-1mm]
PHL~1811          & J215454.9--092331        & 52873 & 1.4    & 354 & 204  & 11.68  &  10.77  & -0.08\pm0.30      &    30   &      3.2:-17:    &    4.4-9.5   &   -9.54: &   -9.79  \\ [-1mm]
PKS~0312--770     & J031201.7--765517        & 17792 & 0.34:  & 239 & 127: & 11.06: &   9.99: &     -0.06:        & \nodata &     \nodata      & $<0.41$      &     .    & <-10.37: \\ [-1mm]
PKS~0312--770     & J031158.5--764855        & 35732 & 2.1:   & 381 & 231: & 11.84: &  10.87: &     +0.10:        & \nodata &     \nodata      & $<1.9$       &     .    & <-10.59: \\ [-1mm]
PKS~0405--123     & 2MASX~J04075411--1214493 & 29050 & 1.2    & 378 & 192  & 11.60  &  10.55  & +0.16\pm0.52      &    75   &      0.31-3.9    &   0.44-2.1   &   -9.96  &  -10.23  \\ [-1mm]
PKS~0405--123     & J040743.9--121209        & 45989 & 0.89   & 197 & 175  & 11.48  &  10.50  &     +0.05         &    44   &     \nodata      & $<2.4$       &     .    & <-10.12  \\ [-1mm]
PKS~0405--123     & J040751.2--121137        & 50127 & 1.8    & 117 & 222  & 11.79  &  10.82  & +0.05\pm1.10      &    34   &     \nodata      & $<2.3$       &     .    & <-10.46  \\ [-1mm]
PKS~1302--102     & NGC~4939                 &  3112 & 3.2    & 261 & 267  & 12.03  &  10.54  &     +0.06         &    61   &      0.57-0.88   &   \nodata    &  -10.59  &     .    \\ [-1mm]
PKS~1302--102     & 2MASX~J13052026--1036311 & 12755 & 2.8    & 227 & 256  & 11.98  &  10.85  &     +0.10         &    34   &     \nodata      &   0.72-1.3   &     .    &  -10.73  \\ [-1mm]
PKS~1302--102     & 2MASX~J13052094--1034521 & 28304 & 3.4    & 353 & 273  & 12.06  &  11.00  &     +0.11         &    32   &   $<0.65$:       &   0.26-0.47  & <-11.19: &  -11.33  \\ [-1mm]
PKS~2155--304     & 2MASX~J21584077--3019271 & 17005 & 1.7    & 425 & 217  & 11.76  &  10.38  &     +0.03         &    65   &     \nodata      &   0.93-2.4   &     .    &  -10.00  \\ [-1mm]
PKS~2155--304     & J215845.1--301637        & 31887 & 2.2:   & 403 & 238: & 11.88: &  10.77: &     +0.09:        & \nodata &     \nodata      &   \nodata    &     .    &     .    \\ [-1mm]
Q~1230+0115       & CGCG~014--054            &  1105 & 0.004  &  70 &  50  &  9.84  &   7.35  & -0.16\pm0.80      & \nodata & $>0.00001$       & $>0.004$     & >-12.21  &  >-9.75  \\ [-1mm]
Q~1230+0115       & SDSS~J123047.60+011518.6 & 23327 & 0.19   &  55 & 105  & 10.81  &   9.63  &     -0.17         &    67   &      0.63-1.5    &   \nodata    &   -9.44  &     .    \\
\enddata

\tablecomments{All masses are in units of $M_{\Sun}$ and specific star formation rate (sSFR) has units of $\mathrm{yr}^{-1}$.}
\tablenotetext{a}{Logarithm of the galaxy's oxygen abundance relative to solar, with $2\sigma$ (95\% confidence) error bars.}
\tablenotetext{b}{This galaxy has no optical emission lines, but \citet{keeney14} estimated its [Fe/H] metallicity using SED modelling of its GALEX+SDSS photometry.}

\end{deluxetable*}

\clearpage

A range of \Ha\ and FUV SFR is listed for each galaxy. The lower value
comes from using the observed \Ha\ and FUV luminosity, corrected for
Galactic foreground extinction, to derive the corresponding SFR using
the calibration of \citet{hunter10}. The extinction correction uses
$E(B-V)$  from \citet{schlafly11} and a \citet{fitzpatrick99}
reddening law with $R_V=3.1$. The upper value attempts to
correct for two additional systematic effects, and is used where possible 
to derive the sSFR values in columns~13 and 14. The first systematic effect 
we correct for is extinction internal to the galaxy. The attenuation at \Ha\ 
is estimated using the galaxy inclination \citep[derived from the galaxy's 
observed axial ratio using the intrinsic  axial ratios for different 
morphological types of][]{masters10} and the attenuation relations of 
\citet{driver08}. The FUV attenuation is estimated from the \Ha\ attenuation 
using the prescription of \citet{calzetti01}. The second systematic effect 
is an aperture correction for galaxies whose \Ha\ SFR is determined 
spectroscopically. In this case we use the \Ha\ curve of growth for CALIFA 
galaxies \citep{iglesias-paramo13} to estimate the amount of \Ha\ emission 
outside the spectroscopic aperture. When measurements of the galaxy's 
half-light radius are not available we compare the galaxy's $r$-band 
magnitude internal to the spectroscopic aperture with its total $r$-band 
magnitude to derive the aperture correction. Comparisons of the two aperture 
correction procedures for the subset of galaxies for which both can 
be derived show that they agree within 10-20\%, and the aperture
corrections of \citet{iglesias-paramo13} have $1\sigma$ uncertainties 
of 5-15\%. 

There are some cases where the spectroscopic aperture covers 
such a small fraction of the galaxy that the aperture correction is very 
large ($>10$); for these galaxies we do not attempt to perform an aperture 
correction and quote only a lower limit to the galaxy SFR that is equal to 
the SFR within the spectroscopic aperture. Fortunately, \Ha\ imaging is available 
for most of the galaxies in our sample with recession velocities $\leq10000$~\kms, 
eliminating the problematical spectroscopic aperture corrections in these cases.

Where two sight lines are listed for the same galaxy, those sight
lines probe the same galaxy at different impact parameters and
position angles relative to the galaxy's disk; e.g., three targets
(HE~0435--5304, HE~0439--5254 and RX~J0439.6--5311) all probe the
nearly edge-on spiral ESO~157--49. Where a single target is listed
more than once, that sight line probes multiple galaxies; e.g., HE~0439--5254 
probes both ESO~157--49 and ESO~157--50.

Unless otherwise noted, narrow-band, redshifted \Ha\ and broad-band images of 
each galaxy were obtained using the SPIcam optical imaging camera at the Apache 
Point Observatory (APO). Typically SDSS $r$-band images were used as the
``off-band'' to produce the pure \Ha\ images even when \Ha\ was
present in the $r$-band \citep[see][for detailed procedure]{keeney13}. 
Broad-band and pure \Ha\ images of both the
targeted and serendipitous galaxies are presented in Appendix~\ref{imaging}
along with observing details (e.g., epoch, exposure time, filters, 
``seeing''). Additionally (unless otherwise indicated below), we
obtained major-axis, long-slit optical spectra of all galaxies using
the Dual-channel Imaging Spectrograph (DIS) at APO. These APO
observations include both low- and high-dispersion spectra unless
adequate low-dispersion spectra were available through the SDSS
spectroscopic database. Some of the serendipitously probed galaxies
are at a distance great enough that the slit spectroscopy provides an
adequate measurement of the total \Ha\ flux. In all other cases the
total \Ha\ flux is determined from the images.

In the Appendix, Section~\ref{indiv:galaxies}, the individual targeted and
serendipitous galaxies are described in some detail both to illustrate
the breadth in galaxy properties probed using the COS, STIS and \fuse\
observations and to make the connection between galaxies and absorbers.

\section{CLOUDY Photo-Ionization Models}
\label{cloudy}

In Paper~1 we performed CLOUDY photo-ionization models \citep{ferland98} of any 
absorber that had metal-line absorption from multiple ionization states of the 
same element (e.g., \SiII\,/\,\SiIII\,/\,\SiIV). Here we model all absorbers in 
Table~\ref{tab:absprop} that meet this criterion, which leads to a slightly 
different list of modeled absorbers than that presented in Paper~1. There are 
five absorbers that we model here but did not model in Paper~1; these absorbers 
all have multi-ion absorption from the same element in Table~\ref{tab:absprop} 
but not in \citet{danforth08}, which provided the serendipitous absorption line 
fits for Paper~1. There are also four absorbers that we modeled in Paper~1 but do 
not model here due to uncertain \HI\ column densities. Specifically, unlike 
Paper~1 we do not model the two low-$z$ absorbers found in the SBS~1108+560 
sight line due to very low S/N in that spectrum at the wavelengths of redshifted 
\lya; the poor quality of the bluest portion of the COS/G130M spectrum is due to 
the presence of an unexpected Lyman limit system at higher redshift that absorbs 
the UV continuum. We do not model the 5444~\kms\ absorber in the PG~0832+251 
sight line because the extremely broad, saturated \lya\ profile 
(Figure~\ref{fig:stackplot}) does not allow us to reliably deconvolve individual 
\HI\ components without higher-order Lyman-series lines (see
Section~\ref{indiv:absorbers:pg0832_5221} for details). Finally, the 
22042~\kms\ absorber in the PHL~1811 sight line is not modeled because 
we find evidence for only \lya\ and possible \SiIII\ absorption at this redshift
in the high-S/N COS spectrum (Figure~\ref{fig:stackplot}), leaving us with too 
few metal lines to meet our modeling criteria.

\begin{deluxetable}{lcccC}
\tablecaption{UVB Model Comparison\label{tab:uvb}}
\tablewidth{0pt}
\tablehead{ \colhead{Model Reference} & \colhead{$\Gamma_0$} & \colhead{$J_0$} & \colhead{$\gamma$} & \colhead{$\delta$} }

\colnumbers
\startdata
\citet{haardt96}          &   4.1  &   1.6   &   3.3   &  -0.8   \\
\citet{shull99}           &   6.3  &   2.5   &   3.9   &  -0.4   \\
\citet{haardt01}          &  10.3  &   3.9   &   3.1   &  -0.9   \\
\citet{haardt05}          &  13.5  &   5.2   &   3.4   &  -0.7   \\
\citet{faucher-giguere09} &   3.8  &   1.4   &   2.9   &  -1.1   \\
\citet{haardt12}          &   2.3  &   0.9   &   4.3   &  -0.1   \\
\citet{kollmeier14}       &  11.5  &   4.4   &   4.3   &  -0.1   \\ 
\citet{shull15}           &   4.6  &   1.7   &   4.4   &  -0.1   \\
\citet{khaire15}          &   4.1  &   1.5   &   4.6   &  +0.1   \\
\citet{gaikwad16}         &   3.9  &   1.4   &   5.0   &  +0.3   \\
\enddata

\end{deluxetable}

\subsection{Effect of Ionizing Background}
\label{cloudy:UVB}

Before delving into the specifics of our photo-ionization models, a brief 
aside is warranted regarding the systematic effects that are coupled to the 
choice of ionizing radiation field. There has been considerable recent 
debate about the strength of the metagalactic UV ionizing background (UVB) at 
$z\sim0$ due to various theoretical models whose \HI\ photo-ionization rates 
vary by a factor of $\sim6$ at $z=0$ 
\citep{haardt96,haardt01,haardt05,haardt12,shull99,shull15,faucher-giguere09,kollmeier14,khaire15,madau15,gaikwad16}.
We examine the nature of these models in Table~\ref{tab:uvb}, whose first 
four columns list: (1) the model reference; (2) the \HI\ photo-ionization rate, 
$\Gamma_{-14}$, at $z=0$ in units of $10^{-14}~{\rm s}^{-1}$; (3) the 
specific intensity of Lyman continuum radiation, $J_{-23}$, at $z=0$ in units of 
$10^{-23}~{\rm erg\,cm^{-2}\,s^{-1}\,Hz^{-1}\,sr^{-1}}$; and (4) a power-law index 
that describes how $\Gamma_{-14}$ and $J_{-23}$ evolve with redshift at $z<0.5$ 
(i.e., $\Gamma_{-14} = \Gamma_0 (1+z)^{\gamma}$ and $J_{-23} = J_0 (1+z)^{\gamma}$). 

\edit1{All of these models, with the exceptions of 
\citet{haardt01,haardt05} and \citet{kollmeier14}, are below observational limits 
on the UVB intensity at $z=0$ 
\citep[and see especially Figure~11 of \citealp{fumagalli17}]{donahue95,adams11}.}
Additionally, all of these models evolve strongly at low redshift 
($\gamma\approx3$-5); however, there is variation in both normalization 
($\Gamma_0 \propto J_0$; see Equation~\ref{eqn:gammaH}) and power-law slope from 
model to model, which makes it non-trivial to compare quantities derived from 
CLOUDY photo-ionization models that assume different underlying UVB prescriptions.

Here we investigate the differences between these models using the cosmological 
simulations of \citet{shull15}, which use a $768^3$ grid that is $50h^{-1}$~Mpc 
on a side. This grid is then irradiated with the \citet{haardt01,haardt05,haardt12}
UVBs to study the relationship between baryon overdensity, $\Delta_{\rm b}$, and 
\HI\ column density for simulated absorbers with $12.5<\log{N_{\rm H\,I}}<15.5$. 
This exercise has been performed by several groups using several different types 
of simulations \citep{dave10,tepper-garcia12,shull15}. While different groups find 
somewhat different relationships between \HI\ column and overdensity, the overall 
normalizations and power-law indices derived for a given ionizing background and 
redshift  are remarkably robust given the other differences between the 
simulations \citep[see discussion in][]{shull15}. 

While \citet{shull15} were primarily interested in comparing the 
$\Delta_{\rm b}$-$N_{\rm H\,I}$ relationship for a given choice of UVB and 
redshift, here we attempt to quantify how different UVB intensities and 
power-law indices will affect the relation. We do so by measuring the 
overdensity in four different redshift bins from $0<z<0.4$ for each 
choice of UVB \citep{haardt01,haardt05,haardt12}. We expect the baryon 
overdensity to vary as
\begin{equation}
\Delta_{\rm b}(z) \equiv \frac{\rho_{\rm b}}{\langle\rho_{\rm b}(z)\rangle} = \Delta_0 (1+z)^{-3} \Gamma_{-14}^{\beta} N_{14}^{\beta},
\end{equation}
where $\rho_{\rm b}$ is the mass density of baryons, 
$\langle\rho_{\rm b}(z)\rangle \propto (1+z)^3$ is the mean baryon density at 
redshift $z$, and $N_{14} = N_{\rm H\,I}/(10^{14}~{\rm cm}^{-2})$. We expect 
$\Gamma_{-14}$ and $N_{\rm H\,I}$ to scale with the same power-law index, $\beta$, 
so that an increase in \HI\ photo-ionization rate will require a corresponding 
decrease in \HI\ column to yield a constant overdensity value. The $(1+z)^{-3}$ 
scaling implies that absorbers of a given column density exposed to a given 
photo-ionization rate exist at a constant physical density ($\rho_{\rm b}$) at 
all redshifts. 

Adding in the redshift dependence of $\Gamma_{-14}$, we find
\begin{equation}
\Delta_{\rm b}(z) = \Delta_0 \Gamma_0^{\beta} N_{14}^{\beta} (1+z)^{\beta\gamma-3},
\end{equation}
where $\Gamma_0 \equiv \Gamma_{-14}(z=0)$ and $\gamma$ are properties of the 
individual UVB model from Table~\ref{tab:uvb}, and $\Delta_0$ and $\beta$ are 
free parameters that hold for all choices of UVB in these simulations. We find 
that $\Delta_0\approx10$ and $\beta\approx\sfrac{2}{3}$ yield values of 
$\Delta_{\rm b}$ that are within 5\% of the measured values for the UVB specified 
by \citet{haardt01,haardt05} and within 20\% of the measured values for the 
\citet{haardt12} UVB.  The final column of Table~\ref{tab:uvb} lists 
$\delta=\beta\gamma-3\approx\sfrac{(2\gamma-9)}{3}$, the power-law index that 
describes the evolution of $\Delta_{\rm b}$ with redshift, for each UVB model. 
Note that some of the UVBs predict very little evolution of $\Delta_{\rm b}$ with 
redshift because they are very close to the $\gamma=\sfrac{9}{2}$ value that 
corresponds to $\delta=0$ when $\beta=\sfrac{2}{3}$.

We can now quantify the effects of the choice of UVB on the outputs and 
derived quantities from our CLOUDY photo-ionization models. The \HI\ 
photo-ionization rate and the specific intensity of Lyman continuum 
radiation (columns~2 and 3 of Table~\ref{tab:uvb}) are related by
\begin{align}
\label{eqn:gammaH}
\Gamma_{\rm H} &= 4\pi \int^{\infty}_{\nu_0} \frac{J_{\nu} \sigma_{\nu}}{h\nu} d\nu   \\*
\Gamma_{-14} &\approx 2.71 \left(\frac{4.4}{\alpha+3}\right) J_{-23}, \nonumber
\end{align}
where $\alpha$ is the power-law slope that characterizes the frequency dependence 
of the UVB. Recent measurements from \hst/COS find $\alpha\approx1.4$ for AGN 
\citep*{shull12,stevans14}, but the assumed value of $\alpha$ varies somewhat for 
the models in Table~\ref{tab:uvb} (from $\alpha\approx1.4$-1.8). 

The baryon overdensity is related to the total number density of hydrogen atoms, 
$n_{\rm H}$:
\begin{align}
n_{\rm H} &= \frac{\rho_{\rm cr}\Omega_{\rm b}(1-Y_p)}{m_{\rm H}}(1+z)^3 \\*
          &= (1.89\times10^{-7}~{\rm cm^{-3}}) (1+z)^3 \Delta_{\rm b}(z), \nonumber
\end{align}
where the product $\rho_{\rm cr}\Omega_{\rm b}$ is the baryon mass density at $z=0$ 
and $Y_p=0.2477$ is the primordial helium abundance \citep*{peimbert07}. The 
ionization parameter, $U$, can then be derived once the number density of 
photons, $n_{\gamma}$, is known:
\begin{align}
n_{\gamma} &= \frac{4\pi J_{\nu}}{hc\alpha} \\*
           &= (5.31\times10^{-8}~{\rm cm}^{-3}) \left(\frac{\alpha+3}{\alpha}\right)  \Gamma_{-14} \nonumber \\*
U &\equiv \frac{n_{\gamma}}{n_{\rm H}} \\*
  &= 0.28 \left(\frac{\alpha+3}{\alpha}\right) (1+z)^{-3} \frac{\Gamma_{-14}}{\Delta_{\rm b}(z)} \nonumber \\*
  &= \frac{0.28}{\Delta_0} \left(\frac{\alpha+3}{\alpha}\right) \Gamma_{-14}^{1-\beta} N_{14}^{-\beta}. \nonumber
\end{align}
The ratio $\sfrac{(\alpha+3)}{\alpha} \approx 3$ for all of the models in 
Table~\ref{tab:uvb}. Using this approximation and substituting the estimated values of 
$\Delta_0\approx10$ and $\beta\approx\sfrac{2}{3}$ from above, we find
\begin{align}
\label{eqn:logU}
\log{U} \approx 8.26 &+ \sfrac{1}{3} \log{\Gamma_{-14}} \\*
                     &- \sfrac{2}{3} \log{(N_{\rm H\,I}\,[{\rm cm}^{-2}])}. \nonumber
\end{align}

There are large uncertainties in the normalization of the UVB at $z\approx0$
(see Table~\ref{tab:uvb}). The most recent UVB models in Table~\ref{tab:uvb} 
suggest that $\Gamma_0\approx4$. However, Figure~2 of \citet{shull15} indicates 
that a somewhat lower value, consistent with the \citet{haardt12} UVB, is a better 
match to the column density distribution function of absorbers with 
$N_{\rm H\,I}>10^{14}~{\rm cm}^{-2}$ than the UVB models of \citet{haardt01} 
or \citet{haardt05}\footnote{\citet{haardt01} and \citet{haardt05} are more 
consistent with the number of absorbers with 
$N_{\rm H\,I}\approx10^{13}~{\rm cm}^{-2}$, but none of the \citeauthor{haardt01} 
models are a good fit to absorbers with $N_{\rm H\,I}=10^{13.5}$ to 
$10^{14}~{\rm cm}^{-2}$.}. Since all but two of the cool CGM absorbers that we 
model have $N_{\rm H\,I}>10^{14}~{\rm cm}^{-2}$ (see Table~\ref{tab:cloudy}), we  
choose to normalize the \HI\ photo-ionization rate in our models to 
$\Gamma_{-14}=2.3$, which is equivalent to the \citet{haardt12} UVB at $z=0$,
and lower than any of the other UVB models in Table~\ref{tab:uvb}. 

Below, we explicitly state the dependence on $\Gamma_{-14}$ for all quantities 
derived from the ionization parameter to facilitate different choices in 
normalization.

\subsection{Model Description and Results}
\label{cloudy:description}

In addition to the above differences in the specific absorbers modeled, we have 
also updated our modeling procedure. In Paper~1 we performed a strict maximum 
likelihood analysis comparing observed and modeled metal-line ratios (i.e., the 
ratio of the observed metal-line column densities to the \HI\ column density), 
adopting the ionization parameter, $U \equiv \sfrac{n_{\gamma}}{n_{\rm H}}$, and 
metallicity, $Z_{\rm abs}$, that maximized the likelihood of simultaneously 
observing all of the modeled metal-line ratios as our ``modeled'' value. 
However, it is usually the case that the metal-line column densities are far 
better constrained than the \HI\ column density, even if the formal uncertainty 
on $N_{\rm H\,I}$ is relatively small. This unhappy circumstance is primarily 
a consequence of the uncertainties inherent in deconvolving the component 
structure of saturated \lya\ profiles when the only \HI\ line available is \lya. 
In Paper~1 we confronted this problem by restricting the ``plausible'' range of 
$N_{\rm H\,I}$ values for a given absorber to those which accommodated a 
single-phase photo-ionization solution with a line-of-sight thickness 
$D_{\rm cl}<\rho$ and metallicity $Z_{\rm abs} \lesssim Z_{\rm gal}$ 
(this procedure is detailed in \citealp{keeney13} and \citealp{davis15}). 

Here we utilize a more principled tactic by assuming a Bayesian prior for the 
absorber ionization parameter and metallicity and adopting the unaltered \HI\ 
column density from the Voigt profile fits\footnote{Since the H\,I column density 
at a particular velocity can be strongly dependent on the number of velocity 
components used to fit saturated \lya\ profiles this does not remove all 
subjective bias from the analysis. We have endeavored to make sensible, 
consistent choices in all of our fits (Section~\ref{absorbers}) but 
in many cases a similar fit quality can be achieved with several combinations 
of the number and location of velocity components.}. Our prior on the absorber 
ionization parameter uses Equation~\ref{eqn:logU} above, normalized to 
$\Gamma_{-14}=2.3$ (the \HI\ photo-ionization rate at $z=0$ for the 
\citet{haardt12} UVB; see Table~\ref{tab:uvb}). There is typically a 
factor of $\sim2$ scatter in the $\Delta_{\rm b}$-$N_{\rm H\,I}$ relation at 
$z\approx0$ \citep[e.g., see Figure~9 of][]{dave10}. We propagate this uncertainty 
along with the uncertainty on $N_{\rm H\,I}$ when determining the uncertainty on 
our prior for $\log{U}$. As in Paper~1, our prior on the absorber metallicity is 
that it is less than or equal to the galaxy metallicity; 
\edit1{if the galaxy metallicity was estimated from the mass-metallicity relations 
of \citet{tremonti04} or \citet{lee06}, our metallicity prior assumes a $1\sigma$ 
uncertainty of 0.5~dex.}

As in \citet{keeney13} and \citet{davis15}, we model the absorbers using a 
plane-parallel grid of CLOUDY models \citep{ferland98} irradiated by the 
\citet{haardt12} UVB at $z=0$. Our models assume solar abundance ratios 
\citep{grevesse10} and vary the absorber metallicity in the range 
$\log{Z_{\rm abs}} = -3$ to 1 in solar units by steps of 0.2~dex, and the 
ionization parameter in the range $\log{U}=-5$ to 1 by steps of 0.2~dex. 

Column densities of \HI\ and all metal lines commonly seen in low-$z$ FUV 
quasar absorption line systems are calculated at each grid point, from 
which model metal-line ratios are calculated. Before comparing these ratios 
with the observed metal-line ratios of our absorption line systems, we 
linearly interpolate the model values to a finer resolution of 0.01~dex in 
both $\log{U}$ and $\log{Z_{\rm abs}}$. For each species, $X$, the 
log-likelihood of a given point in this interpolated grid, $i$, is assumed 
to be
\begin{equation}
\ln{\mathcal{L}^X_i} = -0.5\left(\frac{r^X_i-r^X}{\sigma^X}\right)^2, 
\end{equation}
where $r^X_i$ is the model metal-line ratio (i.e., $N_X/N_{\rm H\,I}$) at 
point $i$, $r^X$ is the observed metal-line ratio, and $\sigma^X$ is 
the uncertainty in the observed metal-line ratio, which takes into account
the uncertainties in both $N_X$ and $N_{\rm H\,I}$. 

For metal-line detections, calculating the metal-line ratio and its associated 
uncertainty is straightforward, but the proper treatment of upper limits is less 
clear. Our approach is to treat the metal-line upper limits as step functions in 
probability; i.e., all column densities less than the $3\sigma$ limit listed 
in Table~\ref{tab:absprop} are equally likely (with $\rm{probability}=1$), as 
are all column densities above the listed limit (with $\rm{probability}=0$). 
However, since we are modeling ratios of metal-line column densities with 
respect to $N_{\rm H\,I}$ we need to take into account uncertainty in the 
\HI\ column density, $\sigma_{\rm H\,I}$. We do this by convolving the 
discontinuous likelihood predicted when $\sigma_{\rm H\,I}=0$ with a Gaussian 
kernel having $\sigma = \sigma_{\rm H\,I}$, which we then use to evaluate $\mathcal{L}^X_i$ for each $r^X_i$ in the interpolated grid. 

Finally, the posterior probability at point $i$ is determined using
\begin{equation}
\ln{P_i} = \ln{\Pi_i} + \sum_x \ln{\mathcal{L}^x_i},
\end{equation}
where $\Pi_i$ is the prior at point $i$ and the sum is performed over all 
species that are modeled. We then identify the interpolated grid points that 
maximize the marginal posterior probability for $\log{U}$ and 
$\log{Z_{\rm abs}}$, and adopt the 68.3\% highest posterior density 
credible interval as the uncertainty in the model parameters.

There are also several quantities that we derive from the model parameters.
The absorber density, $n_{\rm H}$, is given by
\begin{equation}
\label{eqn:nH}
\log{(n_{\rm H}\,[{\rm cm}^{-3}])} = -6.80 + \log{\Gamma_{-14}} - \log{U}.
\end{equation}
The mean pressure in the absorber is defined as 
\begin{align}
\label{eqn:P}
\log{\langle P/k \rangle} &= \log{(n_{\rm H}\,[{\rm cm}^{-3}])} + \log{T} \\*
                          &= -6.80 + \log{T} + \log{\Gamma_{-14}} - \log{U}, \nonumber
\end{align}
where $T$ is the equilibrium temperature ($\approx10^4$~K) output from CLOUDY 
along with the model column densities, and subsequently interpolated onto the same 
fine grid. The neutral fraction, $f_{\rm H\,I}$, is given by
\begin{align}
\label{eqn:fHI}
\log{f_{\rm H\,I}} &= 4.73 - 0.76\log{T} - \log{\Gamma_{-14}} \nonumber \\*
                   &\hspace{3em} + \log{(n_{\rm H}\,[{\rm cm}^{-3}])} \\*
                   &= -2.07 - 0.76\log{T} - \log{U}. \nonumber
\end{align}
\edit1{The \HI\ column density and neutral fraction determine the total hydrogen 
column:
\begin{equation}
\label{eqn:NH}
N_{\rm H} = N_{\rm H\,I}\,\left(1 + f_{\rm H\,I}^{-1}\right) \approx \sfrac{N_{\rm H\,I}}{f_{\rm H\,I}}. 
\end{equation}
}
The absorber line-of-sight thickness, $D_{\rm cl}$, is defined as 
\begin{align}
\label{eqn:Dcl}
\log{(D_{\rm cl}\,[{\rm kpc}])} &= -21.5 + \log{(N_{\rm H}\,[{\rm cm}^{-2}])} \nonumber \\*
                                &\hspace{4em}- \log{(n_{\rm H}\,[{\rm cm}^{-3}])} \\*
                                &= -12.6 + 0.76\log{T} - \log{\Gamma_{-14}} + 2\log{U} \nonumber \\*
                                &\hspace{4em} + \log{(N_{\rm H\,I}\,[{\rm cm}^{-2}])} \nonumber
\end{align}
and its ${\rm hydrogen+helium}$ mass, $M_{\rm cl}$, is given by 
\begin{align}
\label{eqn:Mcl}
\log{(M_{\rm cl}\,[M_{\Sun}])} &= 7.24 + \log{n_{\rm H}} \nonumber \\*
                               &\hspace{3em} + 3\log{(D_{\rm cl}\,[{\rm kpc}])} \\*
                               &= -37.4 + 2.28\log{T} - 2\log{\Gamma_{-14}} \nonumber \\*
                               &\hspace{3em} + 5\log{U} + 3\log{(N_{\rm H\,I}\,[{\rm cm}^{-2}])}. \nonumber
\end{align}
Equation~\ref{eqn:Mcl} assumes a uniform cloud density, a spherical cloud geometry, 
and a primordial helium abundance of $Y_p=0.2477$ \citep{peimbert07}. 

The results of our photo-ionization modeling are summarized in 
Table~\ref{tab:cloudy}, which lists the following information by column: 
(1) the running system number of the absorber from Table~\ref{tab:absprop}; 
(2) the quasar sight line name; (3) the \lya\ absorption velocity in \kms; 
(4) a subjective grade that indicates our relative confidence in the 
photo-ionization model for this absorber (see discussion in 
Section~\ref{ensemble}); (5) the logarithm of the absorber's 
\HI\ column density, in ${\rm cm}^{-2}$; (6) the logarithm of the absorber 
ionization parameter, $U$; (7) the logarithm of the absorber 
metallicity, $Z_{\rm abs}$, in solar units; (8) the logarithm of the mean 
absorber pressure, $\langle P/k \rangle$, in ${\rm cm^{-3}\,K}$; (9) the 
logarithm of the line-of-sight cloud thickness, $D_{\rm cl}$, in kpc; and 
(10) the logarithm of the cloud mass, $M_{\rm cl}$, in $M_{\Sun}$. 
Columns~11-13 of Table~\ref{tab:cloudy} reproduce the \HI\ column density, 
ionization parameter, and absorber metallicity, respectively, for absorbers 
modeled in Table~6 of Paper~1. 

\begin{figure*}[!ht]
\gridline{\fig{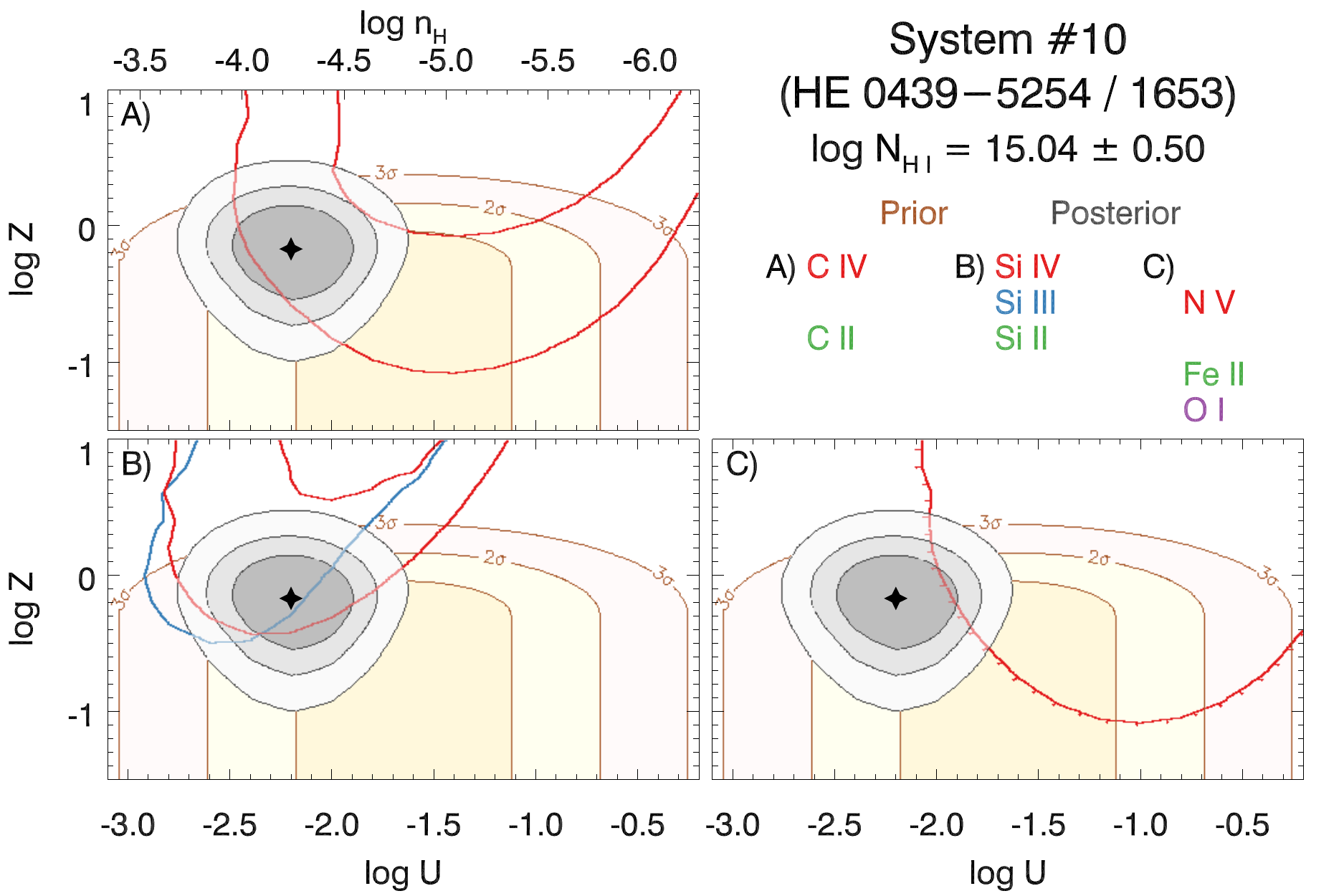}{0.8\textwidth}{5.1: Photo-ionization model of the 1653~\kms\ absorber in the HE~0439--5254 sight line, which is associated with the targeted galaxy ESO~157--49 (Section~\ref{indiv:galaxies:eso157-49}).}}
\gridline{\fig{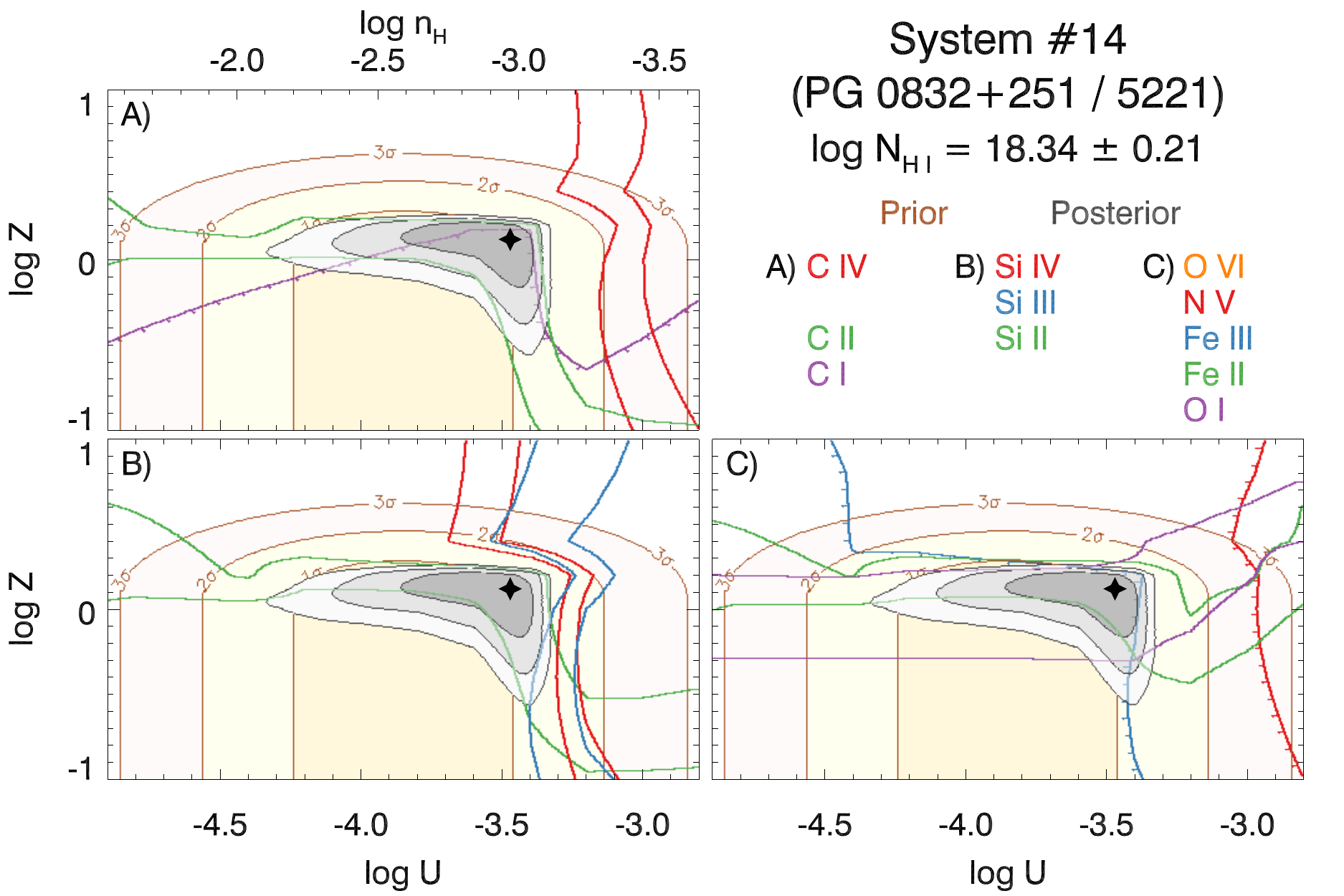}{0.8\textwidth}{5.2: Photo-ionization model of the 5221~\kms\ absorber in the PG~0832+251 sight line, which is associated with the targeted galaxy NGC~2611 (Section~\ref{indiv:galaxies:ngc2611}).}}
\caption{CLOUDY photo-ionization models for all absorbers in Table~\ref{tab:cloudy}.  Metal-line detections are indicated with solid $1\sigma$ contours enclosing the allowable region of parameter space, and $3\sigma$ upper limits are shown with tick marks pointing toward the allowable region of parameter space. Panel~A shows the carbon ions, Panel~B shows the silicon ions, and Panel~C shows everything else, color coded by species. The prior and posterior distributions for a given absorber are identical in all three panels.
The complete figure set (25~images) is available in the online journal.
\label{fig:cloudy}}
\end{figure*}

The photo-ionization models for all absorbers in Table~\ref{tab:cloudy} are shown 
in Figure~\ref{fig:cloudy}. Each model is depicted by three panels; the beige and 
gray contours, which represent the prior and posterior probability distributions
for a given absorber, are identical in all three panels. Panel~``A'' shows all of 
the carbon ions used to constrain the model, Panel~``B'' shows all of the silicon 
ions, and Panel~``C'' shows everything else (sundry ions of oxygen, nitrogen, and 
iron). Metal-line detections are indicated with solid $1\sigma$ contours 
enclosing the allowable region of parameter space, and $3\sigma$ upper limits are 
shown with tick marks pointing toward the allowable region of parameter space. 
Contour lines are color coded to match the legend (e.g., red contours in 
Panel~``A'' signify constraints from \CIV\ measurements). If an ion is listed in 
the legend but not seen in the corresponding panel (e.g., \CII\ in Panel~``A'' for 
System~10, or \SiII\ in Panel~``B'') then the limit is not stringent enough to 
meaningfully constrain the model. The filled star symbol is located at the peaks 
of the marginal distributions for $\log{U}$ and $\log{Z_{\rm abs}}$ (i.e., 
columns~6 and 7 of Table~\ref{tab:cloudy}).

\bigskip
\figsetstart
\figsetnum{5}
\figsettitle{CLOUDY Photo-Ionization Models}

\figsetend

\floattable
\begin{deluxetable*}{clCcCCCCCCCCC}
\rotate
\tablecaption{CLOUDY Modeling Results\label{tab:cloudy}}
\tablewidth{0pt}
\tabletypesize{\scriptsize}
\tablehead{ & & & & & & & & & & \multicolumn{3}{c}{Stocke et~al. (2013)} \\ \colhead{System} & \colhead{Sight Line} & \colhead{$cz_{\rm abs}$} & \colhead{Grade} & \colhead{$\log{N_{\rm H\,I}}$} & \colhead{$\log{U}$} & \colhead{$\log{Z_{\rm abs}}$} & \colhead{$\log{\langle P/k\rangle}$} & \colhead{$\log{D_{\rm cl}}$} & \colhead{$\log{M_{\rm cl}}$} & \colhead{$\log{N_{\rm H\,I}}$} & \colhead{$\log{U}$} & \colhead{$\log{Z_{\rm abs}}$} }

\colnumbers
\startdata
 10 & HE~0439--5254    &    \phn1653 &    C    & 15.04\pm0.50 & -2.20^{+0.20}_{-0.18} & -0.17^{+0.20}_{-0.21} & -0.0^{+0.2}_{-0.3} & +0.8^{+0.6}_{-0.6} & 5.5^{+1.6}_{-1.6} & 15.21\pm0.44 & -2.4^{+0.3}_{-0.2} & +0.1^{+0.9}_{-0.4} \\
 14 & PG~0832+251      &    \phn5221 &    C    & 18.34\pm0.21 & -3.47^{+0.08}_{-0.26} & +0.12^{+0.08}_{-0.11} & +0.9^{+0.3}_{-0.1} & +1.3^{+0.2}_{-0.8} & 8.3^{+0.6}_{-2.1} & 18.48\pm0.17 & -3.5^{+0.1}_{-0.2} & -0.5^{+0.2}_{-0.2} \\
 17 & PG~0832+251      &    \phn5444 & \nodata &    \nodata   &         \nodata       &        \nodata        &      \nodata       &      \nodata       &     \nodata       & 16.39\pm0.91 & -2.4^{+0.4}_{-0.5} & -0.9^{+0.7}_{-0.5} \\
 19 & PMN~J1103--2329  &    \phn1194 &    C    & 14.76\pm0.16 & -2.20^{+0.10}_{-0.16} & +0.35^{+0.12}_{-0.12} & -0.3^{+0.3}_{-0.2} & +0.4^{+0.3}_{-0.4} & 4.1^{+0.8}_{-1.1} & 15.94\pm0.47 & -2.2^{+0.4}_{-0.5} & -0.8^{+0.5}_{-0.4} \\
 21 & RX~J0439.6--5311 &    \phn1674 &    C    & 15.53\pm0.43 & -2.36^{+0.17}_{-0.17} & -0.63^{+0.26}_{-0.25} & +0.2^{+0.2}_{-0.2} & +1.1^{+0.6}_{-0.6} & 6.4^{+1.5}_{-1.6} & 15.41\pm0.42 & -2.6^{+0.4}_{-0.2} & -0.3^{+0.6}_{-0.5} \\
 23 & SBS~1108+560     & \phn\phn654 & \nodata &    \nodata   &         \nodata       &        \nodata        &      \nodata       &      \nodata       &     \nodata       & 17.38\pm0.63 & -3.1^{+0.4}_{-0.4} &  0.0^{+1.0}_{-0.5} \\
 25 & SBS~1108+560     & \phn\phn778 & \nodata &    \nodata   &         \nodata       &        \nodata        &      \nodata       &      \nodata       &     \nodata       & 15.44\pm0.42 & -2.3^{+0.3}_{-0.3} & -0.5^{+0.3}_{-0.3} \\
 26 & SBS~1122+594     &    \phn1221 &    B    & 15.94\pm0.47 & -2.30^{+0.15}_{-0.15} & -0.29^{+0.14}_{-0.21} & +0.1^{+0.2}_{-0.2} & +1.6^{+0.5}_{-0.5} & 7.8^{+1.2}_{-1.4} & 15.92\pm0.42 & -2.5^{+0.4}_{-0.4} & -0.2^{+0.3}_{-0.3} \\
 27 & VII~Zw~244       & \phn\phn715 &    B    & 15.22\pm0.28 & -2.47^{+0.10}_{-0.10} & +0.19^{+0.16}_{-0.14} & +0.0^{+0.2}_{-0.2} & +0.3^{+0.3}_{-0.3} & 4.2^{+0.7}_{-0.8} & 15.81\pm0.26 & -2.8^{+0.1}_{-0.2} & -0.2^{+0.1}_{-0.2} \\
102 & 3C~273           &    \phn1585 &    A    & 15.49\pm0.05 & -3.22^{+0.05}_{-0.02} & -0.48^{+0.13}_{-0.09} & +0.9^{+0.1}_{-0.1} & -0.8^{+0.2}_{-0.1} & 1.5^{+0.5}_{-0.3} & 15.85\pm0.09 & -3.2^{+0.2}_{-0.1} & -0.9^{+0.2}_{-0.2} \\
115 & PG~0953+414      &       42664 &    B    & 13.53\pm0.02 & -1.48^{+0.12}_{-0.12} & +0.01^{+0.09}_{-0.08} & -0.7^{+0.1}_{-0.1} & +0.9^{+0.5}_{-0.4} & 4.9^{+1.2}_{-1.0} &    \nodata   &       \nodata      &       \nodata      \\
123 & PG~1116+215      &       41522 &    A    & 15.95\pm0.03 & -3.09^{+0.04}_{-0.04} & -0.19^{+0.07}_{-0.08} & +0.7^{+0.1}_{-0.1} & -0.2^{+0.2}_{-0.1} & 3.4^{+0.4}_{-0.3} & 16.35\pm0.10 & -3.3^{+0.1}_{-0.1} & -0.3^{+0.1}_{-0.2} \\
126 & PG~1211+143      &       15321 &    A    & 15.54\pm0.06 & -2.74^{+0.13}_{-0.14} & -0.24^{+0.08}_{-0.07} & +0.4^{+0.2}_{-0.3} & +0.2^{+0.6}_{-0.4} & 4.1^{+1.4}_{-1.1} & 15.67\pm0.35 & -2.9^{+0.5}_{-0.3} & -0.5^{+0.3}_{-0.4} \\
130 & PG~1211+143      &       19305 &    B    & 15.31\pm0.04 & -2.20^{+0.05}_{-0.06} & -0.93^{+0.08}_{-0.09} & +0.1^{+0.1}_{-0.1} & +1.2^{+0.2}_{-0.2} & 6.7^{+0.4}_{-0.5} & 15.17\pm0.10 & -2.4^{+0.1}_{-0.2} & -0.9^{+0.1}_{-0.1} \\
132 & PG~1211+143      &       19481 &    D    & 13.29\pm0.17 & -1.24^{+0.25}_{-0.26} & +0.58^{+0.21}_{-0.18} & -0.9^{+0.2}_{-0.2} & +1.1^{+0.9}_{-0.9} & 5.3^{+2.3}_{-2.4} & 13.82\pm0.05 & -2.1^{+0.1}_{-0.1} & -0.2^{+0.2}_{-0.1} \\
137 & PG~1216+069      &       37049 &    B    & 14.57\pm0.05 & -2.02^{+0.05}_{-0.06} & +0.39^{+0.09}_{-0.07} & -0.5^{+0.1}_{-0.2} & +0.5^{+0.1}_{-0.2} & 4.5^{+0.3}_{-0.5} &    \nodata   &       \nodata      &       \nodata      \\
138 & PG~1216+069      &       37138 &    B    & 14.76\pm0.05 & -2.11^{+0.06}_{-0.06} & -0.06^{+0.06}_{-0.07} & -0.2^{+0.1}_{-0.1} & +0.7^{+0.2}_{-0.2} & 5.1^{+0.5}_{-0.5} &    \nodata   &       \nodata      &       \nodata      \\
142 & PG~1259+593      &       13825 &    A    & 15.45\pm0.04 & -2.36^{+0.06}_{-0.06} & -0.61^{+0.06}_{-0.06} & +0.2^{+0.1}_{-0.1} & +1.0^{+0.2}_{-0.2} & 6.1^{+0.5}_{-0.5} & 15.51\pm0.28 & -2.2^{+0.3}_{-0.9} & -1.1^{+0.9}_{-0.3} \\
143 & PG~1259+593      &       13914 &    A    & 14.60\pm0.05 & -1.78^{+0.10}_{-0.12} & -0.46^{+0.09}_{-0.08} & -0.3^{+0.1}_{-0.1} & +1.4^{+0.3}_{-0.4} & 6.6^{+0.8}_{-1.1} & 14.75\pm0.38 & -1.7^{+0.3}_{-1.3} & -0.6^{+0.8}_{-0.5} \\
148 & PHL~1811         &       22042 & \nodata &    \nodata   &         \nodata       &        \nodata        &     \nodata        &      \nodata       &      \nodata      & 14.88\pm0.09 & -2.7^{+0.3}_{-0.2} & -0.3^{+0.2}_{-0.3} \\
149 & PHL~1811         &       23313 &    A    & 15.40\pm0.07 & -3.05^{+0.03}_{-0.02} & -0.15^{+0.06}_{-0.08} & +0.6^{+0.1}_{-0.1} & -0.6^{+0.1}_{-0.1} & 1.9^{+0.4}_{-0.3} & 14.94\pm0.08 & -2.7^{+0.2}_{-0.2} & -0.2^{+0.2}_{-0.2} \\
152 & PHL~1811         &       24226 &    C    & 18.08\pm0.04 & -3.65^{+0.08}_{-0.18} & -0.25^{+0.07}_{-0.07} & +1.2^{+0.3}_{-0.1} & +0.8^{+0.2}_{-0.7} & 6.9^{+0.6}_{-1.8} & 18.00\pm0.50 & -3.5^{+0.3}_{-0.9} & -0.7^{+0.8}_{-1.4} \\
153 & PHL~1811         &       39658 &    D    & 14.61\pm0.01 & -1.37^{+0.19}_{-0.20} & -0.99^{+0.12}_{-0.11} & -0.6^{+0.2}_{-0.2} & +2.3^{+0.6}_{-0.7} & 9.1^{+1.6}_{-1.7} &    \nodata   &       \nodata      &       \nodata      \\
157 & PHL~1811         &       52933 &    B    & 14.84\pm0.04 & -2.37^{+0.12}_{-0.12} & -0.57^{+0.09}_{-0.08} & +0.2^{+0.1}_{-0.1} & +0.4^{+0.4}_{-0.4} & 4.2^{+0.9}_{-0.9} & 14.87\pm0.03 & -2.6^{+0.5}_{-0.5} & -0.5^{+0.4}_{-0.5} \\
170 & PKS~0405--123    &       50059 &    B    & 15.41\pm0.06 & -2.92^{+0.05}_{-0.06} & +0.30^{+0.06}_{-0.07} & +0.2^{+0.1}_{-0.1} & -0.5^{+0.2}_{-0.2} & 2.1^{+0.5}_{-0.6} &    \nodata   &       \nodata      &       \nodata      \\
171 & PKS~0405--123    &       50104 &    B    & 16.45\pm0.02 & -3.16^{+0.03}_{-0.03} & -0.24^{+0.08}_{-0.07} & +0.7^{+0.1}_{-0.1} & +0.2^{+0.1}_{-0.1} & 4.6^{+0.4}_{-0.3} & 16.45\pm0.07 & -3.0^{+0.1}_{-0.1} & +0.1^{+0.2}_{-0.2} \\
175 & PKS~1302--102    &       12655 &    A    & 14.91\pm0.08 & -2.82^{+0.06}_{-0.04} & +0.03^{+0.06}_{-0.06} & +0.4^{+0.1}_{-0.1} & -0.7^{+0.2}_{-0.1} & 1.5^{+0.6}_{-0.4} & 14.83\pm0.17 & -2.8^{+0.1}_{-0.1} & +0.2^{+0.2}_{-0.2} \\
178 & PKS~1302--102    &       28439 &    A    & 17.11\pm0.03 & -3.30^{+0.04}_{-0.04} & -1.26^{+0.14}_{-0.15} & +1.0^{+0.1}_{-0.1} & +0.7^{+0.2}_{-0.1} & 6.1^{+0.4}_{-0.4} & 17.10\pm0.40 & -3.1^{+0.5}_{-0.3} & -1.7^{+0.6}_{-0.4} \\
187 & Q~1230+0115      &       23404 &    C    & 14.77\pm0.07 & -2.16^{+0.11}_{-0.12} & -0.13^{+0.11}_{-0.10} & -0.1^{+0.2}_{-0.2} & +0.6^{+0.3}_{-0.4} & 4.9^{+0.8}_{-1.0} & 15.06\pm0.40 & -2.2^{+0.4}_{-0.7} & -0.2^{+0.4}_{-0.4} \\ [1mm]
\enddata

\end{deluxetable*}

\clearpage

\begin{figure}[!t]
\plotone{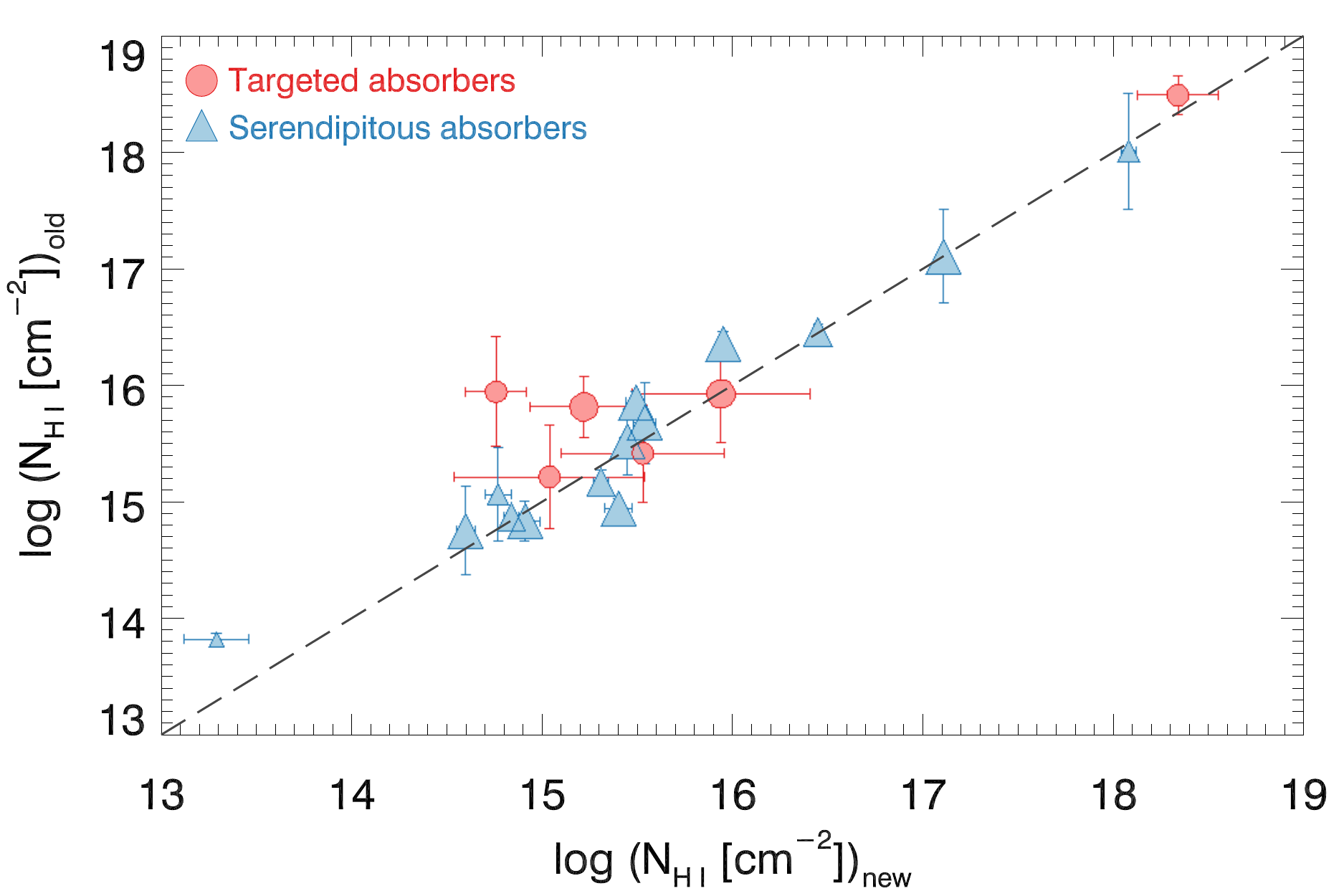}
\plotone{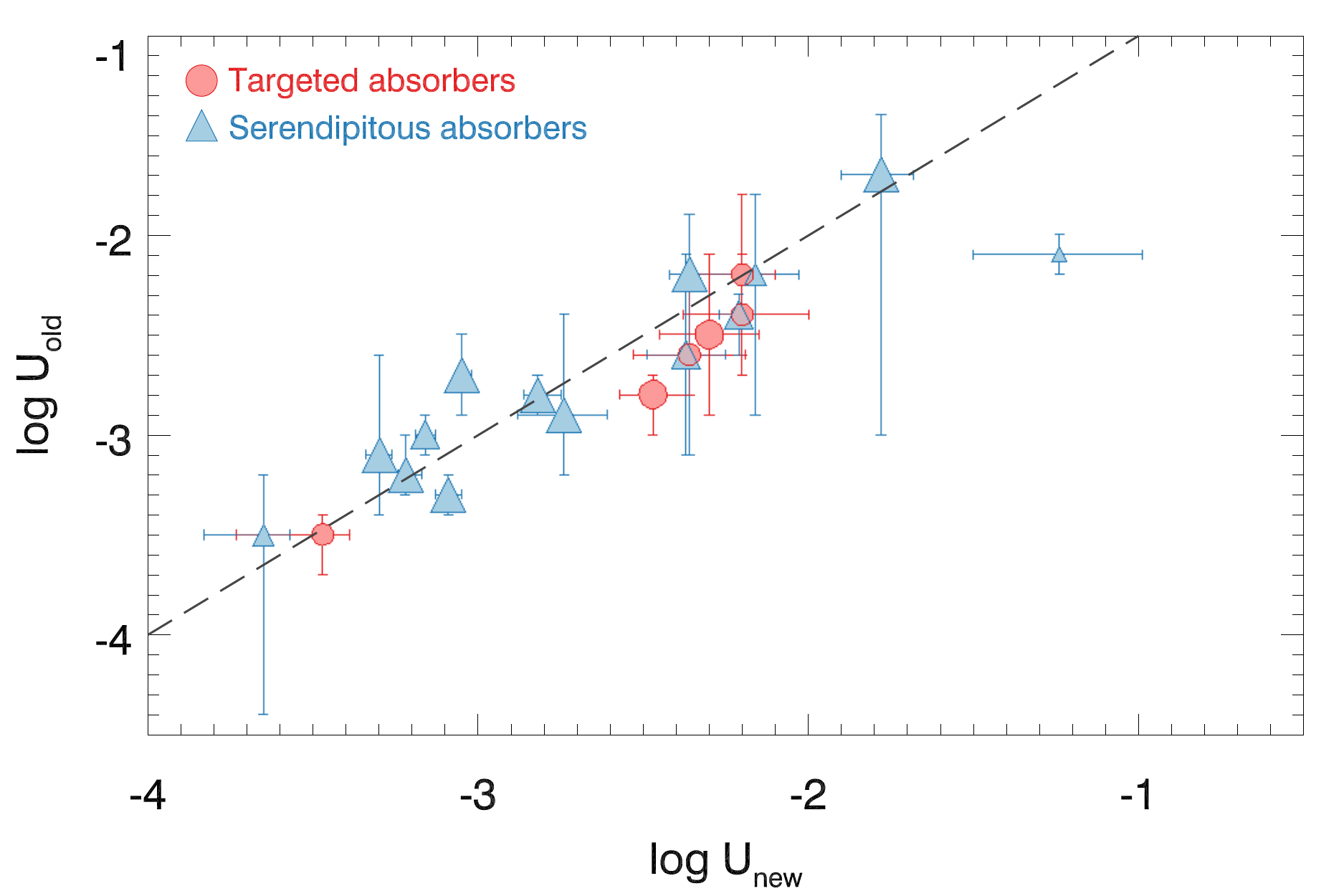}
\plotone{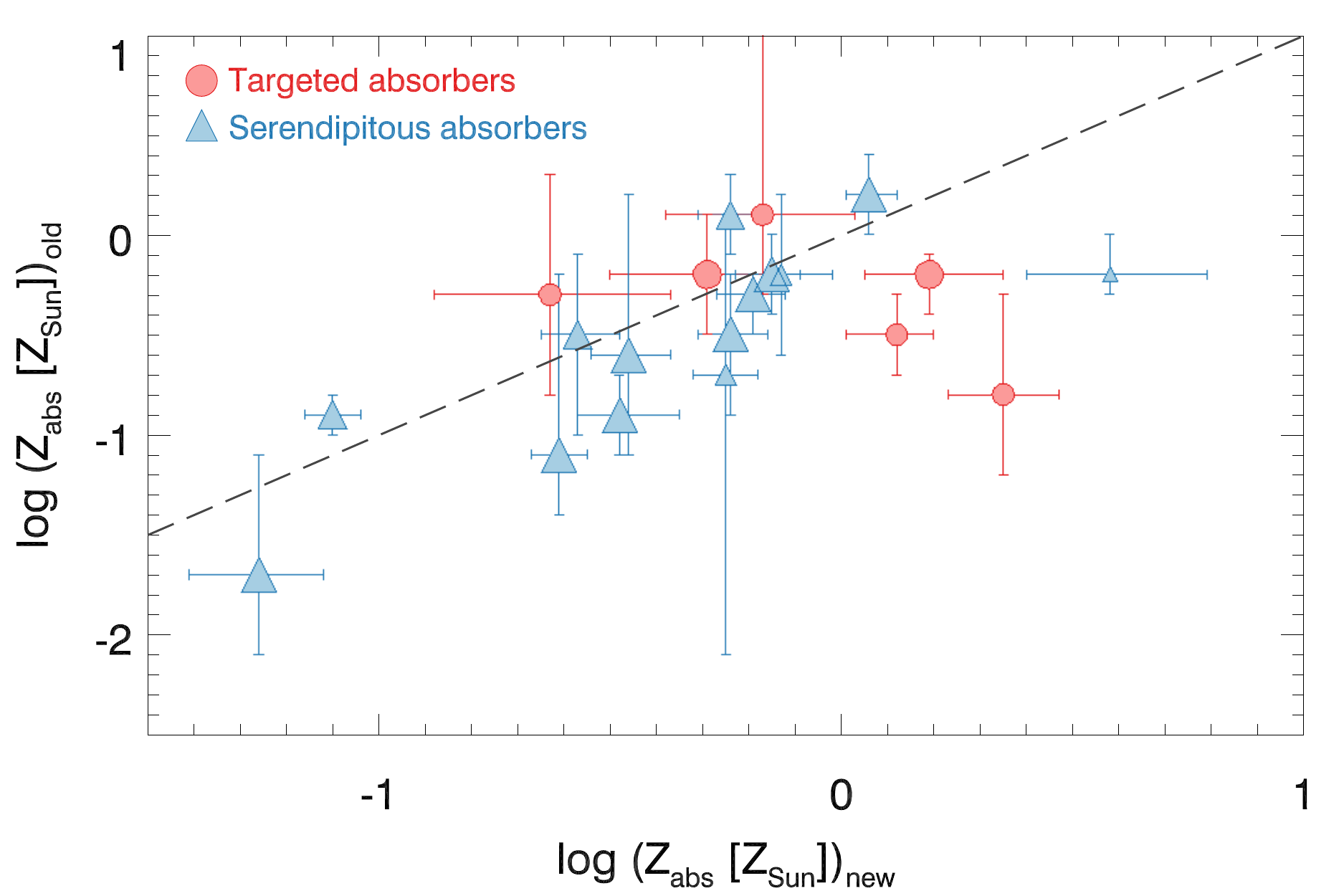}
\caption{\textit{Top:} Comparison of updated values of $\log{N_{\rm H\,I}}$ (``new''; see column~5 of Table~\ref{tab:cloudy}) with the values adopted in Paper~1 (``old''; see column~11 of Table~\ref{tab:cloudy}). \textit{Middle:} Comparison of updated values of $\log{U}$ with the values adopted in Paper~1. \textit{Bottom:} Comparison of updated values of $\log{Z_{\rm abs}}$ with the values adopted in Paper~1. In all panels, the dashed line indicates perfect agreement between the two values. Symbol size is indicative of our confidence in the photo-ionization models (i.e., our absorber grades; see Section~\ref{ensemble}) with higher-confidence absorbers having larger plot symbols. The \HI\ column densities and ionization parameters typically agree well, but our updated absorber metallicities are somewhat higher than the values found in Paper~1.
\label{fig:oldnew}}
\end{figure}

\begin{figure*}
\gridline{\fig{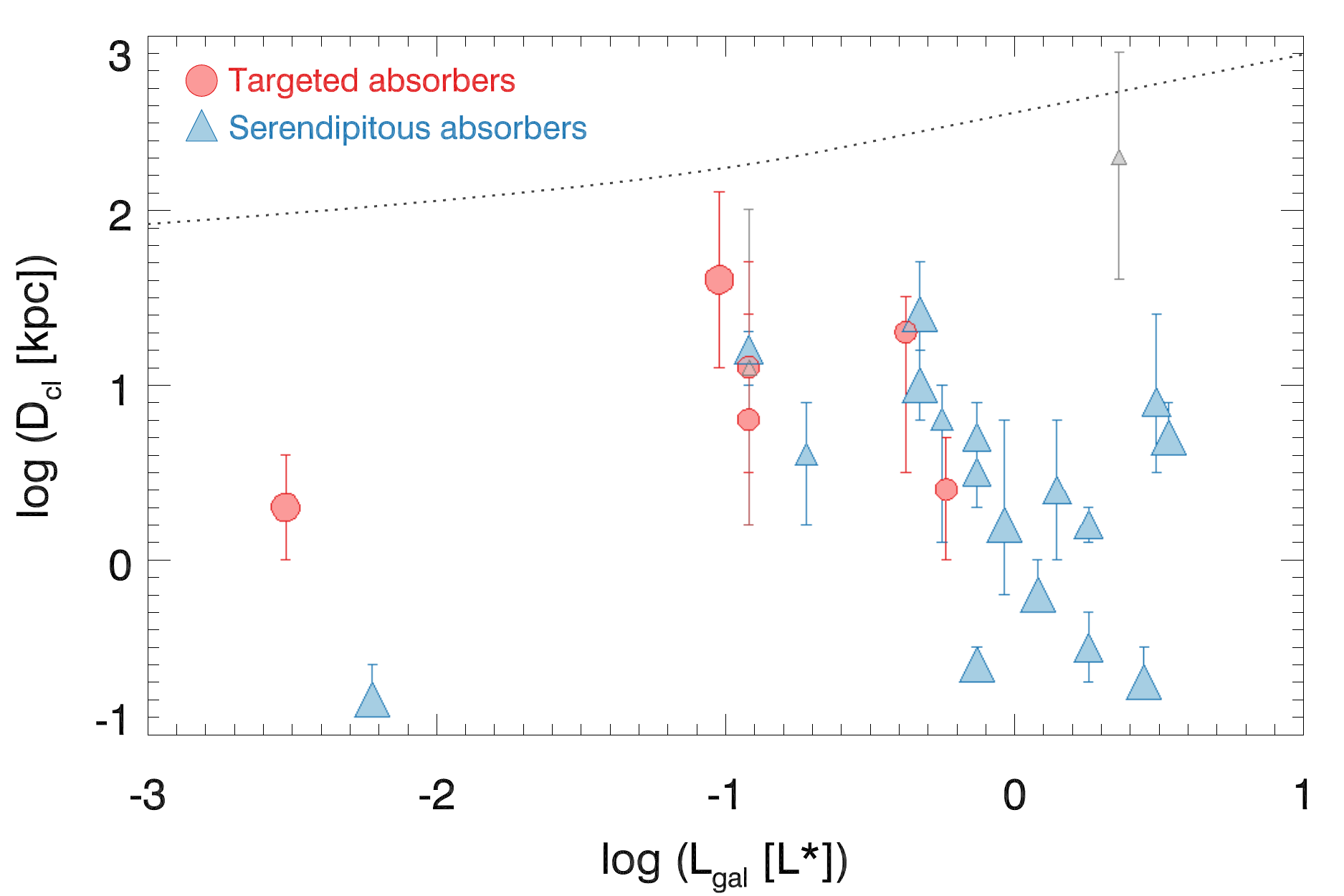}{0.5\textwidth}{(a)}
          \fig{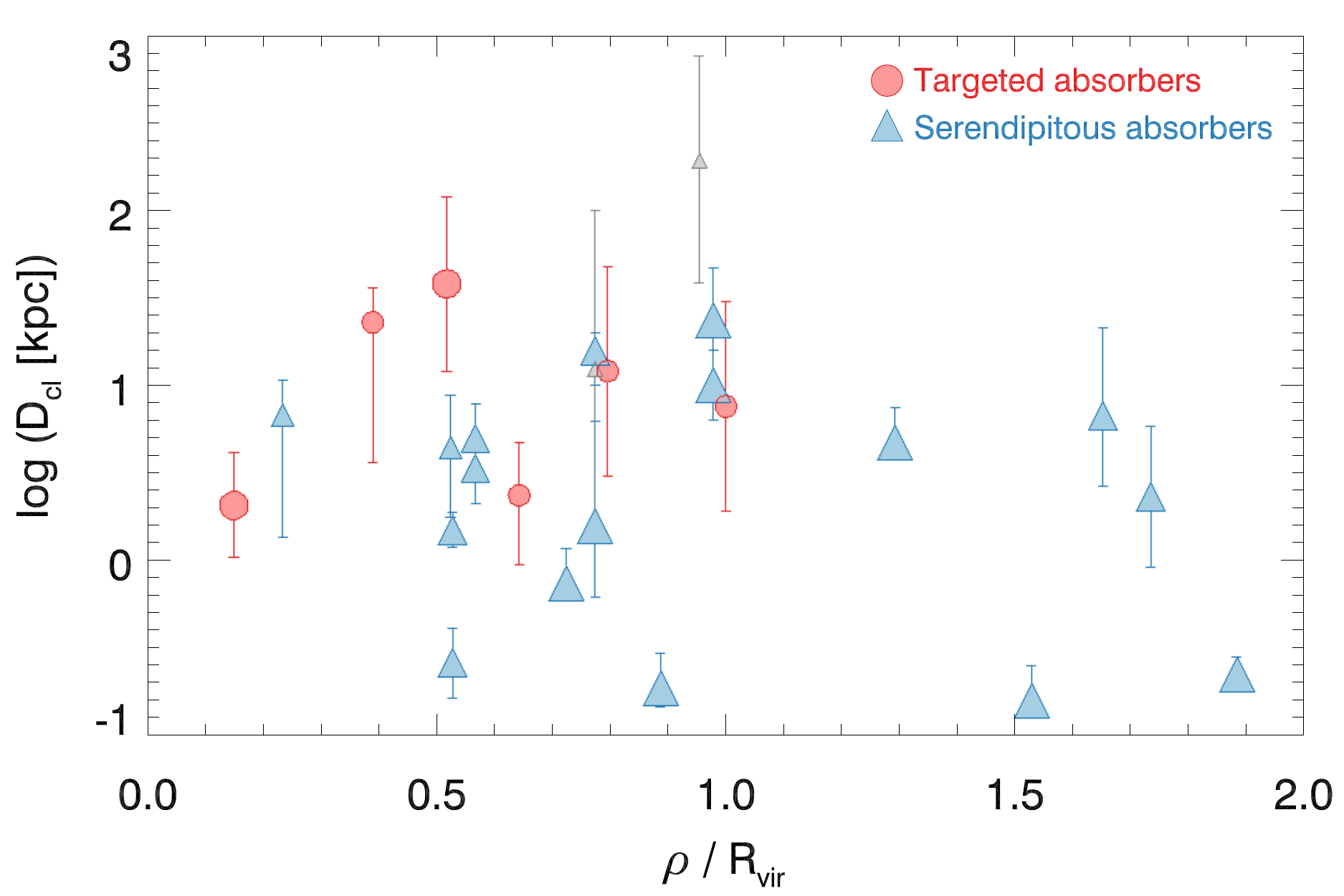}{0.5\textwidth}{(d)}}
\gridline{\fig{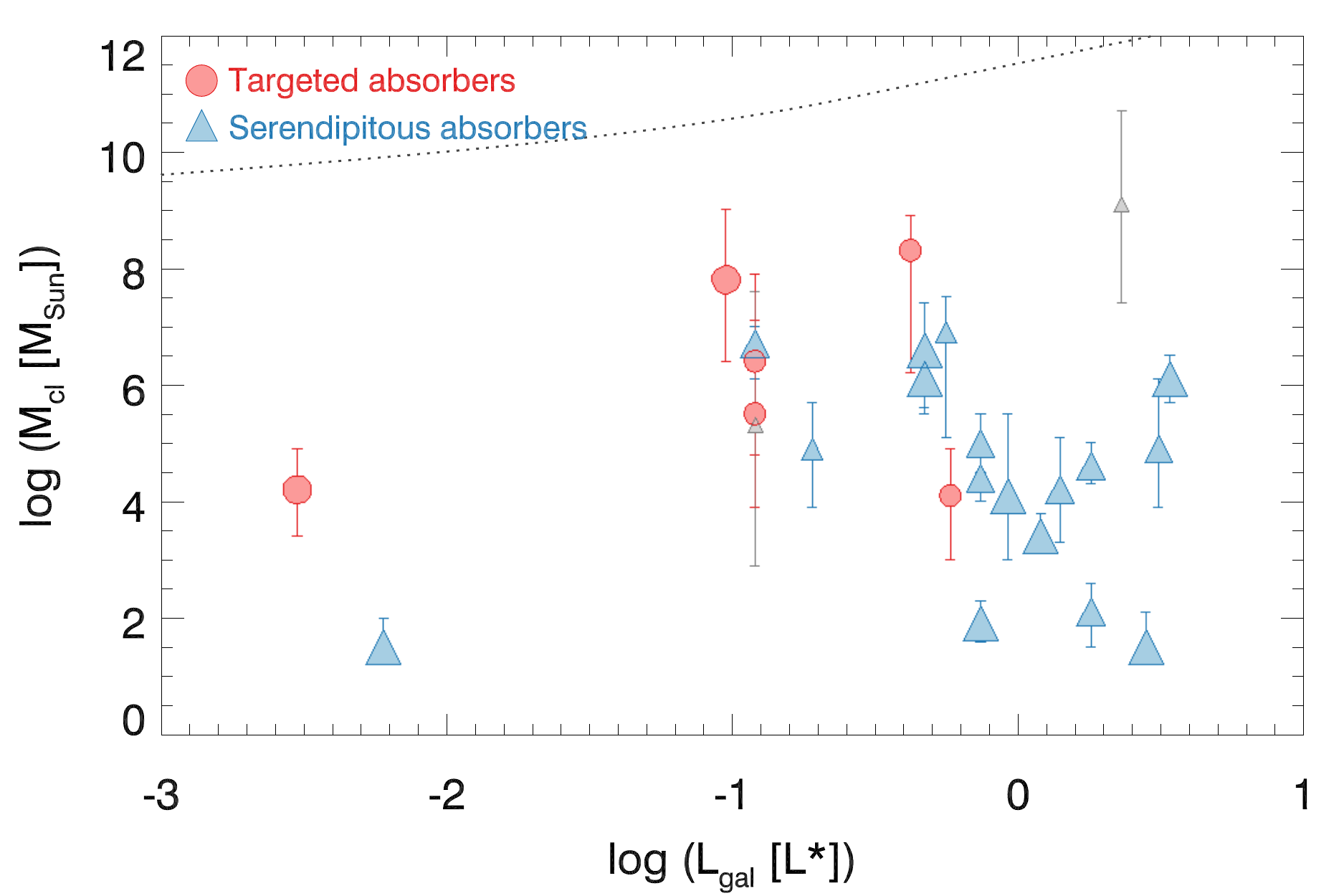}{0.5\textwidth}{(b)}
          \fig{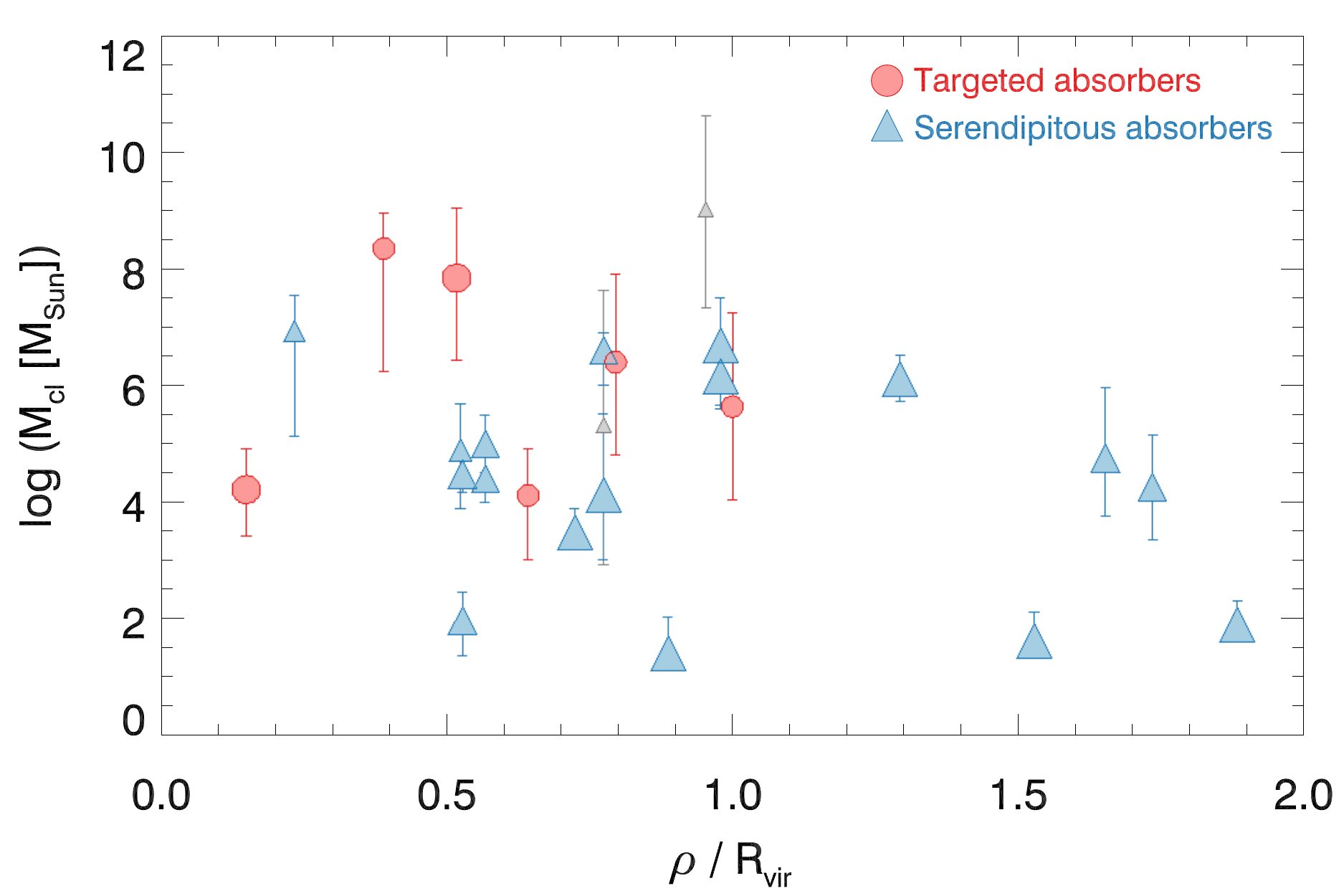}{0.5\textwidth}{(e)}}
\gridline{\fig{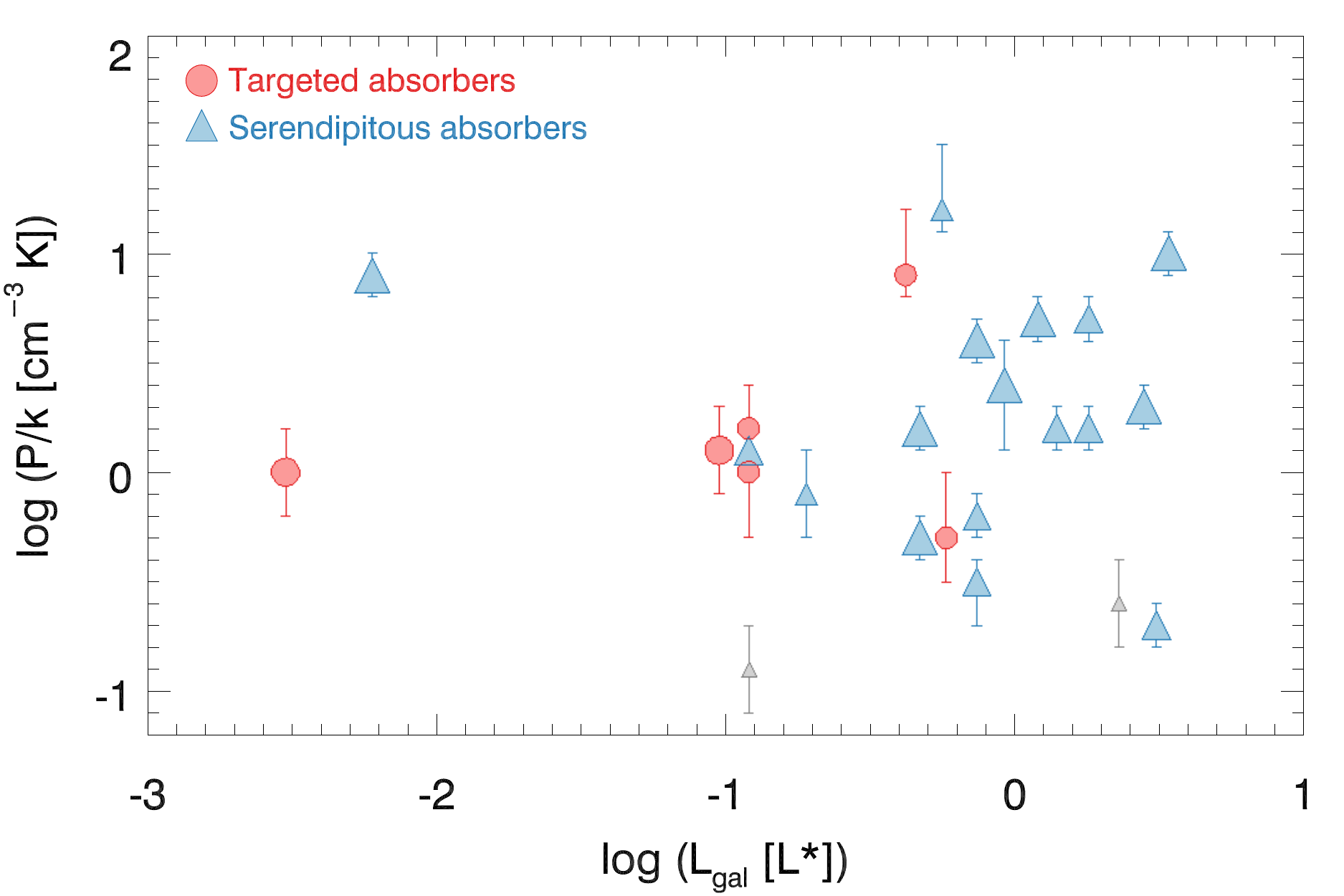}{0.5\textwidth}{(c)}
          \fig{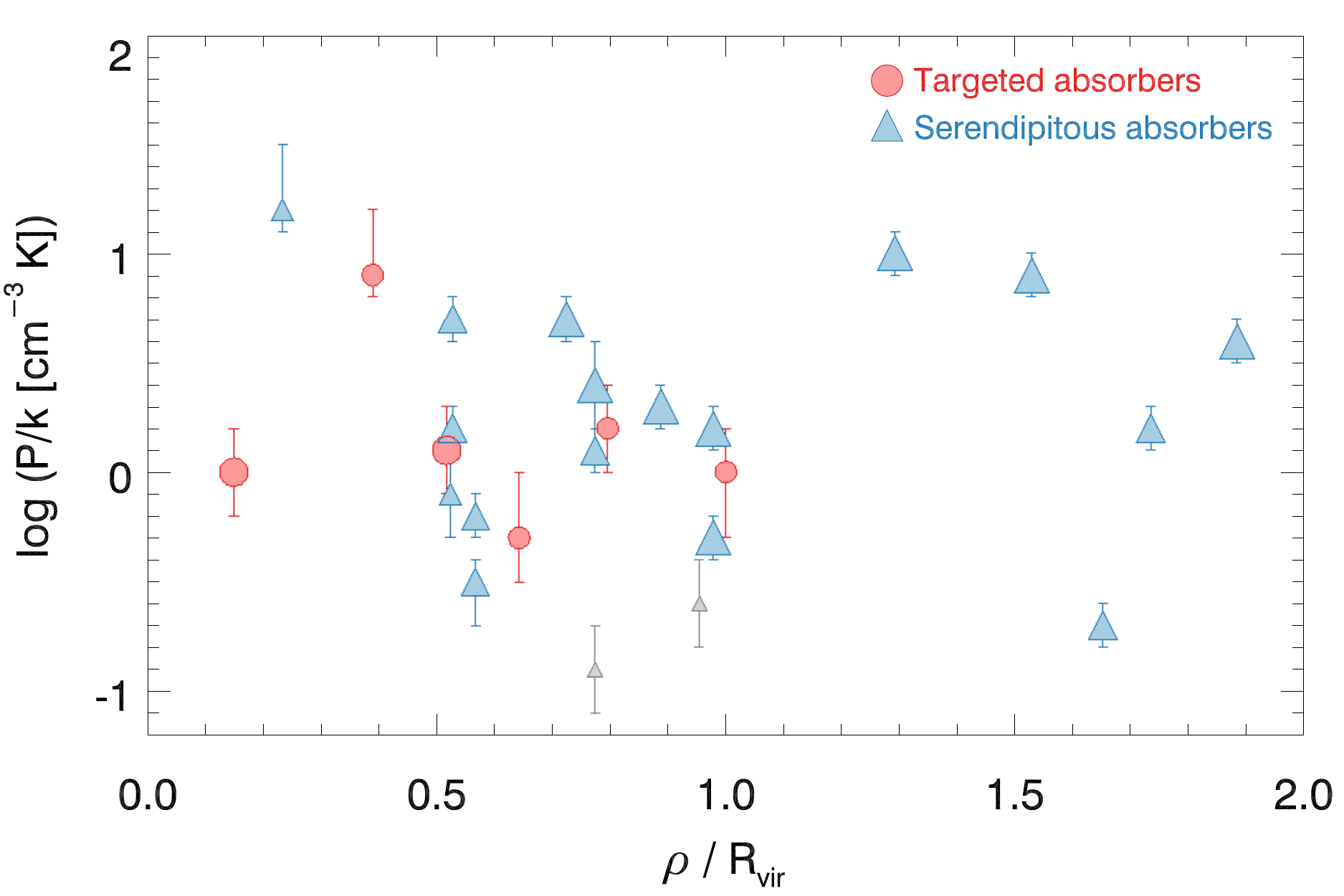}{0.5\textwidth}{(f)}}
\caption{Distributions of $D_{\rm cl}$, $M_{\rm cl}$ and $\langle P/k \rangle$ as a function of nearest galaxy luminosity (Panels~a-c) and normalized impact parameter (Panels~d-f). As in Figure~\ref{fig:oldnew}, symbol size is indicative of our confidence in the photo-ionization models (i.e., our absorber grades) with higher-confidence absorbers having larger plot symbols; the two grade~D absorbers are colored gray. No obvious trends are present in any of the panels, in agreement with Figures~12-14 of Paper~1. The dotted lines in Panels~a-b shows the galaxy's virial diameter and mass, respectively, as a function of luminosity.
\label{fig:comp}}
\end{figure*}

\begin{figure}[!t]
\plotone{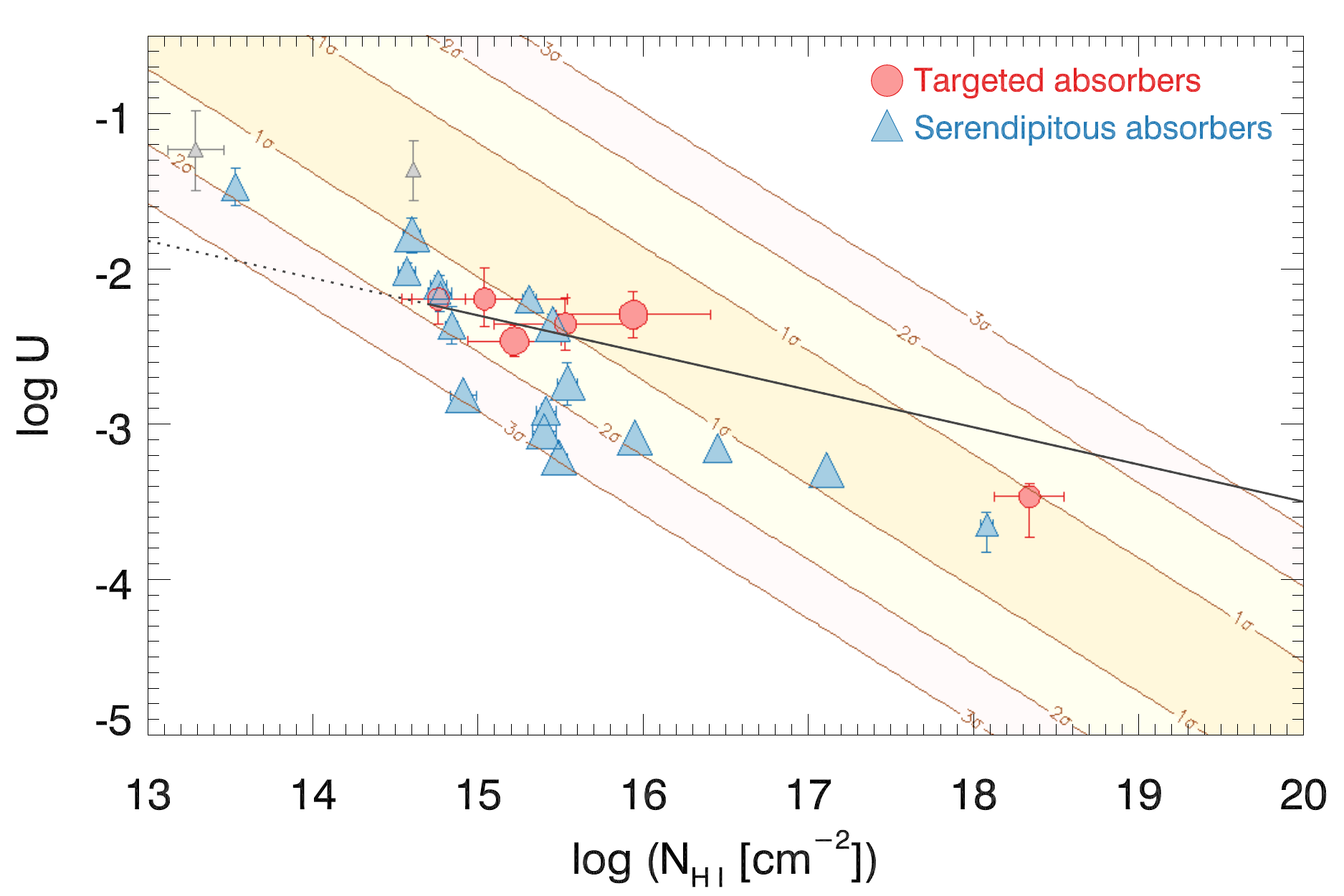}
\plotone{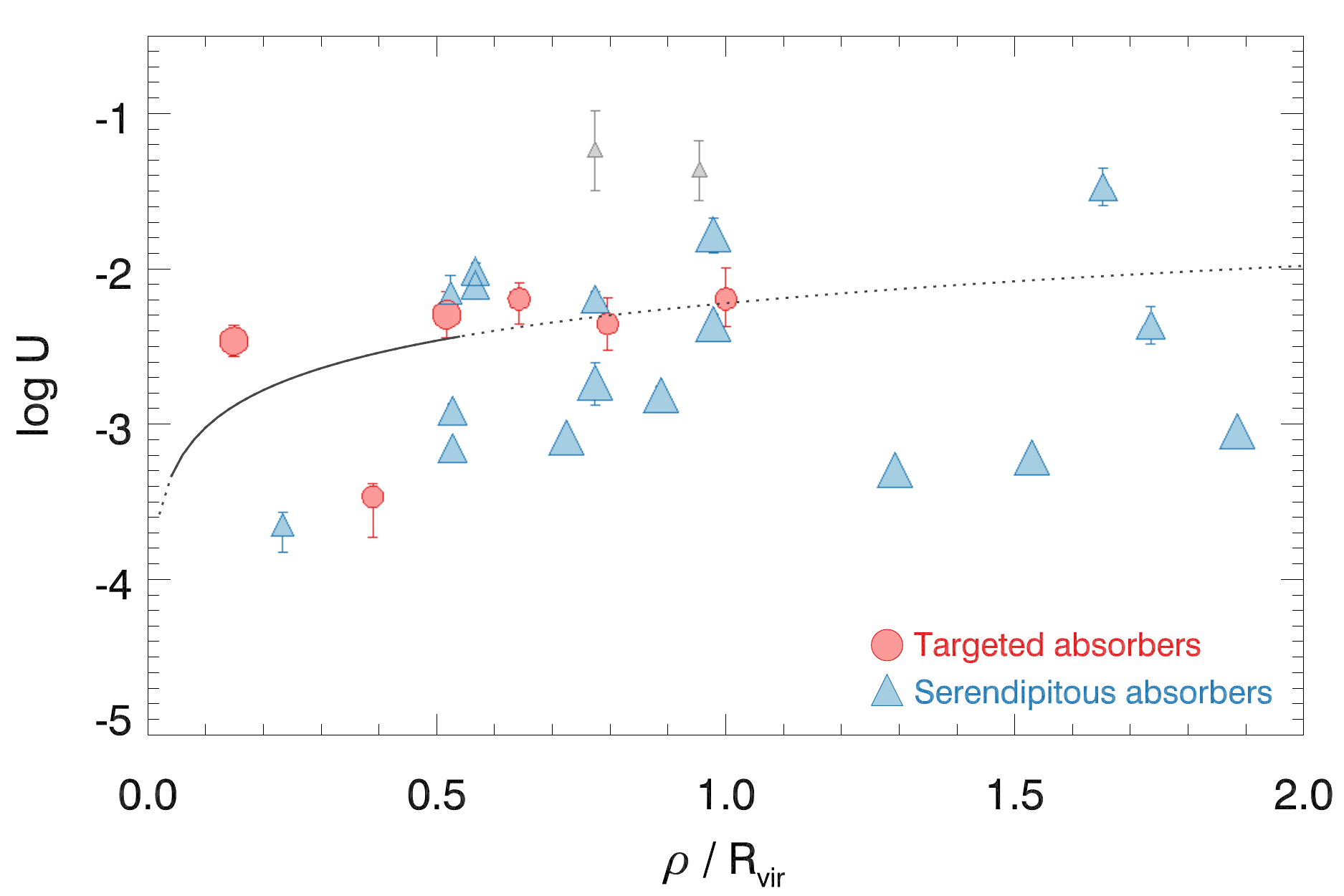}
\plotone{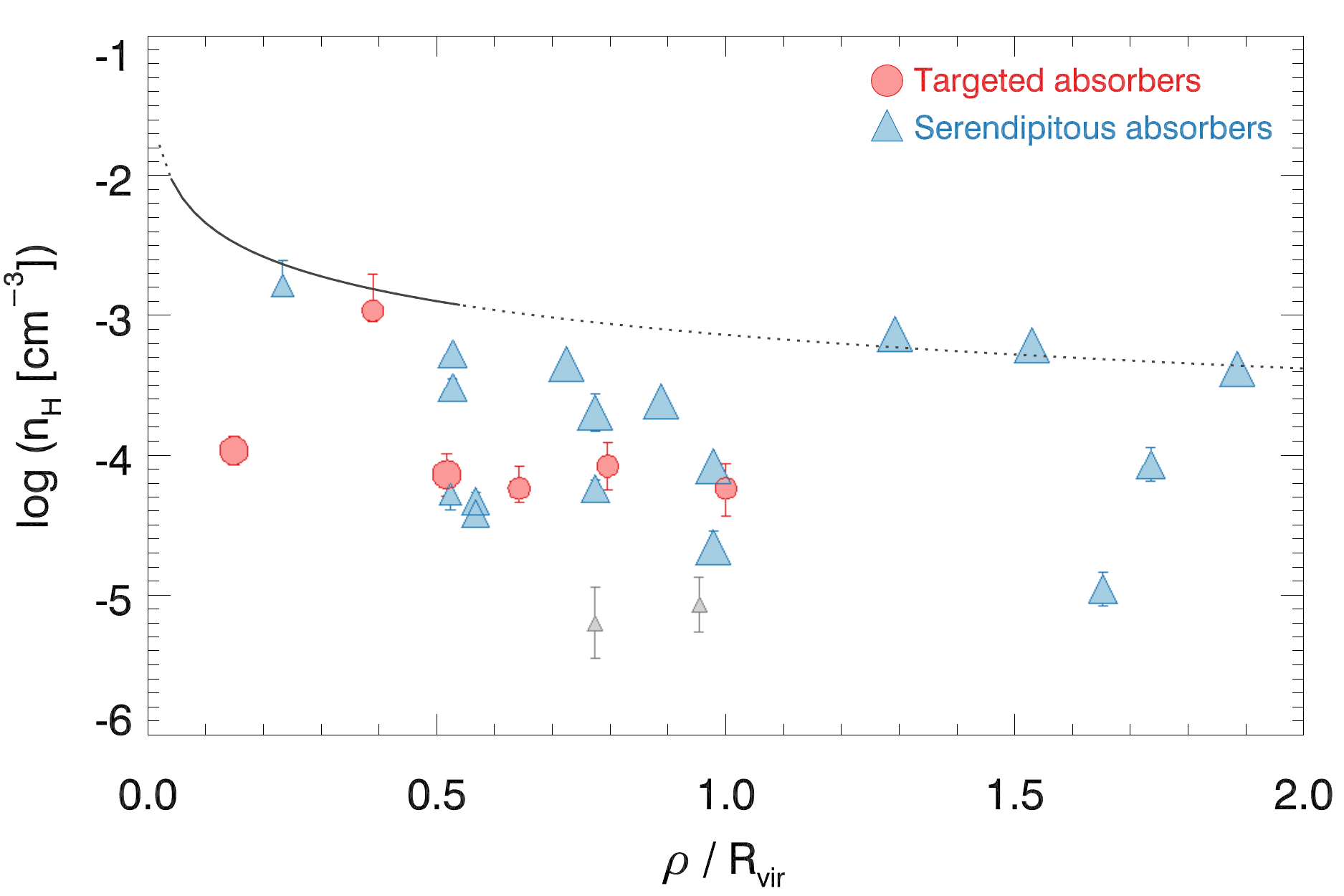}
\plotone{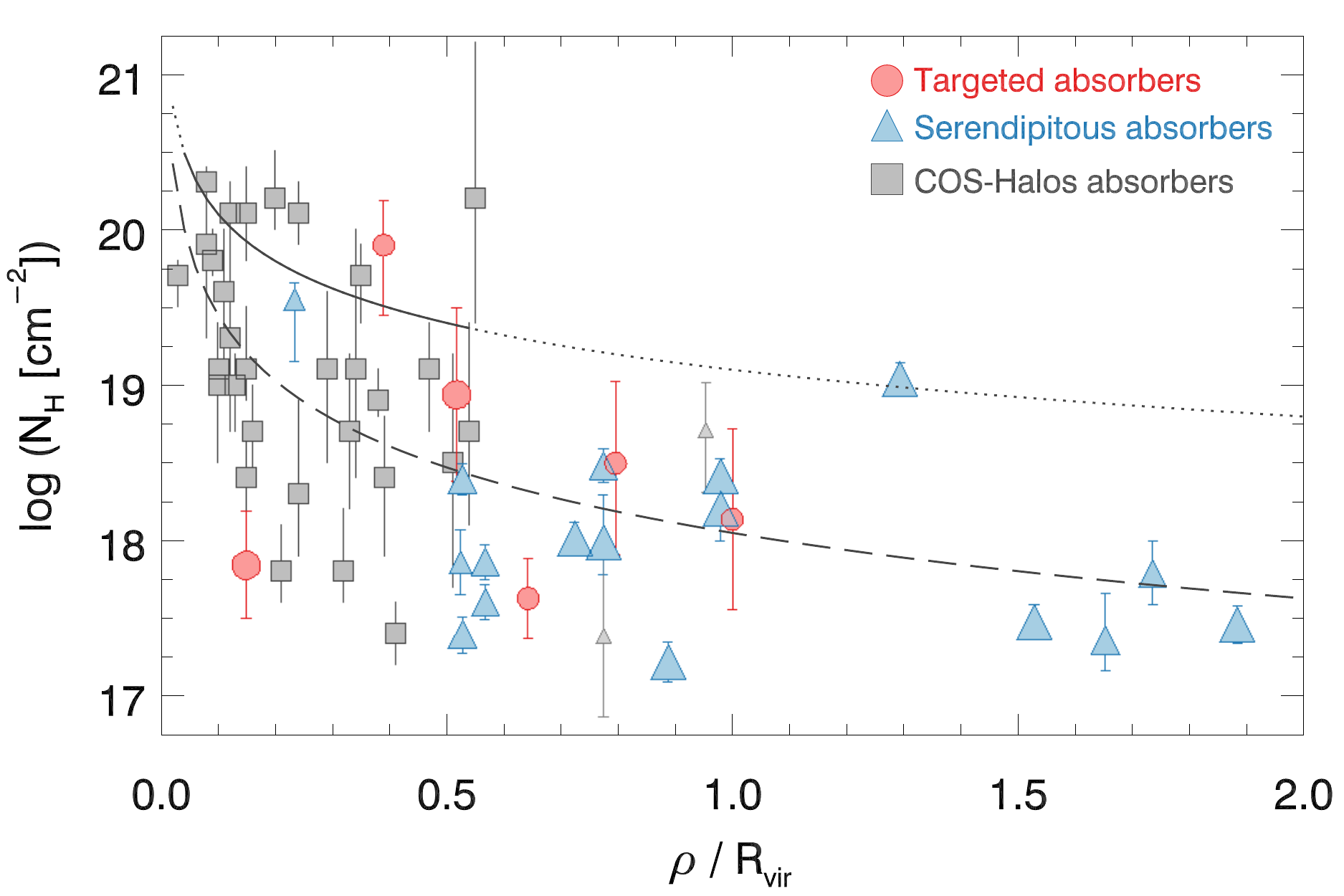}
\caption{Distributions of model parameters as a function of $N_{\rm H\,I}$ and normalized impact parameter for our updated photo-ionization models. These models assume a Bayesian prior for ionization parameter as a function of $N_{\rm H\,I}$, which is shown as 1$\sigma$, 2$\sigma$ and $3\sigma$ contours in the top panel. Each panel has a solid line with a dotted extrapolation that shows the relationship found by \citet{werk14} for COS-Halos absorbers. As in Figure~\ref{fig:oldnew}, symbol size is indicative of our confidence in the photo-ionization models, with higher-confidence absorbers having larger plot symbols. The dashed line in the bottom panel is our power-law fit to the combined targeted, serendipitous, and COS-Halos data, and is described in Section~\ref{cloudy:coshalos}.
\label{fig:werk}}
\end{figure}

In the subsections below we compare the results in Table~\ref{tab:cloudy} with 
those presented in Paper~1 (Section~\ref{cloudy:comp}) and with the 
photo-ionization models of the COS-Halos absorbers 
\citep[Section~\ref{cloudy:coshalos}]{werk14}. Generally speaking, our photo-
ionization models are robust to our choice of Bayesian prior (i.e., running the 
models with and without a prior yield very similar results); individual models 
where this conclusion does not hold are detailed in Section~\ref{indiv:cloudy} of 
the Appendix.

\subsection{Comparison with Photo-Ionization Results of Paper~1}
\label{cloudy:comp}

We find that the values of $N_{\rm H\,I}$, $U$, and $Z_{\rm abs}$ that we 
derive here are generally in good agreement with the values published in 
Paper~1, except that we tend to find somewhat higher absorber metallicities in 
our updated analysis. Plots of the ``old'' values from Paper~1 versus the 
``new'' values from our updated analysis can be found in Figure~\ref{fig:oldnew}.
There are some exceptions to the general agreement in $N_{\rm H\,I}$ and $U$ 
between the two analyses, however, which we detail in the Appendix, Section~\ref{indiv:cloudy},
where appropriate. We also find no clear trends in absorber line-of-sight 
thickness, mass or mean pressure as a function of nearest galaxy luminosity or 
$\rho/R_{\rm vir}$, again consistent with the results of Paper~1. 
Updated plots of these distributions are shown in Figure~\ref{fig:comp}. 
These conclusions hold even if we restrict our analysis to the subset of 
absorbers for which we have the highest confidence in photo-ionization models 
(i.e., grade~A absorbers).

\subsection{Comparison with COS-Halos Photo-Ionization Models}
\label{cloudy:coshalos}

\citet{werk14} presented photo-ionization models of 44 COS-Halos absorbers, 
finding that the circumgalactic gas traced by these absorbers tends to become 
more highly ionized and less dense with increasing distance from the nearest 
galaxy. We compare the results of our updated photo-ionization models to the 
best-fit relations of \citet{werk14} in Figure~\ref{fig:werk}, where the top 
panel shows ionization parameter as a function of $N_{\rm H\,I}$, the middle 
panel shows ionization parameter as a function of normalized impact parameter, 
and the bottom panels show total hydrogen volume and column density, 
respectively, as a function of normalized impact parameter. Our models assume 
a Bayesian prior for $U(N_{\rm H\,I})$ (see Equation~\ref{eqn:logU}), which is 
shown with the filled contours in the top panel. 

\edit1{ All panels have a solid line with a dotted extrapolation that shows the 
best-fit relation from \citet{werk14}. In the top panel, the slope of the 
\citet{werk14} relation differs from the slope assumed by our Bayesian prior 
(see Section~\ref{cloudy:UVB}), but most of our values are in reasonable agreement 
with those of \citet{werk14}. The \citet{werk14} parameterization of ionization 
parameter as a function of normalized impact parameter also provides a reasonable 
description of our data both inside and outside of the region probed by COS-Halos 
($\rho \lesssim 0.5\,R_{\rm vir}$), albeit with large scatter. The \citet{werk14}
fits to the total hydrogen volume and column densities, however, lay well above 
the bulk of our data points; we attribute this difference to the different assumed 
UVB models, a point we will return to in Section~\ref{ensemble:coshalos}.
}

Thus, we conclude that the basic results of our photo-ionization models and those 
of \citet{werk14} are in good agreement. Where we differ is in their 
interpretation. While \citet{werk14} find increasing ionization state and 
decreasing hydrogen density as a function of normalized impact parameter at 
$2\sigma$ significance, our photo-ionization results (Figure~\ref{fig:werk}) are 
consistent with constant values of ionization parameter and total hydrogen density 
(i.e., no change with distance from the nearest galaxy). These slightly differing 
results may be due to the different normalized impact parameter ranges (i.e., 
$\rho/R_{\rm vir} < 0.5$ for \citealt{werk14} and 
$\rho/R_{\rm vir} \approx 0.5$-2.0 for the present sample).

\edit1{
This hypothesis is supported by Figure~\ref{fig:N_HI}, which shows the 
distribution of total CGM $N_{\rm H\,I}$ (i.e., the sum over all velocity 
components associated with a particular galaxy) as a function of normalized 
impact parameter. We have over-plotted the COS-Halos measurements from 
Table~1 of \citet{werk14} for direct comparison with our values. The gray 
squares show the ``adopted'' values of $N_{\rm H\,I}$ and the vertical gray 
lines indicate the full range of values allowed by the data. There is excellent 
agreement in the small range of normalized impact parameter where our sample and 
COS-Halos overlap, and the addition of the COS-Halos values makes the trend of 
increasing $N_{\rm H\,I}$ with decreasing $\rho/R_{\rm vir}$ evident. A 
least-squares fit to the combined targeted, serendipitous, and COS-Halos data 
finds
\begin{equation}
\label{eqn:HIrho}
N_{\rm H\,I} = 10^{14.8\pm0.1}\,\left(\frac{\rho}{R_{\rm vir}}\right)^{-2.7\pm0.3} {\rm cm}^{-2}.
\end{equation}
This parameterization is shown as the dashed line in Figure~\ref{fig:N_HI}. A more 
sophisticated analysis that allows the COS-Halos values of $N_{\rm H\,I}$ to vary 
uniformly over the entire allowable range finds nearly identical results.
}

\edit1{
A similar result is found in the bottom panel of Figure~\ref{fig:werk}, which 
shows total hydrogen column, $N_{\rm H}$, as a function of normalized impact 
parameter for our sample and COS-Halos. Again the ``adopted'' values of 
\citet{werk14} are shown with gray squares, and the full range of $N_{\rm H}$ 
by vertical gray lines. While the correlation is not as strong as in 
Figure~\ref{fig:N_HI}, the bottom panel of Figure~\ref{fig:werk} nevertheless 
suggests that our data and the COS-Halos data can be described by a single power 
law. A least-squares fit to the combined data finds
\begin{equation}
\label{eqn:NHrho}
N_{\rm H} = 10^{18.0\pm0.3}\,\left(\frac{\rho}{R_{\rm vir}}\right)^{-1.4\pm0.3} {\rm cm}^{-2}.
\end{equation}
This parameterization is shown as the dashed line in the bottom panel of 
Figure~\ref{fig:werk}. A more sophisticated analysis that allows the COS-Halos 
$N_{\rm H}$ values to vary over their full allowed range again finds very similar 
results. The power-law slope of this fit is consistent within errors 
with the parameterization of \citet{werk14} for COS-Halos data, but the 
normalization is an order of magnitude lower, a $\sim2\sigma$ discrepancy.
\citet{werk14} justify their use of this high normalization to account for 
saturated \HI\ lines, for which they have only lower limits on $N_{\rm H\,I}$; 
a very recent paper by \citet{prochaska17} presents new observations providing 
better $N_{\rm H\,I}$ values, some but not all of which are higher than the 
lower limits of \citet[see Figure~3 of \citealp{prochaska17}]{werk14}. Despite 
the hint of a discontinuity between our $N_{\rm H}$ values and those of 
COS-Halos at $\rho\sim0.5\,R_{\rm vir}$, independent fits to the two data sets 
do not find statistically significant variations in normalization or power-law 
slope.
}

In addition to the differing impact parameter ranges probed, another clear 
distinction between the COS-Halos sample and ours is that the COS-Halos 
sight lines were chosen to probe $L \approx L^*$ galaxies at $z\approx0.2$, 
whereas our targeted and serendipitous samples probe a wide range of galaxy 
luminosities (see Figure~\ref{fig:Lrho}) at $z\lesssim0.2$. Thus, the fact that 
the targeted and serendipitous absorbers in 
Figures~\ref{fig:oldnew}-\ref{fig:werk} are relatively well-mixed and are 
reasonably fit by the best-fit relations derived from COS-Halos absorbers 
suggests that the luminosity of the nearest galaxy does not have a significant 
effect on the derived circumgalactic absorption properties (see 
Section~\ref{ensemble}).

\section{CGM Ensemble Properties}
\label{ensemble}

In the full complement of $\sim50$ \lya\ absorbers associated with galaxies 
studied here, 28\% of them are found to be multiple velocity component 
absorbers defined as being in a single galaxy CGM by virtue of being within 
$\pm400$~\kms\ of the galaxy redshift. Most of these multiple systems have 
\lya\ lines that are highly saturated and blended, which makes determining their 
basic properties problematical for three reasons. First, deblending of multiple 
components is an uncertain process usually made less subjective by employing a 
$\chi^2$ minimization of a multiple-component Voigt profile fit. Priors sometimes 
are set based on the velocity locations of metal-ion absorptions, which often 
are separated more distinctly than \lya\ since these lines are intrinsically 
narrower. Even this, now rather standard, process 
\citep[e.g.,][]{werk12,savage14, danforth16} assumes that the individual velocity 
components are well-modeled by Voigt profiles, that the metal lines align well in 
velocity with the \HI\ absorptions and that small differences in the resulting 
$\chi^2$ values are meaningful. Nevertheless, for the present analysis we have 
used the procedures outlined in \citet{danforth16} for multiple component fitting. 
This procedure differs from the process employed in Paper~1 by which these 
complexes were assumed to be a single, although complex, absorber associated with 
a single galaxy's CGM. Values and associated uncertainties for the absorption 
line fits are found in Table~\ref{tab:absprop}. 

Secondly, standard analysis procedures for saturated \lya\ lines
\citep[e.g., apparent optical depth;][]{savage91} provide only lower limits 
on \HI\ column density. If higher order Lyman lines are available, a simultaneous
line-fit or curve-of-growth (CoG) analysis can provide a much more secure 
$N_{\rm H\,I}$ value. For the serendipitous sample, CoG analyses are possible for 
most absorbers due to the availability of \fuse\ spectra of these bright AGN 
targets. For the targeted sample, the higher-order \HI\ lines are not available 
(since \fuse\ spectra are not available excepting for a poor, short exposure of 
PG~0832+251) for study and any saturated \lya\ lines yield uncertain 
$N_{\rm H\,I}$ values\footnote{In most cases, a single \lya\ line fit tends to 
under-predict the $N_{\rm H\,I}$ for moderate and stronger \HI\ absorption 
\citep[$\log{N_{\rm H\,I}} \ga 13.5$;][]{danforth10}.}. 
Comparing the physical cloud parameters for targeted absorbers compared to 
serendipitous absorbers finds no large differences between these two samples, but 
the $N_{\rm H\,I}$ values for many absorbers in the targeted sample remain 
uncertain as indicated in Table~\ref{tab:cloudy}. Uncertain values of 
$N_{\rm H\,I}$ are one criterion used to determine the reliability of the results 
from the CLOUDY modeling of these absorbers (see below).     

Thirdly, the basic geometry of these \lya\ complexes is unclear even if the 
line deconvolution is straightforward. Is each velocity component a separate 
cool CGM cloud? This interpretation means that 13 galaxy CGMs probed here have 
3 or 4 clouds found along the line of sight. Different velocity components in 
these complexes may even be associated with different nearby galaxies (see 
Section~\ref{discussion:association}). Or are these differing velocity 
structures within the same CGM cloud? Paper~1 implicitly assumed the latter 
interpretation which led to a modest ``shadowing factor'' of $S=1.4$, in which 
there are few sight lines with multiple clouds in a single CGM but some of these 
clouds have complex velocity distributions. On the other hand, the multi-component 
line deconvolution shown in the COS spectra in Figure~\ref{fig:stackplot} yields a 
larger value of $S=2.0$ by assuming that these velocity components are all 
separate clouds. This shadowing factor ($S$) was introduced and defined in Paper~1 
and is the mean number of discrete clouds found along any one sight line within a 
single galaxy CGM (i.e., the number of clouds ``shadowed'' by another cloud from 
our perspective). The median number of discrete clouds along these sight lines is 
also 2, with 28\% (13) of the CGMs studied having 3 or 4~clouds along the line-of-
sight. For reference, the completely independent evaluation of the shadowing 
factor for the COS-Halos project by \citet{werk14} finds $S=2.4$; i.e., most 
COS-Halos sight lines have multiple detections in a single galaxy CGM. We discuss 
sample differences in the next subsection that might account for the difference 
in shadowing factors obtained between these two studies.

The new \lya\ line deconvolutions performed here and shown in 
Figure~\ref{fig:stackplot} for these complex absorbers could make, at least in
principle, a significant difference in the ensemble CGM properties compared to 
those obtained by Paper~1, including the total filling factor and mass of the cool 
gas in the CGM. The basic procedure to obtain these quantities (see Paper~1 for 
details) is to use those CGM absorbers that have multiple ionization states of the 
same element (e.g., \SiII, \SiIII\ and \SiIV) as representative of the full CGM 
cool cloud population, which may or may not be a true assumption. This is related 
to a basic unstated assumption of both this study and COS-Halos that the 
methodology of using single QSO sight line probes of many galaxies provides a 
statistically accurate picture of the CGM of any one galaxy with similar 
properties (e.g., super-$L^*$, star-forming galaxies). This assumption is 
implicitly made when we calculate ensemble CGM properties in this Section.  
\edit1{ The percentage of absorbers at $\rho \leq R_{\rm vir}$ for which CLOUDY 
modeling is available is high, 12 of 15 (see below), so the physical parameters 
found are a reasonable approximation to the full CGM sample. We take the virial 
radius to be the full extent of the CGM despite some uncertainty \citep{shull14}. 
}

In order to make the best ensemble mass estimate from the current, modest-sized 
sample, we have graded the absorbers A through D in decreasingly well-constrained 
values for ionization parameter ($U$) and thus cloud density, size and mass. 
This letter grade uses both the S/N of the metal-line detections which 
constrain the CLOUDY modeling and the accuracy of the $N_{\rm H\,I}$ value based 
on the availablity and S/N of the higher-order Lyman lines. Grades~A (8~absorbers) 
and B (9~absorbers) include absorbers with high-S/N detections in metal-lines and 
multiple Lyman line detections providing accurate CoG $N_{\rm H\,I}$ values. 
Grade~B absorbers are slightly less well-constrained in $U$ and $N_{\rm H\,I}$ 
than grade~A absorbers by having either lower S/N metal-line detections or 
somewhat inconsistent constraints in the CLOUDY modeling. Grade~C absorbers 
(6~absorbers) have less accurate $N_{H\,I}$ values either due to being \lya-only 
detections (targeted absorbers) or poor CoG solutions that could be due to 
blending of higher order Lyman lines in the \fuse\ band. However, many grade~C 
absorbers have good metal-line detections which yield good constraints on 
ionization parameter, particularly constraining against high values of $\log{U}$ 
(see e.g., the CLOUDY model in Section~\ref{cloudy} for the absorber 
PG~0832+251\,/\,5221). Grade~D absorbers (2~absorbers) lack both accurate 
$N_{\rm H\,I}$ values and also accurate $U$ determinations, with no strong 
constraints against high values of $U$. For example, the absorber 
PHL~1811\,/\,39658 has weak detections of only \CIII\ 977 (4$\sigma$) with \fuse\ 
and the \CIV\ doublet (3.0 and 2.5 $\sigma$) with COS. Not only is $N_{\rm H\,I}$ 
poorly constrained in this case but the very weak and uncertain metal-line 
detections lead to a very uncertain $\log{U}$ value. We do not use the two 
grade~D absorbers in our ensemble mass calculations. Our best estimates for 
ensemble properties use absorbers with grades~A-C, thus maximizing the sample 
size without significantly degrading the quality of the result.
 
Our modeling process assumes a single, homogenous gas phase for the 
metal-enriched clouds, for which a CLOUDY \citep{ferland98} photo-ionization 
model is constructed. These models (see Section~\ref{cloudy}) determine cloud 
density from the resulting value of the ionization parameter assuming a 
meta-galactic UV ionizing spectrum at $z=0$ as specified by \citet{haardt12} 
for all absorbers, since our sample is at low-redshift (see Section~\ref{cloudy} 
for justification of this choice). The \citet{haardt12} UV background best 
reproduces the low-redshift \HI\ column density distribution \citep{danforth16} 
for $\log{N_{\rm H\,I}} > 14$ \citep{shull15}, where almost all CGM absorbers are 
detected. Given the presence of varying ionization states, this modeling relies 
primarily on the lower ions to provide the ionization parameter since the higher 
ionization states may be influenced by some collisional ionization. 

As mentioned in Paper~1, the absorbers in this sample are well outside any 
``proximity distance'' where the ionizing flux which leaks out from the nearby 
galaxy exceeds the meta-galactic ionizing flux from QSOs \citep{giroux97}. This 
calculation assumes that a very high (probably unrealistically high) fraction of 
ionizing radiation escapes from the nearby galaxy 
($\langle f_{\rm esc} \rangle = 5$\%), in order to determine a quite conservative 
estimate for the maximum proximity distance for each absorber based on the nearest 
galaxy's current SFR \citep{giroux97}. Since all of our absorbers are well outside 
this distance, the meta-galactic ionizing flux is assumed to be isotropic in our 
CLOUDY modeling. 

The internal temperature and thus pressure of these clouds is also determined from 
the CLOUDY model (see details in Paper~1 and Section~\ref{cloudy}). The derived 
pressures shown in the bottom panels of Figure~\ref{fig:comp} vary considerably 
around a mean value of $\langle P/k \rangle = 2~{\rm cm^{-3}\,K}$, a value which 
is $\sim0.7$~dex lower than the mean value reported in Paper~1. This is largely 
due to an error in Paper~1 that overestimated the cloud pressure by a factor of 
$\approx5$. 

The calculated cloud density together with the observed \HI\ column 
density and neutral fraction determine the line-of-sight cloud thickness, 
$D_{\rm cl}$. These cloud sizes range from 150~pc to 40~kpc (see 
Table~\ref{tab:cloudy} and Figure~\ref{fig:comp}). The smallest cloud sizes 
are quite close to the minimum size suggested to survive in the CGM over a long 
timescale ($\geq250$~Myr) by \citet{armillotta16}; the largest cloud sizes are 
nearly a factor of ten larger than the largest HVC found near the Milky Way 
\citep[Complex~C;][]{wakker07}. While there is one CLOUDY solution (for the 
absorber PHL~1811\,/\,39658) that suggests a cloud size still larger 
($\sim200$~kpc) this absorber has a very poorly constrained photo-ionization 
model (grade~D). Therefore, we discount this absorber's model and 
suggest a largest cloud size of $\sim40$~kpc (covering about 2\% of the entire CGM 
when viewed from afar). 

Assuming spherical symmetry allows a mass estimate for each cloud modeled. 
Individual cloud mass estimates can be quite uncertain 
($\sigma(M_{\rm cl}) \sim 30$-50\%) due to uncertainties in the 
CLOUDY modeling (including the possible presence of multi-phase gas), to 
uncertainties in the \HI\ column density when the \lya\ is saturated and/or 
blended, and to the substantial sensitivity of the calculated mass to the 
line-of-sight thickness from which it is derived 
($M_{\rm cl} \propto D_{\rm cl}^3$). In this sample, cloud masses range from 
$30~M_{\Sun}$ to $2\times10^8~M_{\Sun}$, the latter value being over two orders 
of magnitude more massive than Complex~C \citep{wakker07}. These cloud mass 
estimates are listed in Table~\ref{tab:cloudy} and shown in Figure~\ref{fig:comp}.

The CLOUDY models for components of the complex absorbers yield smaller 
line-of-sight cloud sizes and masses compared to Paper~1. As inferred by the 
photo-ionization analysis these absorption complexes break up into smaller 
(3-10~kpc) clouds that have estimated cloud masses of $10^5$-$10^{6.5}~M_{\Sun}$. 
This increases the shadowing factor and the filling factor, which increases the 
total CGM ensemble mass. But dividing very massive absorbers into separate clouds 
has the effect of decreasing the total CGM mass estimate because, in the limit of 
unity covering factor, the few most massive clouds dominate the total filling 
factor and the ensemble mass. By this new analysis method for the complex 
absorbers, the total filling factors are 1.5~times greater for all luminosity 
classes than as given in Paper~1 (see their Table~7); e.g., 5-9\% for super-$L^*$ 
galaxies. 

\edit1{
In calculating the ensemble CGM cool gas mass we have followed the procedure 
described in detail in Paper~1 in which the number of clouds ($N_{\rm cl}$) in 
each half-dex size range is determined from the covering factor ($C$), shadowing 
factor ($S$), virial radius of the associated galaxy ($R_{\rm vir}$) 
and cloud size ($R_{\rm cl}$) by the following equation, which was derived in 
Paper~1:
\begin{equation}
N_{\rm cl}= C \, S \, \left(\frac{R_{\rm vir}}{R_{\rm cl}}\right)^2.
\end{equation}
The ensemble mass in each mass bin is then 
$M_{\rm tot} = N_{\rm cl} \, M_{\rm cl}$; for this calculation we have 
assumed that the CGM extends to the virial radius of these galaxies  
\citep[but see][]{shull14} and used the clouds associated with both the 
super-$L^*$ and sub-$L^*$ subsamples (as in Paper~1) as representative of bulk 
CGM properties for galaxies in both of these luminosity subsamples, since the 
covering and shadowing factors for each subsample are nearly identical. While 
the dwarfs have far fewer detected clouds, that subsample also has much lower 
covering factors ($C$) within the virial radius ($C\sim\sfrac{1}{2}$), and 
much larger associated errors. Also as in Paper~1, we have divided the CGM into 
volumes defined by the inner and outer half-radii as the covering factor declines 
slightly between these two impact parameters. Three-quarters of the CGM mass is 
inside $\sfrac{1}{2}\,R_{vir}$. The ensemble cold cloud numbers and mass in each 
luminosity subsample are determined using the mean virial radius for each 
subsample (see Paper~1).
}

The combination of larger filling factors and larger numbers of modest-mass clouds 
(factor of nearly twice more in the $\sim10^5$-$10^{6.5}~M_{\Sun}$ range) leaves 
our best estimate for the ensemble cool CGM mass nearly unchanged at 
$\log{(M/M_{\Sun})} = 10.2\pm0.3$ for super-$L^*$ galaxies, identical (but with 
slightly larger uncertainty) to the value presented in Table~7 of Paper~1. This 
best estimate uses all 23~absorbers with data and CLOUDY modeling quality of A, B 
and C grades. If only the best quality A- and B-grade absorbers are used, this 
sample of 17~absorbers finds $\log{(M/M_{\Sun})} = 10.4\pm0.4$. The larger 
uncertainty is due to the smaller number of large (20-30~kpc), massive 
($>10^{6.5}~M_{\Sun}$) clouds in the full ensemble (3 vs 5). Using only grade~A 
absorbers finds a similar mean mass estimate but with a much larger uncertainty 
due to the smaller sample of massive clouds. 
\edit1{ Almost all of the CGM clouds detected at $\rho \leq R_{vir}$ by this 
survey are modeled in this process; 12 of 15 absorbers have metals and almost all 
of them are modeled in Section~\ref{cloudy}. Since there are no obvious 
differences in basic properties (e.g., $N_{\rm H\,I}$ values) between those few 
unmodeled clouds and the large majority of modeled clouds, the above estimate of 
ensemble CGM cool cloud mass likely is unbiased by the absence of models for these 
few. }

The error budget for this CGM ensemble mass estimate is substantial. Since the 
ensemble mass estimates are dominated by a small number of the largest clouds, 
the sampling errors are large, $\sim50$\%. But then there are also systematic 
errors associated with the CLOUDY modeling itself, due both to uncertainties in 
individual metal-line measurements and to the detailed methodology in constructing 
the best-fit models based on those ratios (see Section~\ref{cloudy}). These errors 
are reflected in the individual uncertainties associated with cloud line-of-sight 
thicknesses and masses recorded in Table~\ref{tab:cloudy}. Despite the statistical 
gain of averaging many modeled clouds, the errors due to modeling uncertainties 
are comparable to the sampling errors, leading to total errors in the ensemble 
mass estimate of 0.3~dex. 

\edit1{ 
Even with this increase in uncertainty, our ensemble super-$L^*$ galaxy CGM 
mass estimate of $\log{(M/M_{\Sun})} =10.2\pm 0.3$ is substantially lower than 
the COS-Halos ``preferred lower limit'' of $\log{(M/M_{\Sun})} \geq 10.8$ 
\citep{werk14} at the $2\sigma$ level. The COS-Halos estimate is presented as a 
firm lower limit on CGM mass due to saturated \HI\ absorption lines in their 
sample. However, a COS-Halos calculation similar to the one performed above and in 
Paper~1 obtains a total CGM cool gas mass of $\log{(M/M_{\Sun})} =10.5$, 
$1\sigma$ higher than the value obtained here and in Paper~1. A more recent study 
of CGM gas structure using a hierarchical, photo-ionized cloud structure to 
explain the various ionization states including \OVI\ finds a total cool gas 
mass of $\log{(M/M_{\Sun})} = 10.1\pm0.1$ \citep{stern16}. The \citet{stern16} 
study also finds covering factors and filling factors for \lya\ and the low ions 
comparable to the values we found in Paper~1 and herein. An even more recent paper
\citep{prochaska17} extends the \citet{werk14} analysis by obtaining better 
$N_{\rm H\,I}$ values for COS-Halos absorbers with highly-saturated \lya. This 
new work suggests a total cool CGM mass of nearly $10^{11}~M_{\Sun}$, at even 
greater variance to the value obtained herein. We critique these various CGM cool 
gas mass estimates in the next subsection.
}

In both the current study and the COS-Halos study, the cool CGM ensemble cloud 
mass is dominated by the few very large clouds with line-of-sight thicknesses of 
$\geq10$~kpc and estimated masses $\geq10^{6.5}~M_{\Sun}$. \citet{werk14} find 
even larger, more massive clouds with estimated line-of-sight thicknesses of up to 
2~Mpc. These cloud sizes are so large that even \citet{werk14} question whether 
inferred sizes can be as large as this, since 2~Mpc is considerably larger than 
the virialized region of an $L^*$ galaxy. Excluding a very poorly-constrained 
(grade~D) CLOUDY solution for one absorber that yields a line-of-sight thickness 
of $\sim200$~kpc, the largest cloud thicknesses we find in the current study are 
20-40~kpc, which yield inferred cloud masses of $10^7$-$10^8~M_{\Sun}$. 

Are even these large cloud sizes reasonable or just the result of unsuspected 
systematics in the CLOUDY modeling?  The recent discovery by \citet{davis15} 
leaves little doubt that large size ($\geq10$-20 kpc) cool CGM clouds can exist. 
\citet{davis15} found \HI\ and/or \CIV\ absorption common to a close triplet of 
QSO sight lines \citep[``The LBQS Triplet'';][]{crighton10} at the redshift of a 
very nearby, star-forming $0.07\,L^*$ spiral. While the physical structure of 
clouds this large is not clear, \citet{davis15} set a firm lower limit of 
$\geq10^6~M_{\Sun}$ on the mass of this cloud. And if the mean ionization 
parameter from the present study is adopted for these absorbers, then the inferred 
mass of this cloud is $\geq10^7~M_{\Sun}$. Further research on the internal, 
physical structure of CGM clouds will provide a much more secure ensemble mass for 
the cool phase of galaxy halos, a work which is now in progress in Cycle~23 of 
\hst\ observations 
\citep[Guest Observer Program \#14127, M.~Fumagalli, PI; see also][]{bowen16}.

\subsection{Systematics and the Estimated Baryon Content of the Cool CGM}
\label{ensemble:coshalos}

The ensemble cool CGM cloud mass determination is important for taking an accurate 
census of spiral galaxy baryons and assessing the status of the ``missing baryon'' 
problem. 
\edit1{ Since the most obvious physical condition for this ``missing'' gas is in 
the hard-to-detect $T=10^5$-$10^{6.5}$~K range 
\citep*[but see][]{savage14,stocke14,werk16,prochaska17}, at present its
amount can best be calculated indirectly by totaling those baryon reservoirs that 
are more easily detectable and assuming a total baryon reservoir set by the 
univeral dark matter to baryon ratio. This hypothesized hotter gas is suggested by 
simulations\citep*[e.g.,][]{klypin01,faerman16} to be massive enough to be the 
dominant baryon reservoir in spiral galaxies with an extent comparable to a small, 
spiral-rich group in which individual star-forming galaxies reside 
\citep{stocke14,faerman16}. }
A hot intra-group medium is also suggested for spiral-rich galaxy groups by 
downward extrapolation from the more massive halos of elliptical-dominated groups 
\citep{mulchaey00} and by some (but not all) observational analyses of the Local 
Group gas \citep[e.g.,][]{anderson11,gupta12, faerman16}.

\edit1{
While data and analysis limitations make all current estimates uncertain to a
factor of $\sim2$, the different results presented in the previous paragraph 
suggest substantial differences in the number of ``missing baryons''. 
The \citet{werk14} COS-Halos ``preferred lower limit'' amounts to $\geq30$\%
of the baryons in spiral galaxies. And the recent \citet{prochaska17} reanalysis 
of the COS-Halos absorbers finds $\approx$ 50\% while the ensemble mass estimated 
here and in Paper~1 yield 10-15\% baryon fraction. The \citet{stern16} formalism 
finds baryon fractions at the $\leq10$\% level while admitting that their assumed 
hierarchical cloud structure minimizes a total mass estimate. While the lower 
limit found by \citet{werk14} and the value quoted by \citet{prochaska17} are 4-5 
times higher than the amount found herein, the COS-Halos galaxies are suggested to 
be in twice-higher-mass halos, so the baryon fractions calculated differ by a 
factor of 2-2.5. Using nominal values (see Table~8 in Paper~1) obtained by other 
studies for the $L^*$ spiral galaxy baryon percentages in stars and gas in the 
spiral disk ($\sim20$\%), hotter \OVI-absorbing gas in the CGM ($\sim6$\%), and 
very hot ($T>10^7$~K) coronal gas ($\leq10$\%), then the \citet{werk14} lower 
limit on cool CGM mass implies that $\geq\sfrac{2}{3}$ of all spiral galaxy 
baryons have been found; i.e, currently detected and identified in emission 
and/or absorption. The most recent \citet{prochaska17} paper suggests that,
between the cool and hot CGM masses, all the baryons in spiral halos may have 
been located. On the other hand, if the value derived herein, in Paper~1, and in 
the \citet{stern16} analysis is used, then $\geq\sfrac{1}{2}$ of spiral galaxy 
baryons remain ``missing''. While these differences are large, they are only 
somewhat greater than the combined statistical errors of these various analyses.
}

\edit1{
While we have quoted statistical errors on our mass estimates, important 
systematic differences exist between the COS-Halos studies and this one. Most 
importantly is the value of the ionization rate assumed to be impinging on these 
clouds because derived cloud densities, and thus cloud masses, are determined 
from the CLOUDY modeling, which uses an assumed meta-galactic ionizing spectrum. 
The use of different values quoted in the most recent works 
\citep*{haardt01,haardt12,kollmeier14,khaire15,shull15,madau15,gaikwad16} can 
add $\geq \pm0.3$~dex of {\it systematic} uncertainty to the total (i.e., 
ionization-corrected) hydrogen column, and a greater systematic uncertainty to 
the computed mass, depending on method used. Additional systematic bias occurs 
when studying absorbers at different redshifts due to the very steep dependence 
of the \HI\ ionization rate ($\Gamma_{\rm H}$) on redshift (see 
Section~\ref{cloudy:UVB}). Our use of the \HI\ photo-ionization rate at $z=0$ as 
compared to a $\sim$25\% larger value at $z=0.05$ (the median redshift of our 
sample) has very little effect on the estimated cloud masses. However, 
\citet{werk14} and \citet{prochaska17} assume a much more intense radiation field 
that amounts to a factor of $\sim$3-4 times larger ionization rate, when corrected 
back to $z=0$ from the mean COS-Halos absorber redshift of $z=0.2$. The 
\citet[][HM12 hereafter]{haardt12} UVB spectrum is assumed for this conversion
from $z=0.2$ to $z=0$. We assert that it is this difference in assumed ionization 
rate (\citet[][HM01 herafter]{haardt01} at $z=0.2$ assumed by \citealt{werk14} and 
\citealt{prochaska17}, and HM12 at $z=0$ assumed herein) that leads to the cloud 
densities in this survey (i.e., largely at $\rho > 0.5\,R_{\rm vir}$) appearing 
to be lower than the COS-Halos cloud densities at slightly smaller impact 
parameters (see Figure~\ref{fig:werk}; bottom two panels). Using the results found 
in \citet{shull15} (see their Table 1 \& Figure 2), for a given $N_H$, the neutral 
fraction is $\sim0.6$~dex higher for the HM12 spectrum than for the HM01 spectrum. 
By Equation~8 in \citet{werk14} and Equation~7 in \citet{prochaska17} this 
$\approx4$ times larger $N_H$ for a given $N_{\rm H\,I}$ translates
directly into a $\approx4$ times larger total CGM cool gas mass. This difference 
can be seen in the bottom panels of Figure~\ref{fig:werk}, in which the 
\cite{werk14} fits to cloud densities are significantly higher than the current 
data; some of the \citet{prochaska17} inferred cloud densities are even higher 
(see their Figure~8, right-hand panels).   
}

\edit1{
Ultimately, the correct value of the \HI\ ionization rate is not known at 
present, particularly at low-$z$ \citep[although see very recent 
work on \Ha\ fluorescence in clouds at $z\approx0$ that may rule out the HM01 
ionization rate;][see also \citealp{donahue95}]{fumagalli17}. 
We justify our use of the HM12 spectrum through the detailed analysis of various 
ionization rates presented in \citet{shull15}, which included a comparison with 
observed column density distributions (see their Figure~2). What is striking 
in this Figure is that the low-\HI\ column density distribution is well-matched 
using the earlier HM01 UVB, but the higher \HI\ column densities are much better 
matched by the HM12 background. While none of the UVB spectra match the 
\textit{slope} of the observed column density distribution, the origin of this 
discrepancy remains a mystery. And what is the best match at 
$\log{N_{\rm H\,I}}>14.5$ is also not clear. But, since the CGM absorbers are all 
at higher column densities, the choice of the HM12 spectrum appears to be the most 
appropriate for CGM studies at this time.  
}

As reported in Paper~1 and herein, none of the present sample of 
absorbers occurs so close to the nearest galaxy that the ionizing flux impinging 
on the cloud becomes dominated by the escaping Lyman continuum radiation from the 
nearby galaxy; i.e., all of the absorbers in the present sample have impact 
parameters larger than the ``proximity distance'' for a rather high assumed escape 
fraction of ionizing photons of 5\% \citep{giroux97}. However, using the SFRs and 
impact parameters for COS-Halos absorbers found in \citet{werk13} we find that 18 
out of 28 of the COS-Halos absorbers lie at impact parameters less than the 
proximity distance for a more plausible escape fraction of 2\%. Both because the 
impact parameter is a projected distance and because the leakage of ionizing 
photons is very likely to be highly anisotropic in star-forming galaxies, only a 
fraction of the 18~absorbers actually may have their photo-ionization rate 
dominated by UV photons from the nearby galaxy. Still, if this occurs even in just 
a few cases the under-estimation of the ionizing flux for these absorbers can lead 
to an under-estimate of the cloud density for the ionization parameter set by the 
metal-line ratios in the absorber. 
\edit1{ If this ``proximity effect'' is present for some COS-Halos absorbers, it 
means that COS-Halos has \textit{under-}estimated the ionization rate in these 
cases. But, even if present, this is likely to be a small effect 
(\citealp{werk14} came to this same conclusion) since the solid angle
illumination of a CGM cloud by the nearby galaxy is modest at CGM cloud 
distances, and much smaller than the isotropic UVB impinging on the cloud. }

\edit1{ 
While there may be differences between these two mass estimates due to 
assumed ionizing radiation field levels, there are definitely differences 
in the radial domain of applicability of these results as described in 
Section~\ref{sample:comparison} and shown in Figures~\ref{fig:Lrho} and 
\ref{fig:coshalos}. Briefly, the COS-Halos study has no data at all at 
$\gtrsim \sfrac{1}{2}\,R_{\rm vir}$, while the present study has 
very little data at $\leq\sfrac{1}{2}\,R_{\rm vir}$ compared with COS-Halos 
(7~absorbers only from this study, all of which have CLOUDY models).
Therefore, one could argue that these studies are largely disjoint and thus 
complementary and cannot be easily compared due to the quite different 
radial distribution of impact parameters. For example, the analysis of 
\citet{prochaska17} suggests that the CGM cool gas mass has converged in the 
COS-Halos data and there is little to no CGM beyond $\rho \geq 0.5R_{\rm vir}$. 
This conclusion is based on a cloud column density distribution which declines 
quickly with impact parameter, truncating at an impact parameter of 160~kpc. 
On the other hand, at larger impact parameters, the present study finds no 
significant correlation between $N_{\rm H\,I}$ and impact parameter. In the 
current sample there are quite a few metal-bearing, high-$N_{\rm H\,I}$ 
absorbers at $\rho \geq 0.5R_{\rm vir}$, including metal-line detections in 
absorbers beyond the virial radius (Figure~\ref{fig:veldist}, and Figures 8 \& 9 
in Paper~1). Paper~1 finds $\approx25$\% of the cool CGM between 
0.5-$1\,R_{\rm vir}$, at variance with the \citet{prochaska17} conclusion. 
}

\edit1{
On the other hand, if the COS-Halos choice of the HM01 ionization is adopted, 
then a total CGM cool gas mass inside the virial radius of slightly greater than 
$10^{11}~M_{\Sun}$ is obtained. If the twice larger halo mass calibration of 
\citet{prochaska11a} is used \citep[see Figure 1 in][]{stocke13}, the baryon 
fraction of the CGM cool gas mass exceeds 50\%. While it is beyond the scope of 
this paper to decide which ionization rate and halo mass calibration are correct, 
we favor a lower value of the CGM cool gas mass both because the HM01 
ionization rate appears to be ruled out by recent low-$z$ \Ha\ fluorescence 
observations \citep{adams11,fumagalli17} and because a larger reservoir of hot, 
compared to cool, gas mass is expected around star-forming galaxies in small 
galaxy groups \citep[][see also \citealp{stocke14} for relevant 
observations]{klypin01,faerman16}. High-S/N \hst/COS observations of QSO probes 
through small galaxy groups are now in progress (cycle~23; J.~Stocke, PI) to 
search for broad, shallow \OVI\ and \lya\ absorption associated with this hotter gas.
}

\edit1{
Since there is only a small overlap in radial domain between our sample and 
COS-Halos, one can argue that to obtain the best total CGM cool gas mass 
estimate, these values should be added. But in this case we need to
renormalize the COS-Halos mass estimate for $\rho < \sfrac{1}{2}\,R_{\rm vir}$ 
to the HM12 ionization rate at $z=0.2$. For the most recent \citet{prochaska17} 
mass estimate this yields: $\log{(M/M_{\Sun})} = 10.4 \pm 0.3$ for the inner  
CGM. This renormalized value is now within the statistical errors of our mass 
estimate here and in Paper 1. For the outer CGM we take 25\% of the CGM calculated 
herein from Paper~1. We make no correction in this calculation for the 
larger halo mass galaxies in the COS-Halos studies compared to the sample herein; 
i.e., we use a total halo mass appropriate for a $2\,L^*$ galaxy from Paper 1 
and here. In this case the total CGM cool gas mass estimate obtained is: 
$\log{(M/M_{\Sun})} = 10.5 \pm 0.3$, $1\sigma$ higher than our estimate. This 
amounts to a cool CGM baryon fraction of $\sim30$\%, twice the value obtained 
from our data alone.
}

\edit1{
In conclusion, while this study updates and slightly revises the ensemble
CGM properties of star-forming galaxies, the new values obtained are nearly 
indistinguishable from the findings of Paper~1. Specifically, for this sample 
alone, the ensemble CGM mass calculation finds the same values as Paper~1 for all 
three luminosity bins. Since the difference between this study and Paper~1 is how 
the multi-component \lya\ absorbers are handled, the cool CGM mass calculation 
appears to be robust with respect to the detailed data analysis process used for 
the multi-component absorber complexes that dominate the CGM cool cloud mass. 
Additionally, the ensemble cool CGM mass of $\log{(M/M_{\Sun})} = 10.2 \pm 0.3$ 
obtained herein and in Paper~1 is identical to within statistical errors with the 
recent result of \citet{stern16}, which uses a much more specific model for cloud 
structure. This suggests that the mass calculation is also robust with respect to 
the detailed structure assumed for these clouds and the cool CGM gas mass 
consititutes a significant, but still minority, contribution (10-15\%) of the spiral galaxy baryon inventory. 
}

\edit1{
However, the COS-Halos study better samples the inner half of the virial radius 
of massive spirals, while the current study better samples the outer half. After 
renormalizing the COS-Halos mass estimate to the less intense ionization rate and 
smaller total halo mass appropriate for our sample, the estimated baryon fraction 
in the cool CGM gas is 30\%, twice the value quoted above. Given all the 
statistical and systematic uncertainties, we consider this baryon fraction to be 
the most accurate estimate currently.
}

\section{Discussion}
\label{discussion}

\subsection{Galaxy-Absorber Correlations}
\label{discussion:correlations}

\subsubsection{Metal-line Detections and Nondetections}
\label{discussion:correlations:metals}

\begin{figure}[!t]
\epsscale{0.86}
\plotone{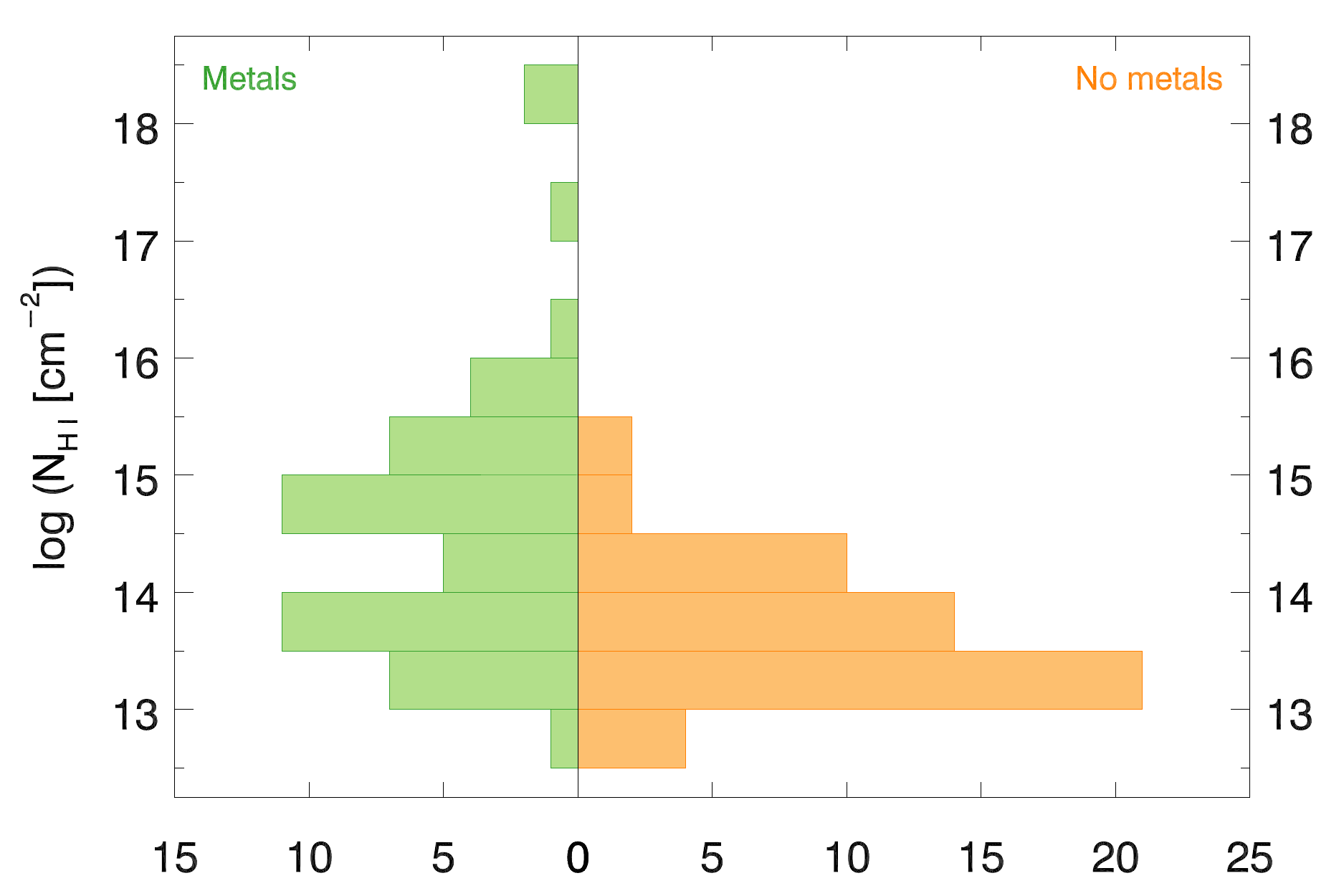}
\plotone{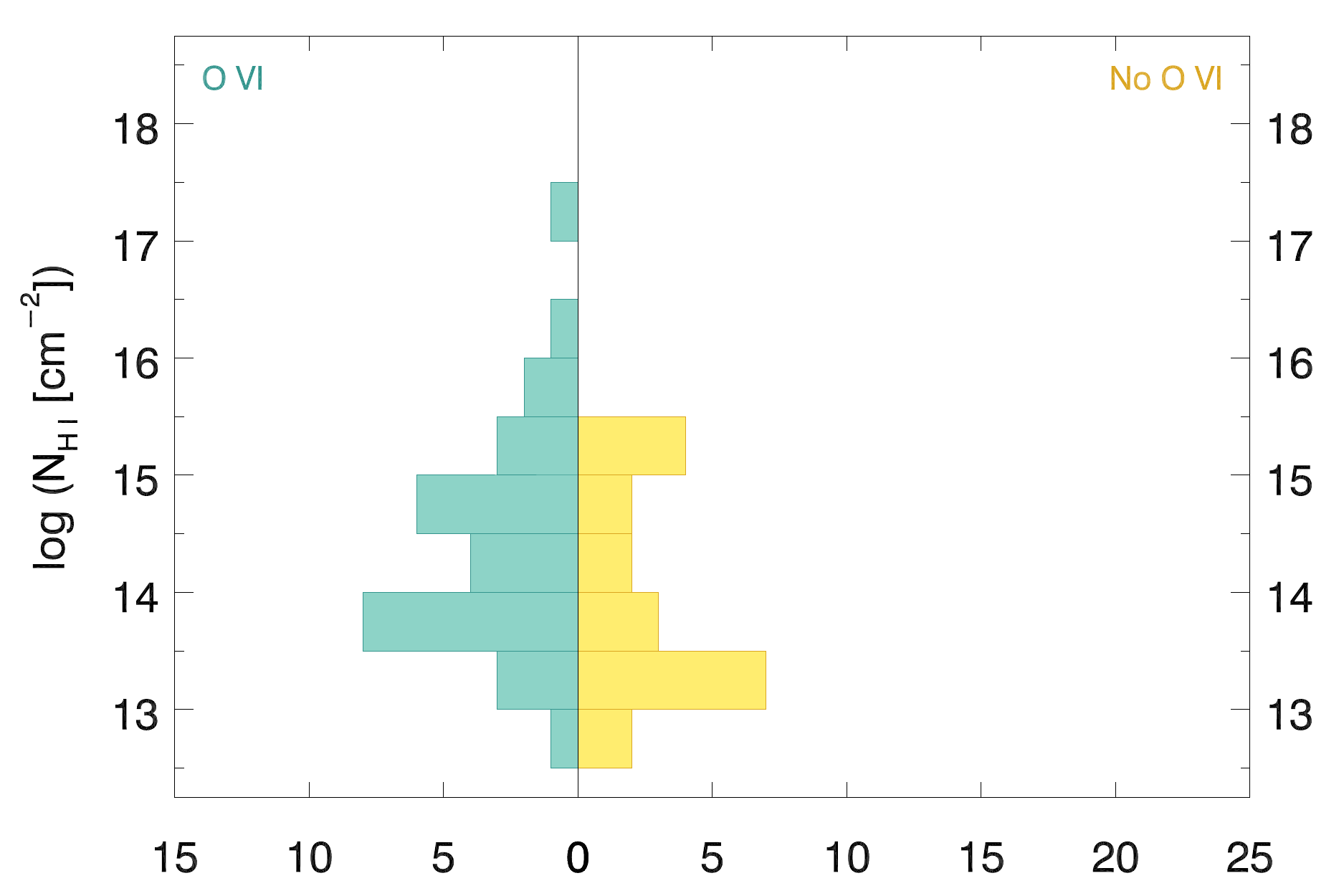}
\caption{Distribution of \HI\ column density for absorbers with and without metals (top) and with and without \OVI\ absorption specifically (bottom). In the top panel no constraints are placed on the strength of the metal-line absorption or lack thereof. In the bottom panel a cutoff of $N_{\rm O\,VI} = 10^{13.2}~{\rm cm}^{-2}$ is used for the presence or absence of \OVI\ absorption (i.e., only \OVI\ absorbers with larger column densities are considered, and \OVI\ must be detectable at the cutoff column density to at least $3\sigma$ significance for \OVI\ non-detections).
\label{fig:HIhist}}
\end{figure}

\begin{figure}[!b]
\epsscale{0.86}
\plotone{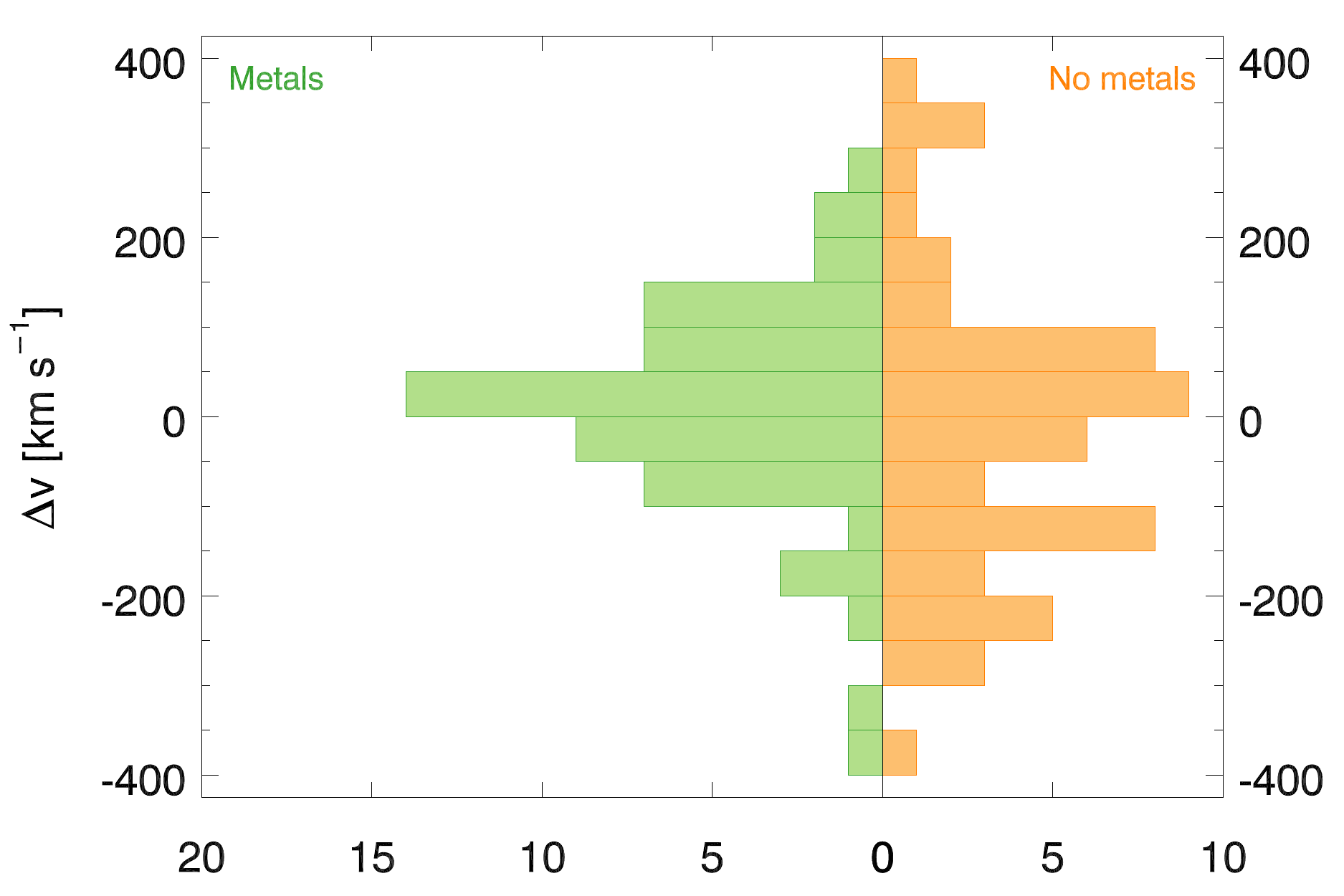}
\plotone{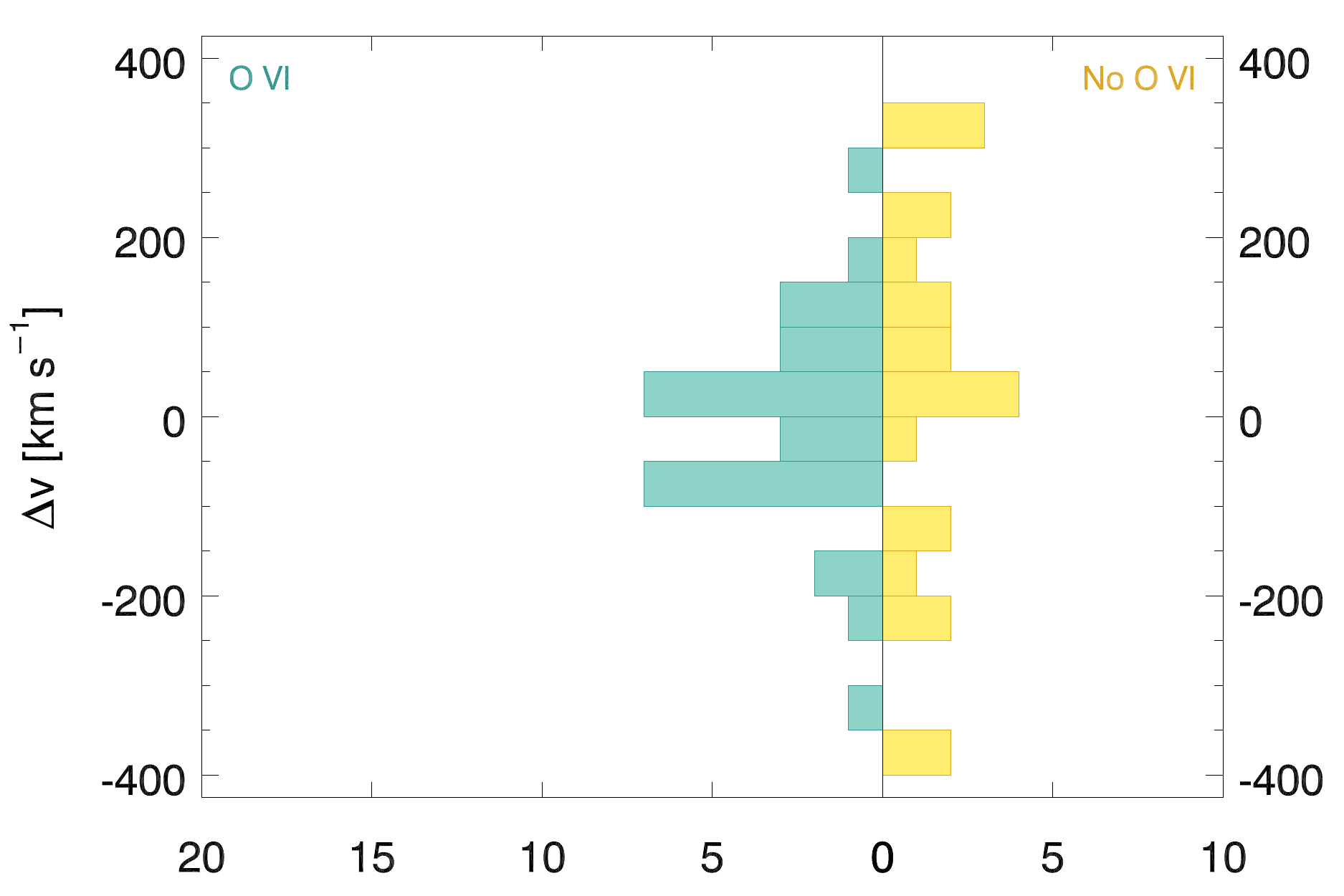}
\caption{Distribution of absorber-galaxy radial velocity difference for absorbers with and without metals (top) and with and without \OVI\ absorption specifically (bottom). In the top panel no constraints are placed on the strength of the metal-line absorption or lack thereof. In the bottom panel a cutoff of $N_{\rm O\,VI} = 10^{13.2}~{\rm cm}^{-2}$ is used for the presence or absence of \OVI\ absorption (i.e., only \OVI\ absorbers with larger column densities are considered, and \OVI\ must be detectable at the cutoff column density to at least $3\sigma$ significance for \OVI\ non-detections).
\label{fig:dvhist}}
\end{figure}

In this Section we examine basic correlations between galaxies and their CGM 
absorbers. Figures~\ref{fig:HIhist} and \ref{fig:dvhist} compare histograms of 
\HI\ column density and absorber-galaxy radial velocity difference, 
respectively, for absorbers with and without metals. The top panel of both 
Figures compares absorbers that exhibit absorption from any metal species with 
those that are metal free (i.e., \HI-only systems) without regard to the 
strength of the metal-line absorption or lack thereof. The bottom panel of 
both Figures compares \OVI\ absorbers specifically to those that do not show 
\OVI\ (but perhaps show absorption from other metal species), both for a more 
direct comparison with the COS-Halos results and because \OVI\ is the metal 
species that is the most sensitive tracer of low-metallicity gas located far
from galaxies \citep{stocke06,stocke07}. To be considered an \OVI\ absorber 
for the purposes of these histograms an absorber must have 
$N_{\rm O\,VI} \geq 10^{13.2}~{\rm cm}^{-2}$. To be an \OVI\ non-detection 
\OVI\ must be detectable (i.e., high quality \fuse\ data must be available or 
the absorber must have sufficient redshift for \OVI\ to be detectable in \hst/COS data) to a $3\sigma$ limit of $N_{\rm O\,VI} < 10^{13.2}~{\rm cm}^{-2}$.

Figure~\ref{fig:HIhist} shows that CGM absorbers with no detected metal lines 
have a lower average \HI\ column than metal-line absorbers. This suggests that 
some of the ``metal-free'' absorbers may have similar metallicity to those with detected metals, but their \HI\ column is so low as to make their associated 
metal lines undetectable in the current spectra; i.e., with greater S/N we 
would expect some of the ``metal-free'' absorbers in the right-hand side of 
Figure~\ref{fig:HIhist} to become metal-line absorbers. Based on the CLOUDY 
models described in Section~\ref{cloudy}, the metal-free absorbers could 
have metallicities $\lesssim0.1\,Z_{\Sun}$. The bottom panel of 
Figure~\ref{fig:HIhist} attempts to correct for this effect by examining 
absorbers with and without \OVI\ above a certain threshold. After this 
correction the average \HI\ column density for CGM absorbers with and 
without \OVI\ are more comparable, although the metal-free absorbers are 
still systematically at lower $N_{\rm H\,I}$. 

If we assume for the moment that the metal-line detections are 
associated with high-metallicity outflows and the nondetections are associated 
with infalling, low-metallicity gas, then the similarity between the velocity 
distributions of \HI\ absorbers with associated metal lines and those without 
(Figure~\ref{fig:dvhist}) suggests that there is no clear kinematic discriminator 
between the two. This speculative conclusion also holds for the \OVI\ and non-
\OVI\ absorbers. However, the relatively high threshold of metallicities 
($\log{(Z/Z_{\Sun})} \geq -1$) accessed by current COS spectroscopy leaves 
unanswered the metallicity of true IGM absorbers far from galaxies since the 
IGM metallicity level found at $z > 2$ is in the range 
$\log{(Z/Z_{\Sun})} \approx -2$ to $-3$ \citep*{schaye03,aguirre04,simcoe04}. 
Much higher S/N COS spectra than those presented here would need to be obtained to 
address this question.

Another issue that Figures~\ref{fig:HIhist} and \ref{fig:dvhist} bring to light 
is systematic uncertainties in the estimate of the CGM mass 
(Section~\ref{ensemble}). The mass of individual CGM 
absorbers can only be estimated if there are sufficient metals present to 
constrain photo-ionization models (see Section~\ref{cloudy} for details), 
so the ``metal-free'' absorbers are de facto assumed to have similar 
sizes and masses to the modeled absorbers. The large number of CGM absorbers 
with low-\HI\ column density and no metals detected suggests that the ensemble 
CGM mass estimate of Section~\ref{ensemble} is likely an 
overestimate of the true value. This increases the discrepancy between our 
CGM cool cloud mass estimate and that of \citet{werk14}. A detailed 
discussion of this point can be found in Section~\ref{ensemble}.

\subsubsection{Targeted and Serendipitous Absorbers}
\label{discussion:correlations:samples}

\begin{figure}[!t]
\plotone{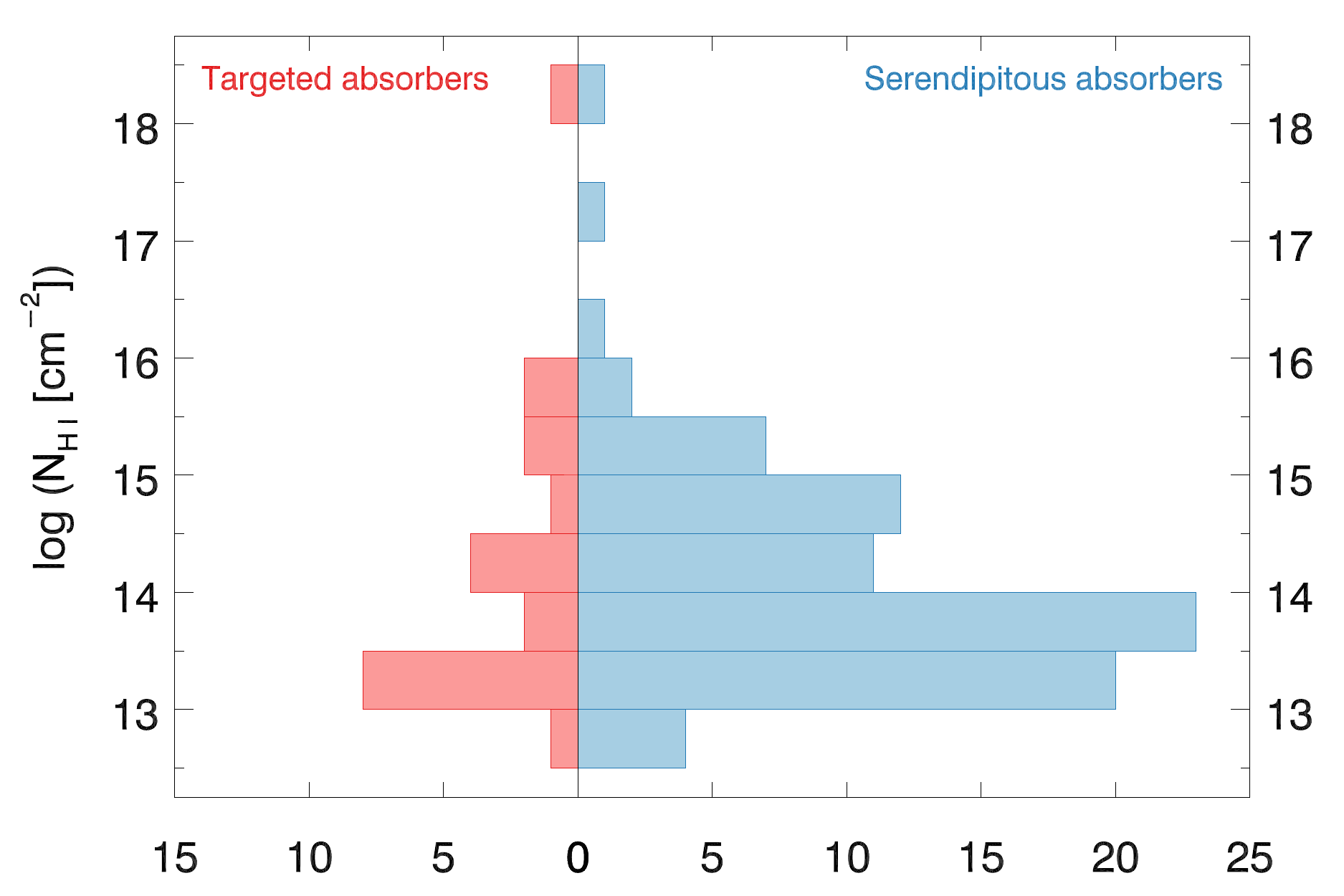}
\plotone{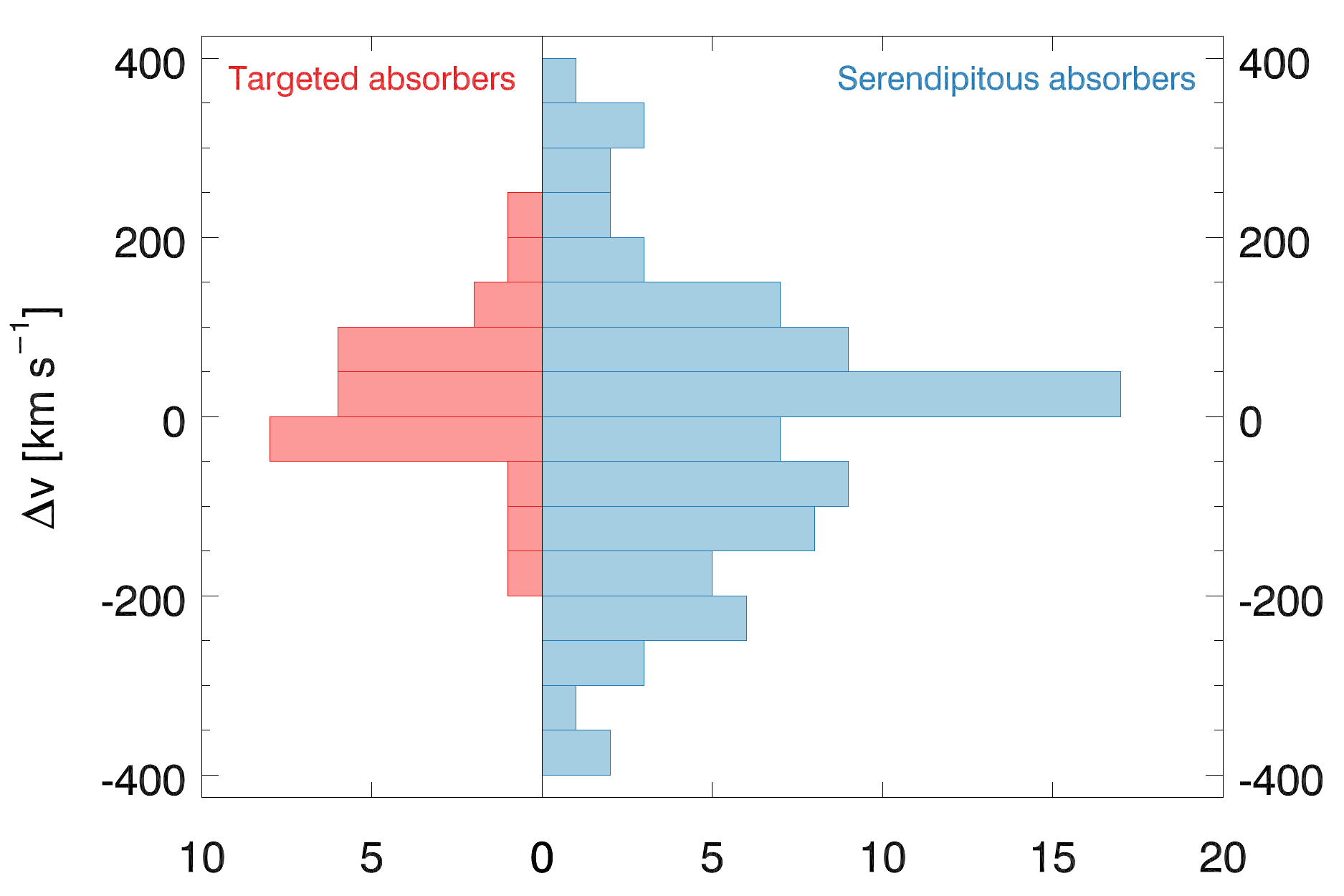}
\caption{Distribution of \HI\ column density (top) and absorber-galaxy radial velocity difference (bottom) for targeted and serendipitous absorbers. There are no significant differences in either quantity between the two samples.
\label{fig:samplehist}}
\end{figure}

Given the differences between the targeted and serendipitous samples, such as 
the higher uncertainties in $N_{\rm H\,I}$ values for targeted absorbers 
(see discussions in Sections~\ref{absorbers} and \ref{cloudy}), it is fair 
to ask whether there are any systematic differences in the CGM properties of 
these galaxies. While Figure~\ref{fig:Lrho} shows that the targeted sample 
preferentially probes lower-luminosity galaxies at smaller impact parameters than 
the serendipitous sample, Figure~\ref{fig:samplehist} shows that there is little 
difference in the distributions of \HI\ column density and absorber-galaxy radial
velocity difference. The targeted sample shows some evidence for having a smaller 
characteristic velocity difference ($|\Delta v| < 200$~\kms) than the 
serendipitous sample, with only $\sim4$\% (1/27) of its absorbers having 
$|\Delta v| > 200$~\kms\ as compared to $\sim23$\% (20/87) for the serendipitous 
one.

\edit1{
One interpretation of Figure~\ref{fig:samplehist} is that the outskirts of 
luminous galaxies look remarkably like the CGM of low-luminosity galaxies, except 
with a larger spread in velocity. This could simply be a consequence of higher 
mass halos having larger velocity dispersions, but it may suggest that the 
luminous serendipitous galaxies possess analogs to Galactic HVCs, which could be 
fossil relics of previous mergers with lower-luminosity companions. In the case of 
Milky Way HVCs, the connection with dwarf satellites is occasionally clear 
\citep[e.g., the Magellanic Stream;][]{d'onghia16}, but often even the largest 
HVCs cannot be associated with a present-day Milky Way satellite \citep[e.g., 
Complex~C;][]{wakker07}.
}

Whenever appropriate in subsequent Figures we have endeavored to plot data points 
from the targeted and serendipitous samples with different plot symbols (circles 
for targeted galaxies/absorbers and triangles for serendipitous ones as in 
previous Figures). In all cases the data from the two different samples are well 
mixed, aside from the aforementioned differences in the host galaxy luminosities 
and impact parameters. This indicates that there are no further systematic 
differences in galaxy or inferred CGM properties between these two samples. 
Therefore, here and in Paper~1 the targeted and serendipitous samples are treated 
as a single sample.

\subsubsection{Galaxy-Absorber Kinematics}
\label{discussion:correlations:kinematics}

\begin{figure}[!t]
\plotone{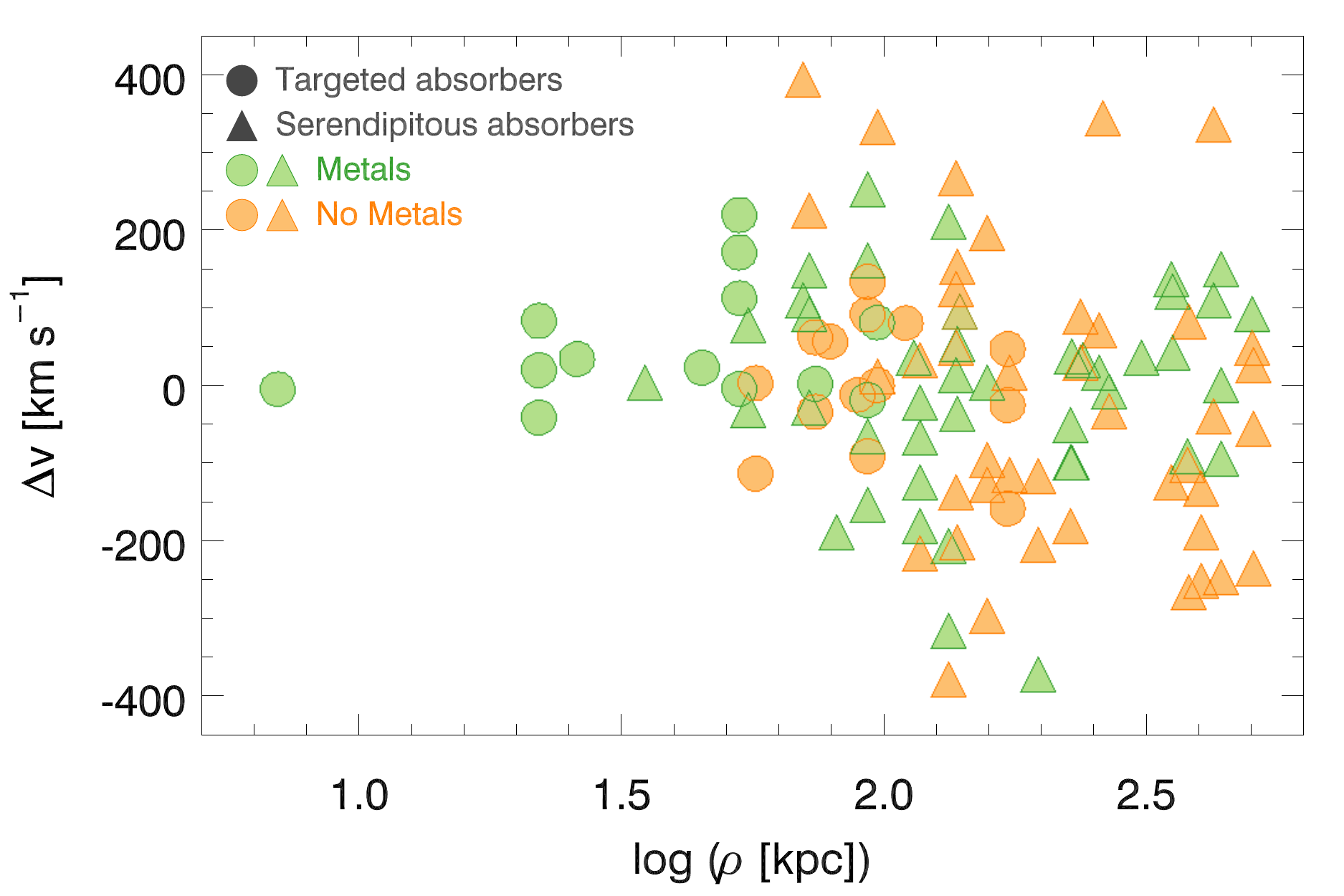}
\plotone{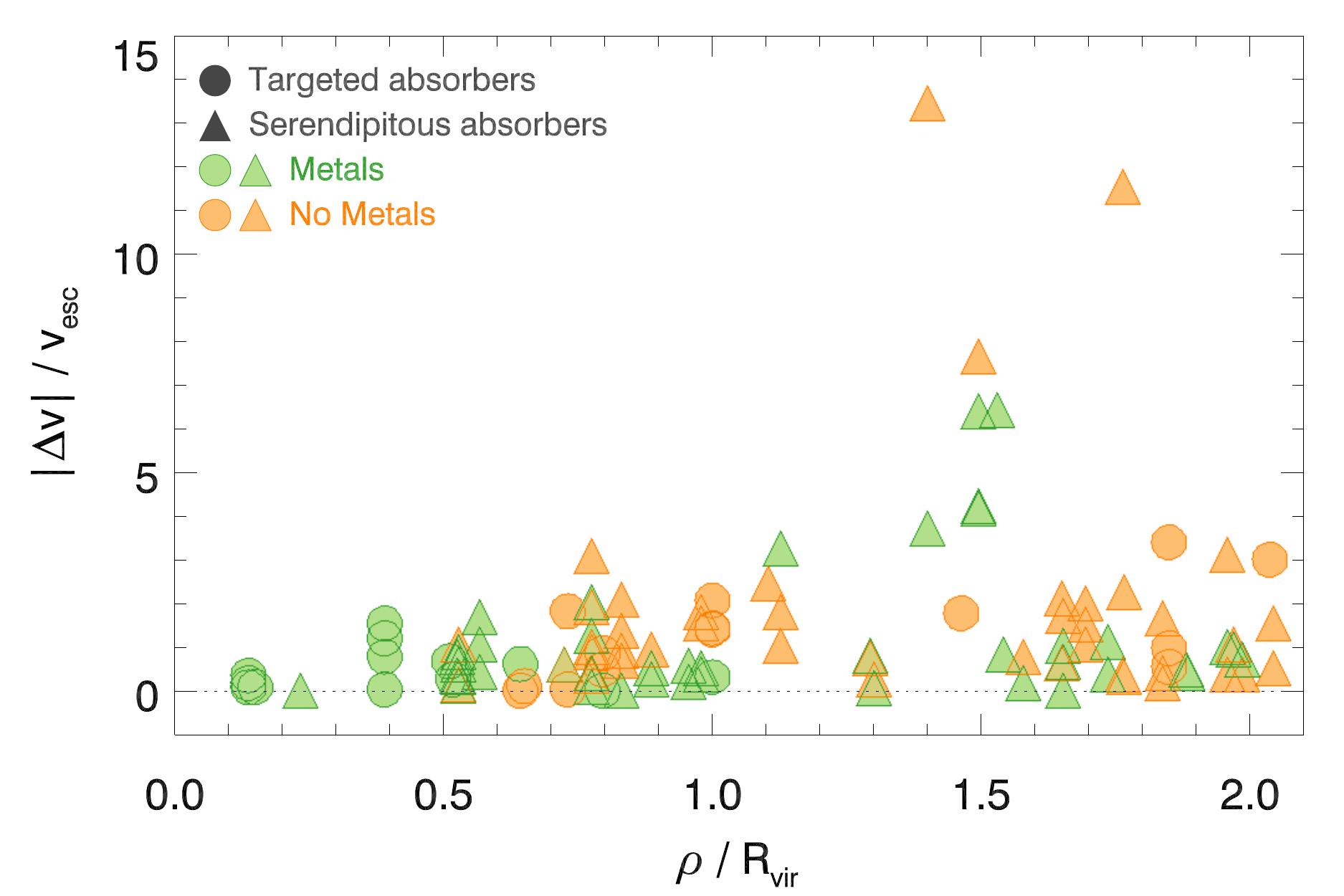}
\caption{Distribution of CGM absorber-galaxy radial velocity difference as a function of impact parameter in directly observable (top) and derived (bottom) units. The absorbers with the largest velocity difference ($|\Delta v| > 5\,v_{\rm esc}$) are all located far from the nearest galaxy ($\rho \gtrsim 1.4\,R_{\rm vir}$).
\label{fig:veldist}}
\end{figure}

\begin{figure}[!t]
\plotone{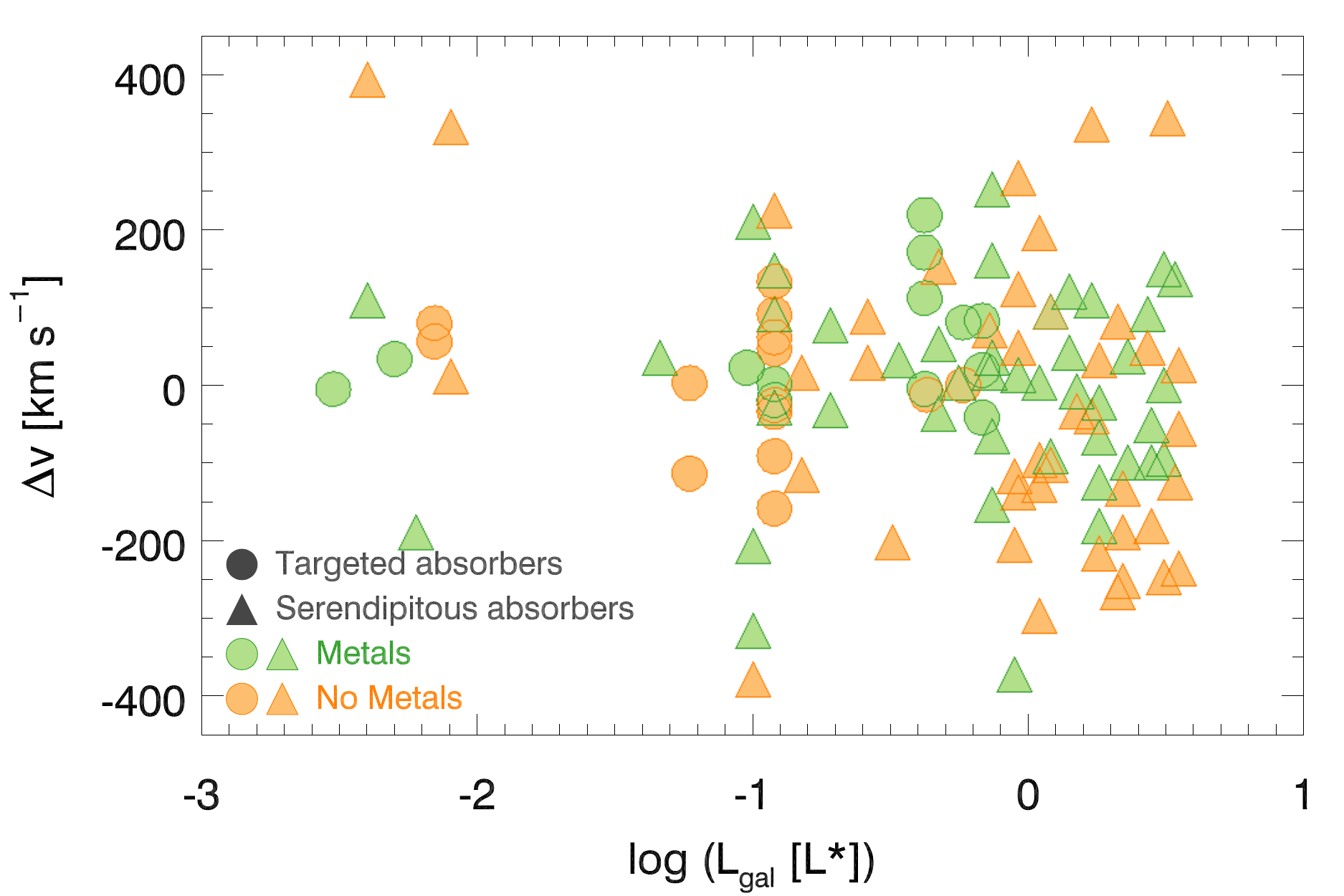}
\plotone{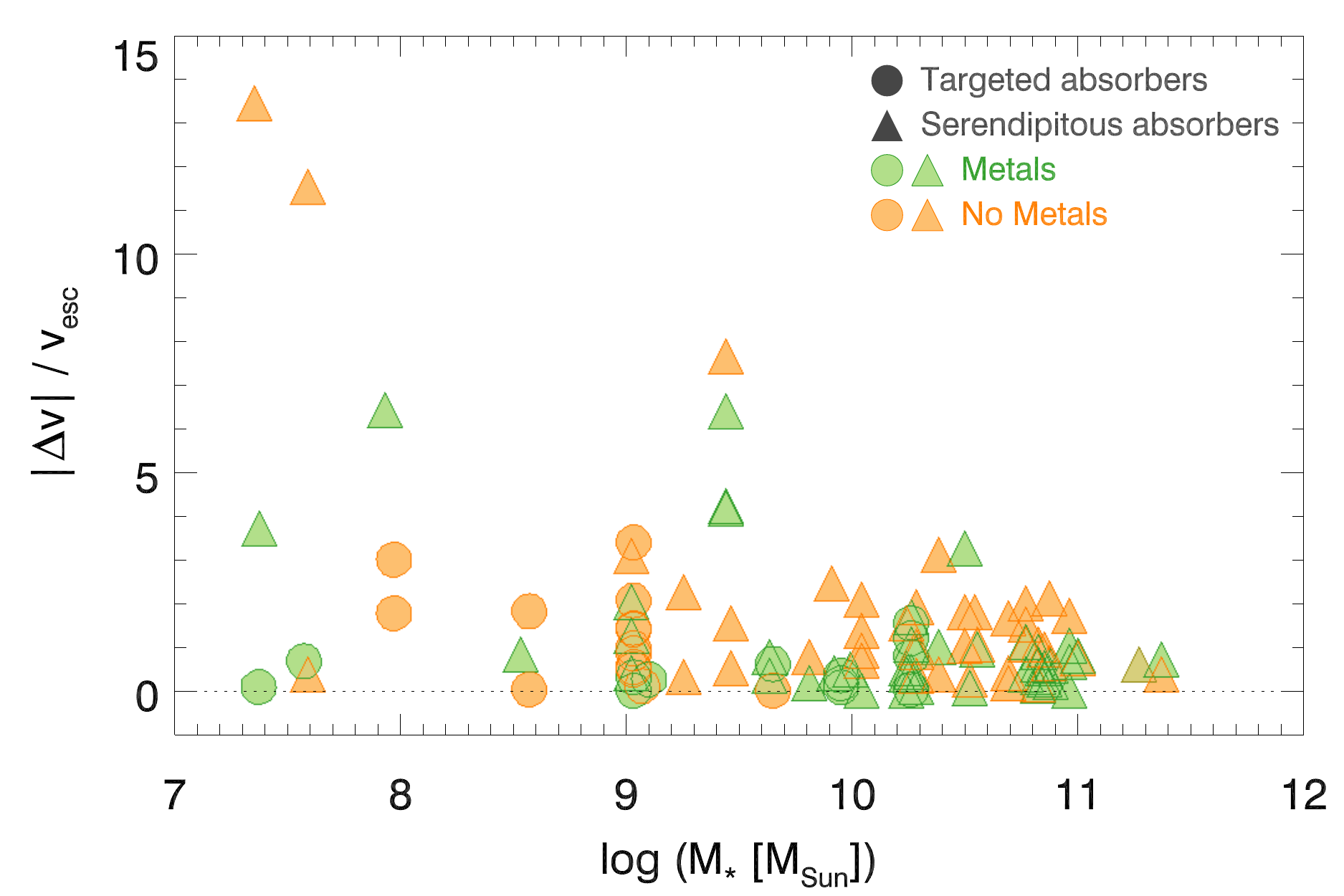}
\caption{Distribution of CGM absorber-galaxy radial velocity difference as a function of galaxy size in directly observable (top) and derived (bottom) units. The absorbers with the largest velocity difference ($|\Delta v| > 5\,v_{\rm esc}$) are all associated with less massive galaxies ($M_* < 10^{10}~M_{\Sun}$).
\label{fig:velsize}}
\end{figure}

\begin{figure}[!t]
\plotone{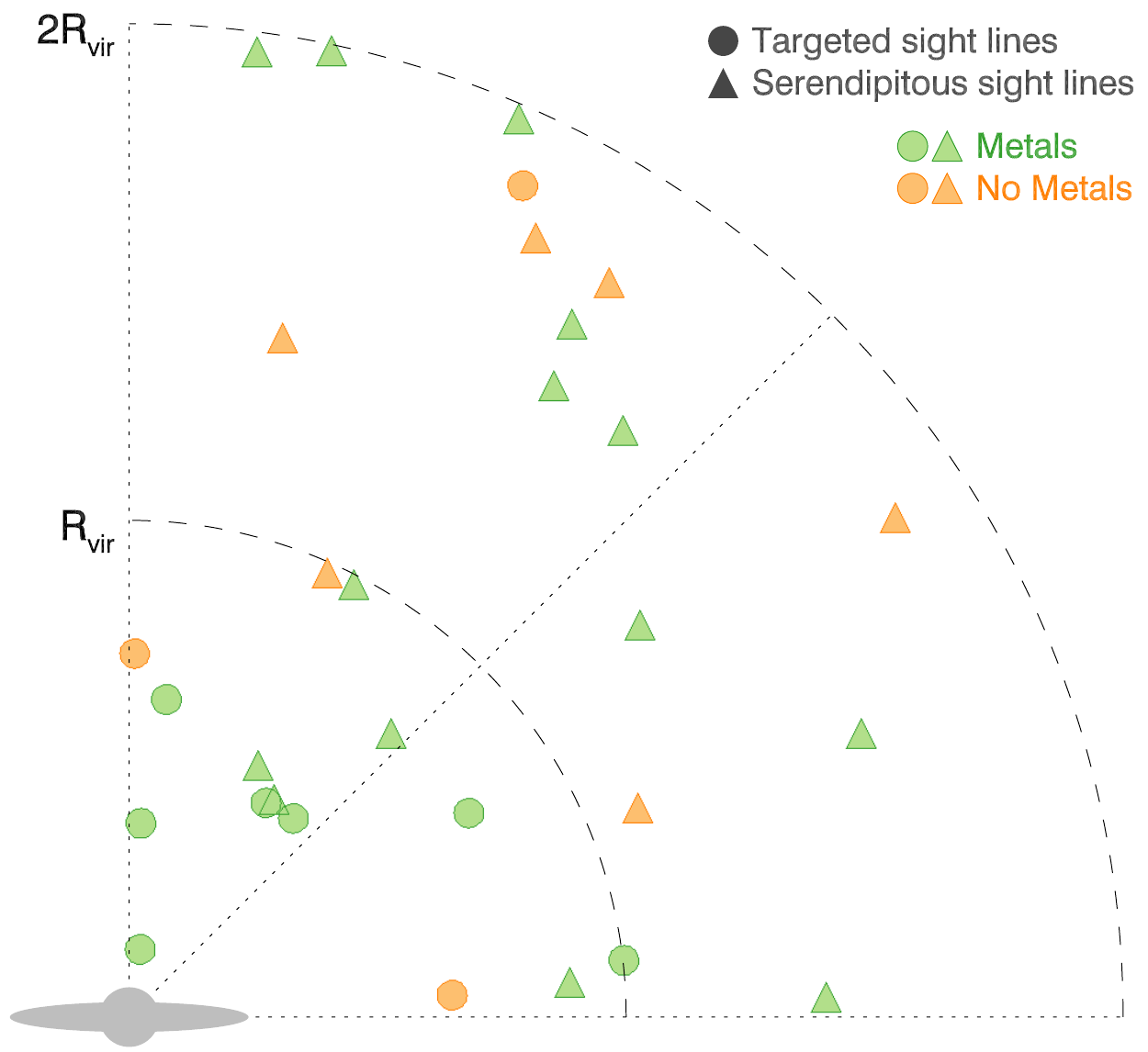}
\caption{Distribution of QSO sight lines with respect to the major axis of the nearest galaxy, where such information is available. Half of the sight lines (15/30) are located within the galaxy's virial radius and 70\% (21/30) are located within $45\degr$ of the galaxy's minor axis. There is no appreciable difference in the fraction of sight lines containing metals as a function of azimuthal angle, but there is some dependence on impact parameter with 80\% (12/15) of sight lines within the galaxy's virial radius containing metals as compared to 60\% (9/15) at larger distances.
\label{fig:posang}}
\end{figure}

Figures~\ref{fig:veldist}-\ref{fig:posang} examine correlations between properties 
of CGM absorbers and their host galaxies. In Sections~4.2 \& 4.4 of Paper~1 we 
investigated basic observables relevant to absorber-galaxy kinematics. While the 
plots here present the data in a different format the overall conclusions here 
and in Paper~1 are the same.

The top panel of Figure~\ref{fig:veldist} shows the galaxy-absorber radial 
velocity difference for metal-line and \HI-only absorbers as a function of 
QSO-galaxy impact parameter. This plot of directly-observable quantities 
shows that there is little difference in the distribution of absorbers with and 
without associated metals, except that virtually all absorbers with 
$\rho \lesssim 50$~kpc have associated metals and velocities close to 
the galaxy velocity. The bottom panel shows the same data plotted in normalized 
units, where the radial velocity difference is displayed as a multiple of 
the escape velocity from the galaxy halo at $R=\rho$ and the impact parameter is 
plotted as a multiple of the galaxy virial radius. While some caution is advisable 
so as not to over-interpret this plot of derived quantities where both axes are 
dividing a projected quantity by a three-dimensional value, both in Paper~1 and 
here we interpret these low impact parameter and low $|\Delta v|$ absorbers 
as recycling gas that is either outflowing or infalling but almost certainly 
remains bound to the associated galaxy.

At larger impact parameters ($\rho \gtrsim 50$~kpc) the $|\Delta v|$ values 
increase significantly. However, this difference may be largely artificial 
because this is also the impact parameter where our overall sample becomes 
dominated by the serendipitous rather than the targeted galaxies. Since the 
serendipitous absorbers are associated with more luminous galaxies (see 
Figure~\ref{fig:Lrho}), the velocity differences are systematically larger
but the $|\Delta v|/v_{\rm esc}$ distributions are not so different; i.e., 
compare the spread in $\Delta v$ at $\rho \approx 100$~kpc in the top panel 
of Figure~\ref{fig:veldist} with the spread in $|\Delta v|/v_{\rm esc}$ at
$\rho/R_{\rm vir} \approx 1$ in the bottom panel.

However, there are a few (5) significant outliers in the bottom panel of 
Figure~\ref{fig:veldist} with $|\Delta v|/v_{\rm esc} > 5$, which are 
certainly not gravitationally bound to their associated galaxy. Whether 
this means that these five are true IGM absorbers, not associated with a single 
galaxy despite their modest impact parameter ($\rho/R_{\rm vir} \approx 1.5$),
or whether these are unbound infalling or ejected clouds, is not clear. 
The three absorbers apparently associated with low mass galaxies 
($\log{M_*}<8$; see Figure~\ref{fig:velsize}) could be associated 
with large-scale structures in their vicinities instead
\citep{rosenberg03,yoon12,keeney14,stocke14}. The same may be the 
case for the multiple-velocity-component absorbers with 
$|\Delta v|/v_{\rm esc} > 5$ apparently associated with galaxies at 
$\log{M_*} \approx 9.4$. The uncertain associations mentioned here
(3C~273\,/\,1585, Mrk~335\,/\,2281, PG~1116+215\,/\,17614 \& 17676, and
Q~1230+0115\,/\,1497) are detailed in Section~\ref{indiv:galaxies} 
of the Appendix (see also Section~\ref{discussion:association}).

Figure~\ref{fig:velsize} is similar to Figure~\ref{fig:veldist} except that 
it plots the luminosity of the host galaxy instead of an absorber's 
distance from it. The top panel of Figure~\ref{fig:velsize} plots 
directly-observable quantities: galaxy-absorber radial velocity difference 
as a function of rest-frame $g$-band galaxy luminosity in $L^*$ units. Again 
we find no clear distinction between CGM absorbers with and without associated 
metals. The bottom panel plots the radial velocity difference as a multiple of 
escape velocity from the galaxy at $R=\rho$ as a function of the host 
galaxy's stellar mass; see Section~\ref{galaxies} for a description of the $M_*$ 
derivation. The dispersion of $|\Delta v|/v_{\rm esc}$ is clearly higher for 
galaxies with $M_* < 10^{10}~M_{\Sun}$ and probably indicates that some of 
these are unbound absorbers.

The general trends found in the bottom panels of Figures~\ref{fig:veldist}
and \ref{fig:velsize} are unsurprising since lower mass galaxies have smaller 
escape velocities and the escape velocity for a given galaxy decreases as the 
distance from the galaxy increases. Thus, an observed value of 
$|\Delta v| < 400$~\kms\ will be a larger multiple of the escape velocity far 
away from less massive galaxies, exactly as found in Figures~\ref{fig:veldist} 
and \ref{fig:velsize}. The handful of absorbers with extremely large peculiar 
velocities ($|\Delta v|/v_{\rm esc} > 5$) are a mixture of metal-line and 
\HI-only absorbers, and all but one have $N_{\rm H\,I} < 10^{14}~{\rm cm}^{-2}$. 
However, the three absorbers with the largest $|\Delta v|/v_{\rm esc}$ values 
are metal-free to a limiting metallicity of $\lesssim10$\% solar. The absorber 
with the largest $|\Delta v|/v_{\rm esc}$ value (Q~1230+0115\,/\,1497) is in 
an area very well surveyed for galaxies to a limit of $<0.01\,L^*$. In 
\citet{rosenberg03} we speculated that this absorber and the 3C~273\,/\,1585 
absorber (at $|\Delta v|/v_{\rm esc}\sim6.5$ in this plot) both are part of a 
large-scale filament of galaxies and gas in this region and not associated with 
any one galaxy \citep[see also][]{keeney14}.

Taken together, these Figures show that there is no easy kinematic or size/mass 
dichotomy between absorbers with associated metal lines and those without. 
Going one step further, this means that we have found no simple kinematic 
diagnostic to distinguish low-metallicity gas accreting onto a galaxy from 
higher-metallicity gas entrained in a galactic wind even though simulations 
strongly imply that both infalling and outflowing gas are pervasive in the CGM 
at $z\sim0$ \citep[e.g.,][]{keres09}. 

This conclusion is bolstered by Figure~\ref{fig:posang}, which shows the 
distribution of CGM sight lines with respect to the host galaxy's major axis for 
the subset of QSO-galaxy pairs where the galaxy's major axis is well-defined. Half 
of the CGM sight lines (15/30) are located within the galaxy's virial radius, and 
we see a modest dependence on impact parameter of the fraction of sight lines 
containing metals, with 80\% (12/15) of the sight lines located within a galaxy's 
virial radius containing metals as opposed to 60\% (9/15) of the sight lines at 
larger distances. However, the fraction of sight lines containing metals as a 
function of azimuthal angle is relatively constant. This is somewhat surprising as 
galactic outflows are found to be bipolar with small opening angles centered on the 
minor axis \citep{veilleux05} and IGM accretion is expected to occur closer to the 
galaxy's major axis \citep{keres09}. 

\citet{bouche12} found a bimodal distribution of CGM absorbers when studying
strong \MgII\ absorbers at $z\sim0.1$. In that study, half of the sight lines 
containing strong \MgII\ absorbers were located near the galaxy's major axis and 
the other half within $30\degr$ of its minor axis. While 70\% (21/30) of our 
CGM sight lines are located closer to the galaxy's minor axis than its 
major axis\footnote{Some of the targeted sight lines were specificially chosen to 
be near the targeted galaxy's minor axis, potentially biasing this result. 
However, we find the same fraction of sight lines closer to the galaxy's minor 
axis in the targeted (7/10) and serendipitous (14/20) samples, suggesting that 
there is no systematic bias between the samples.}, Figure~\ref{fig:posang} shows 
that the azimuthal angle distribution of the sight lines is otherwise rather 
uniform. However, this could be because we are examining all sight lines that show 
CGM absorption, regardless of the \HI\ column density, whereas \citet{bouche12} 
were studying the distribution of strong \MgII\ absorbers, which are known to 
preferentially select Lyman limit systems 
\citep[$N_{\rm H\,I} > 10^{17.3}~{\rm cm}^{-2}$;][]{steidel95}. Our CGM sample 
has too few Lyman limit systems (4) to make a direct comparison with the 
\citet{bouche12} results, but if we limit our sample to only include absorbers 
with $N_{\rm H\,I} > 10^{15}~{\rm cm}^{-2}$ we still do not find a bimodal 
absorber distribution.

\subsection{Are Absorbers Associated Unambiguously with a Single Galaxy?}
\label{discussion:association}

\begin{figure}[!t]
\plotone{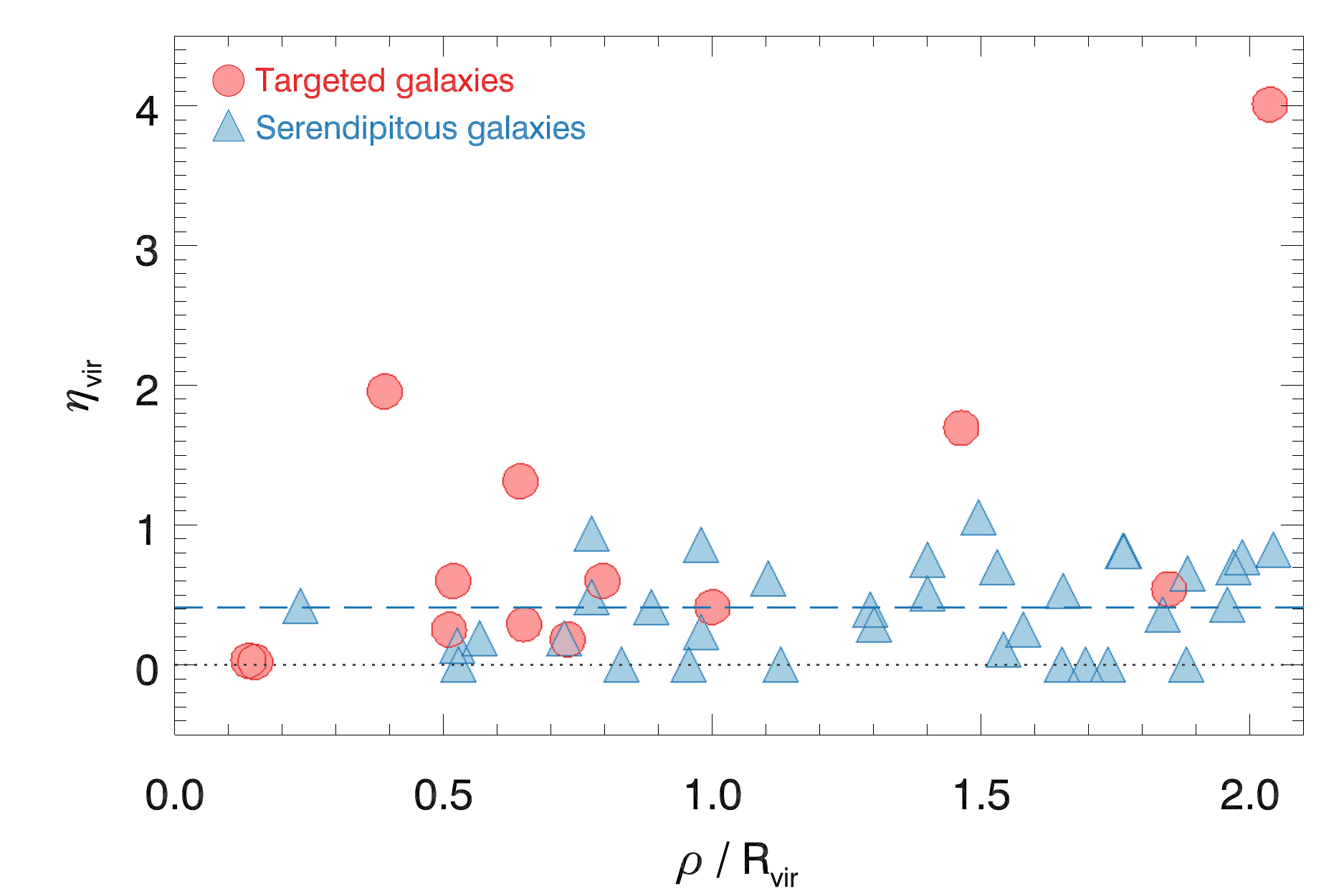}
\plotone{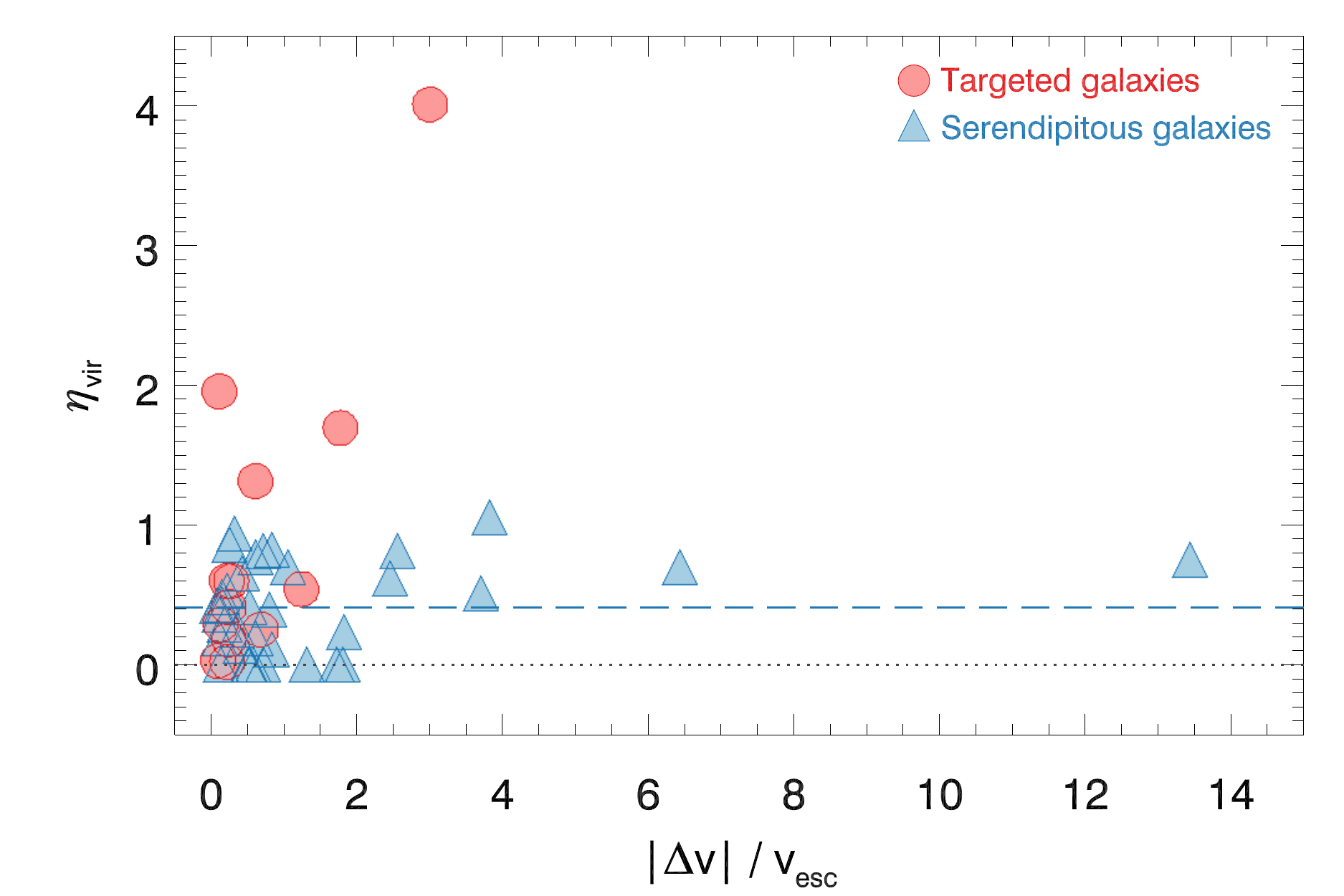}
\caption{Ratio of the normalized impact parameter for the nearest galaxy compared to the normalized impact parameter for the next-nearest galaxy as a function of normalized impact parameter (top) and normalized velocity absorber-galaxy difference (bottom). The dashed horizontal line indicates the median value of $\eta_{\rm vir}$ in the serendipitous sample, which is more homogeneous than the targeted sample. The serendipitous sample shows no clear trend in the top panel and only a slight ($1.4\sigma$) positive correlation in the bottom panel. Small values of $\eta_{\rm vir}$ indicate that the next-nearest galaxy is much further from the absorber than the nearest galaxy.
\label{fig:etavir}}
\end{figure}

We have assumed throughout this Paper that a galaxy-absorber association exists 
if a galaxy is located within a projected distance of $2\,R_{\rm vir}$ of the 
QSO sight line and has a velocity within 400~\kms\ of an \HI\ \lya\ absorber. 
Here we examine the robustness of this assumption by searching for not only the 
nearest galaxy (${\rm ng}$) to the QSO sight line (i.e., those tabulated in 
Tables~\ref{tab:targeted} and \ref{tab:serendipitous}) but the next-nearest
galaxy (${\rm nng}$) as well. The last three columns of Tables~\ref{tab:targeted} 
and \ref{tab:serendipitous} list the ratios of the impact parameters for the 
nearest and next-nearest galaxies 
($\eta_{\rho} = \sfrac{\rho_{\rm ng}}{\rho_{\rm nng}}$), the ratios of their 
normalized impact parameters 
($\eta_{\rm vir} = \sfrac{(\rho/R_{\rm vir})_{\rm ng}}{(\rho/R_{\rm vir})_{\rm nng}}$), 
and the ratios of their normalized absorber-galaxy velocity differences
($\eta_{\Delta v} = \sfrac{(|\Delta v|/v_{\rm esc})_{\rm ng}}{(|\Delta v|/v_{\rm esc})_{\rm nng}}$),
respectively. In all cases a value near zero means that the nearest galaxy 
is significantly closer than the next-nearest galaxy, suggesting a secure 
association. In some cases only one galaxy is located with $\rho<1$~Mpc and 
$|\Delta v|<400$~\kms\ of an absorber (i.e., the maximum extent of our search 
volume; $1~{\rm Mpc} = 2\,R_{\rm vir}$ for a galaxy with $L\approx20\,L^*$ 
according to the prescription of Paper~1), in which case we quote an upper limit 
for $\eta_{\rho}$ but no value for $\eta_{\rm vir}$ or $\eta_{\Delta v}$. 
Figure~\ref{fig:etavir} shows that almost all associations in this sample are 
rather unambiguous since there is no arguably closer galaxy ($\eta_{\rm vir} > 1$) 
in either normalized impact parameter ($\rho/R_{\rm vir}$) or normalized velocity 
difference ($|\Delta v|/v_{\rm esc}$). However, there are a few exceptions in both 
samples.

While the serendipitous sample galaxies were chosen to have $\eta_{\rm vir} < 1$ 
by construction (see Section~\ref{sample} for details), the targeted sample 
galaxies were chosen first with no specific isolation criterion, which could lead 
to some ambiguities. In most cases the targeted galaxy is the closest to the 
absorber, but Figure~\ref{fig:etavir} shows four cases with $\eta_{\rm vir} > 1$,
two at $\rho/R_{\rm vir} \geq 1.5$ and two at much smaller normalized impact 
parameters and velocity differences.

The two targeted absorbers with ambiguous associations ($\eta_{\rm vir} > 1$) at 
large $\rho/R_{\rm vir}$ are related to each other, as the QSO/galaxy pairs were 
chosen specifically to observe two distinct sight lines to probe two galaxies at 
comparable redshifts. It is not surprising that these absorbers have ambiguous 
associations. The 1ES~1028+511 sight line has two \HI\ \lya\ absorbers within 
400~\kms\ of each other and a targeted galaxy that corresponds to each absorber. If 
this were a serendipitous sight line, then the two \lya\ absorbers would be treated 
as probing the CGM of a single galaxy, so for the purposes of the $\eta$ values in 
Table~\ref{tab:targeted} we treat SDSS~J103108.88+504708.7 as the nearest galaxy 
and UGC~5740 as the next-nearest galaxy. Similarly, the galaxy 
SDSS~J103108.88+504708.7 is actually closer to the 1SAX~J1032.3+5051 sight line 
than is UGC~5740 so it is treated as the nearest galaxy there as well. The bottom 
panel of Figure~\ref{fig:etavir} shows that these two absorbers have peculiar 
velocities greater than the escape velocities for these two galaxies, making their 
direct association with either galaxy ambiguous.

The third and fourth ambiguous associations in Figure~\ref{fig:etavir} are targeted 
galaxies in small groups in which a fainter member is closer to the sight line. In 
one case, the luminous starburst galaxy NGC~2611 was the galaxy targeted for 
observation using the PG~0832+251 sight line but subsequent, deeper spectroscopy 
found that it is a member of a small group of galaxies, one of which is significantly 
closer to the sight line than NGC~2611. The case of the PMN~J1103--2329 sight line is 
similar in that it was chosen to probe the starburst galaxy NGC~3511 but the 
lower-luminosity galaxy NGC~3513 is in fact closer to the sight line and has a 
velocity coincident with the 1194~\kms\ \lya\ absorber. NGC~3511 has a velocity 
coincident with the other velocity component in this absorber at 1113~\kms. No 
other targeted associations appear ambiguous.

While the ``typical'' serendipitous galaxy is $\sim2.4$ times closer than the 
next-nearest galaxy ($\eta_{\rm vir}$ = 0.41; dashed horizontal lines in 
Figure~\ref{fig:etavir}) there is large scatter and no trend for galaxies closer to 
the QSO sight line to have smaller $\eta_{\rm vir}$ values (i.e., more secure 
associations). There is also no significant trend ($1.6\sigma$ positive correlation 
as measured by Kendall's tau test) for associations with larger velocity differences 
(see bottom panel of Figure~\ref{fig:etavir}, where we use a single absorption 
velocity equal to the $N_{\rm H\,I}$-weighted mean of all of the associated 
absorption components when the \lya\ absorber is complex). However, in both plots of 
Figure~\ref{fig:etavir}, the $y$-axis is a measure of the robustness of the 
association only on the sky plane, which the original association criteria (i.e., 
choosing the closest galaxy by normalized impact parameter) constrains to be 
$\lesssim1$. Thus, this plot does not measure fully the robustness of the 
serendipitous associations.

\begin{figure}[!t]
\plotone{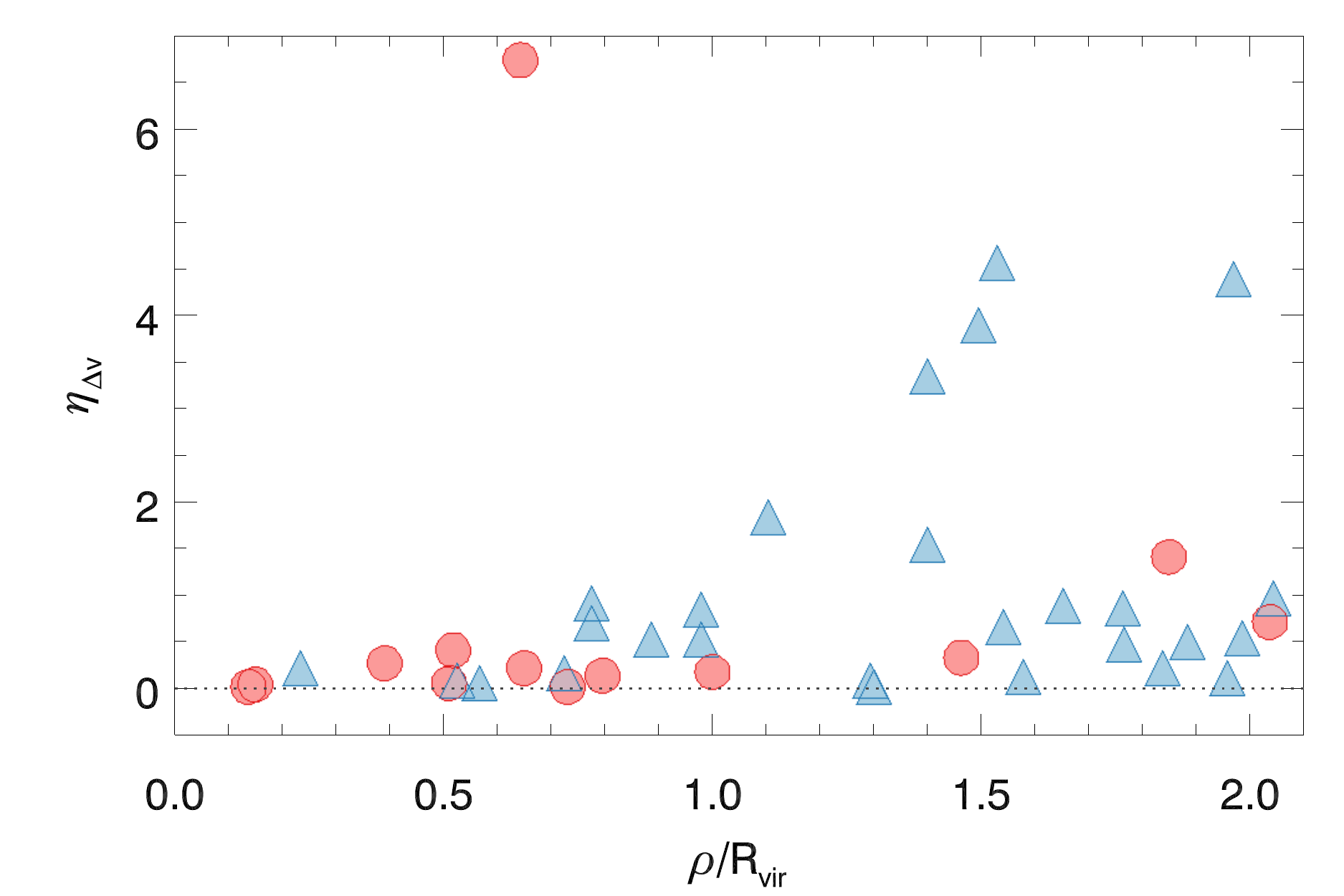}
\caption{The ratio of the normalized absorber-galaxy velocity difference ($|\Delta v|/v_{\rm esc}$) for the nearest galaxy compared to the normalized velocity difference for the next-nearest galaxy as a function of normalized impact parameter ($\rho/R_{\rm vir}$). Small values of $\eta_{\Delta v}$ indicate that the next-nearest galaxy is much farther from the absorber in velocity space than the nearest galaxy. Beyond $\sim1.4$ virial radii, $\sim25$\% of the associations are questionable due to the next-nearest galaxy having a much closer velocity match with the absorber than the nearest galaxy.
\label{fig:etavel}}
\end{figure}

Figure~\ref{fig:etavel} shows the ratio of the normalized velocity differences 
for the nearest and next-nearest galaxies, $\eta_{\Delta v}$, plotted against 
normalized impact parameter, $\rho/R_{\rm vir}$, for the nearest galaxy. As 
before the smaller the $y$-axis value, the more secure the association with 
the individual galaxy we have identified. Here there is a clear trend in which 
there are significantly larger $\eta_{\Delta v}$ values at $\gtrsim1.4$ virial 
radii. This means that while the next-nearest neighbor is farther from the absorber 
on the sky, it is significantly closer in radial velocity difference, making these 
associations more ambiguous.

\clearpage
\begin{deluxetable*}{lcDcccDccl}
\tablecaption{Next-Nearest Galaxy Properties for Passive Galaxy Associations\label{tab:passive}}
\tablewidth{0pt}
\tablehead{ \colhead{Nearest Galaxy} & \colhead{$cz_{\rm ng}$} & \twocolhead{$L_{\rm ng}$} & \colhead{NG Type} & \colhead{$(\rho/R_{\rm vir})_{\rm ng}$}  & \colhead{NNG Type} & \twocolhead{$L_{\rm nng}$}  & \colhead{$\eta_{\rho}$} & \colhead{$\eta_{\Delta v}$} & \colhead{Comments}  \\  & \colhead{(\kms)} & \twocolhead{($L^*$)} & & & & \twocolhead{($L^*$)} }

\decimalcolnumbers
\startdata
SDSS~J122950.57+020153.7   &   1775  &  0.006  &  dSph  &  1.53  &  SBb   &  0.43   &  0.31  &  4.55  &  Dwarf post-starburst; rich group member \\
SDSSJ~111906.68+211828.7   &  41428  &  1.2    &  Sa    &  0.72  &  S0    &  0.92   &  0.22  &  0.16  &  BLA and broad-\OVI\ present; rich group member \\
SDSS~J130101.05+590007.1   &  13862  &  0.47   &  S0    &  0.98  &  dSph  &  0.063  &  1.55  &  0.53  &  Broad-\OVI\ present \\
SDSS~J215517.30--091752.0  &  21951  &  2.7    &  E     &  1.98  &  SBa   &  0.67   &  1.47  &  0.77  &  Small group member; metal-free absorbers \\
2MASX~J13052094--1034521   &  28304  &  3.4    &  E:    &  1.29  &  S:    &  0.33   &  0.91  &  0.08  &  Broad-\OVI\ present \\
\enddata

\end{deluxetable*}

One complication for this plot is that the $N_{\rm H\,I}$-weighted mean absorber 
velocity is used. For some complexes of absorbers, the velocity spread of the 
observed components is large enough that an individual component velocity may 
coincide with the associated absorber velocity even if the mean velocity does not
(see discussion of individual cases in Section~\ref{indiv:galaxies} of the
Appendix). The one targeted absorber/galaxy pair with large 
$\eta_{\Delta v} = 6.7$ is the PMN~J1103--2329\,/\,NGC~3511 association discussed
above, which has one velocity component coincident with the galaxy recession 
velocity and another at $\Delta v = 80$~\kms.

The physical circumstances that create these large values of $\eta_{\Delta v}$ are 
the combination of a small galaxy close to the sight line, which we identify as the 
associated galaxy, and a larger, more luminous and massive galaxy farther away. 
Since the more massive galaxy has a larger estimated escape velocity, its 
$|\Delta v|/v_{\rm esc}$  is smaller. These large $\eta_{\Delta v}$ values in 
Figure~\ref{fig:etavel} are unlikely to be due to incomplete galaxy survey work 
along these sight lines. Two of the serendipitous absorbers with the largest 
$\eta_{\Delta v}$ values are 3C273\,/\,1585 and Q~1230+0115\,/\,1489, which are 
both in regions surveyed completely to well below $0.01\,L^*$. 

This type of ambiguity occurs $\sim25$\% of the time for absorbers with associated 
galaxies at $\rho/R_{\rm vir} \geq 1.4$, but with some lesser ambiguity at 
$1 < \rho/R_{\rm vir} < 1.4$. Although the sample size is small, we conclude that 
claimed associations are robust for absorbers found within the virial radius of 
individual galaxies but become increasingly uncertain beyond one virial radius. 
While the virial radius does not seem to provide any firm physical boundary for the 
CGM \citep[][and see Section 3.1 in Paper~1]{shull14}, operationally it appears to 
be a good, rough estimate for the maximum extent to which individual galaxies can 
be associated with CGM absorbers.

\subsubsection{Do Passive Galaxies Possess Cool CGM Clouds?}
\label{discussion:association:passive}

One of the more intriguing results from the COS-Halos \citep{tumlinson11} study 
of the cool gas CGM of low-redshift galaxies is that luminous (see 
Figure~\ref{fig:coshalos}), passive galaxies with very low sSFR have \HI\ \lya\ 
absorption in their halos \citep{thom12}. 
\edit1{ Unlike star-forming galaxies in the COS-Halos study, most of these galaxy halos are not detected in \OVI, the targeted metal ion \citep{tumlinson11}. However, some are detected in low-ionization absorption \citep[e.g., \MgII, \SiII, \CII;][]{werk13}, }
and \MgII\ absorption has been detected in luminous early-type galaxies at 
intermediate redshifts in some cases \citep{zahedy16}. Both the presence and 
origins of this cool gas are uncertain and problematical since in-falling cool CGM 
gas clouds are thought to fuel new star formation in the galaxy disk and winds 
produced by recent star formation are thought to create outflowing and recycling 
CGM clouds in galaxy halos. Neither seems to be the case for passive galaxies.

In the current serendipitous sample there are only five apparently associated 
galaxies which have very low-sSFR (${\rm sSFR} < 10^{-11}~{\rm yr^{-1}}$; 
see Figure~\ref{fig:coshalos}) based on the non-detection of \Ha\ in our galaxy 
survey data \citep{keeney17}. If the cool gaseous halos of early- and late-type 
galaxies are similar we would expect that a comparable number of bright early-type 
and late-type galaxy halos would be detected since they are comparably numerous 
in the low-$z$ universe based on the SDSS luminosity functions of red and blue 
galaxies shown in \citet{montero-dorta09}. But, instead, only $\sim15$\% (5/35) 
of the associated galaxies in our sample are \lya\ detections in very low-sSFR 
galaxies. No very low-sSFR galaxies were targeted by the COS GTO observations.

The primary concern mentioned in Paper~1 about the potential association of 
low-sSFR galaxies with absorbers is that early-type systems are typically found 
in rich groups or clusters. This makes it likely that other possible 
associated galaxies with high sSFR can be nearby. However, this does not diminish
concerns as to why this gas does not fall into the passive galaxy stimulating star
formation. 
\edit1{The five absorbers in Table~\ref{tab:absprop} that are associated 
with very low-sSFR galaxies are listed in Table~\ref{tab:passive} keyed by the
associated galaxy name. Both the basic properties of the associated galaxy and the
next nearest galaxy (NNG) are listed with column headings as defined in 
Table~\ref{tab:ser_galprop}. The first entry (SDSS~J122950.57+020153.7) is a dwarf 
post-starburst galaxy near the 3C~273 sight line, which has been extensively 
modeled and discussed by \citet{stocke04} and \citet{keeney14}. Since this galaxy 
is quite different from the others in Table~\ref{tab:passive}, it will not be 
discussed further here.
}

\edit1{
Three of the remaining associations are at $\rho < 1.4\,R_{\rm vir}$ for which our 
analysis above suggests that associations should be secure. Two of these 4 are 
identified as being members of small galaxy groups and there is a broad \OVI\ + 
BLA absorber that has been identified as group gas near SDSS~J111906.68+211828.7 
\citep{stocke14}. Two other absorbers with passive galaxy associations possess 
broad \OVI\ absorption. In two of the four cases the NNG is also an early-type 
galaxy so that for these two the only association ambiguity is between an 
individual passive galaxy (either the nearest galaxy or the NNG) or with an 
entire group of galaxies. The last two absorbers in Table~\ref{tab:passive} have 
the largest impact parameters and have NNGs which are late-type star forming 
galaxies which could be the associated galaxy; e.g., these two have  
$\eta_{\rho} \geq 1$ but are not so well-matched in velocity with the absorber. 
For these two the concerns of Paper~1 seem valid that an alternative late-type 
galaxy association is plausible.
}

\edit1{
But based on the two firmest associations in Table~\ref{tab:passive} 
(SDSS~J111906.68+211828.7 and SDSS~J130101.05+590007.1) we confirm the result 
of \citet{thom12} that \lya\ absorption is associated with some early-type, 
low-$z$ galaxies. Unlike \citet{thom12} we find metal absorption, including \OVI,
associated with at least one \HI\ velocity component in all but one case in 
Table~\ref{tab:passive}. The two-component \lya-only absorber 
PHL~1811\,/\,21998, 22042 is almost 2 virial radii from the nearest luminous 
galaxy, SDSS~J215517.30--091752.0, and could be primordial infall onto this 
galaxy and/or others in the region.
}

\edit1{
While the very low, \textit{current} sSFR of these galaxies makes them typical of 
early-type galaxies in the local universe, only one of these four luminous passive 
galaxies has sensitive \galex\ upper limits, confirming that this galaxy 
(SDSS~J215517.30--091752.0) is fully passive. For 2MASX~J13052094--1034521, 
weak \galex\ UV detections require a longer-term sSFR of $\log{\rm sSFR} = -11.3$; 
the other two galaxies have no \galex\ observations. 
} If these two and the 16 COS-Halos passive galaxies of \citet{thom12} all lack 
significant UV flux, then these \lya-detected passive galaxies are rather typical 
red galaxies in which star formation has largely, if not completely, ceased. And 
yet they possess cool gas clouds in their halos. Near-UV imaging with \hst/WFC3 can 
measure their longer-term star formation rate and resolve this outstanding 
question. Similar observations of the \citet{thom12} galaxies can determine if 
CGM-selected passive galaxies had star forming events too long ago to show \Ha\ 
emission. \lya\ absorption-selected passive galaxies can provide new insights into 
the evolutionary progress of galaxies across the ``green valley''. Their further 
study is warranted for understanding how star formation is quenched in galaxies.

Three of the five absorbers in this sample are complexes of two or three \lya\ 
velocity components; at least one of these components is metal-rich in each case. 
These three have strong, broad \OVI\ which is likely to be ``warm'' ($T>10^5$~K) 
gas associated with the entire galaxy group \citep{stocke14}. One of these 
(PG~1116+215\,/\,41428) was described in \citet{stocke14} as a ``warm'' absorber 
in a small galaxy group. The warm gas in these groups is postulated to be interface 
gas between the cool, photo-ionized gas clouds in the CGM and a hotter intra-group 
medium in the process of developing in these galaxy groups \citep{stocke14}. 
Similar ``warm'' absorption could be present in the \citet{thom12} absorbers 
associated with passive galaxies but the COS-Halos spectra have insufficient 
signal-to-noise to detect the broad, shallow \OVI\ and \lya\ that we predict is 
present.

The hypothesis which arises out of this analysis is that the hot intra-group gas 
may be inhibiting the accretion of these cool gas clouds onto the associated 
galaxy so that no on-going star formation is induced by CGM gas in these 
cases. The source of this cool CGM gas could be either a past star formation event 
in the passive galaxy or a wind from a lower luminosity, even dwarf galaxy like 
in the case of the 3C~273 dwarf post-starburst galaxy. After being ejected from 
a group galaxy, the cool gas clouds in small galaxy groups may not accrete easily 
onto passive galaxies in their vicinity. They remain ``warm'' 
($5<\log{T}<6.5$)\footnote{All temperatures reported in logarithms 
have units of Kelvin.} due to being shock-ionized by collisions with other CGM 
clouds or through contact with the hotter intra-group medium which is in the 
process of developing in these groups. New \hst/COS observations of nearby galaxy 
groups can be used to test this hypothesis (\hst\ Cycle~23 project \#14277; J. 
Stocke, PI).

\subsection{CGM Properties and SFR Correlations}
\label{discussion:SFR}

Because there is a strong dichotomy in CGM properties between star-forming and 
``passive'' galaxies in the COS-Halos study \citep{tumlinson11,thom12}, it might 
be expected that correlations between basic CGM properties and SFR/sSFR would be 
present in our sample. Since one source of CGM gas is ejected and/or recycling gas 
produced by current or longer-term star formation in the disk, CGM/galaxy disk 
correlations would be expected. On the other hand, QSO probes through the CGM of 
individual galaxies are anecdotal in that they sample only a specific pencil beam 
through each galaxy's gaseous halo; i.e., while the CGM of disk galaxies with high 
and low SFRs might possess quite different bulk CGM properties, individual sight 
lines may not reflect accurately such differences. Further, there may be timing 
differences between the observed SFR and the gaseous content of the CGM so that 
current CGM properties may be correlated with past, rather than current, star-forming 
activity.

At high-$z$, the \CII* $\lambda$1335.7~\AA\ transition can derive the sSFR of DLA 
host galaxies \citep*{wolfe03a, wolfe03b}, but these values were not used 
individually but rather as averages in redshift bins over many systems to verify 
the SFR vs. redshift relationship \citep[a.k.a. the ``Madau plot'';][]
{madau96,steidel99}. Therefore, while correlations between CGM absorption 
properties and associated galaxy SFR would be interesting, if no correlation is 
present we cannot conclude that the CGMs of high- vs low-SFR galaxies are similar.

\begin{figure}[!t]
\plotone{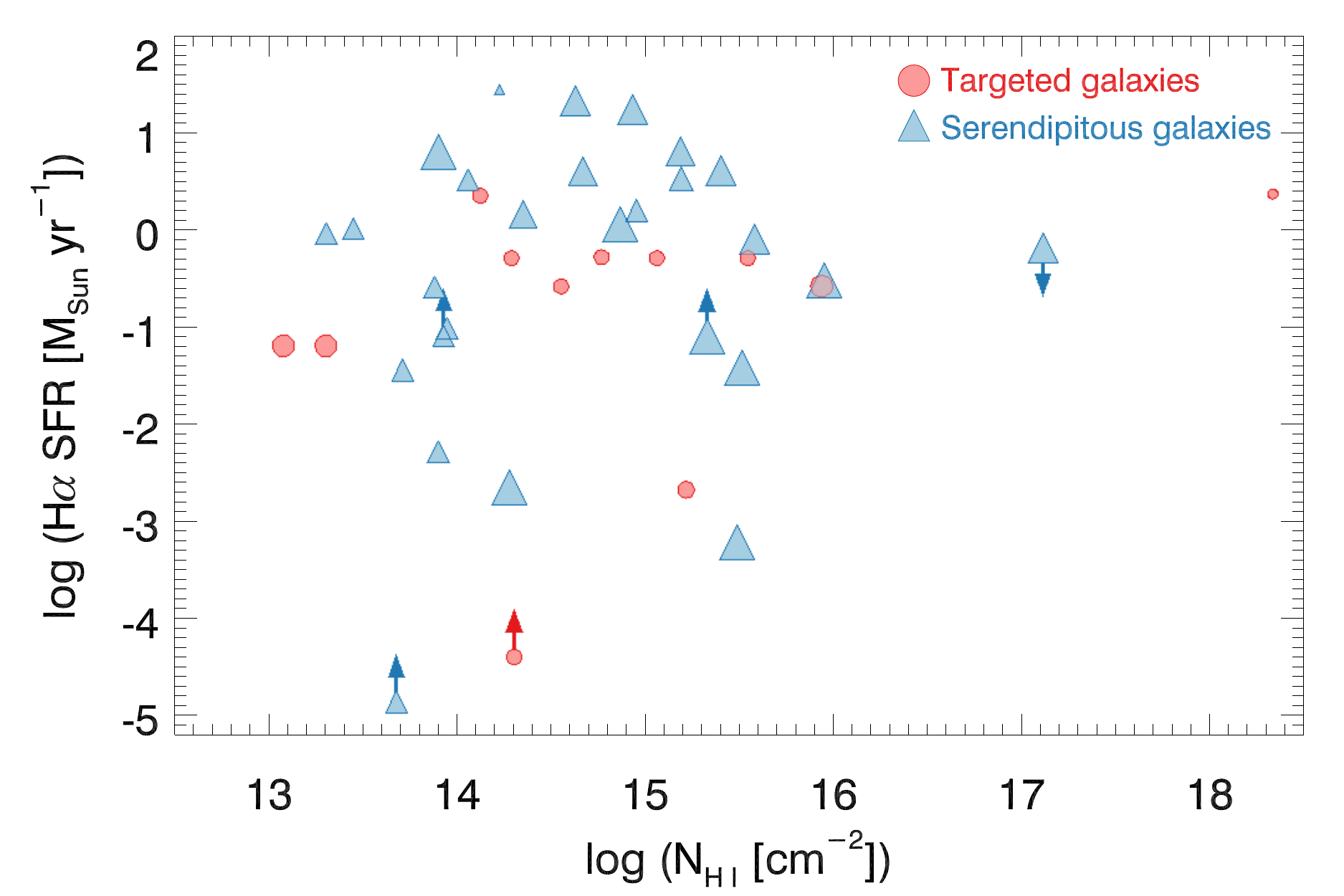}
\plotone{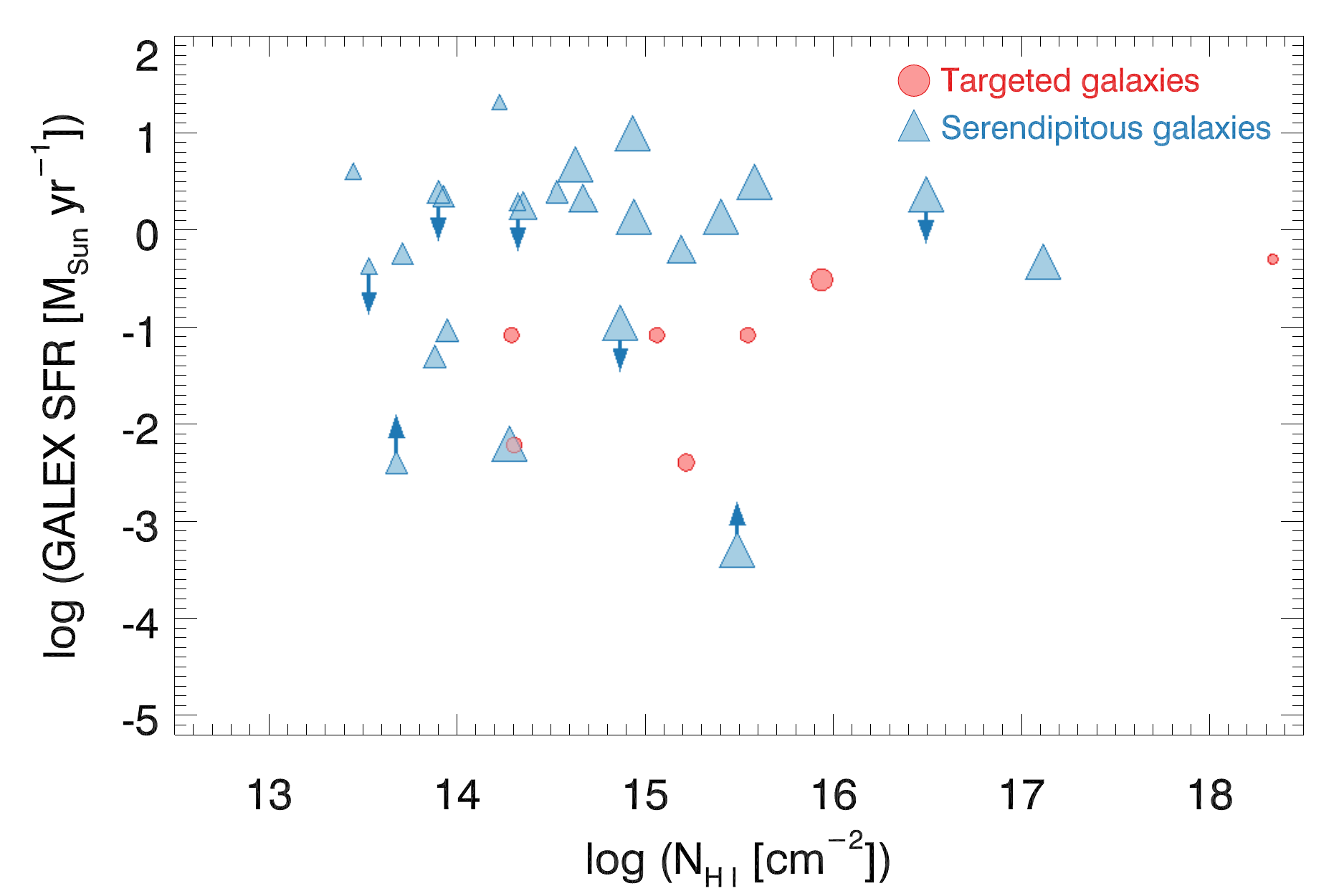}
\caption{Galaxy SFR as measured by \Ha\ (top) and \galex\ FUV continuum (bottom) luminosity as a function of total CGM \HI\ column density. No statistically significant correlation is evident.
\label{fig:sfr_HI}}
\end{figure}

\begin{figure}[!t]
\plotone{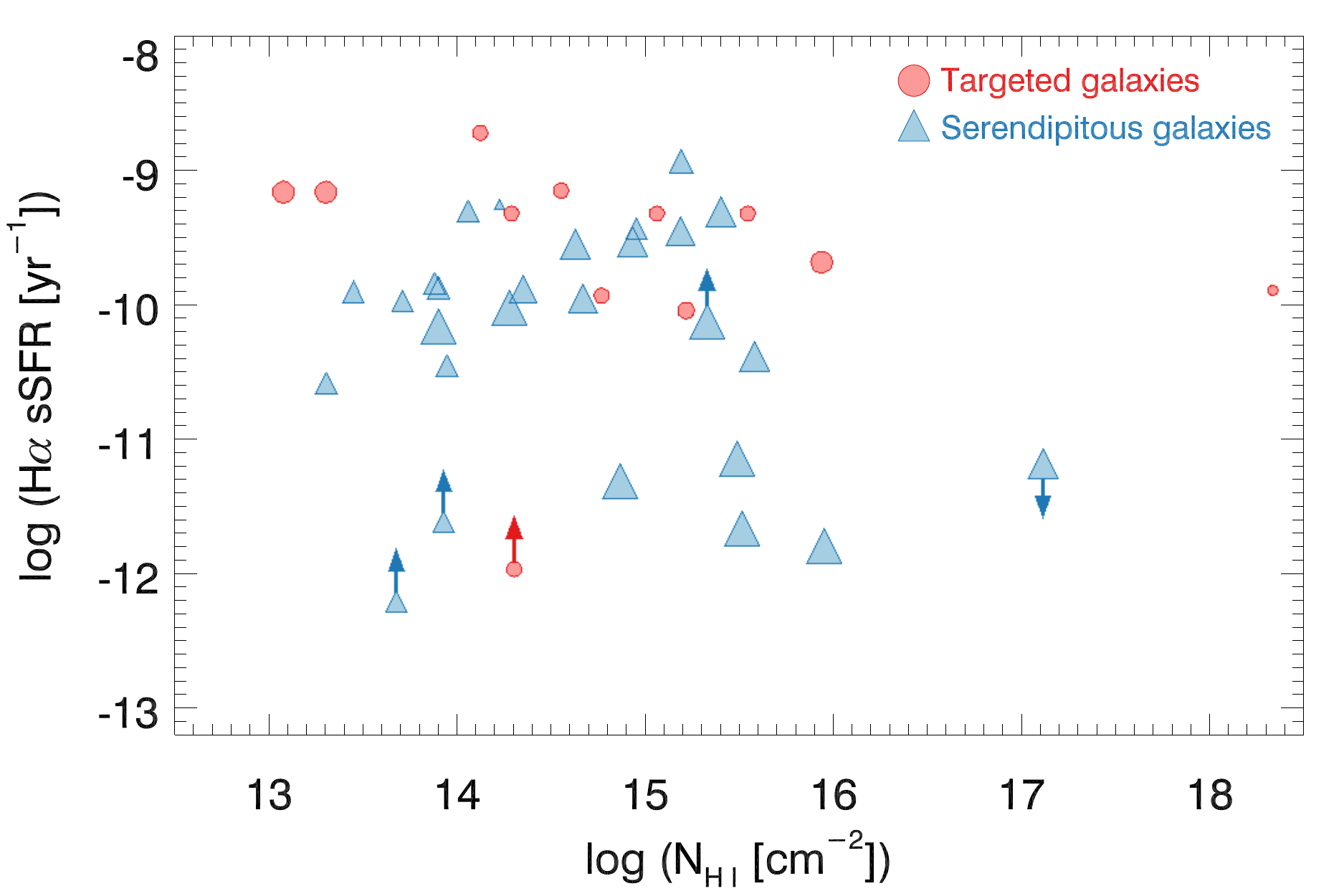}
\plotone{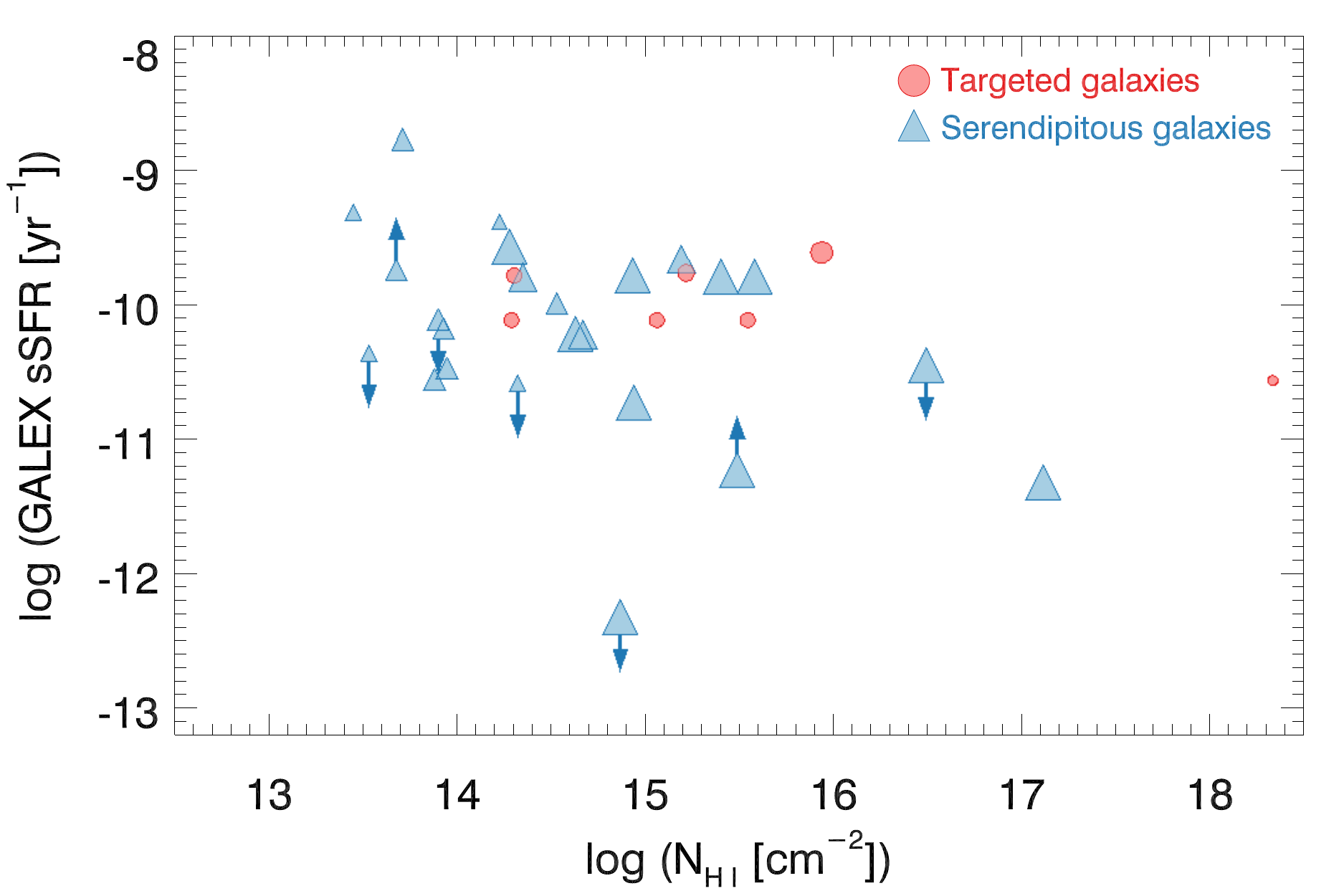}
\caption{Galaxy sSFR as measured by \Ha\ (top) and \galex\ FUV continuum (bottom) luminosity as a function of total CGM \HI\ column density. No statistically significant correlation is evident.
\label{fig:ssfr_HI}}
\end{figure}

Figure~\ref{fig:sfr_HI} shows the relationship between SFR determined by \Ha\ 
line emission (top) and by \galex\ FUV continuum emission (bottom) with 
$\log{N_{\rm H\,I}}$. No correlations are found except perhaps a very shallow 
decrease in $\log{N_{\rm H\,I}}$ with increasing \Ha\ SFR. A similar absence of 
correlations are found by using sSFR (Figure~\ref{fig:ssfr_HI}). Because our 
sample shows no trend of \HI\ column density with impact parameter, we make no 
re-normalization of the column density data to a specific impact parameter. 
However, we have used the cumulative $N_{\rm H\,I}$ values for complex 
absorbers described in the previous paragraph, thus summing all of the CGM 
clouds along each sight line.

While the \HI\ column density is one of a handful of direct observables, we also 
searched for correlations between SFR and the two model parameters emerging 
from the CLOUDY modeling: ionization parameter and absorber metallicity. While this 
greatly reduces our sample size (23 absorbers have good quality CLOUDY model fits 
but not all of these have good SFR estimates), a case can be made that these 
derived CGM quantities may be more closely tied to current or recent SFR. We are 
left with $\sim12$ sample points only and no correlations between nearest galaxy 
SFR and either ionization parameter or metallicity is present. We conclude that 
individual sight line CGM properties cannot be used as indicative of the entire 
CGM condition in any one galaxy probed.

\section{Conclusions}
\label{conclusion}

In this paper we have presented the basic data used to determine specific CGM cool 
cloud properties and ensemble cool ($T\sim10^4$~K) CGM gas mass estimates 
associated with low-$z$ galaxies as first presented in Paper~1. We refer the 
reader to \citet{savage14} and \citet{stocke14} for a detailed discussion of the 
\OVI\ absorption and the ``warm'' gas phase since in this paper and Paper~1 we 
concentrate on the cool, photo-ionized gas.

These data include the \HI\ and metal-line absorption (Section~\ref{absorbers}) 
seen in FUV spectra of bright AGN taken with \hst/COS, \hst/STIS and \fuse. 
Detailed line fits and tabulated column densities are presented in 
Figure~\ref{fig:stackplot} and Table~\ref{tab:absprop}. The sample analyzed herein 
includes a ``targeted'' set of sight-line/galaxy pairs observed by the COS GTO 
team using \hst/COS with the G130M and G160M gratings. A second sample of 
``serendipitous'' sight-line/galaxy pairs uses \hst/COS, \hst/STIS and \fuse\ FUV 
spectra of very bright AGN whose detected \HI\ absorptions at $z \leq 0.2$ lie 
within $\sim2$ virial radii of a foreground galaxy in our extensive database of 
$\sim700$ galaxies located $<1$~Mpc from these sight lines. These samples nicely 
complement the two samples of galaxies constituting the COS-Halos and COS-Dwarfs 
surveys \citep{tumlinson11, bordoloi14}. Basic data for the associated galaxies 
also are presented in Section~\ref{galaxies}, including redshifts, luminosities, 
metallicities and star-formation rates determined both by \Ha\ imaging (see 
Appendix~\ref{imaging}) and by \galex\ UV imaging, as well as inferred properties 
such as virial radius, stellar mass and halo mass (see 
Tables~\ref{tab:targ_galprop} \& \ref{tab:ser_galprop} in Section~\ref{galaxies}). 
Single-phase, homogeneous CLOUDY photo-ionization models are presented in 
Section~\ref{cloudy}; when multi-phase gas is present we attempt to model only the 
coolest phase detected (i.e., the lowest ions).

\clearpage
The following major results are obtained:
\begin{enumerate}
\item We find that associations between absorbers and an individual, nearest 
galaxy are robust, and almost always unambiguous, \textit{if the sight line lies 
within the virial radius of the galaxy}. These associations become less secure at 
larger impact parameters and are sometimes ambiguous at $\rho > 1.4\,R_{\rm vir}$, 
where the absorbers are often more appropriately linked to several nearby galaxies 
or to an entire small galaxy group (see Section~\ref{discussion:association}).

\item We find no evidence for increasing ionization parameter or declining \HI\ 
column density, line-of-sight cloud thickness, cloud mass, or cloud pressure with 
increasing impact parameter. These results differ from the trends found by 
\citet{werk14} from the COS-Halos sample, but are not inconsistent because these 
two studies probe different ranges of impact parameter: $\rho < 0.5\,R_{\rm vir}$ 
for COS-Halos and $0.5\,R_{\rm vir} < \rho < 2\,R_{\rm vir}$ for this study. The 
associated galaxies also differ since the COS-Halos galaxies were selected to be 
both luminous and isolated while the current sample has no such restrictions. 

\item We find no correlations between current or longer term (i.e., $\leq1$~Gyr) 
SFR and CGM absorber parameters like $N_{\rm H\,I}$. This is expected since it 
seems quite unlikely that a single, pencil-beam probe of the CGM can adequately 
constrain its bulk properties.

\edit1{
\item We have found at least two, and possibly four, serendipitously discovered absorbers with \HI\ and metal-lines associated with passive galaxies (Section~\ref{discussion:association:passive}), in support of the COS-Halos discoveries of such unexpected associations \citep{thom12}. Since the presence of cool gas in passive galaxy halos is unexpected, these absorption-line detected, passive galaxies deserve further study.
}

\edit1{
\item Using the metal-rich absorbers detected in these \hst/COS and \hst/STIS spectra as a fair sampling of the CGM of low-$z$ galaxies, we confirm the ensemble CGM cool cloud mass obtained by Paper~1 of $\log{(M/M_{\Sun})} = 10.2\pm0.3$. Since we have employed a different line-fitting method than used in Paper~1, this result suggests that these mass estimates are robust to the details of the data analysis. 
}

\edit1{
\item We summarize the statistical understanding of the cool, photo-ionized CGM obtained from this study of a diverse set of low-$z$ galaxies as follows. The cool cloud population of a typical $L \geq L^*$, star-forming galaxy consists of a few thousand individual clouds at sizes of $>1$~kpc (there are likely $>10,000$ smaller clouds with sizes $200~{\rm pc} \leq D_{\rm cl} < 1~{\rm kpc}$) that fill $\sim5$-9\% of the CGM volume inside the virial radius of these galaxies. The 300-500 largest ($\geq20$~kpc), most massive ($\geq10^7~M_{\Sun}$) clouds dominate the ensemble CGM mass but are detected only $\gtrsim20$\% of the time along random sight lines through the volume inside the virial radius. Thus, the number of these most massive clouds in any absorption-line probe of the CGM is small, which leads to sparse sampling statistics and large uncertainties in ensemble mass ($\pm0.3$~dex estimated herein for the super-$L^*$ sample). Lower mass star-forming galaxies have similar CGM properties (scaled down by galaxy mass), although dwarfs have lower cloud covering factors, as detailed in Paper~1.
}

\edit1{
\item While the statistical error budget on the determination of the CGM cool gas mass is large (factor of $\sim2$) , both for this study and COS-Halos, the \citet{werk14} ``preferred lower limit'' and the latest COS-Halos value of nearly $10^{11}~M_{\Sun}$ \citep{prochaska17} suggest a very different picture of the CGM than the current result. Much of the difference in these values can be attributed to the considerably different low-$z$ ionization rate assumed; this study assumed \citet{haardt12} while COS-Halos assumes the higher values of \citet{haardt01}. While it is beyond the scope of this paper to determine which of these two ionization rates (if either) are correct for CGM absorbers, we have instead corrected the latest COS-Halos CGM cool gas mass estimate to the \citet{haardt12} ionization rate and then used both results in the radial regime for which each best-samples the CGM (COS-Halos at $\rho \leq 0.5\,R_{\rm vir}$ and the present study outside that radius). By this method, we obtain a ``best-value'' for the CGM cool gas mass of $\log{(M/M_{\Sun})} = 10.5\pm0.3$, amounting to $\sim30$\% of the total baryon inventory of $L \geq L^*$ spiral galaxies. 
}
\end{enumerate}

We further suggest that many of the \OVI\ absorbers, cataloged but not studied 
herein, are likely detections of ``warm'' or ``warm-hot'' interface gas at 
$T \approx10^5$-$10^{6.5}$~K \citep{savage14, stocke14}. While this 
\OVI-absorbing gas probably does not contain sufficient baryons to account for 
all those ``missing'', the abundance of these absorbers at $z\sim0$ 
\citep[$dN/dz\approx4$;][]{stocke14} attests to the likely presence of large 
amounts of even hotter ($T \approx 10^6$-$10^7$~K) gas that may permeate small, 
low-redshift, spiral-rich galaxy groups. Despite \OVI\ being a minority tracer of 
gas at those temperatures, the lack of high-resolving-power spectrographs at soft 
X-ray wavelengths means that \hst/COS remains the most sensitive way to detect 
such gas currently through broad, shallow \OVI\ and \lya\ absorbers.

\acknowledgements

\edit1{We wish to thank the anonymous referee for comments that stimulated us to improve the quality of the final manuscript.}
This work was supported by NASA grants NNX08AC146 and NAS5-98043 to
the University of Colorado at Boulder for the \hst/COS project. BAK
and JTS gratefully acknowledge additional support from NSF grant
AST1109117. This research has made use of the NASA/IPAC Extragalactic
Database (NED) which is operated by the Jet Propulsion Laboratory,
California Institute of Technology, under contract with the National
Aeronautics and Space Administration. This research has also made use
of the ``K-corrections calculator'' service available at \url{http://kcor.sai.msu.ru/}. We thank J.~Rosenberg and E.~Ryan-Weber, 
who assisted in selecting the targeted QSO/galaxy pairs in this study, 
and J. Moloney for assistance with the cosmological simulations of 
\citet{shull15}.

\facilities{HST (COS, STIS), ARC (SPIcam, DIS), Sloan, GALEX, KPNO:2.1m, CTIO:0.9m}

\appendix

\section{Notes on Individual Cases}
\label{indiv}

\subsection{Absorption-Line Fits}
\label{indiv:absorbers}

In the descriptions below each absorber discussed is
identified by the sight line target name/recession velocity (\kms).

\subsubsection{1ES~1028+511\,/\,967}
\label{indiv:absorbers:1es1028_967}

Paper~1 listed a column density of $\log{N_{\rm H\,I}} =
17.21^{+0.22}_{-3.20}$  for this absorber, but Table~\ref{tab:absprop}
lists $b\approx20$~\kms\ and $\log{N_{\rm H\,I}} \approx 14.3$. The
column densities formally overlap to  within the quite large
uncertainties listed in Paper~1 but our re-analysis finds that
$\log{N_{\rm H\,I}}>17$ requires $b<8$~\kms, which cannot  be resolved
using COS data with a velocity resolution of 17~\kms.

\subsubsection{FBQS~J1010+3003\,/\,1264 \& 1380}
\label{indiv:absorbers:f1010_1380}

Similar to the case above, Paper~1 lists $\log{N_{\rm H\,I}} =
17.79^{+0.11}_{-3.48}$ and Table~\ref{tab:absprop} lists
$b\approx30$~\kms\  and $\log{N_{\rm H\,I}}\approx14.5$. Here the
\lya\ absorption is located  near Ly$\gamma$ absorption at
$z=0.2546,0.2549$ \citep{danforth16} and whether  a
high-column-density solution is allowed is somewhat dependent on
assumed  component structure. We have adopted a two-component,
low-column-density solution here, matching the fits of 
\citet{danforth16}, as the lack of any metals associated
with an absorber with $\log{N_{\rm H\,I}}\sim18$ would be remarkable.

\subsubsection{HE~0439--5254\,/\,1581, 1653, 1763 \& 1805}
\label{indiv:absorbers:he0439_1653}

This is another case where the assumed component structure
matters. Paper~1 and  \citet{keeney13} fit a total of two component in
this region, with the strong one  centered at 1662~\kms\ with
$\log{N_{\rm H\,I}} = 14.36^{+0.13}_{-0.07}$. Here we  fit a total of
four components, with one BLA to account for the very red wing at
$\lambda>1223.2$~\AA\ and a weak metal-free component at
1581~\kms. This allows for a narrower, higher column density component
at 1653~\kms\ than found in  Paper~1, which is preferred by the
photo-ionization model in Section~\ref{indiv:cloudy:he0439_1653}.

\subsubsection{PG~0832+251\,/\,5221, 5337, 5396 \& 5444}
\label{indiv:absorbers:pg0832_5221}

This system shows extremely broad, saturated \lya\ absorption with
asymmetric  wings, suggesting multiple velocity components. In Paper~1
we fit this profile  with two components at 5227 and 5425~\kms, the
first of which had a column  density of $\log{N_{\rm
H\,I}}\approx18.5$. While a \fuse\ spectrum exists for this sight line
that shows \OVI\ absorption at $\sim5221$~\kms, its S/N is quite low
and cannot help constrain the \HI\ component structure. Here we are
basing our component structure on the metal-line velocity components,
which  show four consistent components in low-ionization species. We
adopt these  component velocities for \HI\ and hold them fixed while
fitting the \lya\ profile.  Unfortunately, the \lya\ profile is
completely saturated for the two interior  components so we are unable
to constrain their Voigt profile parameters with any confidence. The
outermost components can be constrained by the blue and red
wings of the \lya\ profile, but the parameters for the 5441~\kms\
component are still quite uncertain. The specified \lya\ parameters 
are consistent with the low S/N \lyb\ detection in the \fuse\ data.

\subsubsection{PMN~J1103--2329\,/\,1113 \& 1194}
\label{indiv:absorbers:p1103_1194}

Paper~1 fits this profile with a single velocity component but we use
two to better fit the blue wing of the profile. The column densities of 
the 1194~\kms\ component are consistent to within uncertainties regardless
of which component structure is adopted.

\subsubsection{RX~J0439.6--5311\,/\,1638, 1674 \& 1734}
\label{indiv:absorbers:rxj0439_1674}

Paper~1 \citep[and][]{keeney13} fit this profile with a
single \lya\  component at 1672~\kms, while we fit it here with
multiple components, including  one BLA, to better fit asymmetries in
the wings of the \lya\ line profile. These  components are heavily
blended so the inferred Voigt profile parameters are quite
uncertain. As with the HE~0439--5254 absorption system at similar
velocity  (Section~\ref{indiv:absorbers:he0439_1653}), the higher
column density afforded  by the multi-component fit is preferred by
the photo-ionization models of  Section~\ref{indiv:cloudy:rxj0439_1674}.

\subsubsection{SBS 1108+560\,/\,654, 715 \& 778}
\label{indiv:absorbers:sbs1108_654}

This sight line presents a unique challenge because a Lyman-limit
system (LLS) at  $z=0.4634$ dramatically reduces the observed flux at
$\lambda\lesssim1340$~\AA.  Above this cutoff, the spectrum exhibits
quite high S/N, while below it the S/N  is effectively zero in some
regions. As with the PG~0832+251 system in
Section~\ref{indiv:absorbers:pg0832_5221} we use the associated
low-ionization metals to constrain the \HI\ component structure,  
and attempt to fit the \lya\ profile
holding these velocities fixed. However, the very poor S/N in this
region of the spectrum, combined with the proximity to the  Galactic
DLA absorption at such low redshift, renders us unable to meaningfully
constrain the \HI\ column density for this absorber. In Paper~1 we did
report \HI\  column densities for two velocity components at 665 \&
778~\kms, but our  re-analysis cannot support those values.

\subsubsection{SBS~1122+594\,/\,1221}
\label{indiv:absorbers:sbs1122_1221}

The column density listed for this absorber in Paper~1 ($\log{N_{\rm 
H\,I}} =  17.71^{+0.35}_{-2.85}$) agrees with the value in
Table~\ref{tab:absprop} to  within the large uncertainties. The higher
column density does not fit the blue  wing of the \lya\ profile as
well as the lower column density does, but  both are acceptable fits
to the data.

\subsubsection{VII~Zw~244\,/\,715}
\label{indiv:absorbers:pg0838_715}

As with the previous absorber, we report a lower column density than
Paper~1, but  both solutions are acceptable fits to the data. We
prefer the lower column  density solution in this case because it
provides a better fit to the red wing of the \lya\ profile.

\subsubsection{3C~273\,/\,1019}
\label{indiv:absorbers:3c273_1019}

As discussed in \citet{savage14} and \citet{stocke14} this absorber includes a
BLA and associated broad \OVI\ absorption. \citet{stocke14} and \citet{yoon12} 
identify this absorber with a very rich foreground group on the southern
outskirts of the Virgo Cluster.

\subsubsection{3C~273\,/\,1585}
\label{indiv:absorbers:3c273_1585}

This absorber was analyzed extensively by \citet{tripp02} and
\citet{sembach01} based on STIS and \fuse\ spectroscopy, respectively.
The \HI\ fit in Table~\ref{tab:absprop} finds a slightly higher
$b$-value and  lower column density than those of \citet{sembach01}
and our metal-line  values also differ slightly, owing primarily to
the increased S/N of the COS spectrum we analyze.  Detailed studies of
the associated galaxy by \citet{stocke04} and \citet {keeney14} make a plausible
case for the association of this absorber with a dwarf post-starburst
galaxy in a galaxy filament near the Virgo Cluster, although ambient
filament gas is also probably necessary to understand its detailed
properties \citep{keeney14}. Gas at a comparable recession velocity
was found in the nearby sight line Q~1230+0115; a discussion of the
physical sizes of these absorbers can be found in
\citet{rosenberg03}. Single-phase, homogeneous photo-ionization models
of this absorber by \citet{keeney14} differ only slightly from the
original \citet{tripp02} model in being at slightly higher metallicity
($\sim15$\% solar abundance) and larger size ($\sim130$~pc), primarily
based on better upper limits on \SiIV\ and \CIV\ in the COS data. The
photo-ionization models presented in Section~\ref{indiv:cloudy:3c273_1585},
which use an  updated methodology, suggest a higher metallicity of
$35\pm10$\% solar abundance  and a line-of-sight cloud size of
50-100~pc, but given the systematic  uncertainties in photo-ionization
modeling the three models agree remarkably well.

\subsubsection{H~1821+643\,/\,36139, 36307, 36339, 36439, \& 36631}
\label{indiv:absorbers:h1821_36339}

This absorption complex was analyzed by \citet{tripp01} who concluded
that the higher redshift component contains a BLA and an associated
broad  \OVI\ line that is almost certainly due to
collisionally-ionized gas at  $\log{T}\approx5.5$ 
\citep[see also][]{savage14}. Our simultaneous  fit to the \lya\ and \lyb\
profiles suggests that there are three blended  components (two BLAs)
with velocities of 36307, 36339, \& 36439~\kms. This three-component 
solution agrees with the fits of \citet{danforth16}, although Paper~1 and 
\citet{savage14} only fit two components to this profile.

\subsubsection{PG~0953+414\,/\,42512, 42664, 42759 \& 42907}
\label{indiv:absorbers:pg0953_42664}

First studied by \citet{savage02} using rather low S/N STIS data, the
COS spectrum obtained by the COS GTOs has much higher S/N
\citep{savage14}. The new spectrum reveals a total of four components
in \lya, two of which align with rather strong \OVI\ lines \citep[see
Figure~1 in][]{savage14}. \citet{savage14} find that the \lya + \OVI\
absorbers are consistent with cool, photo-ionized gas. In this
compilation we identify two independent \HI\ systems, each associated
with \OVI\ absorption (see Table~\ref{tab:absprop}).

\subsubsection{PG~1116+215\,/\,17614, 17676, 17786 \& 18202}
\label{indiv:absorbers:pg1116_17676}

This system was most recently studied by \citet{savage14} who found
three \HI\ components in \lya\ and two \OVI\ absorbers, one of which
aligns well with one of the \HI\ components. The $b$-values of these
lines suggests photo-ionized gas. According to \citet{savage14} the
other metal line absorptions (\CIV\ and \OVI) do not align well with
the weak \HI\ component allowing no firm conclusions to be drawn about
the physical conditions of this gas. We also fit the \lya\ with three
components and find that, using our wavelength solution, the
$cz=17786$~\kms\ \HI\ aligns reasonably well with the metal
lines. Under the assumption that the \HI\ and metal-line absorption
arise in the same gas, the line widths suggest temperatures consistent
with photo-ionized gas although these detections are weak enough that
there is a considerable range in suggested temperatures.

\subsubsection{PG~1116+215\,/\,41522 \& 41522}
\label{indiv:absorbers:pg1116_41522}

This absorber has been studied extensively in \citet{danforth10},
\citet{savage14} and \citet{stocke14}. While all three papers suggest
that a BLA is present which aligns well with the broad
\OVI\ absorption, \citet{savage14} fit the \lya\ line with three
components and find a significantly smaller $b$-value than the other
analyses. This lower value suggests photo-ionized gas, while the BLA
fit made by \citet{stocke14} found collisionally-ionized gas at
$T\approx400,000$~K. Here we fit the \HI\ lines with two components, 
consistent with the approach of \citet{stocke14}. The two velocity 
components, one broad and one narrow, have best-fit velocities that 
are coincident (see Table~\ref{tab:absprop}).

\subsubsection{PG~1211+143\,/\,15170, 15321, 15357, 15431 \& 15574}
\label{indiv:absorbers:pg1211_15321}

This absorption was studied in detail using the STIS data by
\citet{tumlinson05}. This complex includes several velocity components
which align well with the velocities of nearby galaxies in a rich
group of galaxies. As shown in the plot of the 15321~\kms\ absorber
there is a very broad \OVI\ absorption detected only in the weaker
line of the doublet due to obscuring Galactic lines at
this redshift. A BLA is almost certainly present in the highly
saturated \lya\ profile, which we and \citet{tumlinson05} fit as three
components, only one of which has associated metals.

\subsubsection{PG~1211+143\,/\,19305, 19424, 19481 \& 19557}
\label{indiv:absorbers:pg1211_19305}

This absorption complex was also studied in detail by
\citet{tumlinson05} and it also contains a very strong, very broad
\OVI\ line. \citet{tumlinson05} decomposed this complex into four
components, including a BLA, but our 
simultaneous fit to all available \HI\ lines from STIS and \fuse\ do
not require a BLA (although at least one of the components has 
$b>35$~\kms). In this case these
absorptions have only one nearby galaxy to which they are most likely
associated, although the galaxy survey in this region is not
exceptionally deep ($M_r < -19.5$; $L > 0.2\,L^*$) 
since it is due entirely to the SDSS.

\subsubsection{PG~1216+069\,/\,37049, 37138, 37363 \& 37455}
\label{indiv:absorbers:pg1216_37049}

Paper~1 lists only one velocity component at 37091~\kms\ for this
sight line,  but the COS spectrum clearly shows two components in
\lya\ and \lyb\ near  this velocity as well as two more components
$\sim300$~\kms redward, all of which  have associated metals.

\subsubsection{PG~1259+593\,/\,13825, 13914 \& 14014}
\label{indiv:absorbers:pg1259_13825}

Both of these absorbers contain strong \OVI\ absorption which
\citet{savage02} and \citet{savage14} find to be consistent with
photo-ionization. A third component at 14014~\kms\ is clearly present
in the COS data but was not listed in Paper~1.

\subsubsection{PHL~1811\,/\,24226}
\label{indiv:absorbers:phl1811_24226}

This absorber is the strong LLS studied in detail by
\citet{jenkins03}, who also presented a detailed discussion of the luminous, 
early-type associated galaxy. \citet{savage14} deconvolved a BLA from the
highly-saturated \lya\ absorption that corresponds to a very broad
\OVI\ line, yielding an approximate temperature of 300,000~K in
collisionally-ionized gas. We fit only a single velocity component
to the \lya\ profile and find no strong evidence for associated 
\OVI\ absorption, differing from the \citet{savage14} analysis. 
However, line blending in the \fuse\ data preclude an entirely 
consistent \HI\ solution that satisfies the constraints of all of 
the lines, so the detailed component structure is less certain for 
this absorber than for other absorbers with comparable data. 
Nonetheless, the AGN continuum in the \fuse\ data is completely absorbed 
shortward of the Lyman limit at $cz=24226$~\kms, implying that the 
total \HI\ column among all components is $>10^{17.9}~{\rm cm}^{-2}$, 
consistent with our best-fit value of $\log{N_{\rm H\,I}}=18.08\pm0.04$.

\subsubsection{PHL~1811\,/\,39658 \& 39795}
\label{indiv:absorbers:phl1811_39658}

The absorption tabulated here has a strong, aligned \HI\ +
\OVI\ absorber whose $b$-values suggest cool, photo-ionized gas
\citep{savage14}. But \citet{savage14} also find another three
\OVI\ velocity components, one of which (shifted by
$\sim270$~\kms\ relative to the strong, photo-ionized absorber) lacks
detectable \lya\ absorption, implying a very hot, collisionally-ionized
plasma at $\log{T} > 5.7$ \citep{stocke14}. We fit a total of three \OVI\ 
components and confirm the lack of \HI\ associated with the redmost 
component centered at 39930~\kms.

\subsubsection{PHL~1811\,/\,52914 \& 52933}
\label{indiv:absorbers:phl1811_52914}

This absorber was found by \citet{savage14} to have two \lya\ velocity
components, neither of which align with an observed
\OVI\ absorber. \citet{stocke14} reanalyzed this spectrum and found
that, while \lya\ absorption exists at the \OVI\ velocity (see 
Figure~\ref{fig:stackplot}), the inferred $b$-values of \HI\ and \OVI\ 
suggest cool, photo-ionized gas.

\subsubsection{PKS~0405--123\,/\,28947 \& 28958}
\label{indiv:absorbers:pks0405_28945}

\citet{savage14} fit the \lya\ absorption with a two-component model,
one narrow and one BLA, which we adopt here. Weak \OVI\ found at this 
velocity in the \fuse\ spectrum can be either cool, photo-ionized gas or warm,
collisionally-ionized gas since the velocity separation of the two
\lya\ components is comparable to the uncertainties in registering the
\fuse\ data to the COS data \citep[see also][]{stocke14}.

\subsubsection{PKS~0405--123\,/\,45617, 45783 \& 45871}
\label{indiv:absorbers:pks0405_45410}

Paper~1 lists only two velocity components for this system, but our simultaneous
fit to the \lya\ and \lyb\ profiles suggests that there are three components
total. The bluer system from Paper~1 is now comprised of two weak, blended components.

\subsubsection{PKS~0405--123\,/\,49910, 49946, 50001, \\ 50059, 50104 \& 50158}
\label{indiv:absorbers:pks0405_50104}

The very well-known, strong LLS absorber at 
50104~\kms\ was first studied in \citet{chen00} and \citet{prochaska04}, 
and the much higher S/N COS spectrum is described in detail in \citet{savage10}. 
The latter authors fit this absorption complex with multiple components, 
several of which contain broad \OVI\ \citep{tripp08, thom08} at temperatures
consistent with collisionally-ionized gas. At
$\approx -280$~\kms\ relative to the LLS is an \OVI-only system with
no associated \HI\ absorption requiring a very hot temperature
\citep[$\log{T}> 6.1$;][]{savage10}. The decrement at the Lyman limit in 
the \fuse\ spectrum of this absorber suggests a total \HI\ column among 
all velocity components of $\log{N_{\rm H\,I}}=16.63\pm0.02$, which is
consistent with the total \HI\ column ($\log{N_{\rm H\,I}}=16.49\pm0.21$) 
found by the Voigt profile fits of \citet{savage14}, which we adopt here.

\subsubsection{PKS~1302--102\,/\,12573, 12655 \& 12703}
\label{indiv:absorbers:pks1302_12655}

Again Paper~1 lists only two velocity components for this system but  
simultaneously fitting the \lya\ and \lyb\ profiles indicates that
the redward component of Paper~1 contains two narrower, blended components.

\subsubsection{PKS~2155--304\,/\,16965, 17113 \& 17340}
\label{indiv:absorbers:pks2155_16964}

This absorption complex was first studied in detail by \citet{shull98} and 
\citet*{shull03} using GHRS data. In the latter publication the
\OVI\ detection is discussed in the context of a possible \OVIII\ detection 
using {\sl Chandra} and {\sl XMM}. Here we model this complex with two 
strong components and a much weaker third one. The \OVI\ detection discussed 
by \citet{shull03} is redshifted relative to the $cz= 17112$~\kms\ \HI\ by 
$\approx45$~\kms.

\subsubsection{Q~1230+0115\,/\,23294 \& 23404}
\label{indiv:absorbers:q1230_23404}

Paper~1 lists only one \HI\ component at this velocity, but two are clearly 
present in the COS data.

\subsection{Galaxy Properties}
\label{indiv:galaxies}

In Sections~\ref{indiv:galaxies:ugc5740}-\ref{indiv:galaxies:ugc4527}, the targeted 
galaxies are discussed in the order in which they appear in 
Table~\ref{tab:targ_galprop}. Then the serendipitous galaxies are discussed in 
Sections~\ref{indiv:galaxies:sdssj1228+01}-\ref{indiv:galaxies:sdssj1230+01} in the 
order in which they appear in Table~\ref{tab:ser_galprop}.

\subsubsection{UGC~5740}
\label{indiv:galaxies:ugc5740}

This dwarf spiral is observed at intermediate inclination and probed
by both the 1ES~1028+511 and also the 1SAX~J1032.3+5051 sight lines at impact
parameters of 79-110~kpc (1.5-$2.0\,R_{\rm vir}$). It is the only targeted 
galaxy other than ESO~157--49 to be probed by multiple sight
lines or at $\rho>R_{\rm vir}$. Its recession velocity of $649\pm4$~\kms\ 
is derived from \HI\ 21-cm emission \citep{schneider92} and its \Ha\ flux 
has been measured by \citet{kennicutt08}, whose value we adopt. All 
distance-dependent quantities (e.g., luminosity, stellar mass, SFR) for 
this galaxy assume a distance of 18.5~Mpc \citep{tully88,sorce14} rather 
than the Hubble-flow distance. This galaxy was not detected by \galex\ so 
we are unable to estimate its FUV SFR. See Section~\ref{discussion:association}
for a discussion of the absorber associations for this and the following galaxy.

\subsubsection{SDSS~J103108.88+504708.7}
\label{indiv:galaxies:sdssj1031+50}

This dwarf irregular galaxy is probed by the 1ES~1028+511 sight line
at an  impact parameter of 26~kpc ($0.5\,R_{\rm vir}$). Its SDSS
spectrum indicates it has a recession velocity of $934\pm7$~\kms\ and
shows weak \Ha\ emission, which is used to derive its \Ha\ SFR. The 
\Ha\ SFR is quoted as a lower limit because the aperture correction for 
this very low-$z$ galaxy is quite large \citep{iglesias-paramo13}.

\subsubsection{UGC~5478}
\label{indiv:galaxies:ugc5478}

This dwarf irregular galaxy is probed by the FBQS~J1010+3003 sight
line with an  impact parameter of 57~kpc ($0.7\,R_{\rm vir}$). Its
recession velocity of $1378\pm5$~\kms\ is derived from \HI\ 21-cm
emission \citep{devaucouleurs91}. The \Ha\ SFR is derived from 
narrowband images and the internal extinction correction is calculated 
from the Balmer decrement. All distance-dependent quantities assume a distance 
of 23.4~Mpc \citep{tully88}.

\subsubsection{ESO~157--49}
\label{indiv:galaxies:eso157-49}

This edge-on spiral is probed by the HE~0435--5304, HE~0439--5254, and
RX~J0439.6--5311 sight lines at impact parameters of 74-172~kpc
(0.8-$1.8\,R_{\rm vir}$) and was studied in detail by
\citet{keeney13}.  Associated metal lines are found in the closer
sight lines but only  \lya\ is detected toward HE~0435--5304. ATCA
\HI\ 21-cm emission maps  with 4~\kms\ velocity resolution and \Ha\
images from the CTIO 0.9-m were used to study its \HI\ morphology and
\Ha\ SFR, respectively (see \citealp{keeney13} for details). We adopt
the HIPASS recession velocity of  $1673\pm7$~\kms\ \citep{meyer04},
and the metallicity is derived using the N2 index of
\citet{pettini04}. Since SDSS magnitudes are not available  for this
galaxy we estimate its $g$- and $i$-band magnitudes from the broadband
($B,R,I$) magnitudes listed in the NASA Extragalactic Database (NED)
using the photometric conversions of \citet{jester05}. The  galaxy's
stellar mass is then calculated from these derived values. All
distance-dependent quantities  for this galaxy assume a distance of
$23.9\pm0.2$~Mpc \citep{willick97,tully09}.

\subsubsection{ESO~157--50}
\label{indiv:galaxies:eso157-50}

This galaxy, also an edge-on spiral, is located near ESO~157--49 on
the sky but has a higher redshift and luminosity. It is probed by the
HE~0439--5254 sight line at an impact parameter of 89~kpc
($0.65\,R_{\rm vir}$) and was again studied by \citet{keeney13} using
ATCA \HI\ 21-cm maps and CTIO 0.9-m \Ha\ images. Its  recession
velocity of $3874\pm12$~\kms\ is from HIPASS \citep{meyer04}, the
galaxy metallicity is derived using the N2 index of \citet{pettini04},
and the  stellar mass is calculated from NED ($B,R,I$) photometry
converted to SDSS $g$-  and $i$-band magnitudes using the relations of
\citet{jester05}.

\subsubsection{NGC~2611}
\label{indiv:galaxies:ngc2611}

This nearly edge-on spiral is probed by the PG~0832+251 sight line at an impact 
parameter of 53~kpc ($0.4\,R_{\rm vir}$).  Its recession velocity of 
$5226\pm11$~\kms\ is derived from \HI\ 21-cm emission \citep{springob05}. 
The associated \lya\ absorption profile in the COS data is exceptionally broad 
and saturated, easily a LLS, but without higher-order Ly-series lines we
cannot deconvolve the precise \HI\ columns associated with most of
the velocity components present in the metal-line data (see
Section~\ref{indiv:absorbers:pg0832_5221}) The galaxy metallicity
is derived from our high-resolution DIS spectrum, which shows significant
\Hb\ absorption that must be accounted for to accurately measure
the \Hb\ emission flux. As discussed in Section~\ref{discussion:association}, 
this galaxy is actually a member of a small group of galaxies, one of which is 
considerably closer to the QSO sight line though it is also considerably fainter 
than NGC~2611.

\subsubsection{NGC~3511}
\label{indiv:galaxies:ngc3511}

This galaxy is a nearly edge-on spiral that is probed by
the  PMN~J1103--2329 sight line at an impact parameter of 97~kpc
($0.64\,R_{\rm vir}$).  Its recession velocity of $1114\pm6$~\kms\
comes from HIPASS \citep{meyer04}, and \Ha\ images of this galaxy were
obtained with the KPNO 2.1-m telescope. We calculate the stellar mass
using SDSS $g$- and $i$-band magnitudes derived from NED $B,V,R,I$
photometry \citep{jester05}. All distance-dependent quantities for
this galaxy assume a distance of 13.6~Mpc (median of 18
redshift-independent distances listed by NED).

As discussed in Section~\ref{discussion:association}, the association
of this absorber with NGC~3511 is unclear. While the recession velocity of 
NGC~3511 matches the lower of two velocity components in the absorber, the 
higher velocity component is a close match to the recession velocity of 
another group member, NGC~3513, which is also closer to the sight line.
Notably, if this absorber were in our serendipitous sample we would 
associate it with NGC~3513 rather than NGC~3511. The sight line is, however, 
projected along the minor axis of the low-level starburst galaxy NGC~3511.

\subsubsection{M~108}
\label{indiv:galaxies:m108}

This intermediate inclination spiral is the highest luminosity galaxy in the
targeted sample,  with an \HI\ 21-cm-derived recession velocity of
$696\pm1$~\kms\ \citep{springob05}.  It is probed by the SBS~1108+560
sight line at an impact parameter of 22~kpc  ($0.14\,R_{\rm vir}$),
which also makes it the closest probe of a galaxy's CGM  (in terms of
virial radii) in either sample. Unfortunately, a LLS at $z=0.4634$
precludes us from analyzing the \HI\ absorption in this system (see
Section~\ref{indiv:absorbers:sbs1108_654}). Furthermore, the galaxy's 
large size on the sky prevent the use of our standard ground-based methods. 
SDSS pipeline photometry is not reliable for such large sources, so we convert 
the broadband ($B,V,I$) magnitudes listed in NED to SDSS $g$- and $i$-band values 
\citep{jester05} to calculate the galaxy's stellar mass. We do not have images 
with a large enough FOV to capture all of the galaxy's \Ha\ flux, so we adopt 
the value of \citet{kennicutt08}.  The galaxy metallicity in
Table~\ref{tab:targ_galprop} is the average of individual
measurements of several \HII\ regions from SDSS and our own DIS
spectroscopy.  The FUV luminosity is from \citet*{rifatto95} and 
uncorrected for intrinsic extinction, and thus highly uncertain.
All distance-dependent quantities for this galaxy assume a distance of 
10.7~Mpc  (median of 9 redshift-independent distances listed by NED).


\subsubsection{IC~691}
\label{indiv:galaxies:ic691}

This dwarf irregular galaxy is probed by the SBS~1122+594 sight line
at an impact  parameter of 45~kpc ($0.5\,R_{\rm vir}$) and was studied
in detail by \citet{keeney06}. \Ha\ images of this galaxy come from
the KPNO 2.1-m telescope, its recession velocity of
$1199\pm7$~\kms\ is derived from \HI\ 21-cm emission
\citep{devaucouleurs91}, and its intrinsic extinction is estimated
from its observed Balmer decrement as described in \citet{calzetti01}. 
\galex\ photometry for this galaxy is from  \citet{hao11}, and all 
distance-dependent quantities assume a distance of 23.7~Mpc \citep{tully88}.

\subsubsection{UGC~4527}
\label{indiv:galaxies:ugc4527}

This low surface brightness dwarf galaxy is the faintest galaxy in either 
sample and is located only 7~kpc ($0.15\,R_{\rm vir}$) from the VII~Zw~244
sight line, making it the closest probe of a galaxy's CGM (in terms
of impact parameter, second closest in terms of  virial radii) in
either sample. Its recession velocity of $721\pm6$~\kms\ is derived
from \HI\ 21-cm emission \citep{schneider92}, but the luminosity, halo
mass, and stellar mass of this galaxy are very uncertain because
there is insufficient optical photometry available in NED to reliably
estimate its SDSS $g$- and $i$-band magnitudes.  However, this galaxy
is clearly forming stars as evidenced by its weak \Ha\ and FUV
emission.

\subsubsection{SDSS~J122815.96+014944.1}
\label{indiv:galaxies:sdssj1228+01}

This star-forming dwarf is located 70~kpc ($1.4\,R_{\rm vir}$) from the 3C~273 
sight line. Its metallicity is derived from emission lines present in the SDSS spectrum, and its \Ha\ SFR is derived from narrowband images.

\subsubsection{SDSS~J122950.57+020153.7}
\label{indiv:galaxies:sdssj1229+02}

This dwarf post-starburst galaxy, located 81~kpc ($1.5\,R_{\rm vir}$) from the 
3C~273 sight line, was studied in detail by \citet{stocke04} and \citet{keeney14}.
We adopt the FUV SFR and metallicity of \citet{keeney14}, who use SED modeling of 
the galaxy's \galex+SDSS photometry to estimate its metallicity due to the lack 
of any optical emission lines in low- or high-resolution optical spectra. The 
\Ha\ SFR comes from narrowband imaging. Despite the large uncertainties, this galaxy 
has the lowest metallicity estimate of all galaxies in this sample. The absorber is 
at high $|\Delta v|/v_{\rm esc} > 5$, contains metal absorption, and may have been 
ejected from this galaxy. Alternately, the absorber may be associated with a large-
scale structure or rich galaxy group in this vicinity 
\citep{rosenberg03,yoon12,stocke14}. The absorber association and ``passive'' nature 
of this galaxy are discussed further in Section~\ref{discussion:association:passive}.

\subsubsection{Mrk~892}
\label{indiv:galaxies:mrk892}

This galaxy was part of the ``alternate'' sample of Paper~1, and is probed by 
the 3C~351 sight line at an impact parameter of 173~kpc ($1.8\,R_{\rm vir}$).
Its \Ha\ SFR and metallicity are derived from SDSS emission-line fluxes. The 
lenticular morphology and bright nucleus of this galaxy suggest that it may 
harbor an AGN, but SDSS classifies it as a star-forming galaxy 
using emission line diagnostics \citep{kewley06}, an interpretation which is 
supported by the galaxy's strong \galex\ FUV emission.

\subsubsection{SDSS~J182202.70+642138.8}
\label{indiv:galaxies:sdssj1822+64}

This $L^*$ galaxy is probed by the H~1821+643 sight line at an impact parameter
of 157~kpc ($0.8\,R_{\rm vir}$). Its recession velocity is adopted from 
\citet{tripp98} but the only optical spectrum we have in hand for this galaxy 
is from WIYN/HYDRA and does not cover the \Ha\ emission line. 
\edit1{Thus, we cannot directly measure the galaxy's metallicity; we use the 
mass-metallicity relationship of \citet{lee06} to estimate it instead. We 
attempt to estimate the galaxy's }
\Ha\ SFR by bootstrapping from \Hb\ as follows: (1) we estimate the intrinsic 
extinction at \Ha\ using the galaxy's observed morphology and inclination angle 
(\Ha\ attenuation $\sim1.2$); (2) we determine the Balmer decrement 
\citep[$\sim3.2$ assuming an intrinsic value of 2.87;][]{calzetti01} that yields 
the same \Ha\ attenuation as our morphological estimate; (3) we multiply the 
observed \Hb\ flux by this derived Balmer decrement to estimate the \Ha\ flux that 
would have been present in the galaxy spectrum. Since this is a spectroscopic 
estimate we apply our usual aperture correction procedure (a factor of $\sim2.0$ 
in this case). A similar procedure is followed for several galaxies below for which 
we have optical spectra that cover \Hb\ but not \Ha.

\subsubsection{SDSS~J000529.16+201335.9}
\label{indiv:galaxies:sdssj0005+20}

This dwarf irregular galaxy is located 97~kpc ($1.8\,R_{\rm vir}$) from the 
Mrk~335 sight line. Its recession velocity is derived from \HI\ 21-cm emission 
maps \citep{vangorkom96} and its long-slit optical spectrum shows weak \Ha\ and 
marginal \Hb\ emission. We have estimated its metallicity using the observed 
emission line fluxes. The \Ha\ SFR comes from narrowbnad imaging, but have not 
applied our usual intrinsic extinction correction due to the large uncertainties 
in the observed Balmer decrement and galaxy morphology; thus, the tabulated \Ha\ 
SFR is highly uncertain.

\subsubsection{NGC~6140}
\label{indiv:galaxies:ngc6140}

This intermediate-inclination galaxy was part of the ``alternate'' sample of 
Paper~1 and is probed by the Mrk~876 sight line at an impact parameter of 
257~kpc ($1.6\,R_{\rm vir}$). Its recession velocity of $908\pm1$~\kms\ is 
derived from \HI\ 21-cm emission \citep{springob05}, its stellar mass is 
calculated from NED ($B,V,I$) photometry converted to SDSS $g$- and $i$-band 
magnitudes using the relations of \citet{jester05}, and all distance-dependent
quantities for this galaxy assume a distance of 18.6~Mpc \citep{tully88}.
The galaxy metallicity is estimated from an APO/DIS spectrum taken on 
2016 Mar 20 and the \Ha\ SFR is taken from the \Ha\ imaging of
 \citet{sanchez-gallego12}, who have corrected for [\NII] contamination
and internal extinction.

\subsubsection{SDSS~J095638.90+411646.1}
\label{indiv:galaxies:sdssj0956+41}

This luminous spiral is located 438~kpc ($1.7\,R_{\rm vir}$) from the PG~0953+414
sight line. Its \Ha\ SFR and metallicity are derived from 
SDSS emission-line fluxes. The intrinsic extinction estimate derived from the 
face-on morphological correction predicts an \Ha\ attenuation ($\sim1.3$) that is 
approximately half the value inferred from the observed Balmer decrement for this 
galaxy. If the large spread in \lya\ velocity components for the associated 
absorber are separate clouds, some components may be associated with another 
nearby galaxy (see Section~\ref{discussion:association}).

\subsubsection{SDSS~J111905.51+211733.0}
\label{indiv:galaxies:sdssj1119+21a}

This galaxy is probed by the PG~1116+215 sight line at an impact parameter of 
133~kpc ($1.5\,R_{\rm vir}$). Its recession velocity is from \citet{prochaska11b}, 
but the only optical spectrum we have for this galaxy is from WIYN/HYDRA and 
does not cover the \Ha\ region. As in 
Section~\ref{indiv:galaxies:sdssj1822+64}, we bootstrap an \Ha\ SFR 
estimate from the observed \Hb\ emission flux but 
\edit1{cannot directly measure the galaxy's metallicity. We use the estimate 
the galaxy metallicity using the mass-metallicity relation of \citet{lee06} 
instead.}
The four-component \HI+\OVI\ absorber near this galaxy has two components at 
$|\Delta v|/v_{\rm esc}>5$ (see Section~\ref{discussion:association}). Some or all 
of these absorptions may be associated instead with a galaxy group that includes 
this galaxy \citep{stocke14}.

\subsubsection{SDSS~J111906.68+211828.7}
\label{indiv:galaxies:sdssj1119+21b}

This intermediate-inclination galaxy is located 139~kpc ($0.7\,R_{\rm vir}$) 
from the PG~1116+215 sight line. Its SDSS spectrum shows weak Balmer emission, 
from which its \Ha\ SFR and metallicity are derived. The 
metallicity estimate uses the N2 index of \citet{pettini04} because no 
[\OIII] emission is observed. This galaxy has the highest metallicity 
estimate of all galaxies where such estimates are available, and a sSFR low 
enough to be classified as passive by the COS-Halos definition 
\citep[$<10^{-11}~{\rm yr}^{-1}$;][]{tumlinson11}. At least some of this absorption
may be due to a galaxy group to which this galaxy belongs \citep{stocke14}.

\subsubsection{IC~3061}
\label{indiv:galaxies:ic3061}

This very low-$z$, edge-on spiral is projected 138~kpc ($1.1\,R_{\rm vir}$) from 
the quasar PG~1211+143. Its recession velocity of $2136\pm1$~\kms\ is derived from 
\HI\ 21-cm emission measurements \citep{springob05} and its metallicity is derived 
from SDSS emission-line fluxes. All distance-dependent quantities for this galaxy 
assume a distance of 42.0~Mpc (median of 10 redshift-independent distances listed by 
NED). The FUV magnitude is from \citet{rifatto95} and its SFR calibration is highly 
uncertain. While another nearby Virgo Cluster galaxy has a closer velocity match with 
the absorber (see Table~\ref{tab:serendipitous}), the impact parameter to IC~3061 is 
quite small and the association is deemed secure.

\subsubsection{SDSS~J121409.55+140420.9}
\label{indiv:galaxies:sdssj1214+14a}

This intermediate-inclination spiral is probed by the PG~1211+143 sight line at 
an impact parameter of 137~kpc ($0.8\,R_{\rm vir}$). Its \Ha\ SFR and 
metallicity are derived from emission lines in the SDSS spectrum.
Its \Ha\ SFR is treated as uncertain, however, because emission line diagnostics
\citep{kewley06} indicate a significant AGN contribution.

\subsubsection{SDSS~J121413.94+140330.4}
\label{indiv:galaxies:sdssj1214+14b}

This galaxy is located 72~kpc ($0.8\,R_{\rm vir}$) from the PG~1211+143 sight line
and has no FUV photometry available. Its \Ha\ SFR and metallicity are estimated
from an APO/DIS spectrum taken on 2016~Feb~6.

\subsubsection{SDSS~J121930.86+064334.4}
\label{indiv:galaxies:sdssj1219+06a}

This starburst galaxy, probed by the PG~1216+069 sight line at an impact parameter
of 505~kpc ($1.8\,R_{\rm vir}$), is the most luminous galaxy ($3.5\,L^*$) in 
either sample and the galaxy probed at the largest impact parameter. Its \Ha\ 
SFR and metallicity are derived from SDSS emission-line fluxes. Its \Ha\ SFR 
is treated as uncertain, however, because its very broad emission lines indicate 
a strong AGN contribution.

\subsubsection{SDSS~J121923.43+063819.7}
\label{indiv:galaxies:sdssj1219+06b}

This galaxy is projected 93~kpc ($0.6\,R_{\rm vir}$) from the quasar PG~1216+069.
Its recession velocity is from \citet{prochaska11b}, but the only optical spectrum 
in hand for this galaxy is from WIYN/HYDRA and does not cover the \Ha\ region. 
As before, we bootstrap an \Ha\ SFR estimate from the observed \Hb\ emission 
\edit1{and estimate the galaxy metallicity using the relation of 
\citet{tremonti04}.}

\subsubsection{UGC~8146}
\label{indiv:galaxies:ugc8146}

This edge-on dwarf spiral is probed by the PG~1259+593 sight line at an impact
parameter of 114~kpc ($1.5\,R_{\rm vir}$). Its recession velocity of 
$668\pm1$~\kms\ is derived from \HI\ 21-cm emission maps \citep{springob05}
and its metallicity is derived from SDSS emission-line fluxes. The \Ha\ SFR is 
derived from narrowband images. All distance-dependent quantities assume a 
distance of 18.5~Mpc (median of 15 redshift-independent distances listed by NED).

\subsubsection{SDSS~J130101.05+590007.1}
\label{indiv:galaxies:sdssj1301+59}

This intermediate-inclination spiral is projected 138~kpc ($1.0\,R_{\rm vir}$) 
from the quasar PG~1259+593. The SDSS spectrum of this galaxy shows very weak 
\Ha\ and \Hb\ emission, from which the \Ha\ SFR and metallicity are derived. 
The metallicity uses the N2 index of \citet{pettini04} because no [\OIII] 
emission is observed. This galaxy has a sSFR low enough to be classified as 
passive by the COS-Halos definition 
\citep[$<10^{-11}~{\rm yr}^{-1}$;][]{tumlinson11}, but there is no compelling 
reason to doubt this association (see Section~\ref{discussion:association:passive}).

\subsubsection{SDSS~J215456.65--091808.6}
\label{indiv:galaxies:sdssj2154-09}

This galaxy is probed by the PHL~1811 sight line at an impact parameter of 
269~kpc ($1.3\,R_{\rm vir}$). Its \Ha\ SFR and metallicity are derived from 
SDSS emission-line fluxes. The \Ha\ SFR is listed as a lower limit due to the 
large aperture correction for this galaxy.

\subsubsection{SDSS~J215517.30--091752.0}
\label{indiv:galaxies:sdssj2155-09}

This galaxy was part of the ``alternate'' sample of Paper~1, and is probed by the 
PHL~1811 sight line at an impact parameter of 502~kpc ($2.0\,R_{\rm vir}$). 
Its \Ha\ SFR and metallicity are derived from SDSS emission-line fluxes. This 
galaxy has the largest stellar mass of any studied herein and a sSFR low enough 
to be classified as passive \citep{tumlinson11}. There are no other luminous 
galaxies close to the absorber in this field, so the association with this 
passive galaxy seems reasonable despite the large impact parameter (see 
Section~\ref{discussion:association:passive}).

\subsubsection{J215447.5--092254}
\label{indiv:galaxies:j2154-09a}

This galaxy is projected 309~kpc ($1.3\,R_{\rm vir}$) from the PHL~1811 sight 
line, and was part of the ``alternate'' sample of Paper~1. We adopt the recession 
velocity of \citet{prochaska11b} but the \Ha\ SFR and metallicity are derived 
from an AAT/AA$\Omega$ spectrum with uncertain flux calibration.  
We flag the \Ha\ SFR as uncertain but not the metallicity as the O3N2 index 
relies on ratios of emission line fluxes in narrow spectral regions (specifically, 
[\OIII]$\,\lambda5007/\Hb$ and [\NII]$\,\lambda6583/\Ha$).
The stellar mass was derived from SDSS $g,r,i$ images obtained with 
the MOSAIC imager on the CTIO Blanco 4-m telescope.

\subsubsection{J215450.8--092235}
\label{indiv:galaxies:j2154-09b}

This galaxy is also part of the ``alternate'' sample of Paper~1, and is 
probed by the PHL~1811 sight line at an impact parameter of 237~kpc 
($2.0\,R_{\rm vir}$). Its recession velocity is adopted from 
\citet{prochaska11b} but its \Ha\ SFR and metallicity are derived from 
an AAT/AA$\Omega$ spectrum as with the previous galaxy. Similarly, the 
galaxy's stellar mass is derived from CTIO/MOSAIC imaging in the 
SDSS $g,r,i$ bands. In addition to being at the limit of this survey 
($\rho=2.0\,R_{\rm vir}$), this galaxy has the highest sSFR of any galaxy 
studied herein.

\subsubsection{2MASS~J21545996--0922249}
\label{indiv:galaxies:2massj2154-09}

This galaxy is projected 35~kpc ($0.2\,R_{\rm vir}$) from the PHL~1811 sight 
line and its recession velocity of $24223\pm45$~\kms\ is adopted from 6dF 
\citep{jones05}. An AAT/AA$\Omega$ spectrum is available for this galaxy but 
shows no emission lines, so we are unable to estimate its \Ha\ SFR. Its stellar 
mass was derived from CTIO/MOSAIC imaging in the SDSS $g,r,i$ bands,
\edit1{and its metallicity was estimated using the relation of \citet{tremonti04}.}


\subsubsection{J215506.5--092326}
\label{indiv:galaxies:j2155-09}

This luminous starburst galaxy is probed by the PHL~1811 sight line at an impact 
parameter of 228~kpc ($1.0\,R_{\rm vir}$) and has the highest \Ha\ SFR of any 
galaxy in either sample. Its recession velocity is adopted from 
\citet{prochaska11b}, its \Ha\ SFR and metallicity are derived from an 
AAT/AA$\Omega$ spectrum and its stellar mass is determined using CTIO/MOSAIC 
$g,r,i$ imaging.

\subsubsection{J215454.9--092331}
\label{indiv:galaxies:j2154-09c}

This starburst galaxy has the highest redshift and largest FUV SFR of any galaxy 
studied herein. It is located 354~kpc ($1.7\,R_{\rm vir}$) from the PHL~1811 sight 
line and its recession velocity is adopted from \citet{prochaska11b}. Its \Ha\ 
SFR and metallicity are derived from an AAT/AA$\Omega$ spectrum and its stellar 
mass is measured from CTIO/MOSAIC $g,r,i$ images.


\subsubsection{J031201.7--765517}
\label{indiv:galaxies:j0312-76}

This galaxy is from the ``alternate'' sample of Paper~1, and is probed by the 
PKS~0312--770 sight line at an impact parameter of 239~kpc ($1.9\,R_{\rm vir}$).
Its recession velocity is from \citet{prochaska11b}, but we have no optical 
spectroscopy in hand to estimate its \Ha\ SFR or metallicity. Furthermore, 
its luminosity, halo mass, and stellar mass are highly uncertain because 
there is insufficient optical photometry available in NED to reliably 
estimate its SDSS $g$- and $i$-band magnitudes.
\edit1{Nevertheless, we use the relation of \citet{lee06} to estimate the 
galaxy metallicity from its ill-constrained stellar mass.}

\subsubsection{J031158.5--764855}
\label{indiv:galaxies:j0311-76}

This galaxy is also from the ``alternate'' sample of Paper~1 and is located 
381~kpc ($1.6\,R_{\rm vir}$) from the PKS~0312--770 sight line. Its recession 
velocity is from \citet{prochaska11b}, but we have no optical spectrum available 
to measure its \Ha\ SFR or metallicity. As with the previous galaxy, there is 
insufficient optical photometry available in NED to reliably estimate its 
luminosity, halo mass \& stellar mass.
\edit1{We have estimated the galaxy metallicity using the relation of 
\citet{tremonti04}.}




\subsubsection{2MASX~J04075411--1214493}
\label{indiv:galaxies:2masxj0407-12}

This $L^*$ galaxy is probed by the PKS~0405--123 sight line at an impact 
parameter of 378~kpc ($2.0\,R_{\rm vir}$). Its recession velocity of 
$29050\pm3$~\kms\ is adopted from \citet{johnson13} and its stellar mass is 
derived from CTIO/MOSAIC imaging in the SDSS $g,r,i$ bands. Its \Ha\ SFR is 
bootstrapped from the \Hb\ flux as described in 
Section~\ref{indiv:galaxies:sdssj1822+64} because no \Ha\ coverage is 
available in our CTIO/HYDRA spectrum, and is listed as a lower limit due 
to the large intrinsic extinction correction for this galaxy. A low-resolution
APO/DIS spectrum with no flux calibration taken on 2012 Sep 16 shows emission 
from \Ha\ and [\NII] but not \Hb\ and [\OIII]; therefore, the 
galaxy's metallicity is estimated using the N2 index calibration of 
\citet{pettini04}. The lack of \Hb\ and [\OIII] emission lines in 
the APO spectrum is not surprising, however, since they are located in regions 
of greatly reduced sensitivity due to the instrumental dichroic. The next-nearest
galaxy is physically closer to the sight line and a better match to the 
$N_{\rm H\,I}$-weighted mean velocity of the two \lya\ components, making this 
association uncertain.

\subsubsection{J040743.9--121209}
\label{indiv:galaxies:j0407-12a}

This galaxy is projected 197~kpc ($1.1\,R_{\rm vir}$) from the quasar 
PKS~0405--123. Its recession velocity of $45989\pm3$~\kms\ is adopted from 
\citet{johnson13} and is $\sim270$~\kms\ larger than the value in Paper~1, 
which used the redshift reported by \citet{chen01}. Its stellar mass is derived 
from CTIO/MOSAIC $g,r,i$ images, and an APO/DIS spectrum taken on 2012~Sep~23
shows no optical emission lines.
\edit1{Thus, we use the relation of \citet{tremonti04} to estimate the galaxy 
metallicity.}


\subsubsection{J040751.2--121137}
\label{indiv:galaxies:j0407-12b}

This luminous, intermediate-inclination galaxy is responsible for the LLS in the 
PKS~0405--123 spectrum first studied by \citet{chen00}. It is projected 117~kpc 
($0.5\,R_{\rm vir}$) from the sight line, and we now adopt the galaxy redshift 
of \citet{johnson13} rather than the value from \citet{chen09} as in Paper~1.
The galaxy's stellar mass is estimated from CTIO/MOSAIC images in the SDSS 
$g,r,i$ bands. The galaxy metallicity and \Ha\ SFR are estimated from an APO/DIS 
spectrum obtained on 2016~Oct~29. The metallicity is derived using the N2 index of
\citet{pettini04} because the \Hb\ and [\OIII] emission lines are affected 
by the instrumental dichroic (a low-S/N CTIO/HYDRA spectrum of this galaxy has no 
\Ha\ coverage and no \Hb\ emission).


\subsubsection{NGC~4939}
\label{indiv:galaxies:ngc4939}

This luminous, intermediate-inclination spiral is projected 261~kpc 
($1.0\,R_{\rm vir}$) from the quasar PKS~1302--102. Its recession velocity 
of $3112\pm2$~\kms\ is derived from \HI\ 21-cm emission maps \citep{springob05}.
We calculate its stellar mass using NED ($B,V,R,I$) photometry converted to 
SDSS $g$- and $i$-band magnitudes using the relations of \citet{jester05}. 
All distance-dependent quantities assume a distance of 39.3~Mpc (median of 18 
redshift-independent distances listed by NED). No FUV photometry or optical 
spectroscopy are available for this galaxy, but a catalog of \Ha\ + [\NII]
fluxes for 250 \HII\ regions in the galaxy is available in 
\citet{tsvetanov95}. To calculate the galaxy's \Ha\ SFR we have summed the 
fluxes from all of the individual \HII\ regions and assumed that 
[\NII]$/\Ha = \sfrac{1}{3}$.
\edit1{The galaxy metallicity is estimated using the relation of 
\citet{tremonti04}.}


\subsubsection{2MASX~J13052026--1036311}
\label{indiv:galaxies:2masxj1305-10a}

This galaxy is probed by the PKS~1302--102 sight line at an impact parameter
of 227~kpc ($0.9\,R_{\rm vir}$). Its recession velocity of $12755\pm45$~\kms\ 
is adopted from 6dF \citep{jones05}, but we have no optical spectra in hand 
from which to measure its \Ha\ SFR or metallicity. Its stellar mass is estimated
from the $B,R$ photometry of \citet{prochaska11b} using the relations of 
\citet{jester05} to convert to SDSS $g$- and $i$-band magnitudes.
\edit1{The galaxy metallicity is estimated using the relation of 
\citet{tremonti04}.}


\subsubsection{2MASX~J13052094--1034521}
\label{indiv:galaxies:2masxj1305-10b}

This massive galaxy is located 353~kpc ($1.3\,R_{\rm vir}$) from the 
PKS~1302--102 sight line, and its recession velocity of $28304\pm45$~\kms\ 
is adopted from 6dF \citep{jones05}. Unlike the previous galaxy, we do have an 
optical spectrum in hand, which shows no emission lines; thus we are unable to 
measure the galaxy's metallicity 
\edit1{(it is estimated using the relation of \citet{tremonti04}),}
but we are able to set a spectroscopic limit on the \Ha\ SFR. We flag the \Ha\ 
SFR limit as uncertain because we do not know if any \Ha\ emission is present 
outside of the spectroscopic aperture. The galaxy's stellar mass is derived from 
$B,R$ photometry \citep{prochaska11b} converted to SDSS $g,i$ magnitudes using the 
conversions of \citet{jester05}, and its sSFR is low enough to be passive by the 
COS-Halos definition \citep{tumlinson11}, but there is no compelling reason to 
doubt this association (see Section~\ref{discussion:association:passive}).




\subsubsection{2MASX~J21584077--3019271}
\label{indiv:galaxies:2masxj2158-30}

This galaxy is probed by the PKS~2155--304 sight line at an impact parameter of 
425~kpc ($2.0\,R_{\rm vir}$). Its recession velocity of $17005\pm31$~\kms\ is 
adopted from 6dF \citep{jones05} but we have no additional optical spectrum 
from which to measure its \Ha\ SFR or metallicity. Its stellar mass is calculated 
from $B,R$ photometry \citep{prochaska11b} converted to SDSS $g,i$ magnitudes 
\citep{jester05}.
\edit1{The galaxy metallicity is estimated using the relation of 
\citet{tremonti04}.}




\subsubsection{J215845.1-301637}
\label{indiv:galaxies:j2158-30}

This luminous galaxy is located 403~kpc ($1.8\,R_{\rm vir}$) from the PKS~2155--304
sight line, and was part of the ``alternate'' sample of Paper~1. Its recession 
velocity is adopted from \citet{mclin03} and \citet{yao10}, but we have no 
optical spectrum from which to measure its \Ha\ SFR or metallicity. Further, its
luminosity, halo mass, and stellar mass are highly uncertain because there is 
insufficient optical photometry in NED to reliably estimate its SDSS $g$- and 
$i$-band magnitudes.
\edit1{The galaxy metallicity is estimated using the relation of 
\citet{tremonti04}.}

\subsubsection{CGCG~14--54}
\label{indiv:galaxies:cgcg14-54}

This dwarf irregular galaxy, located 70~kpc ($1.4\,R_{\rm vir}$) from the 
Q~1230+0115 sight line, is from the ``alternate'' sample of Paper~1 and 
has the lowest stellar mass of any galaxy studied herein. Its recession 
velocity of $1105\pm5$~\kms\ is derived from \HI\ 21-cm emission, 
and its \Ha\ SFR and metallicity are derived from SDSS emission-line fluxes.
The \Ha\ SFR is listed as a lower limit due to the large aperture correction for
this galaxy. All distance-dependent quantities assume a distance of 
9.6~Mpc \citep{karachentsev13}. The associated absorber has 
$|\Delta v|/v_{\rm esc} > 10$ and no associated metals. We speculate that this 
absorber is not associated with a single galaxy \citep[see also][]{rosenberg03}.

\subsubsection{SDSS~J123047.60+011518.6}
\label{indiv:galaxies:sdssj1230+01}

This galaxy is probed by the Q~1230+0115 sight line at an impact parameter of 
55~kpc ($0.5\,R_{\rm vir}$). Its recession velocity of $23327\pm18$~\kms\ 
is measured from a WIYN/HYDRA spectrum that does not cover the \Ha\ region; this 
is $\sim200$~\kms\ bluer then the redshift quoted in Paper~1. The updated redshift
aligns much more closely with the \lya\ absorption velocity and is derived 
from a high-quality emission line spectrum; the previous redshift was derived from 
a low-S/N spectrum contaminated by moonlight \citep{mclin03} and is deemed 
spurious. As before when we lack \Ha\ coverage, we bootstrap an \Ha\ SFR estimate 
from the observed \Hb\ flux 
\edit1{and estimate the galaxy metallicity using the relation of \citet{lee06}.}

\subsection{Photo-Ionization Models}
\label{indiv:cloudy}

Below, each modeled absorber is discussed in turn, with the model ``grade'' 
(A, B, C, D) assessing the quality of the absorption-line data and the CLOUDY 
constraints noted in brackets after each absorber identification (see 
Section~\ref{ensemble} and Table~\ref{tab:cloudy}).

\subsubsection{HE~0439--5254\,/\,1653 [C]}
\label{indiv:cloudy:he0439_1653}

With detections in only \HI, \CIV, \SiIII\ and \SiIV, this targeted absorber 
(System~10 in Table~\ref{tab:absprop}) barely meets our threshold for CLOUDY 
modeling. Nonetheless, we find a single-phase solution consistent with all of 
the metal-line detections above, as well as the upper limits set by the 
non-detection of \CII, \NV, etc. Since our $N_{\rm H\,I}$ value is consistent 
with that of Paper~1 we find a very similar $\log{U}$ value, but our updated 
model metallicity is somewhat lower (although still consistent with the Paper~1 
value to within the uncertainties).

This is a rare case where our assumed prior effects the model's preferred values. 
A pure maximum likelihood analysis that assumes no prior predicts 
$\log{U}=-2.30^{+0.19}_{-0.18}$ and $\log{Z_{\rm abs}}=+0.16^{+0.44}_{-0.34}$ 
for this absorber.  The ionization parameter preferred by the maximum 
likelihood analysis is close to the value in Table~\ref{tab:cloudy}, so
the effect of the prior is to (unsurprisingly) nudge the absorber metallicity 
toward the galaxy metallicity (see Table~\ref{tab:targ_galprop}).

\subsubsection{PG~0832+251\,/\,5221 [C]}
\label{indiv:cloudy:pg0832_5221}

This targeted absorber (System~14 in Table~\ref{tab:absprop}) has a plethora of 
metal-line detections, but has perhaps the most uncertain $N_{\rm H\,I}$ value of 
any tabulated in Table~\ref{tab:cloudy} (see 
Section~\ref{indiv:absorbers:pg0832_5221} for details). However, if its 
\HI\ column is taken at face value then this is the strongest \HI\ absorber that
we attempt to model. Our updated model attempts to reproduce the metal-line ratios 
from the observed \OI, \CII, \SiII, and \FeII\ detections, as well as the \CI\ and 
\FeIII\ upper limits; the region of highest posterior probability is mildly 
inconsistent with the \CI\ upper limit. The contours for the \SiIII, \SiIV, and 
\CIV\ detections and the \NV\ upper limit are also shown, but are not used to 
constrain the model. The \CIV\ detection in particular is clearly inconsistent with 
the solution derived from the lower ions, exemplifying the multi-phase nature of 
this absorber. This updated model finds a similar ionization parameter but higher 
metallicity than the model of Paper~1.

\subsubsection{PG~0832+251\,/\,5444 [---]}
\label{indiv:cloudy:pg0832_5444}

This absorber (System~17 in Table~\ref{tab:absprop}) was modeled in Paper~1 but we 
do not model it here. There are sufficient metal-line detections to attempt to 
model it, but the \HI\ column is too uncertain because this velocity component is 
located in the trough of an extremely strong \lya\ absorber (see 
Figure~\ref{fig:stackplot} and Section~\ref{indiv:absorbers:pg0832_5221}).

\subsubsection{PMN~J1103--2329\,/\,1194 [C]}
\label{indiv:cloudy:p1103_1194}

This targeted absorber (System~19 in Table~\ref{tab:absprop}) was detected in 
\SiIII, \SiIV, \CIV\ and \NV, and has stringent upper limits on \CII\ and \SiII. 
We fit the \lya\ profile with two velocity components (see 
Section~\ref{indiv:absorbers:p1103_1194}) and find a considerably lower 
\HI\ column density than in Paper~1; consequently, our updated CLOUDY model 
finds a similar ionization parameter but a much higher metallicity than the 
model of Paper~1. Our photo-ionization model cannot simultaneously reproduce 
the metal-line ratios for all of the detected ions, so it uses the \SiIII\ and 
\SiIV\ detections (and \CII\ and \SiII\ limits) to constrain the model, ignoring 
the \CIV\ and \NV\ detections. This is another case where our imposed prior on 
absorber metallicity affects the model's preferred values. A pure maximum 
likelihood analysis with no prior finds $\log{U} = -2.27^{+0.14}_{-0.15}$ and 
$\log{Z_{\rm abs}} = +0.48^{+0.16}_{-0.15}$.

\subsubsection{RX~J0439.6--5311\,/\,1674 [C]}
\label{indiv:cloudy:rxj0439_1674}

As with the HE~0439--5254 absorber that probes the same targeted galaxy 
(Section~\ref{indiv:cloudy:he0439_1653}), this absorber (System~21 in 
Table~\ref{tab:absprop}) is detected only in \SiIII, \SiIV\ and \CIV, but 
has stringent limits on \CII, \SiII\ and \NV. Our updated model assumes a similar 
$N_{\rm H\,I}$ value to that of Paper~1 and finds a comparable ionization parameter 
but a lower metallicity than the model of Paper~1, although all of the values 
overlap within uncertainties. All of this absorber's metal-line detections and 
limits can be reproduced with this single-phase model.

\subsubsection{SBS~1108+560\,/\,654 [---]}
\label{indiv:cloudy:sbs1108_654}

This absorber (System~23 in Table~\ref{tab:absprop}) was modeled in Paper~1 and 
has many metal-line detections, but a higher redshift LLS (see 
Section~\ref{indiv:absorbers:sbs1108_654} for details) reduces the continuum 
near \lya\ for this absorber to such a degree that we cannot reliably constrain the 
\HI\ column. Thus, we do not attempt to model this absorber here.

\subsubsection{SBS~1108+560\,/\,778 [---]}
\label{indiv:cloudy:sbs1108_778}

This absorber (System~25 in Table~\ref{tab:absprop}) was modeled in Paper~1 but 
we do not do so here for the reasons elaborated in Section~\ref{indiv:cloudy:sbs1108_654}
above.

\subsubsection{SBS~1122+594\,/\,1221 [B]}
\label{indiv:cloudy:sbs1122_1221}

We have higher confidence in \HI\ column density for this absorber (System~26 in 
Table~\ref{tab:absprop}) than the values of the previous two targeted absorbers 
because of its relatively simple \lya\ profile. It has associated metal-line 
detections in \CII, \CIV, \SiII, and \SiIV, and stringent limits in \SiII\ and \NV. 
The CLOUDY model simultaneously reproduces the metal-line ratios inferred from all 
of these species and finds very similar values for the ionization parameter and 
absorber metallicity as the model of Paper~1. This absorber has the largest 
line-of-sight thickness and second-largest cloud mass of all the targeted absorbers 
that we are able to model, which may be related to its associated galaxy being a 
dwarf starburst \citep{keeney06}.

\subsubsection{VII~Zw~244\,/\,715 [B]}
\label{indiv:cloudy:viizw244_715}

This absorber (System~27 in Table~\ref{tab:absprop}) also has a relatively 
simple \lya\ profile, with a correspondingly higher confidence in the \HI\ 
column density as compared to other targeted absorbers. It is detected in 
\CII, \CIV, \SiII, \SiIII\ and \SiIV, and has a stringent \NV\ upper limit. 
Our updated CLOUDY model is able to simultaneously accommodate all of the 
metal-line constraints with a single-phase solution. We assume a considerably 
lower \HI\ column than in Paper~1 and thus find subsequently higher values for 
the ionization parameter and absorber metallicity than previously.

\subsubsection{3C~273\,/\,1585 [A]}
\label{indiv:cloudy:3c273_1585}

This serendipitous absorber (System~102 in Table~\ref{tab:absprop}) has been 
studied extensively by various groups (see 
Section~\ref{indiv:absorbers:3c273_1585} for details) and has an 
unusual ionization pattern in that it is detected in several ions (\CII, 
\SiII\ and \SiIII) with low ionization potentials but not in any species with 
intermediate or high ionization potentials (e.g., \CIII, \CIV, \SiIV, \NV, \OVI).
Our updated CLOUDY model accommodates all of the metal-line constraints with a 
single-phase solution, although the \CIII\ upper limit that we adopt from 
\citet{tilton12} is somewhat incompatible with the \CII\ column density we measure 
in the COS spectrum. Since we assume a smaller \HI\ column than either Paper~1 or 
\citet{tripp02} we find a larger metallicity than they do but our ionization 
parameter is similar to theirs.

\subsubsection{PG~0953+414\,/\,42664 [B]}
\label{indiv:cloudy:pg0953_42664}

This serendipitous absorber (System~115 in Table~\ref{tab:absprop}) is one of only 
two absorbers with $N_{\rm H\,I}<10^{14}~{\rm cm}^{-2}$ that we attempt to model. 
It is detected in \CIII\ and \CIV, which we use to constrain our model, as well as 
\NV\ and \OVI, which we don't. This absorber was not modeled in Paper~1 because 
\citet{danforth08} only lists associated \CIII\ and \OVI; the \CIV\ and \NV\ are 
only detectable in the COS spectrum (see Figure~\ref{fig:stackplot}).

\subsubsection{PG~1116+215\,/\,41522 [A]}
\label{indiv:cloudy:pg1116_41522}

This absorber (System~123 in Table~\ref{tab:absprop}) is detected in many metal 
ions. We use the \CII, \SiII\ and \SiIII\ detections to constrain the model, 
along with the limits on \OI\ and \FeIII. The higher-ionization metal detections 
do not agree with the lower-ionization constraints, emphasizing the multi-phase 
nature of this absorber. We use a lower \HI\ column than Paper~1 and find a somewhat higher ionization parameter and an entirely consistent absorber metallicity.

\subsubsection{PG~1211+143\,/\,15321 [A]}
\label{indiv:cloudy:pg1211_15321}

This absorber (System~126 in Table~\ref{tab:absprop}) is also detected in many 
ions. We use the \CII, \SiII\ and \SiIII\ detections to constrain the model, 
but not the higher-ionization detections in \CIII, \CIV, \SiIV, \NV, and \OVI. 
We assume an \HI\ column consistent with the value from Paper~1 and find 
consistent values of ionization parameter and metallicity for this multi-phase 
absorber.

\subsubsection{PG~1211+143\,/\,19305 [B]}
\label{indiv:cloudy:pg1211_19305}

This absorber (System~130 in Table~\ref{tab:absprop}) is detected in \SiIII, 
\CIII, \CIV, and \OVI. We use all but the \OVI\ to constrain our model and find 
a comparable ionization parameter and metallicity to Paper~1 using an \HI\ column 
a bit larger than they assumed.

\subsubsection{PG~1211+143\,/\,19481 [D]}
\label{indiv:cloudy:pg1211_19481}

This serendipitous absorber (System~132 in Table~\ref{tab:absprop}) has the 
lowest \HI\ column of any we attempt to model here, and consequently the 
highest absorber metallicity. It is detected in only \HI, \CIII\ and \CIV, but our 
\HI\ column is considerably smaller than the value assumed in Paper~1. This causes 
our ionization parameter and metallicity to differ more from the values in Paper~1 
than any other absorber for which a comparison can be made. The poorly constrained 
values for \HI\ column density and ionization parameter lead to this absorber being 
assigned grade~D.

\subsubsection{PG~1216+069\,/\,37049 [B]}
\label{indiv:cloudy:pg1216_37049}

This absorber (System~137 in Table~\ref{tab:absprop}) and the following were 
listed as a single system in Paper~1, where it was not modeled because 
\citet{danforth08} list only \SiII, \CIII\ and \OVI\ at this redshift. 
The high-S/N COS spectrum (Figure~\ref{fig:stackplot}) reveals that there 
are in fact two \HI\ components here, along with another two $\sim300$~\kms\ 
redward, as well as metal-line absorption from \SiIII, \SiIV, \CIV, \NV, 
and \OVI. All of these species except for \NV\ ang \OVI\ are used to constrain
our model, along with \CIII\ from \fuse\ and stringent limits on \CII\ and \SiII.

\subsubsection{PG~1216+069\,/\,37138 [B]}
\label{indiv:cloudy:pg1216_37138}

This absorber (System~138 in Table~\ref{tab:absprop}) was not modeled in Paper~1 
for the reasons described in Section~\ref{indiv:cloudy:pg1216_37049}. It is detected
in the same ions as the absorber above and its model was constrained in the same 
way.

\subsubsection{PG~1259+593\,/\,13825 [A]}
\label{indiv:cloudy:pg1259_13825}

This absorber (System~142 in Table~\ref{tab:absprop}) is detected in \CIII, \CIV, 
\SiIII, \SiIV, and \OVI. All but the \OVI\ are used to constrain our model, which 
assumes a nearly identical \HI\ column as Paper~1 and predicts values for the 
ionization parameter and absorber metallicity that are consistent those of Paper~1, 
thanks to the rather large uncertainties quoted therein.

\subsubsection{PG~1259+593\,/\,13914 [A]}
\label{indiv:cloudy:pg1259_13914}

This absorber (System~143 in Table~\ref{tab:absprop}) is detected in \CIII\ and
\CIV, which are used to constrain the model, and \OVI, which is not. This model 
assumes a similar value of $N_{\rm H\,I}$ as Paper~1 and finds values of the 
ionization parameter and metallicity consistent with those quoted in Paper~1.

\subsubsection{PHL~1811\,/\,22042 [---]}
\label{indiv:cloudy:phl1811_22042}

This absorber (System~148 in Table~\ref{tab:absprop}) was modeled in Paper~1 
using \SiIII\ and \SiIV\ column densities from \citet{danforth08}. The high-S/N
COS spectrum (Figure~\ref{fig:stackplot}) shows no indication of \SiIV\ 
absorption, however, so we do not model this absorber here.

\subsubsection{PHL~1811\,/\,23313 [A]}
\label{indiv:cloudy:phl1811_23313}

This absorber (System~149 in Table~\ref{tab:absprop}) has a peculiar 
ionization signature similar to the 3C~273 absorber we model in 
Section~\ref{indiv:cloudy:3c273_1585} in so far as it is detected in 
\SiII, \SiIII\ and \CIII\ but no higher-ionization species. Our updated 
model assumes a higher \HI\ column than in Paper~1, and subsequently prefers 
a lower ionization parameter, but the two models find consistent metallicities.

\subsubsection{PHL~1811\,/\,24226 [C]}
\label{indiv:cloudy:phl1811_24226}

This absorber (System~152 in Table~\ref{tab:absprop}) is detected in many metal 
species and has the largest $N_{\rm H\,I}$ value of any serendipitous absorber. 
It is also remarkable in its simple component structure for such a high column 
density system; all of the \HI\ and metal-line detections have a single velocity 
component except for \CIV, which has two. Our model finds evidence for a multi-phase 
absorber whose \SiIV, \CIII\ and \CIV\ detections do not match the solution preferred 
by the lower ions (detections in \OI, \CII, \SiII, and \FeII, with a stringent limit 
on \CI). We find a ionization parameter and metallicity values consistent with those 
of Paper~1.

\subsubsection{PHL~1811\,/\,39658 [D]}
\label{indiv:cloudy:phl1811_39658}

This absorber (System~153 in Table~\ref{tab:absprop}) is detected only in 
\CIII\ and \CIV, which we use to constrain our model, and \OVI, which we don't. 
It was not modeled in Paper~1 because \citet{danforth08} found only \CIII\ and 
\OVI\ absorption at this redshift; the \CIV\ only becomes apparent in the high-S/N
COS spectrum (Figure~\ref{fig:stackplot}). Due to its high ionization parameter
and low metallicity this absorber has the largest line-of-sight thickness and mass 
of any absorber we attempt to model. However, due to its very weak metal-line 
detections and correspondingly large uncertainty in ionization parameter this 
model is grade~D.

\subsubsection{PHL~1811\,/\,52933 [B]}
\label{indiv:cloudy:phl1811_52933}

This absorber (System~157 in Table~\ref{tab:absprop}) is detected in \CIII, 
\SiIII\ and \SiIV, and has stringent limits on \CII, \SiII, \NV, and \OVI. 
We find a single-phase solution that accommodates all of these constraints and 
is entirely consistent with the solution of Paper~1.

\subsubsection{PKS~0405--123\,/\,50059 [B]}
\label{indiv:cloudy:pks0405_50059}

This absorber (System~170 in Table~\ref{tab:absprop}) was not modeled in Paper~1
because \citet{danforth08} assumed a different component structure for this 
complex absorber than \citet{savage14}. Our model assumes the \citet{savage14} 
values for \HI\ and finds a multi-phase absorber whose \CIII\ and \OVI\ detections 
are inconsistent with the solution that matches the \CII, \SiII\ and \SiIII\ 
detections.

\subsubsection{PKS~0405--123\,/\,50104 [B]}
\label{indiv:cloudy:pks0405_50104}

This multi-phase absorber (System~171 in Table~\ref{tab:absprop}) is 
well-studied by several groups (see 
Section~\ref{indiv:absorbers:pks0405_50104} for details) and detected 
in many ions. As with the previous absorber, we adopt the \HI\ component 
structure of \citet{savage14}. Our model cannot simultaneously meet the 
constraints of the high- and low-ionization species, so we choose to model the 
low ion detections (\CII, \SiII, \SiIII, and \FeIII). Our updated model finds a 
lower metallicity for this absorber than the value reported in Paper~1.

\subsubsection{PKS~1302--102\,/\,12655 [A]}
\label{indiv:cloudy:pks1302_12655}

This is another multi-phase absorber (System~175 in Table~\ref{tab:absprop}) 
detected in many ions. Our model is entirely consistent with that of Paper~1 and 
uses the low- and intermediate-ionization detections (\CII, \CIII, \SiII, \SiIII) 
to constrain the model, but not the higher ionization species (\CIV\ and \OVI).

\subsubsection{PKS~1302--102\,/\,28439 [A]}
\label{indiv:cloudy:pks1302_28439}

This is another absorber (System~178 in Table~\ref{tab:absprop}) seen mostly 
in low ions. In this regard it is similar to the 3C~273 absorber studied above 
(Section~\ref{indiv:cloudy:3c273_1585}) except that it does show \OVI\ absorption, 
which we do not use to constrain the model. Our updated model simultaneously 
satisfies the constraints from \CII, \CIII, \SiII, and \SiIII\ detections, as 
well as limits from \CIV, \SiIV, \OI, \FeIII, and \NV. Even though the updated 
absorber metallicity is higher than found in Paper~1, it is the lowest of any 
absorber we model here.

\subsubsection{Q~1230+011\,/\,23404 [C]}
\label{indiv:cloudy:q1230_23404}

Our updated model for this multi-phase absorber (System~187 in 
Table~\ref{tab:absprop}) is entirely consistent with the 
model of Paper~1, although we have less confidence in the \HI\ column density than 
is the case for the typical serendipitous absorber. The \SiIII, \SiIV\ and \CIII\ 
detections are used to constrain the model, along with the limits on \SiII\ and 
\CII, but not the higher-ionization detections of \CIV\ and \NV.

\section{Ground-Based Galaxy Images}
\label{imaging}

As discussed in Section~\ref{galaxies}, we endeavored to derive \Ha\ SFRs from 
narrowband \Ha\ images for all galaxies with recession velocities $\leq10000$~\kms, 
where spectroscopic aperture corrections are large and unreliable. All ten targeted 
galaxies are at these low redshifts, as well as eight of the serendipitous 
galaxies. \Ha\ images were available in the literature for four of these galaxies: 
UGC~5470 \citep{kennicutt08}, M~108 \citep{kennicutt08}, NGC~6140 
\citep{sanchez-gallego12}, and NGC~4939 \citep{tsvetanov95}. We obtained 
narrowband \Ha\ images for 12 others, which are detailed in Table~\ref{tab:ha}. 
Table~\ref{tab:ha} lists the following information by column: (1) galaxy name; (2) 
telescope where images were obtained; (3) instrument used to obtain images; (4) 
date of observation; (5) broadband and narrowband filters used for galaxy imaging; 
(6) exposure time per filter, in units of kiloseconds; and (7) the seeing in each 
filter, in units of arcseconds.

\clearpage
\floattable
\begin{deluxetable*}{llllccc}
\tablecaption{Journal of \Ha\ Imaging Observations\label{tab:ha}}
\tablewidth{0pt}
\tabletypesize{\small}
\tablehead{ \colhead{Galaxy} &  \colhead{Telescope} & \colhead{Instrument} & \colhead{Date} & \colhead{Filters} & \colhead{Exposure} & \colhead{Seeing} \\ & & & & & \colhead{(ksec)} & \colhead{(arcsec)}}

\colnumbers
\startdata
UGC~5478                 & KPNO 2.1-m & T2KA   & 2004 Apr 16    & $B$, $R$,         KP1563  & 0.6, 0.5,       2.4 & 0.9, 0.8,      1.0 \\
ESO~157--49              & CTIO 0.9-m & CFCCD  & 2003 Aug 22-23 & $B$, $R$,         6600/75 & 3.0, 1.8,       9.0 & 2.3, 1.8,      1.5 \\
ESO~157--50              & CTIO 0.9-m & CFCCD  & 2003 Aug 24-25 & $B$, $R$,         6649/75 & 1.8, 1.8,      10.8 & 2.8, 2.5,      1.6 \\
NGC~2611                 &  APO 3.5-m & SPIcam & 2013 Mar 15    &      $r$, 645/10,  665/8  &      0.1, 2.7,  3.6 &      1.3, 1.3, 1.1 \\
NGC~3511                 & KPNO 2.1-m & T2KA   & 2004 Apr 13    & $B$, $R$,         KP1563  & 0.6, 0.5,       2.4 & 2.0, 1.9,      2.0 \\
IC~691                   & KPNO 2.1-m & T2KA   & 2003 Mar  4, 6 & $B$, $R$, KP1565, KP1563  & 0.5, 1.7, 2.4,  2.4 & 1.5, 1.8, 1.5, 1.5 \\
UGC~4527                 & KPNO 2.1-m & T2KA   & 2004 Apr 13-16 & $B$, $R$,         KP1563  & 0.9, 0.7,       2.4 & 1.1, 2.2,      1.6 \\
SDSS~J122815.96+014944.1 &  APO 3.5-m & SPIcam & 2013 Mar 15    &      $r$,          657/8  &      0.1,       1.2 &      1.2,      1.2 \\
SDSS~J122950.57+020153.7 &  APO 3.5-m & SPIcam & 2011 May  4    &      $R$,          661/8  &      1.8,       4.5 &      1.4,      1.2 \\
SDSS~J000529.16+201335.9 &  APO 3.5-m & ARCTIC & 2016 Nov  1    &      $r$, 657/3,   660/3  &      0.1, 6.0,  6.0 &      0.9, 0.7, 0.6 \\
IC~3061                  &  APO 3.5-m & SPIcam & 2009 Jan 20    &           645/10,  661/8  &           1.8,  2.4 &           2.2, 2.1 \\
UGC~8146                 &  APO 3.5-m & SPIcam & 2008 Feb  2    &           645/10,  657/8  &           1.4,  1.8 &           1.0, 0.9 \\
\enddata

\end{deluxetable*}

\begin{figure}[!t]
\gridline{\fig{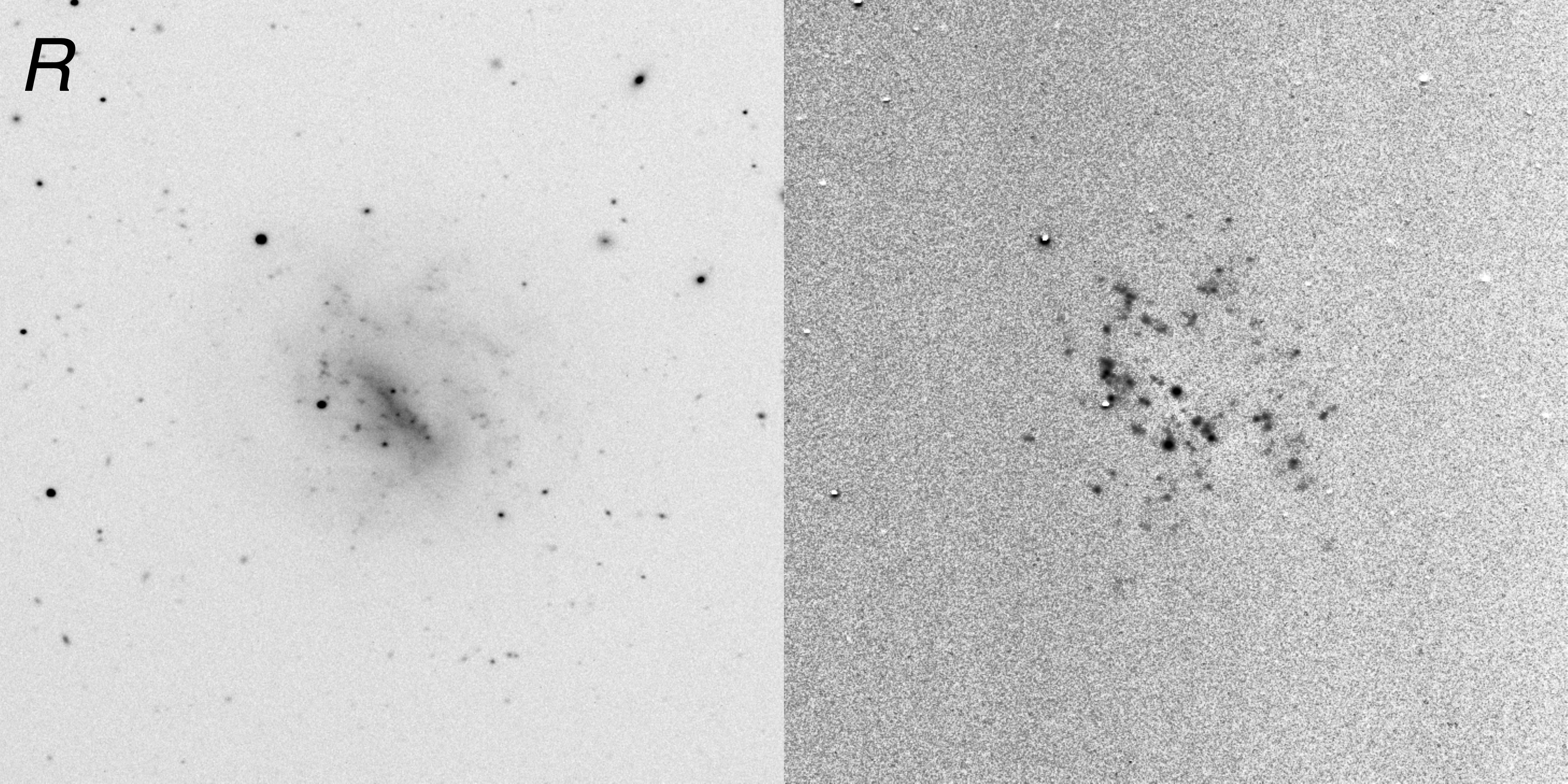}{0.45\textwidth}{B1.1: Continuum and emission-line images for UGC~5478 (Section~\ref{indiv:galaxies:ugc5478}).}}
\gridline{\fig{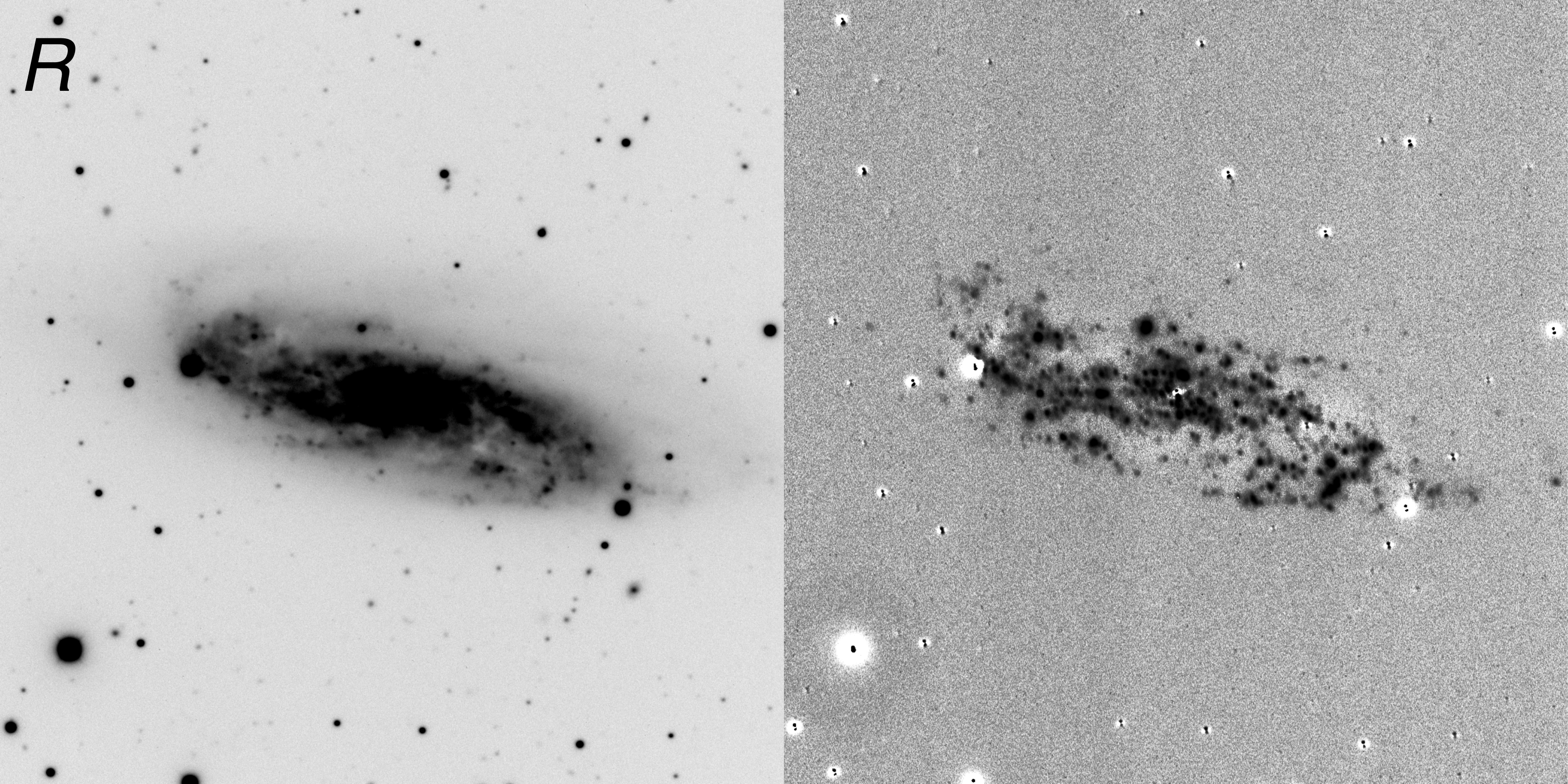}{0.45\textwidth}{B1.5: Continuum and emission-line images for NGC~3511 (Section~\ref{indiv:galaxies:ngc3511}).}}
\caption{Continuum (left) and \Ha\ + [\NII] (right) images of targeted and serendipitous galaxies. All images are oriented north-up, east-left and have a field-of-view of $25\times25~{\rm kpc}^2$ (except for UGC~4527 and SDSS~J122815.96+014944.1, which have a field-of-view of $15\times15~{\rm kpc}^2$). The name of the continuum filter for each galaxy is also labeled in the left-hand panel. Residual structure at the locations of point sources in the right-hand panel (e.g., white rings around black cores) are the result of seeing differences between the continuum and narrowband \Ha\ images. The complete figure set (12~images) is available in the online journal.
\label{fig:images}}
\end{figure}

One complication that sometimes arises is a mismatch in seeing between the 
continuum image and the narrowband \Ha\ image. The narrowband image still 
contains some stellar continuum flux (i.e., it is not a pure emission-line image) 
that must be estimated from the off-band continuum image. We do this by scaling 
the sky level in the continuum image to match that of the narrowband image, then 
subtracting this scaled continuum image from the narrowband image to create a 
pure emission-line image, from which the galaxy's \Ha\ SFR is derived. If the 
continuum image and the narrowband image have significantly different seeing 
(e.g., NGC~3511; see Figure~\ref{fig:images}), there will be residual artifacts 
in the the emission-line image from point sources that are not cleanly subtracted 
due to mismatched point-spread functions. However, since the galaxies themselves 
are much larger on the sky than point sources at these very low redshifts, seeing 
differences do not have a significant effect on the estimated continuum flux 
distributions.

Figure~\ref{fig:images} shows continuum and emission-line images for all galaxies 
listed in Table~\ref{tab:ha}. These images are all oriented north-up, east-left, 
and have a field-of-view of $25\times25~{\rm kpc}^2$, except for UGC~4527 and 
SDSS~J122815.96+014944.1, which have a field-of-view of $15\times15~{\rm kpc}^2$. 
The narrowband \Ha\ filters employed are all wide enough that the emission-line 
images (right-hand panels of Figure~\ref{fig:images}) include contributions from 
both \Ha\ and [\NII]. We use optical spectra of the galaxy, where available, to 
estimate the contribution of [\NII] to the emission-line image. These spectra 
cover the nuclear regions of the galaxy and do not allow us to account for any 
variations in \Ha/[\NII] for other regions of the galaxy. If no galaxy spectrum is 
available, we use the procedure of \citet{kennicutt08} to statistically estimate 
the galaxy's \Ha/[\NII] ratio.

\clearpage
Some of our galaxies (e.g., NGC~2611) were observed through a narrow- or 
intermediate-width, off-band continuum filter that does not include any 
\Ha\ + [\NII] emission from the galaxy, while others (e.g., UGC~5478) used a 
broad off-band continuum filter that does include some contamination from the 
galaxy. Continuum subtraction in the former case is straightforward, but requires 
some care in the latter case; we use the procedure of \citet{kennicutt08} to 
statistically correct for any galaxy emission present in the continuum filter. 
The name of the off-band continuum filter used for each galaxy is labeled in 
that galaxy's continuum image (left-hand panel) in Figure~\ref{fig:images}.

The \Ha\ SFR for the galaxy is derived from the emission-line images in 
Figure~\ref{fig:images}, corrected for [\NII] emission in the narrowband 
filter as well as any \Ha\ emission that is present in the continuum filter. 
The SFR is derived from the galaxy's \Ha\ luminosity using the conversion of 
\citet{hunter10}, and is listed in column~11 of Tables~\ref{tab:targ_galprop} 
and \ref{tab:ser_galprop}. The lower value in the tabulated SFR range is the 
value measured directly from the images, after correcting for Galactic 
foreground extinction, and the higher value employs a correction for internal 
extinction in the galaxy, as described in Section~\ref{galaxies}.

\Ha\ images for three of our galaxies have been previously published. The 
images for IC~691 were published in \citet{keeney06}, and those for ESO~157--49
and ESO~157--50 were published in \citet{keeney13}. We reproduce these images 
in Figure~\ref{fig:images} for completeness but adopt the previously published 
SFR values, corrected to use the \citet{hunter10} SFR conversion when necessary
for consistency with the rest of the sample.

\bigskip
\figsetstart
\figsetnum{B1}
\figsettitle{Continuum and Emission-Line Images}

\figsetgrpstart
\figsetgrpnum{B1.1}
\figsetgrptitle{UGC~5478}
\figsetplot{./fig_images_ugc5478.pdf}
\figsetgrpnote{Continuum and emission-line images for UGC~5478 (Section~\ref{galaxies:targeted:ugc5478}).}
\figsetgrpend

\figsetgrpstart
\figsetgrpnum{B1.2}
\figsetgrptitle{ESO~157--49}
\figsetplot{./fig_images_eso157-49.pdf}
\figsetgrpnote{Continuum and emission-line images for ESO~157--49 (Section~\ref{galaxies:targeted:eso157-49}).}
\figsetgrpend

\figsetgrpstart
\figsetgrpnum{B1.3}
\figsetgrptitle{ESO~157--50}
\figsetplot{./fig_images_eso157-50.pdf}
\figsetgrpnote{Continuum and emission-line images for ESO~157--50 (Section~\ref{galaxies:targeted:eso157-50}).}
\figsetgrpend

\figsetgrpstart
\figsetgrpnum{B1.4}
\figsetgrptitle{NGC~2611}
\figsetplot{./fig_images_ngc2611.pdf}
\figsetgrpnote{Continuum and emission-line images for NGC~2611 (Section~\ref{galaxies:targeted:ngc2611}).}
\figsetgrpend

\figsetgrpstart
\figsetgrpnum{B1.5}
\figsetgrptitle{NGC~3511}
\figsetplot{./fig_images_ngc3511.pdf}
\figsetgrpnote{Continuum and emission-line images for NGC~3511 (Section~\ref{galaxies:targeted:ngc3511}).}
\figsetgrpend

\figsetend


\begin{thebibliography}

\bibitem[Adams et~al.(2011)]{adams11} Adams, J. J., Uson, J. M., Hill, G. J., \& MacQueen, P. J. 2011, \apj, 728, 107

\bibitem[Adelberger et~al.(2005)]{adelberger05} Adelberger, K. L., Shapley, A. E., Steidel, C. C., et~al. 2005, \apj, 629, 636

\bibitem[Aguirre et~al.(2004)]{aguirre04} Aguirre, A., Schaye, J., Kim, T.-S., et~al. 2004, \apj, 602, 38

\bibitem[Alam et~al.(2015)]{alam15} Alam, S., Albareti, F. D., Allende Prieto, C., et~al. 2015, \apjs, 219, 12

\bibitem[Anderson \& Bregman(2011)]{anderson11} Anderson, M. E. \& Bregman, J. N. 2011, \apj, 737, 22

\bibitem[Armillotta et~al.(2016)]{armillotta16} Armillotta, L., Werk, J. K., Prochaska, J. X., Fraternali, F., \& Marinacci, F. 2016, \mnras, submitted (arXiv:1608.05416)

\bibitem[Asplund et~al.(2009)]{asplund09} Asplund, M., Grevesse, N., Sauval, A. J., \& Scott, P. 2009, \araa, 47, 481

\bibitem[Binney \& Tremaine(1987)]{binney87} Binney, J. \& Tremaine, S. Galactic Dynamics (Princeton: Princeton Univ. Press), pp. 567-575

\bibitem[Bordoloi et~al.(2014)]{bordoloi14} Bordoloi, R., Tumlinson, J., Werk, J. K., et~al. 2014, \apj, 796, 136

\bibitem[Bouch\'e et~al.(2012)]{bouche12} Bouch\'e, N., Hohensee, W., Vargas, R., et~al. 2012, \mnras, 426, 801

\bibitem[Bowen et~al.(2016)]{bowen16} Bowen, D. V., Chelouche, D., Jenkins, E. B., et~al. 2016, \apj, 826, 50


\bibitem[Calzetti(2001)]{calzetti01} Calzetti, D. 2001, \pasp, 113, 1449

\bibitem[Chen \& Prochaska(2000)]{chen00} Chen, H.-W. \& Prochaska, J. X. 2000, \apj, 543, L9

\bibitem[Chen et~al.(2001)]{chen01} Chen, H.-W., Lanzetta, K. M., Webb, J. K., \& Barcons X. 2001, \apj, 559, 654

\bibitem[Chen \& Mulchaey(2009)]{chen09} Chen, H.-W. \& Mulchaey, J. S. 2009, \apj, 701, 1219

\bibitem[Chiappini et~al.(2001)Chiappini, Matteucci \& Romano]{chiappini01} Chiappini, C, Matteucci, F. \& Romano, D. 2001, \apj, 554, 1044

\bibitem[Chilingarian et~al.(2010)Chilingarian, Melchior, \& Zolotukhin]{chilingarian10} Chilingarian, I. V., Melchior, A.-L., \& Zolotukhin, I. Y. 2010, \mnras, 405, 1409

\bibitem[Chilingarian \& Zolotukhin(2012)]{chilingarian12} Chilingarian, I. V. \& Zolotukhin, I. Y. 2012, \mnras, 419, 1727

\bibitem[Chomiuk \& Povich(2011)]{chomiuk11} Chomiuk, L. \& Povich, M. S. 2011, \aj, 142, 197

\bibitem[Collins et~al.(2004)Collins, Shull, \& Giroux]{collins04} Collins, J. A., Shull, J. M., \& Giroux, M. L. 2004, \apj, 605, 216

\bibitem[Collins et~al.(2009)Collins, Shull, \& Giroux]{collins09} Collins, J. A., Shull, J. M., \& Giroux, M. L. 2009, \apj, 705, 962

\bibitem[Crighton et~al.(2010)]{crighton10} Crighton, N. H. M., Morris, S. L., Bechtold, J., et~al. 2010, \mnras, 402, 1273

\bibitem[Danforth \& Shull(2005)]{danforth05} Danforth, C. W. \& Shull, J. M. 2005, \apj, 625, 555

\bibitem[Danforth \& Shull(2008)]{danforth08} Danforth, C. W. \& Shull, J. M. 2008, \apj, 679, 194

\bibitem[Danforth et~al.(2010)]{danforth10} Danforth, C. W., Keeney, B. A., Stocke, J. T., Shull, J. M., \& Yao, Y. 2010, \apj, 720, 976

\bibitem[Danforth et~al.(2016)]{danforth16} Danforth, C. W., Keeney, B. A., Tilton, E. M., et~al. 2016, \apj, 817, 111

\bibitem[Dav\'e et~al.(2010)]{dave10} Dav\'e, R., Oppenheimer, B. D., Katz, N., Kollmeier, J. A., \& Weinberg, D. H. 2010, \mnras, 408, 2051

\bibitem[Davis et~al.(2015)]{davis15} Davis, J. D., Keeney, B. A., Danforth, C. W., \& Stocke, J. T. 2015, \apj, 810, 92

\bibitem[de~Vaucouleurs et~al.(1991)]{devaucouleurs91} de~Vaucouleurs, G., de~Vaucouleurs, A., Corwin, H. G., Jr., et~al. 1991, Third Reference Catalogue of Bright Galaxies, Version~3.9

\bibitem[Diehl et~al.(2006)]{diehl06} Diehl, R., Halloin, H., Kretschmer, K., et~al. 2006, \nat, 439, 45

\bibitem[Donahue et~al.(1995)Donahue, Aldering, \& Stocke]{donahue95} Donahue, M., Aldering, G., \& Stocke, J. T. 1995, \apj, 450, L45

\bibitem[D'Onghia \& Fox(2016)]{d'onghia16} D'Onghia, E. \& Fox, A. J. 2016, \araa, 54, 363

\bibitem[Driver et~al.(2008)]{driver08} Driver, S. P., Popescu, C. C., Tuffs R. J., et~al. 2008, \apjl, 678, L101

\bibitem[Faerman et~al.(2016)Faerman, Sternberg, \& McKee]{faerman16} Faerman, Y., Sternberg, A., \& McKee, C. F. 2016, \apj, submitted (arXiv:1602.00689)

\bibitem[Faucher-Gigu\`ere et~al.(2009)]{faucher-giguere09} Faucher-Gigu\`ere, C.-A., Lidz, A., Zaldarriaga, M., \& Hernquist, L. 2009, \apj, 703, 1416

\bibitem[Ferland et~al.(1998)]{ferland98} Ferland, G. J., Korista, K. T., Verner, D. A., et~al. 1998, \pasp, 110, 761

\bibitem[Fitzpatrick(1999)]{fitzpatrick99} Fitzpatrick, E. L. 1999, \pasp, 111, 63

\bibitem[Fukugita et~al.(1995)Fukugita, Shimasaku, \& Ichikawa]{fukugita95} Fukugita, M., Shimasaku, K., \& Ichikawa, T. 1995, \pasp, 107, 945

\bibitem[Fumagalli et~al.(2017)]{fumagalli17} Fumagalli, M., Haardt, F., Theuns, T., et~al. 2017, \mnras, in press (arXiv:1702.04726)

\bibitem[Garcia et~al.(1992)]{garcia92} Garcia, A. M., Bottinelli, L., Garnier, R., Gouguenheim, L., \& Paturel, G. 1992, \aaps, 96, 435

\bibitem[Gaikwad et~al.(2016)]{gaikwad16} Gaikwad, P., Srianand, R., Choudhury, T. R., \& Khaire, V. 2016, \mnras, submitted (arXiv:1610.06572)

\bibitem[Giroux \& Shull(1997)]{giroux97} Giroux, M. L. \& Shull, J. M. 1997, \aj, 113, 1505

\bibitem[Green et~al.(2012)]{green12} Green, J. C., Froning, C. S., Osterman, S. et~al. 2012, \apj, 744, 60

\bibitem[Grevesse et~al.(2010)]{grevesse10} Grevesse, N., Asplund, M., Sauval, A. J., \& Scott, P. 2010, \apss, 328, 179

\bibitem[Gupta et~al.(2012)]{gupta12} Gupta, A., Mathur, S., Krongold, Y., Nicastro, F., \& Galeazzi, M. 2012, \apj, 756, L8

\bibitem[Haardt \& Madau(1996)]{haardt96} Haardt, F. \& Madau, P. 1996, \apj, 461, 20

\bibitem[Haardt \& Madau(2001)]{haardt01} Haardt, F. \& Madau, P. 2001, in Clusters of Galaxies and the High Redshift Universe Observed in X-Rays, ed. D. M. Neumann \& J. T. V. Tran, 64

\bibitem[Haardt \& Madau(2005)]{haardt05} Haardt, F. \& Madau, P. 2005, unpublished spectra in 2005~August update to \citet{haardt01} and included in the photo-ionization code CLOUDY

\bibitem[Haardt \& Madau(2012)]{haardt12} Haardt, F. \& Madau, P. 2012, \apj, 746, 125

\bibitem[Hao et~al.(2011)]{hao11} Hao, C.-N., Kennicutt, R. C., Jr., Johnson, B. D., et~al. 2011, \apj, 741, 124

\bibitem[Hinshaw et~al.(2013)]{hinshaw13} Hinshaw, D., Larson, G., Komatsu, E., et~al. 2013, \apjs, 208, 19

\bibitem[Hunter et~al.(2010)Hunter, Elmegreen, \& Ludka]{hunter10} Hunter, D. A., Elmegreen, B. G., \& Ludka, B. C. 2010, \aj, 139, 447

\bibitem[Iglesias-P\'aramo et~al.(2013)]{iglesias-paramo13} Iglesias-P\'aramo, J., V\'ilchez, J. M., Galbany, L., et~al. 2013, \aap, 553, L7

\bibitem[Jenkins et~al.(2003)]{jenkins03} Jenkins, E. B., Bowen, D. V., Tripp, T. M., et~al. 2003, \aj, 125, 2824

\bibitem[Jester et~al.(2005)]{jester05} Jester, S., Schneider, D. P., Richards, G. T., et~al. 2005, \aj, 130, 873

\bibitem[Johnson et~al.(2013)Johnson, Chen, \& Mulchaey]{johnson13} Johnson, S. D., Chen, H.-W., \& Mulchaey, J. S. 2013, \mnras, 434, 1765

\bibitem[Jones et~al.(2005)]{jones05} Jones, D. H., Saunders, W., Read, M., \& Colless, M. 2005, \pasa, 22, 277

\bibitem[Karachentsev \& Nasonova(2013)]{karachentsev13} Karachentsev, I. D. \& Nasonova, O. G. 2013, \mnras, 429, 2677

\bibitem[Keeney et~al.(2006)]{keeney06} Keeney, B. A., Stocke, J. T., Rosenberg, J. L., Tumlinson, J., \& York, D. G. 2006b, \aj, 132, 2496

\bibitem[Keeney et~al.(2012)]{keeney12} Keeney, B. A., Danforth, C. W., Stocke, J. T., France, K., \& Green, J. C. 2012, \pasp, 124, 830

\bibitem[Keeney et~al.(2013)]{keeney13} Keeney, B. A., Stocke, J. T., Rosenberg, J. L., et~al. 2013, \apj, 765, 27

\bibitem[Keeney et~al.(2014)]{keeney14} Keeney, B. A., Joeris, P., Stocke, J. T., Danforth, C. W., \& Levesque, E. M. 2014, \aj, 148, 103

\bibitem[Keeney et~al.(2017)]{keeney17} Keeney, B. A., Stocke, J. T., Pratt, C., et~al. 2017, in prep.

\bibitem[Kennicutt et~al.(2008)]{kennicutt08} Kennicutt, R. C., Jr., Lee, J. C., Funes, S. J. J. G., Sakai, S., \& Akiyama, S. 2008, \apjs, 178, 247

\bibitem[Kere\u{s} \& Hernquist(2009)]{keres09} Kere\u{s}, D. \& Hernquist, L. 2009, \apjl, 700, L1

\bibitem[Kewley et~al.(2006)]{kewley06} Kewley, L. J., Groves, B., Kauffmann, G., \& Heckman, T. 2006, \mnras, 372, 961

\bibitem[Khaire \& Srianand(2015)]{khaire15} Khaire, V. \& Srianand, R. 2015, \mnras, 451, L30

\bibitem[Klypin et~al.(2001)]{klypin01} Klypin, A., Kravtsov, A. V., Bullock, J. S. \& Primack, J. R. 2001, \apj, 554, 903

\bibitem[Kollmeier et~al.(2014)]{kollmeier14} Kollmeier, J. A., Weinberg, D. H., Oppenheimer, B. D., et~al. 2014, \apj, 789, L32

\bibitem[Kriss(2011)]{kriss11} Kriss, G. A. 2011, COS Instrment Science Report 2011-01(v1), Improved Medium Resolution Line Spread Functions for COS FUV Spectra (Baltimore: STScI)

\bibitem[Larson(1972)]{larson72} Larson, R. B. 1972, \mnras, 157, 121

\bibitem[Lee et~al.(2006)]{lee06} Lee, H., Skillman, E. D., Cannon, J. M., et~al. 2006, \apj, 647, 970

\bibitem[Lehner \& Howk(2011)]{lehner11} Lehner, N. \& Howk, J. C. 2011, Science, 334, 955

\bibitem[Lehner et~al.(2015)]{lehner15} Lehner, N., Howk, J. C., \& Wakker, B. P. 2015, \apj, 804, 79 

\bibitem[Madau et~al.(1996)]{madau96} Madau, P., Ferguson, H. C., Dickinson, M. E., et~al. 1996, \mnras, 283, 1388

\bibitem[Madau \& Haardt(2015)]{madau15} Madau, P. \& Haardt, F. 2015, \apj, 813, L8

\bibitem[Martin et~al.(2015)]{martin15} Martin, C. L., Dijkstra, M., Henry, A., et~al. 2015, \apj, 803, 6

\bibitem[Masters et~al.(2010)]{masters10} Masters, K. L., Nichol, R., Bamford, S., et~al. 2010, \mnras, 404, 792

\bibitem[McGaugh et~al.(2000)]{mcgaugh00} McGaugh, S. S., Schombert, J. M., Bothun, G. D., \& de Blok, W. J. G. 2000, \apjl, 533, L99

\bibitem[McLin(2003)]{mclin03} McLin, K. M. 2003, Ph.D. dissertation, Univ. of Colorado Boulder

\bibitem[Meyer et~al.(2004)]{meyer04} Meyer, M. J., Zwaan, M. A., Webster, R. L., et~al. 2004, \mnras, 350, 1195

\bibitem[Montero-Dorta \& Prada(2009)]{montero-dorta09} Montero-Dorta, A. D. \& Prada, F. 2009, \mnras, 399, 1106

\bibitem[Moster et~al.(2013)Moster, Naab, \& White]{moster13} Moster, B. P., Naab, T. \& White, S. D. M. 2013, \mnras, 428, 3121

\bibitem[Mulchaey(2000)]{mulchaey00} Mulchaey, J. S. 2000, \araa, 38, 289

\bibitem[Pachat et~al.(2016)]{pachat16} Pachat, S., Narayanan, A., Muzahid, S., et~al. 2016, \mnras, 458, 733

\bibitem[Peimbert et~al.(2007)Peimbert, Luridiana, \& Peimbert]{peimbert07} Peimbert, M., Luridiana, V., \& Peimbert, A. 2007, \apj, 666, 636

\bibitem[Penton et~al.(2004)Penton, Stocke, \& Shull]{penton04} Penton, S. V., Stocke, J. T., \& Shull, J. M. 2004, \apjs, 152, 29

\bibitem[Pettini \& Pagel(2004)]{pettini04} Pettini, M. \& Pagel, B. E. J. 2004, \mnras, 348, L59

\bibitem[Prochaska et~al.(2004)]{prochaska04} Prochaska, J. X., Chen, H.-W., Howk, J. C., Weiner, B. J., \& Mulchaey, J. 2004, \apj, 617, 718

\bibitem[Prochaska et~al.(2011a)]{prochaska11a} Prochaska, J. X., Weiner, B., Chen, H.-W., Mulchaey, J., \& Cooksey, K. 2011a, \apj, 740, 91

\bibitem[Prochaska et~al.(2011b)]{prochaska11b} Prochaska, J. X., Weiner, B., Chen, H.-W., Cooksey, K., \& Mulchaey, J. 2011b, \apjs, 193, 28

\bibitem[Prochaska et~al.(2017)]{prochaska17} Prochaska, J. X., Werk, J. K., Worseck, G., et~al. 2017, \apj, in press (arXiv:1702.02618)

\bibitem[Robitaille \& Whitney(2010)]{robitaille10} Robitaille, T. P. \& Whitney, B. A. 2010, \apjl, 710, L11

\bibitem[Rifatto et~al.(1995)Rifatto, Longo, \& Capaccioli]{rifatto95} Rifatto, A., Longo, G., \& Capaccioli, M. 1995, \aaps, 114, 527

\bibitem[Rosenberg et~al.(2003)]{rosenberg03} Rosenberg, J. L., Ganguly, R., Giroux, M. L., \& Stocke, J. T. 2003, \apj, 597, 677

\bibitem[Rudie et~al.(2013)]{rudie13} Rudie, G. C., Steidel, C. C., Shapley, A. E., \& Pettini, M. 2013, \apj, 769, 146

\bibitem[S\'anchez-Gallego et~al.(2012)]{sanchez-gallego12} S\'anchez-Gallego, J. R., Knapen, J. H., Wilson, C. D., et~al. 2012, \mnras, 422, 3208

\bibitem[Savage \& Sembach(1991)]{savage91} Savage, B. D. \& Sembach, K. R. 1991, \apj, 379, 245

\bibitem[Savage et~al.(2002)]{savage02} Savage, B. D., Sembach, K. R., Tripp, T. M., \& Richter, P. 2002, \apj, 564, 631

\bibitem[Savage et~al.(2010)]{savage10} Savage, B. D., Narayanan, A., Wakker, B. P., et~al. 2010, \apj, 721, 960

\bibitem[Savage et~al.(2014)]{savage14} Savage, B. D., Kim, T.-S., Wakker, B. P., et~al. 2014, \apjs, 212, 8

\bibitem[Schaye et~al.(2003)]{schaye03} Schaye, J., Aguirre, A., Kim, T.-S., et~al. 2003, \apj, 596, 768

\bibitem[Schlafly \& Finkbeiner(2011)]{schlafly11} Schlafly, E. F. \& Finkbeiner, D. P. 2011, \apj, 737, 103

\bibitem[Schneider et~al.(1992)]{schneider92} Schneider, S. E., Thuan, T. X., Mangum, J. C., \& Miller, J. 1992, \apjs, 81, 5

\bibitem[Sembach et~al.(1995)]{sembach95} Sembach, K. R., Savage, B. D., Lu L., \& Murphy, E. 1995, \apj, 451, 616

\bibitem[Sembach et~al.(2001)]{sembach01} Sembach, K. R., Howk, J. C., Savage, B. D., Shull, J. M., \& Oegerle, W. R. 2001, \apj, 561, 573

\bibitem[Sembach et~al.(2003)]{sembach03} Sembach, K. R., Wakker, B. P., Savage, B. D., et~al. 2003, \apjs, 146, 165

\bibitem[Shull et~al.(1998)]{shull98} Shull, J. M., Penton, S. V., Stocke, J. T., et~al. 1998, \aj, 116, 2094

\bibitem[Shull et~al.(1999)]{shull99} Shull, J. M., Roberts, D., Giroux, M. L., Penton, S. V., \& Fardal, M. A. 1999, \aj, 118, 1450

\bibitem[Shull et~al.(2003)Shull, Tumlinson, \& Giroux]{shull03} Shull, J. M., Tumlinson, J., \& Giroux, M. L. 2003, \apj, 594, L107

\bibitem[Shull et~al.(2009)]{shull09} Shull, J. M., Jones, J. R., Danforth, C. W., \& Collins, J. A. 2009, \apj, 699, 754

\bibitem[Shull et~al.(2011)]{shull11} Shull, J. M., Stevans, M., Danforth, C., et~al. 2011, \apj, 739, 105

\bibitem[Shull et~al.(2012)Shull, Stevans, \& Danforth]{shull12} Shull, J. M., Stevans, M. L., \& Danforth, C. W. 2012, \apj, 752, 162

\bibitem[Shull(2014)]{shull14} Shull, J. M. 2014, \apj, 784, 142

\bibitem[Shull et~al.(2015)]{shull15} Shull, J. M., Moloney, J., Danforth, C. W., \& Tilton, E. M. 2015, \apj, 811, 3

\bibitem[Simcoe et~al.(2004)Simcoe, Sargent, \& Rauch]{simcoe04} Simcoe, R. A., Sargent, W. L. W., \& Rauch, M. 2004, \apj, 606, 92

\bibitem[Sorce et~al.(2014)]{sorce14} Sorce, J. G., Tully, R. B., Courtois, H. M., et~al. 2014, \mnras, 444, 527

\bibitem[Springob et~al.(2005)]{springob05} Springob, C. M., Haynes, M. P., Giovanelli, R., \& Kent, B. R. 2005, \apjs, 160, 149

\bibitem[Steidel(1995)]{steidel95} Steidel, C. C. 1995, in QSO Absorption Lines, ed. G. Meylan (Garching: Springer), 139

\bibitem[Steidel et~al.(1999)]{steidel99} Steidel, C. C., Adelberger, K. L., Giavalisco, M., Dickinson, M., \& Pettini, M. 1999, \apj, 519, 1

\bibitem[Stern et~al.(2016)]{stern16} Stern, J., Hennawi, J. F., Prochaska, J. X., \& Werk, J. K. 2016, \apj, 830, 87

\bibitem[Stevans et~al.(2014)]{stevans14} Stevans, M. L., Shull, J. M., Danforth, C. W., \& Tilton, E. M. 2014, \apj, 794, 75

\bibitem[Stocke et~al.(2004)]{stocke04} Stocke, J. T., Keeney, B. A., McLin, K. M., et~al. 2004, \apj, 609, 94

\bibitem[Stocke et~al.(2006)]{stocke06} Stocke, J. T., Penton, S. V., Danforth, C. W., et~al. 2006, \apj, 641, 217

\bibitem[Stocke et~al.(2007)]{stocke07} Stocke, J. T., Danforth, C. W., Shull, J. M., Penton, S. V., \& Giroux, M. L. 2007, \apj, 671, 146

\bibitem[Stocke et~al.(2013)]{stocke13} Stocke, J. T., Keeney, B. A., Danforth, C. W., et~al. 2013, \apj, 763, 148

\bibitem[Stocke et~al.(2014)]{stocke14} Stocke, J. T., Keeney, B. A., Danforth, C. W., et~al. 2014, \apj, 791, 128

\bibitem[Taylor et~al.(2011)]{taylor11} Taylor, E. N., Hopkins, A. M., Baldry, I. K., et~al. 2011, \mnras, 418, 1587

\bibitem[Tepper-Garcia et~al.(2012)]{tepper-garcia12} Tepper-Garcia, T., Richter, P., Schaye, J., et~al. 2012, \mnras, 425, 1640

\bibitem[Thom et~al.(2008)]{thom08} Thom, C. \& Chen, H.-W. 2008, \apjs, 179, 37

\bibitem[Thom et~al.(2012)]{thom12} Thom, C., Tumlinson, J., Werk, J. K., et~al. 2012, \apj, 758, L41

\bibitem[Tilton et~al.(2012)]{tilton12} Tilton, E. M., Danforth, C. W., Shull, J. M., \& Ross, T. L. 2012, \apj, 759, 112

\bibitem[Tremonti et~al.(2004)]{tremonti04} Tremonti, C. A., Heckman, T. M., Kauffmann, G., et~al. 2004, \apj, 613, 898

\bibitem[Tripp et~al.(1998)Tripp, Lu, \& Savage]{tripp98} Tripp, T. M., Lu, L., \& Savage, B. D. 1998, \apj, 508, 200

\bibitem[Tripp et~al.(2001)]{tripp01} Tripp, T. M., Giroux, M. L., Stocke, J. T., Tumlinson, J., \& Oegerle, W. R. 2001, \apj, 563, 724

\bibitem[Tripp et~al.(2002)]{tripp02} Tripp, T. M., Jenkins, E. B., Williger, G. M., et~al. 2002, \apj, 575, 697

\bibitem[Tripp et~al.(2008)]{tripp08} Tripp, T. M., Sembach, K. R., Bowen, D. V., et~al. 2008, \apjs, 177, 39

\bibitem[Tsvetanov \& Petrosian(1995)]{tsvetanov95} Tsvetanov, Z. I. \& Petrosian, A. R. 1995, \apjs, 101, 287

\bibitem[Tully(1988)]{tully88} Tully, R. B., 1988, Nearby Galaxies Catalog (Cambridge:Cambridge)

\bibitem[Tully et~al.(2009)]{tully09} Tully, B. R., Rizzi, L., Shaya, E. J., et~al. 2009, \aj, 138, 323

\bibitem[Tumlinson et~al.(2005)]{tumlinson05} Tumlinson, T., Shull, J. M., Giroux, M. L., Stocke, J. T. 2005, \apj, 620, 95

\bibitem[Tumlinson et~al.(2011)]{tumlinson11} Tumlinson, J., Thom, C., Werk, J. K., et~al. 2011, Science, 334, 948

\bibitem[Tumlinson et~al.(2013)]{tumlinson13} Tumlinson, J., Thom, C., Werk, J. K., et~al. 2013, \apj, 777, 59

\bibitem[van~Gorkom et~al.(1996)]{vangorkom96} van~Gorkom, J. H., Carilli, C. L., Stocke, J. T., Perlman, E. S., \& Shull, J. M. 1996, \aj, 112, 1397

\bibitem[Veilleux et~al.(2005)Veilleux, Cecil, \& Bland-Hawthorn]{veilleux05} Veilleux, S., Cecil, G., \& Bland-Hawthorn, J. 2005, \araa, 43, 769

\bibitem[Wakker et~al.(2007)]{wakker07} Wakker, B. P., York, D. G., Howk, J. C., et~al. 2007, \apj, 670, L113

\bibitem[Werk et~al.(2012)]{werk12} Werk, J. K., Prochaska, J. X., Thom, C., et~al. 2012, \apjs, 198, 3

\bibitem[Werk et~al.(2013)]{werk13} Werk, J. K., Prochaska, J. X., Thom, C., et~al. 2013, \apjs, 204, 17

\bibitem[Werk et~al.(2014)]{werk14} Werk, J. K., Prochaska, J. X., Tumlinson, J., et~al. 2014, \apj, 792, 8

\bibitem[Werk et~al.(2016)]{werk16} Werk, J. K., Prochaska, J. X., Cantalupo, S., et~al. 2016, \apj, submitted (arXiv:1609.00012)

\bibitem[Willick et~al.(1997)]{willick97} Willick, J. A., Courteau, S., Faber, S. M., et~al. 1997, \apjs, 109, 333

\bibitem[Wolfe et~al.(2003a)Wolfe, Prochaska, \& Gawiser]{wolfe03a} Wolfe, A. M., Prochaska, J. X., \&  Gawiser, E. 2003a, \apj, 593, 215

\bibitem[Wolfe et~al.(2003b)Wolfe, Gawiser, \& Prochaska]{wolfe03b} Wolfe, A. M., Gawiser, E., \& Prochaska, J. X. 2003b, \apj, 593, 235

\bibitem[Yao et~al.(2010)]{yao10} Yao, Y., Wang, Q. D., Penton, S. V., et~al. 2010, \apj, 716, 1514

\bibitem[Yoon et~al.(2012)]{yoon12} Yoon, J. H., Putman, M. E., Thom, C., Chen, H.-W., \& Bryan, G. L. 2012, \apj, 754, 84

\bibitem[Zahedy et~al.(2016)]{zahedy16} Zahedy, F. S., Chen, H.-W., Rauch, M., Wilson, M. L., \& Zabludoff, A. 2016, \mnras, 458, 2423

\end{thebibliography}
\end{document}